\documentclass[a4paper,fleqn,usenatbib,useAMS]{mnras}

\usepackage{graphicx}	
\usepackage{amsmath}	
\usepackage{amssymb}	
\usepackage{multicol}        
\usepackage{bm}		
\usepackage{pdflscape}	
\usepackage{caption}
\usepackage{natbib}
\usepackage{longtable,lscape}
\usepackage{rotating}
\usepackage{multirow,multicol}

\usepackage[space]{grffile}  
\usepackage{threeparttable}
\usepackage[normalem]{ulem} 

\def\Kepler{\textit{Kepler}} 

\usepackage[T1]{fontenc}
\usepackage{ae,aecompl}

\usepackage{newtxtext,newtxmath}

\title[Clustered Planetary Systems]{Architectures of Exoplanetary Systems. I: A Clustered Forward Model for Exoplanetary Systems around \Kepler's FGK Stars}

\author[He, Ford, and Ragozzine]{Matthias Y. He$^{1,2,3,4}$\thanks{Contact e-mail: \href{mailto:myh7@psu.edu}{myh7@psu.edu}}, 
Eric B. Ford$^{1,2,3,4}$, and Darin Ragozzine$^{5}$ \\
$^{1}$Department of Astronomy and Astrophysics, 525 Davey Laboratory, The Pennsylvania State University, University Park, PA 16802, USA\\
$^{2}$Center for Exoplanets and Habitable Worlds, 525 Davey Laboratory, The Pennsylvania State University, University Park, PA 16802, USA\\
$^{3}$Center for Astrostatistics, 525 Davey Laboratory, The Pennsylvania State University, University Park, PA 16802, USA\\
$^{4}$Institute for CyberScience, The Pennsylvania State University, University Park, PA 16802, USA\\
$^{5}$Department of Physics and Astronomy, N283 ESC, Brigham Young University, Provo, UT 84602, USA\\
}

\date{Accepted ?. Received ?; in original form ?}

\pubyear{2019}

\begin{document}
\label{firstpage}
\pagerange{\pageref{firstpage}--\pageref{lastpage}}
\maketitle

\begin{abstract}
Observations of exoplanetary systems provide clues about the intrinsic distribution of planetary systems, their architectures, and how they formed. We develop a forward modelling framework for generating populations of planetary systems and ``observed'' catalogues by simulating the \Kepler{} detection pipeline (\texttt{SysSim}).
We compare our simulated catalogues to the \Kepler{} DR25 catalogue of planet candidates, updated to include revised stellar radii from Gaia DR2.
We constrain our models based on the observed 1D marginal distributions of orbital periods, period ratios, transit depths, transit depth ratios, transit durations, transit duration ratios, and  transit multiplicities.  
Models assuming planets with independent periods and sizes do not adequately account for the properties of the multiplanet systems.
Instead, a clustered point process model for exoplanet periods and sizes provides a significantly better description of the \Kepler{} population, particularly the observed multiplicity and period ratio distributions.
We find that $0.56^{+0.18}_{-0.15}$ of FGK stars have at least one planet larger than $0.5 R_\oplus$ between 3 and 300 d. Most of these planetary systems ($\sim 98\%$) consist of one or two clusters with a median of three planets per cluster.
We find that the \Kepler{} dichotomy is evidence for a population of highly inclined planetary systems and is unlikely to be solely due to a population of intrinsically single planet systems.
We provide a large ensemble of simulated physical and observed catalogues of planetary systems from our models, as well as publicly available code for generating similar catalogues given user-defined parameters.
\end{abstract}

\begin{keywords}
methods: statistical -- planetary systems -- planets and satellites: detection, fundamental parameters, terrestrial planets -- stars: statistics
\end{keywords}



\section{Introduction} \label{Introduction}

Within the past decade, NASA's \Kepler{} mission \citep{B2010, B2011a, B2011b, B2013} has discovered thousands of exoplanets and hundreds of multiplanet systems around Sun-like stars. Within these systems, there is an abundance of short period planets (i.e. with orbits much smaller than that of Earth) and tightly packed multiple planets \citep{La2011, Li2011b, Li2014, R2014}. The wealth of transit detections generated by \Kepler{} and the relative homogeneity of the sample allows for the study of exoplanet systems as a whole, enabling statistical exploration of the population of exoplanets detectable by \Kepler{}.

Multitransiting planetary systems are especially valuable because they serve as crucial tests for our models of planetary formation, their resulting architectures, and their subsequent evolution and stability \citep{RH2010, Fo2014, WF2015}. The abundance of systems with many transiting planets indicates that systems with small mutual inclinations are common, as small mutual inclination angles between planets in the same system are required to explain how a single observer can see so many planets in a transiting configuration. These observations contribute to our theories of planet formation, supporting the picture that planets likely formed in relatively flat, gaseous discs. Previous studies have explored the degree of coplanarity required to explain the multitude of multitransiting systems, focusing almost exclusively on the observed multiplicity and the transit duration ratio distributions \citep{La2011, Li2011b, FM2012, F2012, J2012, TD2012, WSS2012, F2014}. However, one emergent puzzle from the studies of observed transiting multiplicities is the apparent excess of single transiting systems, which has led some authors to speculate the existence of a second population of intrinsic singles or highly inclined multiplanet systems, a so-called ``\Kepler{} dichotomy'' \citep{Li2011b, J2012, HM2013, BJ2016}.

It is unclear how the planet multiplicities and their orbital inclinations mesh with other observed properties of the transiting population, such as the distributions of their orbital periods and period ratios. The period ratios of adjacent planet pairs in multiplanet systems, which is a direct measure of their physical separations and thus stability, have been studied \citep{F2014, SH2015}. The period ratios range from close to unity to over several dozen, with 1.172 being the smallest observed period ratio in the KOI-1665 system \citep{Li2011a}. Smaller period ratios of 1.038 (in KOI-284) and 1.065 (in KOI-2248) are now interpreted as due to transiting planets with very similar periods around different stars in a binary system \citep{Li2011a, Li2014}. These studies also find relative excesses of (apparently) adjacent planet pairs with period ratios near (slightly larger than) first-order mean-motion resonances (MMRs), namely near the 3:2 and 2:1 ratios \citep{F2014, SH2015}.  Inferring the true rate of near resonant systems is difficult, due to the interaction of the geometric joint transit probability, detection efficiency, and the unknown distribution of mutual inclinations and eccentricities.  

A few studies have also attempted to understand the size distribution of planets in multiplanet systems, by probing their relative radii using transit depth ratios in order to minimize the effects of our uncertainties in the stellar radii. For example, \citet{C2013} found that for planet pairs with planets larger than $\sim 3 R_\oplus$, the outer planet tends to be larger than the inner one, although they did not observe this trend for planets $\lesssim 3 R_\oplus$. In addition, planet sizes appear to be highly correlated, as evidenced by the peaked nature of the adjacent radii ratio distribution \citep{W2018a}, and this clustering extends to planet masses \citep{M2017}. While the mechanisms of photoevaporation \citep{OW2013, F2017, OW2017, vE2017, C2018} or core-powered mass-loss from formation \citep{GSS2016, GSS2018, GS2018} have been proposed to explain these features in the distributions of radii and radii ratios, complex observational biases limit our ability to distinguish models or understand the relative contribution of physical and observational effects. To further illustrate this point, most recently \citet{Z2019} has challenged the statistical significance of the correlations reported by \cite{C2013} and \citet{W2018a}. \citet{Z2019} found that testing the significance of correlations by bootstrap re-sampling with cuts on the signal-to-noise ratios (as opposed to planet size) resulted in smaller and less significant correlations for sizes of planets in one system, uniformity of spacing, and preference for the outer planet to be larger than the inner planet.  They suggest that a better way to study such correlations would be via forward modelling the detection and selection processes using a model for the intrinsic distribution of planetary properties.  This study takes that approach and provides a fresh perspective on the significance of correlations in orbital period, planet size, and uniformity of spacing.

There is an overwhelming wealth of information in the observed \Kepler{} population of exoplanets, especially encoded in the multiplicity, period ratio, and transit depth and duration ratio distributions.  Exploratory analyses are difficult to interpret due to complex geometric and sensitivity biases in the \Kepler{} detection pipeline. Earlier works such as those discussed above attempted to study these features but had fewer detections to work with, a more limited understanding of the \Kepler{} pipeline's detection efficiency, greater uncertainties in the stellar properties, and intractable biases in the planet catalogue due to human-influenced vetting of planet candidates. In this work, we are able to address these concerns by making use of several recent analyses and updates to the \Kepler{} stellar and planetary properties. We use the final \Kepler{} DR25 catalogue \citep{T2018}, which was generated using a completely automated procedure for the vetting of planet candidates using the \textit{Kepler Robovetter}. We rely and build on the Exoplanets Systems Simulator (``SysSim''), which makes use of several additional DR25 data products describing \Kepler's detection pipeline \citep{BC2017a, BC2017b, BC2017c, C2017, Co2017}, and is described in \citet{H2018, H2019}. Finally, we adopt the improved stellar properties from the European Space Agency's Gaia mission \citep{Gaia2018}, and planet candidates around a clean sample of FGK dwarfs as defined by \citet{H2019}.

In this paper, we outline a new framework for simulating catalogues of observed transiting planets in \S\ref{Methods}, which we use to explore several statistical, yet physically motivated, models for the intrinsic distribution of planetary systems.
In particular, we explore a marked, clustered point process for generating planetary systems.  For each planetary system, we attempt to generate the planets by first drawing a period and radius scale for each cluster and then drawing orbital periods and planet sizes for each planet in a cluster conditioned on the cluster's period and radius scales. These models are described in \S\ref{Models}-\S\ref{Obs_pipeline}.
We define a set of summary statistics, a distance function to compare models to data, and a subset of the \Kepler{} DR25 catalogue we use to constrain our models, in \S\ref{Obs_compare}.
We describe the optimization procedure used to explore the multidimensional parameter space of each model in \S\ref{Optimization}.
In \S\ref{GP}, we provide a brief review of the statistical machinery of Gaussian processes, and describe how we adapt it to serve as an ``emulator'' for our models and compute the credible regions for each model parameter using Approximate Bayesian Computation.
The results for the credible regions of the parameters for each model are presented and discussed in \S\ref{Model_params}. 
The results of each of our models, for both the observed and physical (intrinsic) distributions of planetary systems, and a direct comparison between them are described in \S\ref{Discussion}. We compute some estimates for the fraction of stars with planets and the number of planets per star or planetary system in \S\ref{Planet_rates}.
A discussion of the \Kepler{} dichotomy is presented in \S\ref{Dichotomy}.
We discuss how our models can be improved in \S\ref{secFuture}, and summarize our main results and encourage the use of our code and model catalogues in \S\ref{Conclusions}.

\section{Methods} \label{Methods}

The models described in this paper are developed in the framework of the Exoplanets Systems Simulator, which we refer to as ``SysSim''. The SysSim codebase is written in the Julia language \citep{B2014} and can be installed as the ExoplanetsSysSim.jl package \citep{F2018b}. Our models can be accessed at \url{https://github.com/ExoJulia/SysSimExClusters}.

Our general framework for performing an (approximate) Bayesian analysis can be summarized with the following steps:
\begin{description}
 \item \textbf{Step 0: Define a statistical description for the intrinsic distribution of exoplanetary systems} \\
  We present three separate statistical models in this paper (\S\ref{Models}). Each model is described by a set of physically motivated model parameters.
 \item \textbf{Step 1: Generate an underlying population of exoplanetary systems (\textit{physical catalogue}) from a given model} \\
  Each model involves randomly drawing stars from the \Kepler{} stellar catalogue and populating them with planets. Planet radii and periods are drawn. Then, masses are assigned from the radii and a rejection-sampling algorithm is applied to only keep planetary systems that are physical.
 \item \textbf{Step 2: Generate an observed population of exoplanetary systems (\textit{observed catalogue}) from the \textit{physical catalogue}} \\
  The systems are oriented by assigning orbits to each planet and the SysSim forward model simulating the \Kepler{} detection pipeline is applied to generate a catalogue of observed, transiting planets along with their ``measured'' properties.
 \item \textbf{Step 3: Compare the simulated \textit{observed catalogue} with the \Kepler{} data} \\
  A set of summary statistics are computed for both the simulated observed and \Kepler{} data, and a distance function is computed to compare the two catalogues.
 \item \textbf{Step 4: Optimize the distance function to find the best-fitting model parameters} \\
 Steps 0--3 are repeated by passing the distance function into an optimization algorithm in order to explore the parameter space and attempt to find the best-fitting model parameters.
 \item \textbf{Step 5: Explore the posterior distribution of model parameters using a Gaussian Process (GP) emulator} \\
  For each model, a GP model is trained on a set of points (model parameters with computed distances) in order to form an emulator for the full forward model. The emulator is used to quickly characterize the parameter space.
 \item \textbf{Step 6: Compute credible intervals for model parameters and simulated catalogues using Approximate Bayesian Computing} \\
  Model parameters are drawn from prior distributions, and used to draw an emulated distance from the GP emulator. If the emulated distance is less than a specified distance threshold, we accept the set of model parameters.
  The credible regions are reported based on the accepted sample of model parameters, and we present and analyze the simulated physical and observed catalogues that resulted in the distance from the full forward model being less than the distance threshold.
\end{description}

We describe each of these steps in full detail in the following subsections.

\subsection{Models for generating planetary systems} \label{Models}

\begin{figure}
\includegraphics[scale=0.28,trim={2.5cm 1cm 2.5cm 0.5cm},clip]{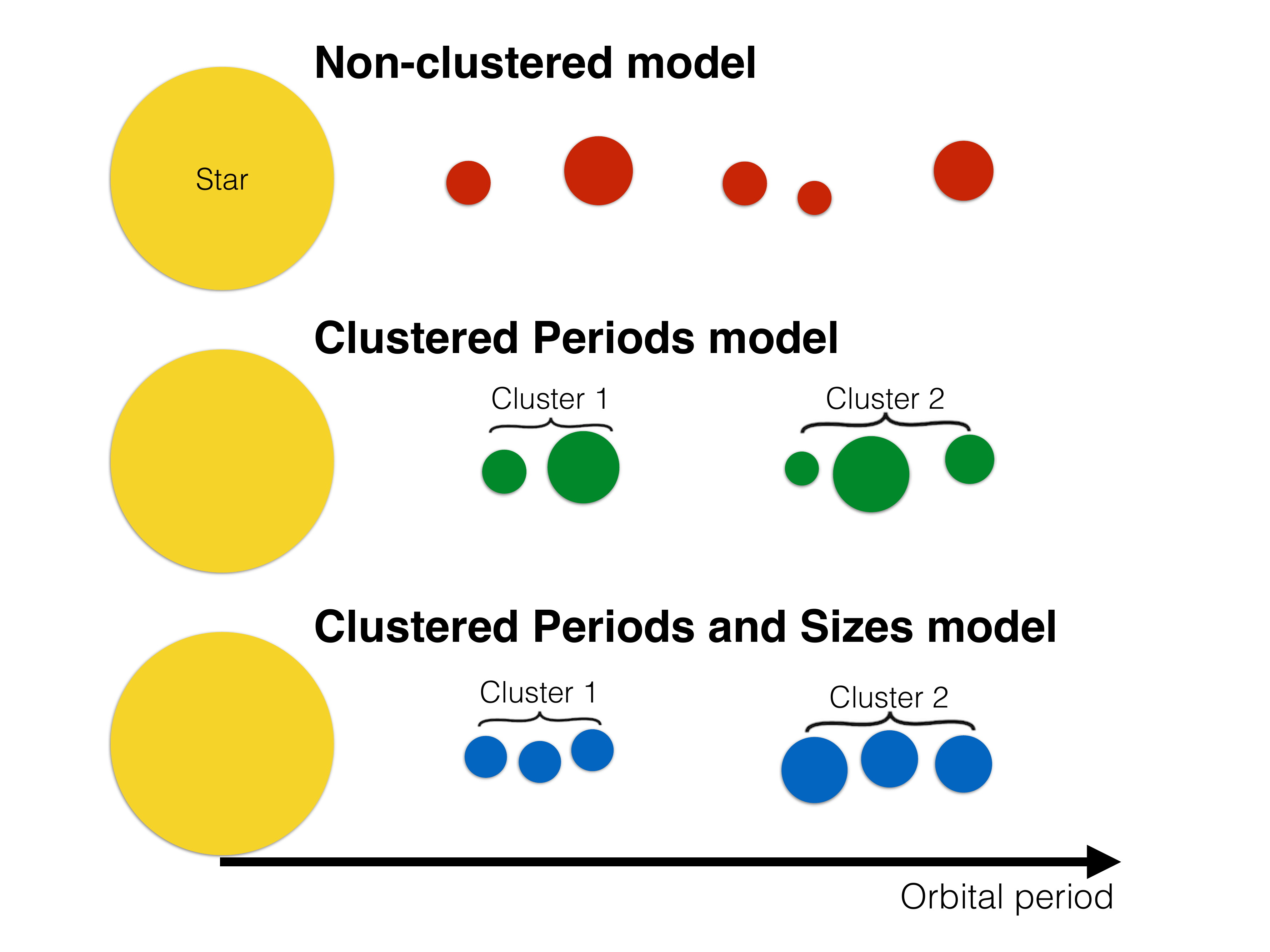}
\caption{Cartoon illustration of our three models, drawn along an arbitrary orbital period axis. In the non-clustered model, the number of planets is drawn from a Poisson distribution, and the orbital periods and planet radii are drawn independently from a power law and a broken power law, respectively. In the clustered models, the number of clusters is drawn from a Poisson distribution (two clusters are shown here for illustrative purposes). Similarly, the number of planets per cluster is drawn from a zero-truncated Poisson distribution. For the clustered periods model, the orbital periods of the planets are clustered together, but their sizes are still drawn independently. For the clustered periods and sizes model, the radii of planets in the same cluster are also similar. The models do not enforce any correlation between periods and planet radii.
This paper provides a forward model for drawing planetary systems from each of the three models and simulating \Kepler{} observations of such planetary systems. In \S\ref{Discussion}, we show that the two clustered models provide a significantly better match to the \Kepler{} observations than the non-clustered model.}
\label{fig:models_cartoon}
\end{figure}

We describe three separate models, starting with a baseline \textbf{non-clustered} model, extending to a \textbf{clustered periods} model, and then a \textbf{clustered periods and sizes} model.  First, we give a brief overview of all three models.  Then we provide details about the method for drawing each of the properties under the different models.  Results from the non-clustered model shown in \S\ref{Discussion}, will provide empirical support for motivating use of the clustered models.

In the non-clustered model, we first draw a number of planets for each star and then draw orbital periods and planet sizes independently for each planet in a system, using a simple power law for period and a broken power law for radius.  
In both of the clustered models, we first draw a number of ``clusters'' of planets for each star.  For each cluster, we draw a number of planets and a period scale. In the clustered periods and sizes model, we also draw a radius scale for each cluster. Then, the periods and sizes of the planets are drawn from distributions centred on the period scale and the radius scale, respectively (i.e. conditioned on the properties of their parent cluster). Thus, the properties of planets are explicitly correlated with those of other planets from the same cluster. This leads to closely spaced planets often having strong correlations, but more widely spaced planets having weaker correlations.
We show a cartoon illustration of our three models in Figure \ref{fig:models_cartoon}.

\subsubsection{Numbers of planets}

We will constrain our models based on the observed multiplicity distribution (as described in \S\ref{Summary_stats}) and use the results to address both the overall rate of planets per star and the fraction of stars with at least one planet in this study.  Therefore, it is important to consider the process for assigning the number of planets to each star.

For the non-clustered model, we draw the number of planets in each system from a Poisson distribution, $N_p \sim {\rm Poisson}(\lambda_p)$. In the clustered models we first draw a number of clusters from a Poisson distribution, $N_c \sim {\rm Poisson}(\lambda_c)$. Then, we draw a number of planets for each cluster from a zero-truncated Poisson (ZTP) distribution, $N_p \sim {\rm ZTP}(\lambda_p)$, where the probability mass function is given by:
\begin{align}
 g(k | \lambda) = \frac{\lambda^k}{(e^\lambda - 1)k!}. \label{eq_ZTP}
\end{align}
The choice of the ZTP is simply to avoid drawing empty clusters with no planets.
\footnote{The mean of a ZTP-distributed random variable $X$ with parameter $\lambda$ is given by $E[X] = \frac{\lambda}{1-e^{-\lambda}}$.}

The way in which zero-planet systems are drawn is an important consideration for our models. In our non-clustered model, assigning a star no planets occurs when $N_p = 0$ is drawn from the Poisson distribution for the number of planets per system, while in our clustered models, this occurs only when $N_c = 0$ is drawn from the Poisson distribution for the number of clusters per system.
Both the number of clusters and planets per cluster are truncated to not exceed maximum values, $N_c \leq N_{c,\rm max}$ and $N_p \leq N_{p,\rm max}$, respectively. We set $N_{c,\rm max} = 10$ and $N_{p,\rm max} = 15$ in our clustered models, and $N_{p,\rm max} = 20$ in our non-clustered model, unless stated otherwise in order to avoid generating systems with extremely large numbers of planets, as these are computationally expensive.

In addition to truncation effects, the true mean rates of planets per system (non-clustered model), clusters per system and planets per cluster (clustered models) are somewhat less than the values suggested by the parameters $\lambda_p$ and $\lambda_c$ due to our stability criteria and rejection sampling algorithm (described in \S\ref{Stability} and \S\ref{Procedure}). Thus, while the values of $\lambda_p$ (non-clustered) and $\lambda_c \lambda_p$ (clustered) serve as approximations for the mean number of planets per star, they should not be overinterpreted. We use the true rates of planets per star for more detailed calculations in \S\ref{Discussion}.

\subsubsection{Orbital Periods}
\label{sec:OribtalPeriods}
\begin{description}
 \item \textbf{Non-clustered model:} \\
 Orbital periods are drawn independently from a simple power-law distribution:
 \begin{align}
  f(P) &\propto {P}^{\alpha_P}, \quad P_{\rm min} \leq P \leq P_{\rm max}, \label{eq_P}
 \end{align}
 where $f(P)$ is the probability density function (PDF) and $\alpha_P$ is the power-law index for the period distribution. 
 We choose $P_{\rm min} = 3$ d and $P_{\rm max} = 300$ d.

 \item \textbf{Clustered periods and clustered periods and sizes models:} \\
 The orbital period of each planet is drawn conditionally on the period scale for its parent cluster, $P_c$. For each cluster we draw a trial $P_c$ from a simple power law, and the planets in each cluster have trial periods drawn from a log-normal distribution conditioned on $P_c$ with a cluster width that is scaled to the number of planets:
 \begin{align}
  f(P_c) &\propto {P_c}^{\alpha_P} \label{eq_Pc} \\
  P'_i &\sim {\rm Lognormal}(0, N_p\sigma_P) \label{eq_P_clusP} \\
  P_i &=  P_c P'_i,\quad P_{\rm min} \leq P_c \leq P_{\rm max}
 \end{align}
 where $P'_i$ are the unscaled periods, $P_i$ are the true periods, $N_p$ is the number of planets in the cluster, and $\sigma_P$ is the cluster width scale factor, per planet in the cluster.
 The trial periods are accepted or rejected based on a heuristic for orbital instability, as described in \S\ref{Procedure}.\footnote{In practice, we first draw unscaled periods $P'_i \sim {\rm Lognormal}(0, N_p\sigma_P)$, truncated between $P'_{\rm min} = \sqrt{P_{\rm min}/P_{\rm max}}$ and $P'_{\rm max} = \sqrt{P_{\rm max}/P_{\rm min}}$ (this is to avoid drawing unscaled periods such that maximum period ratio is greater than $P_{\rm max}/P_{\rm min}$, which would have no chance fitting between $P_{\rm min}$ and $P_{\rm max}$ regardless of $P_c$). We then draw a cluster period scale $P_c$ from Equation (\ref{eq_Pc}) (truncated between $P_{\rm min}/{\rm min}\{P'_i\}$ and $P_{\rm max}/{\rm max}\{P'_i\}$) to give $P_i = P'_i{P_c}$, checking that the drawn value of $P_c$ allows the planets in the cluster to be stable with all other planets in the system; see \S\ref{Procedure} for a detailed outline of this process.}
 
\end{description}

\subsubsection{Planet sizes}
\begin{description}
 \item \textbf{Non-clustered model and clustered periods model:} \\
 Planet radii are drawn independently from a broken power law:
 \begin{align}
  f(R_p) &\propto
  \begin{cases}
   {R_p}^{\alpha_{R1}}, & R_{p,\rm min} \leq R_p \leq R_{p,\rm break} \label{eq_brokenR} \\
   {R_p}^{\alpha_{R2}}, & R_{p,\rm break} < R_p \leq R_{p,\rm max}
  \end{cases},
 \end{align}
 where $f(R_p)$ is the PDF, and $\alpha_{R1}$ and $\alpha_{R2}$ are the broken power-law indices for the planet radius distribution below and above $R_{p,\rm break}$, the break radius, respectively. 
 Our choice of a broken power law for the radius distribution is motivated by previous studies suggesting that there is an observed break at around 2--3$R_\oplus$, with a rise in occurrence down to $\sim 2 R_\oplus$ and a plateau below that \citep{Y2011, H2012, PMH2013b}. 

 \item \textbf{Clustered periods and sizes models:} \\
The radius of each planet is drawn conditionally on the radius scale for its parent cluster, $R_{p,c}$.  The cluster radius scales $R_{p,c}$ are drawn from a broken power law, while the planets in each cluster have their radii drawn from a log-normal distribution conditioned on $R_{p,c}$ with a fixed scale for the cluster width $\sigma_R$:
 \begin{align}
  f(R_{p,c}) &\propto
  \begin{cases}
   {R_{p,c}}^{\alpha_{R1}}, & R_{p,\rm min} \leq R_{p,c} \leq R_{p,\rm break} \label{eq_brokenRc} \\
   {R_{p,c}}^{\alpha_{R2}}, & R_{p,\rm break} < R_{p,c} \leq R_{p,\rm max}
  \end{cases}. \\
  R_{p,i} &\sim {\rm Lognormal}(R_{p,c}, \sigma_R). \label{eq_R_clusR}
 \end{align}
\end{description}

For all three models, we limit our analyses to $R_{p,\rm min} = 0.5 R_\oplus$ and $R_{p,\rm max} = 10 R_\oplus$, as the distribution of larger or smaller planets is not well constrained by \Kepler{} observations.  
 
\subsubsection{Planet masses}

Planets vary significantly in structure and composition \citep{WL2012, LF2014, CK2016, WRF2016}. We adopt a non-parametric mass--radius relation from \citet{NWG2018} in order to assign planet masses probabilistically given their drawn planet radii. This mass-radius relation involves a series of Bernstein polynomials used to model the joint distribution of mass and radius based on a sample of 127 \Kepler{} exoplanets with RV or TTV masses, and is more flexible than simpler, power law mass--radius relations.
To accelerate computations, we pre-compute and use a $1001\times1001$ look-up table for the mass--radius relation.

\subsubsection{Eccentricities}
We draw the orbital eccentricities from a Rayleigh distribution, $e \sim {\rm Rayleigh}(\sigma_e)$, where the Rayleigh parameter ($\sigma_e$) defines the scale for eccentricities, following previous works (e.g. \citealt{M2011,F2014}).

\subsubsection{Mutual inclinations} \label{Incl}

Finally, we allow for two populations of planetary systems with separate distributions of mutual inclinations between planets. We use Rayleigh distributions for both populations, as the Rayleigh distribution has been used to describe mutual inclinations in previous works (e.g. \citealt{FW2009, Li2011b, FM2012,DLC2016}). Thus, for a fraction $f_{\sigma_{i,\rm high}}$ of systems we draw from a broad distribution of mutual inclinations while for the remaining $1 - f_{\sigma_{i,\rm high}}$ fraction of systems we draw from a narrower distribution:
\begin{align}
 i_m \sim
 \begin{cases}
  {\rm Rayleigh}(\sigma_{i,\rm high}), & u < f_{\sigma_{i,\rm high}} \label{eq_incl} \\
  {\rm Rayleigh}(\sigma_{i,\rm low}), & u \geq f_{\sigma_{i,\rm high}}
 \end{cases},
\end{align}
where $u \sim {\rm Unif}(0,1)$ and $\sigma_{i,\rm low} \leq \sigma_{i,\rm high}$.

In addition, we check whether each planet is near a 2:1, 3:2, 4:3, or 5:4 MMR with another planet, and draw $i_m \sim {\rm Rayleigh}(\sigma_{i,\rm low})$ for planets that are. For our purposes, we regard adjacent planet pairs as ``near an MMR'' if their period ratio is within 5\% exterior to the MMR. This amounts to period ratios between 1.5 and 1.575 as near the 3:2 MMR and period ratios between 2 and 2.1 as near the 2:1 MMR, for example. Our motivation for forcing planets near MMRs to have more coplanar orbits than other planets is to explore whether this alone can produce the apparent relative excesses of planets with period ratios just wide of MMRs \citep{F2014, SH2015}, or whether it is essential to generate more planets near MMRs than what is drawn naturally from our model for orbital periods (see \S\ref{sec:OribtalPeriods}). Since planets with more coplanar orbits are more likely to transit together than highly mutually inclined planets, our model has the effect of producing slightly more observed planet pairs near MMRs than planet pairs with arbitrary period ratios.

To test the robustness of our conclusions to this special treatment of the near-MMR planets, we also explore a model without the lowering of mutual inclinations for such planets, using our clustered periods and sizes model. We refer to this model as the ``alternative MMR inclinations'' model for the remainder of the paper. The results of this model will be primarily discussed in \S\ref{Model_params} and \S\ref{Dichotomy}.

We leave $f_{\sigma_{i,\rm high}}$, $\sigma_{i,\rm high}$, and $\sigma_{i,\rm low}$ as free parameters of the models.  

\subsubsection{Stability criteria} \label{Stability}

We test whether planetary systems are likely to be long-term Hill stable after drawing their periods.  When a planetary system is identified as likely unstable, then it is discarded and we attempt to redraw orbital periods.  Our instability criterion is based on the spacing between adjacent planets ($\Delta$) normalized by the mutual Hill radius ($R_H$), which is given by:
\begin{align}
 R_H = \bigg(\frac{a_{\rm in} + a_{\rm out}}{2}\bigg)\bigg[\frac{m_{\rm in} + m_{\rm out}}{3 M_\star}\bigg]^{1/3}, \label{eq_mhill}
\end{align}
where $a_{\rm in}$, $a_{\rm out}$ are the semimajor axes and $m_{\rm in}$, $m_{\rm out}$ are the masses of the inner and outer planets, respectively, and $M_\star$ is the mass of the stellar host. 
We define an instability criterion that is parametrized in terms of $R_H$:
\begin{align}
 \Delta = \frac{a_{\rm out}(1-e_{\rm out}) - a_{\rm in}(1+e_{\rm in})}{R_H}. \label{eq_Nhill}
\end{align}
To avoid generating planetary systems likely to be unstable, we require $\Delta \geq \Delta_c$ where $\Delta_c$ is the minimum separation (held fixed). For two circular and coplanar orbits, the minimum separation required to be Hill stable (i.e., no close encounters) is given by $\Delta > 2\sqrt{3} \simeq 3.46$ \citep{G1993}. \citet{CWB1996} used numerical simulations to conclude that systems with equal-mass planets separated by $\Delta < 10$ are virtually always unstable. \citet{PW2015} find that $\Delta \gtrsim12$ is required for long-term stability of planets in multiplanet systems given certain assumptions about the masses and eccentricity distribution for the \Kepler{} exoplanets. 
In our early exploratory analyses where $\Delta_c$ was treated as a free parameter, we found that our models preferred smaller values, $\Delta \lesssim 10$. Therefore, we set $\Delta_c = 8$ for the remainder of this paper unless otherwise stated. 

\subsubsection{Summary of the free parameters of our models}

Our non-clustered model has a total of 10 free parameters: $\lambda_p$, $\alpha_P$, $\alpha_{R1}$, $\alpha_{R2}$, $R_{p,\rm break}$, $\sigma_e$, $\Delta_c$, $f_{\sigma_{i,\rm high}}$, $\sigma_{i,\rm high}$, and $\sigma_{i,\rm low}$. The clustered periods model has an additional two parameters $\lambda_c$ and $\sigma_P$ (the cluster width in log-period per planet in the cluster) to give 12 free parameters, while the clustered periods and sizes model adds yet another parameter $\sigma_R$ (the cluster width in log-radius) for a total of 13 free parameters. In our early exploratory analysis we found that the break radius $R_{p,\rm break}$ and the minimum separation $\Delta_c$ were not well constrained by \Kepler{} observations (see \S\ref{Optimization} for more details). In particular, $R_{p,\rm break}$ can take on a wide range of values and the models seem to prefer smaller values of $\Delta_c \lesssim 10$. Thus, in order to improve the efficiency of the fitting procedure, we reduce the number of free parameters and thus the size of the parameter space by setting these parameters to $R_{p,\rm break} = 3 R_\oplus$ and $\Delta_c = 8$ based on previous studies (e.g., \citealt{PMH2013b}). This reduces the number of free parameters for each model by two. A list of all the parameters of our models used in this paper is provided in Table \ref{tab:params}.

\begin{table*}
\centering
\caption{List describing the important model parameters, and how they are used, in this paper.}
\begin{tabular}{ l | p{9cm} | c | l }
 \hline
 \hline
 Parameter & Definition of parameter & Equation & Relevant quantity/distribution \\
 \hline
 $N_{\rm stars,sim}$ & Number of simulated stars & - & - \\
  $\lambda_c$ & Mean number of clusters per star (before rejection sampling) & - & $N_c \sim {\rm Poisson}(\lambda_c)$ \\
 $N_{c,\rm max}$ & Maximum number of clusters per star & - & $N_c \leq N_{c,\rm max}$ \\
 $\lambda_p$ & Mean number of planets per system (non-clustered model) or cluster (clustered models) (before rejection-sampling) & - & $N_p \sim {\rm Poisson}(\lambda_p)$ or ${\rm ZTP}(\lambda_p)$ \\
 $N_{p,\rm max}$ & Maximum number of planets per cluster & - & $N_p \leq N_{p,\rm max}$ \\
 $\alpha_P$ & Power law index for distribution of periods (non-clustered model) or period scales (clustered models) & (\ref{eq_P}), (\ref{eq_Pc}) & $P_c \sim f(P_c) \propto {P_c}^{\alpha_P}$ \\
 $P_{\rm min}$ & Minimum period (days) & (\ref{eq_P}), (\ref{eq_Pc}) & $P = P'P_c \geq P_{\rm min}$ \\
 $P_{\rm max}$ & Maximum period (days) & (\ref{eq_P}), (\ref{eq_Pc}) & $P = P'P_c \leq P_{\rm max}$ \\
 $\alpha_{R1}$ & Power law index for planetary radius distribution for $R_p \leq R_{p,\rm break}$ & (\ref{eq_brokenR}), (\ref{eq_brokenRc}) & $R_{p,c} \sim f(R_p) \propto {R_p}^{\alpha_{R1}}, R_p \leq R_{p,\rm break}$ \\
 $\alpha_{R2}$ & Power law index for planetary radius distribution for $R_p > R_{p,\rm break}$ & (\ref{eq_brokenR}), (\ref{eq_brokenRc}) & $R_{p,c} \sim f(R_p) \propto {R_p}^{\alpha_{R2}}, R_p > R_{p,\rm break}$ \\
 $R_{p,\rm break}$ & Break radius for planetary radii ($R_\oplus$) & (\ref{eq_brokenR}), (\ref{eq_brokenRc}) & $R_{p,\rm min} < R_{p,\rm break} < R_{p,\rm max}$ \\
 $R_{p,\rm min}$ & Minimum planetary radius ($R_\oplus$) & (\ref{eq_brokenR}), (\ref{eq_brokenRc}) & $R_p \geq R_{p,\rm min}$ \\
 $R_{p,\rm max}$ & Maximum planetary radius ($R_\oplus$) & (\ref{eq_brokenR}), (\ref{eq_brokenRc}) & $R_p \leq R_{p,\rm max}$ \\
 $\sigma_P$ & Scale factor, per planet in the cluster, for the cluster (unscaled) period distribution & (\ref{eq_P_clusP}) & $P' \sim {\rm Lognormal}(0, {N_p}\sigma_P)$ \\
 $\sigma_R$ & Scale factor for the cluster radius distribution & (\ref{eq_R_clusR}) & $R_p \sim {\rm Lognormal}(R_{p,c}, \sigma_R)$ \\
 $\sigma_e$ & Scale for orbital eccentricities & - & $e \sim {\rm Rayleigh}(\sigma_e)$ \\
 $\Delta_c$ & Minimum allowed separation between adjacent planets in mutual Hill radii & - & $\Delta \geq \Delta_c$ \\
 $f_{\sigma_{i,\rm high}}$ & Fraction of systems with relatively high mutual inclinations, $\sigma_{i,\rm high}$ & (\ref{eq_incl}) & - \\
 $\sigma_{i,\rm high}$ & Scale for mutual inclinations for systems with high mutual inclinations (deg) & (\ref{eq_incl}) & $i_m \sim {\rm Rayleigh}(\sigma_{i,\rm high})$ \\
 $\sigma_{i,\rm low}$ & Scale for mutual inclinations for systems with low mutual inclinations (deg) & (\ref{eq_incl}) & $i_m \sim {\rm Rayleigh}(\sigma_{i,\rm low})$ \\
 $N_{\rm attempts}$ & Maximum number of attempts for re-sampling periods and period scales for each cluster & - & - \\
 \hline
 \hline
\end{tabular}
\label{tab:params}
\end{table*}

\subsection{Procedure for generating a \textit{physical catalogue}} \label{Procedure}

We outline the procedure for generating an underlying population of planetary systems (a \textit{physical catalogue}) from the clustered periods and sizes model as follows. The procedure for the other two models are analogous (we explain how to modify the procedure after these steps).
\begin{enumerate}
 \item Set a number of target stars $N_{\rm stars,sim}$ for each simulated catalogue, typically equal to the number of \Kepler{} targets being used as observational constraints.
 \item Set a value for each of the model parameters.
 \item For each target, assign stellar properties drawn from the \Kepler{} stellar catalogue (updated based on Gaia DR2, as described in \citealt{H2019}, and allowing for the uncertainties in the stellar properties).  Stellar properties include radius and mass which affect the observed transit properties, as well as the one-sigma depth function, window function, and contamination that are necessary for the planet detection model \citep{H2019}.
 \item Draw a number of clusters in the system, $N_c \sim {\rm Poisson}(\lambda_c)$. Re-sample until $N_c \leq N_{c,\rm max}$.
 \item For each cluster:
 \begin{enumerate}
  \item Draw a number of planets in the cluster from a zero-truncated Poisson (ZTP) distribution, $N_p \sim {\rm ZTP}(\lambda_p)$. Re-sample until $N_p \leq N_{p,\rm max}$.
  \item Draw a radius for each planet in the cluster: first, draw a characteristic radius $R_{p,c}$ for the cluster according to Equation (\ref{eq_brokenRc}). If $N_p = 1$, the radius of the cluster's one planet is simply $R_p = R_{p,c}$. If $N_p > 1$, draw a radius for each of the cluster's planets, $R_{p,i} \sim {\rm Lognormal}(R_{p,c}, \sigma_R)$, where $i = 1,...,N_p$ (the log is base-$e$). Draw their masses using the mass--radius relation from the non-parametric model in \citet{NWG2018}.
  \item Draw an orbital eccentricity $e \sim {\rm Rayleigh}(\sigma_e)$ and argument of periastron $\omega \sim {\rm Unif}(0,2\pi)$ for each planet in the cluster.
  \item Draw orbital periods for each planet in the cluster:  If $N_p = 1$, assign an unscaled period of $P' = 1$. If $N_p > 1$, draw their unscaled periods $P'_i \sim {\rm Lognormal}(0, N_p\sigma_P)$, where $i = 1,...,N_p$ (the log is base-$e$), and sort them in increasing order. Check if $\Delta = [a_{i+1}(1-e_{i+1}) - a_i(1+e_i)]/R_H(i,i+1) \geq \Delta_c$ for all $i$ in the cluster. Re-sample the unscaled periods $P'_i$ until this condition is satisfied or the maximum number of attempts $N_{\rm attempts}$ is reached.\footnote{We adopt a value of $N_{\rm attempts} = 100$, for the purposes of computational efficiency. This is necessary to prevent certain (e.g. extremely populated but compact; large $N_p$ with a small set value of $\sigma_P$) clusters from significantly increasing the computational time for simulating a \textit{physical catalogue}. On the other hand, if we attempt the drawing of planets in each cluster only a few times (or even just once), the rejection-sampling algorithm tends to produce too few planetary systems. Thus, a modest number of attempts per cluster provides a compromise between these two extremes. Also, a consequence of this procedure is that the true mean rates of clusters and planets per cluster in our samples are somewhat less than $\lambda_c$ and $E[N_p]$, respectively.} If the case is the latter, discard the cluster.  Draw a period scale factor $P_c$ (d) according to Equation (\ref{eq_P}) and multiply each planet's unscaled periods by the period scale for its parent cluster: $P_i = {P'_i}P_c$. 
 \item Test for stability:  Check if $\Delta \geq \Delta_c$ for all adjacent planet pairs in the entire system, including planets from previously drawn clusters.  If the system is identified as likely unstable, redraw $P_c$ for the current cluster until this condition is satisfied or until the maximum number of attempts $N_{\rm attempts}$ is reached.  If the latter case occurs, discard the cluster.
\end{enumerate}

 \item For each system, draw an inclination angle (relative to the plane of the sky) for the reference plane of the system isotropically, $\cos{i_{\rm ref}} \sim {\rm Unif}(0,1)$.  For each system, draw a number $u \sim {\rm Unif}(0,1)$. If $u < f_{\sigma_{i,\rm high}}$, set $\sigma_i = \sigma_{i,\rm high}$; otherwise, set $\sigma_i = \sigma_{i,\rm low}$. Assign an orbit to each planet in the system:
 \begin{enumerate}
  \item Compute the period ratios $\mathcal{P} = P_{i+1}/P_i$ of all adjacent planet pairs in the system. For each planet, check if it is near any MMRs with any adjacent planet by checking if $\mathcal{P}_{\rm mmr} \leq \mathcal{P} \leq 1.05\mathcal{P}_{\rm mmr}$ for any $\mathcal{P}_{\rm mmr}$ in $\{2, 1.5, 4/3, 1.25\}$ (aside from the inner- and outer-most planet, each planet is part of two adjacent planet pairs).
  If the planet is not near any MMRs with another adjacent planet, draw a mutual inclination angle (relative to the reference plane) $i_m \sim {\rm Rayleigh}(\sigma_i)$; otherwise, draw a mutual inclination angle $i_m \sim {\rm Rayleigh}(\sigma_{i,\rm low})$ regardless of whether $\sigma_{i,\rm high}$ or $\sigma_{i,\rm low}$ was assigned to the non-resonant planets of the system.
  \item For each planet's orbit, draw an angle of ascending node $\Omega \sim {\rm Unif}(0,2\pi)$ in the reference plane.
  \item Compute the inclination angle $i$ (relative to the plane of the sky) for each planet's orbit using the spherical law of cosines, $\cos{(i)} = \cos{(i_{\rm ref})}\cos{(i_m)} + \sin{(i_{\rm ref})}\sin{(i_m)}\cos{(\Omega)}$.
 \end{enumerate}
\end{enumerate}

The procedure is very similar for the other models. For the clustered periods model, instead of drawing a characteristic radius $R_{p,c}$ for each cluster as in Step 5(b), we simply assign planet radii by drawing them directly from Equation (\ref{eq_brokenR}). In the case of the non-clustered model, Steps 4--5 are simplified to drawing a number of planets $N_p \sim {\rm Poisson}(\lambda_p)$ and directly drawing their periods and radii by Equations (\ref{eq_P}) and (\ref{eq_brokenR}).\footnote{The non-clustered model could be considered as a special case of the clustered models by setting the number of planets per cluster to one, $N_p = 1$, instead of drawing from a ZTP distribution in Step 4(a). If there is exactly one planet per ``cluster'', then the properties of planets in the same planetary system are effectively drawn independently.  In this case, the total number of planets in any system is $N_c$. 
In practice, for the non-clustered model, we draw periods for all the $N_p$ planets simultaneously and accept or reject all periods at once, so as to improve computational efficiency and to minimize artefacts near $P_{\rm min}$ and $P_{\rm max}$ for systems with many planets.}

\subsection{Procedure for generating an \textit{observed catalogue}} \label{Obs_pipeline}

The underlying population of planetary systems is then used to generate an \textit{observed catalogue} of exoplanets using a procedure which simulates the observational pipeline employed by \Kepler{}. This procedure is the same regardless of the choice of model used for generating the \textit{physical catalogue}. The full pipeline that is implemented in SysSim as described in \citet{H2018, H2019} incorporates many \Kepler{} DR25 data products, including tabulated window functions, one sigma depth functions, and a detection efficiency derived from analysing pixel-level transit injections.  We briefly summarize the main considerations here. 

First, we reduce the number of planets by only keeping (in the observed catalogue) planets that transit.\footnote{We use SysSim in the single-observer mode, and do not make use of CORBITS for sky-averaging \citep{BR2016}.  Working in single-observer model allows our forward model to reproduce the variations in the observed \Kepler{} catalogue due to the finite number of targets (see \citealt{H2019}) and avoids issues with two-planet correlations that are not calculated by CORBITS.}

This amounts to requiring that the impact parameter $b$ is less than $1+R_p/R_\star$,
\begin{align}
 b = \frac{a\cos{i}}{R_\star}\frac{1 - e^2}{1 + e\sin{\omega}} < 1+R_p/R_\star. \label{eq_b}
\end{align}
Next, we select a subset weighted by their detection probabilities. 
We require at least three transits to be observed by the \Kepler{} mission based on window functions provided as part of \Kepler{} DR25, as described in \citet{H2019}.  A planet is labelled as detected if a random number $u \sim {\rm Unif}(0,1)$ is less than the planet's detection probability (conditioned on it transiting), as calculated by assuming the joint detection and vetting efficiency model described in \citet{H2019}.  

For each of the detected transiting planets, we compute the true transit depths accounting for limb darkening and the true transit durations according to \citet{K2010}, equation (15) therein:
\begin{align}
 t_{\rm dur} &= \frac{P}{\pi}\frac{\rho_c^2}{\sqrt{1-e^2}}\sin^{-1}{\bigg(\frac{\sqrt{1-b^2}}{a_R\rho_c\sin{i}}\bigg)} \label{eq_tdur} \\
 &\simeq{} \frac{P}{\pi{a_R}}\frac{\sqrt{1-e^2}}{1+e\sin{\omega}}\sqrt{1-b^2}, \label{eq_tdur_approx} \\
 \rho_c &\equiv \frac{1-e^2}{1+e\sin{\omega}},
\end{align}
where $t_{\rm dur}$ is the full-width half-maximum transit duration and $a_R \equiv a/R_\star$ is the semimajor axis in units of the stellar radius. The one-term analytic expression in Equation (\ref{eq_tdur}) only assumes that the planet--star separation does not change during the entire transit event and neglects the planet size \citep{K2010}.  
%
%
Using this definition grazing transits (i.e., $1\le b<1+R_p/R_\star$) have zero duration.  

Finally, we add measurement noise to the true period, transit depth, and transit durations and compile the ``observed'' properties of the detected transiting planets into a simulated observed catalogue.  
Some of the simulations used for exploring parameter space and training the emulator (see \S\ref{Optimization}-\S\ref{GP}) use a simplified noise model where the fractional uncertainties in orbital periods, transit depths, and transit durations are held fixed.  However, the final simulations for constraining model parameters and inferring the distributions of observed and physical properties use a diagonal version of the transit noise model of \citet{PR2014}, which is based on a trapezoidal transit model and accounts for the finite integration time.

\subsection{Observational comparisons} \label{Obs_compare}

Next, we compare the simulated \textit{observed catalogues} of exoplanets derived from our models to the actual observed population of exoplanets by the \Kepler{} mission.
In principle, one could attempt to generate simulated catalogues that precisely match all planet properties to those of the actual \Kepler{} catalogue. 
Of course, the odds of generating such a catalogue are minuscule, even if one could use the perfect model for the underling exoplanet population due to the stochastic nature of the model.  
Instead, we identify a set of summary statistics that encode the most physically important properties of the distributions of planetary systems observed by \Kepler{} in \S\ref{Summary_stats}.  
Next, we define a distance function that quantifies the degree of similarity between the summary statistics for the two catalogues in \S\ref{Distance}.
We describe a procedure for identifying sets of parameters for our physical models that approximately minimize the distance between simulated and observed catalogues (allowing for inevitable shot noise due to the stochastic nature of the model) in \S\ref{Optimization}.
Finally, we constrain the model parameters using approximate Bayesian Computing (ABC).
Given the cost of the full model, we train a Gaussian process emulator to approximate the distribution of the distances computed from our forward model in \S\ref{GP}.  
We obtain samples from the ABC posterior by drawing trial model parameters based on the GP emulator, computing the distance with the full SysSim forward model, and accepting parameter values that result in a small distance between the simulated observed catalogue and the \Kepler{} DR25 catalogue.

\subsubsection{Summary statistics} \label{Summary_stats}

Our procedure for generating \textit{observed catalogues} yields an observed catalogue with a ``measured'' orbital period $P$, transit duration $t_{\rm dur}$, and transit depth $\delta$ for each observed planet.  
A good forward model must result in a similar number of detected planets, as well as a similar number of systems with $m$ detected planets.  
Additionally, we want our models to reproduce the observed distributions of orbital periods, transit depths and transit durations.  
Finally, we want our forward models to produce planetary systems that have realistic correlations between the orbital periods and sizes of planets within the same system.  
Therefore, we compute the following summary statistics for each observed catalogue:
\begin{enumerate}
 \item the overall rate of observed planets relative to the number of target stars, $f = N_{p,\rm tot}/N_{\rm stars}$, where $N_{p,\rm tot}$ and $N_{\rm stars}$ are the total numbers of observed planets and stars in the catalogue, respectively,
 \item the observed multiplicity distribution, $\{N_m\}$, where $N_m$ is the number of systems with $m$ observed planets and $m = 1,2,3,...$,
 \item the observed orbital period distribution, $\{P\}$,
 \item the observed distribution of period ratios of apparently adjacent planets, $\{\mathcal{P}\}$, where $\mathcal{P} = P_{i+1}/P_i$,
 \item the observed transit depth distribution, $\{\delta\}$,
 \item the observed distribution of the transit depth ratios of apparently adjacent planets, $\{\delta_{i+1}/\delta_i\}$,
 \item the observed transit duration distribution, $\{t_{\rm dur}\}$,
 \item the observed distribution of (period-normalized) transit duration ratios of apparently adjacent planets near mean motion resonances, $\{\xi_{\rm res}\}$, and
 \item the observed distribution of (period-normalized) transit duration ratios of apparently adjacent planets not near mean motion resonances, $\{\xi_{\rm non-res}\}$.
\end{enumerate}

Previously published studies have attempted to match a subset of these summary statistics, but not all at once.  Each summary statistic is most sensitive to one or two model parameters, but typically have weaker dependencies on other model parameters.

For example, the transit duration distribution is useful for characterizing the orbital eccentricities \citep{FQV2008}. Moreover, the period-normalized transit duration ratio $\xi$ (also known as the orbital-velocity normalized transit duration ratio; \citealt{F2014}) is a useful probe of the orbital properties of exoplanets in multitransiting systems. It can be shown that the ratio of the period to the transit duration cubed is proportional to the stellar density, $P/t_{\rm dur}^3 \propto \rho_\star$ \citep{SM2003}. Thus, for two planets transiting the same star, it is useful to consider the ratio of their period-normalized transit durations. This quantity has been used to test whether two planets really transit the same star; if they do, $\xi$ should be near unity \citep{Li2012, F2014}. Following \citet{S2010} and \citet{F2014}, we define $\xi$ as:
\begin{align}
 \xi = \bigg(\frac{t_{\rm dur, in}}{t_{\rm dur, out}}\bigg)\bigg(\frac{P_{\rm out}}{P_{\rm in}}\bigg)^{1/3}. \label{eq_xi}
\end{align}
More importantly for our purposes, the distribution of $\xi$ encodes information about the orbital eccentricities and impact parameters (i.e., inclination angles) of the transiting planets \citep{F2014, M2016phd}. It is useful to transform this quantity by taking the logarithm, $\log{\xi}$, so that negative values imply $\xi < 1$ and positive values denote $\xi > 1$. For planets in circular, coplanar orbits around the same star, $\log{\xi}$ must be non-negative (if they both transit at the equator, $b = 0$ and $\xi = 1$; otherwise, the inner planet must have a smaller impact parameter $b$ than the outer planet, and thus a longer period-normalized transit duration, $\xi > 1$). Deviating from coplanarity by increasing the mutual inclination angles results in a lower skewness (i.e., greater symmetry around 0) for the distribution of $\log{\xi}$, because the impact parameters become more randomized as the outer transiting planets need not necessarily have larger values of $b$ than the inner planets. For eccentric orbits, the distribution of $\log{\xi}$ becomes more spread out (i.e., more extreme values of $\xi$ are more common) with increasing eccentricities because the velocities of the planets during transit become more randomized, depending on whether the transits occur near periastron or apastron.

Our forward model involves assigning systems to one of two separate mutual inclination scales and assigning planets near MMRs to follow the smaller mutual inclination distribution ($i_m \sim {\rm Rayleigh}(\sigma_{i,\rm low})$; see Step 6(a) in \S\ref{Procedure}).
In order to constrain the inclinations of both populations, we include two summary statistics based on the $\xi$ distribution: $\{\xi_{\rm res}\}$ calculated using only observed planet pairs apparently near a MMR (i.e. based on the observed periods) and $\{\xi_{\rm non-res}\}$ calculated using the remaining planet pairs (i.e., planets pairs not apparently near any MMRs).

\subsubsection{Distance function} \label{Distance}

ABC is a powerful method for performing inference on models where it is impractical to write an explicit likelihood, such as the case for studying multiplanet systems observed by \Kepler{} \citep{H2018}.  
In ABC, one must define a distance function to quantify how different two catalogues are.  
This distance is a function of only the summary statistics for each catalogue and goes to zero when the two summary statistics are identical.  
For simple analytic distributions, one can identify sufficient summary statistics for which one can rigorously prove that the ABC posterior approaches the true posterior as a distance threshold goes to zero. 
In practice, ABC is most useful for complex problems like ours, for which it is impractical to identify sufficient statistics.  
In these cases, the ABC posterior will be broader than the true posterior, since some draws may reasonably reproduce the summary statistics, but differ in some way that did not contribute to the distance function.

Having identified summary statistics that are both observable and physically significant in \S\ref{Summary_stats}, we proceed to define a distance function for each summary statistic.  
For most of our summary statistics, we use either the two-sample Kolmogorov--Smirnov (KS; \citealt{K1933, S1948}) distance or a rescaled variant of the two-sample Anderson--Darling (AD; \citealt{AD1952, P1976}) distance. 
A component distance of zero would mean that the marginal distributions for each of these summary statistics are identical.   
The use of KS distances rewards distributions that agree best near the median of the marginal distribution, while the use of AD distances is more sensitive to differences in the tails of the marginal distributions.  
We perform calculations using both, primarily as a sensitivity test.  
By comparing results using two different distance functions, we can check whether any of our conclusions are sensitive to the choice of distance function.

For our total distance, we take a linear combination of individual distance terms for each summary statistic:
\begin{align}
 \mathcal{D}_W &= \left[ \sum_{i'}{w_{i'} \left\|\mathcal{D}_{i'}\right\|}^{\alpha_D} \right]^{1/\alpha_D}, \label{eq_dist_general}
\end{align}
where $\mathcal{D}_W$ is the total weighted distance, we set $\alpha_D =1$, and $w_{i'} = 1/\hat{\sigma}(\mathcal{D}_{i'})$ is the weight (defined below) of the $i'^{\rm th}$ distance term, $\mathcal{D}_{i'}$.
In principle, one could have chosen another value of $\alpha_D$, corresponding to the Euclidean normal or maximum norm.  Either of these would result in a total distance that is more sensitive to the most discrepant summary statistic than our choice of $\alpha_D=1$, which was informed by our preliminary exploratory analyses.

We want a total distance function that takes into account each of the observed marginal distributions of the \Kepler{} population described in \S\ref{Summary_stats}, as well as the overall rate of planets per observed star, $f_{\rm Kepler}$.  We aim to assign weights to the individual distances that reflect the precision with which they were measured by \Kepler{}.  
Therefore, we specialize Equation \ref{eq_dist_general} to:
\begin{align}
 \mathcal{D}_W &= \frac{D_f}{\hat{\sigma}(D_f)} + \frac{D_{\rm mult}}{\hat{\sigma}(D_{\rm mult})} + \sum_{i=1}^{7} \frac{\mathcal{D}_i}{\hat{\sigma}(\mathcal{D}_i)}, \label{eq_dist}
\end{align}
where $D_f$ is the distance between the two rates of observed planets, $D_{\rm mult}$ is the distance of the multiplicity distributions, $\mathcal{D}_i$ is the distance between the distributions of the $i^{\rm th}$ observable, and $\hat{\sigma}(\mathcal{D})$ is the estimated root-mean-square (RMS) of the distance $\mathcal{D}$ given a ``perfect'' model (i.e., comparisons of simulated catalogues to a reference simulated catalogue with the same model parameters). 
The indices for the observables run from $i = 1$ to $i = 7$, denoting the following distributions of measured properties in order: period $\{P\}$, period ratio $\{\mathcal{P}\}$, transit depth $\{\delta\}$, transit depth ratio $\{\delta_{i+1}/\delta_i\}$, transit duration $\{t_{\rm dur}\}$, and period-normalized transit duration ratio for near-resonance $\{\xi_{\rm res}\}$ and not-near-resonance $\{\xi_{\rm non-res}\}$ planet pairs.

{\em Rate of Observed Planets:} 
Since the rate of observed planets is a scalar, the distance function is simply, $D_f = | f_{\rm sim} - f_{\rm Kepler} |$.

{\em Multiplicity:}  
The observed multiplicities $\left\{N_m\right\}$ is a vector of integers.  If each system had the same probability to be observed as an $m$-planet system, then it would be a multinomial distribution.
Formally, $\left\{N_m\right\}$ is not drawn from a multinomial since each system has a different set of probabilities for being observed as an $m$-planet system.  
Nevertheless, the theory of multinomial distribution is useful for identifying an appropriate distance function for the observed multiplicities.  
The Cressie--Read power divergence (CRPD) statistic \citep{CR1984} is commonly used in comparing multinomial distributions like the multiplicity distribution and is known to be more robust than other choices like $\chi^2$ for cases like ours where there is a large dynamic range between the values in each category and one of the categories ($O_5$) often has values less than 5.\footnote{\citet{Li2011a} use the exact multinomial statistic, but using this statistic is intractable for our case. After significant testing of different statistics, CRPD was shown to be the best approximation of the exact statistic for distributions similar to ours.}
Therefore, for $D_{\rm mult}$, we adopt the CRPD statistic which is given by:
\begin{align}
 \rho_{\rm CRPD} = \frac{9}{5}\sum_j O_j \bigg[{\bigg(\frac{O_j}{E_j}\bigg)}^{2/3} - 1\bigg], \label{eq_CRPD}
\end{align}
where $O_j$ is the number of ``observed'' systems according to one of our models and $E_j$ is the number of expected systems in the $j^{th}$ bin based on the rate of such systems in the \Kepler{} data set.
The indices $j$ label the bins corresponding to 1, 2, 3, 4, and $5+$ observed planets.
Note that by definition, this statistic only sums over bins $j$ for which $E_j \neq 0$ (as it is formally impossible to have a match between the observed and expected distributions when $E_j = 0$ and $O_j > 0$). These caveats are not significant for our analysis since our distance function $D_{\rm mult}$ meets the primary goal of clearly favouring better matches to the \emph{Kepler} data. 
\footnote{We also considered switching the interpretation of $O_j$ and $E_j$, i.e. let $O_j$ denote the number of actual observed \Kepler{} systems and $E_j$ denote the number of expected observed systems given a model.  During exploratory analyses, we found that $\rho_{\rm CRPD}$ would occasionally give infinite values if a simulated catalogue included zero 5+ planet systems.  Eliminating that term from the sum is not viable, as this can result in negative values, which violates the non-negative property of a distance function and would lead to favouring models with fewer multiplanet systems than were observed by \Kepler{}.} 

{\em Continuous Distributions:}  
For remaining $\mathcal{D}_i$ terms, we perform the full analysis with two different distance functions for comparing samples of continuous random variables: the two-sample Kolmogorov--Smirnov (KS) distance, and a rescaled variant of the two-sample Anderson--Darling (AD) distance. For two finite samples of sizes $n$ and $m$, described by empirical distribution functions $F_n(x)$ and $G_m(x)$, the two-sample KS and AD distances are defined as:
\begin{align}
 \mathcal{D}_{\rm KS} &= {\rm max}| F_n(x) - G_m(x) |, \label{eq_KS} \\
 \mathcal{D}_{\rm AD} &\equiv \frac{nm}{N} \int_{-\infty}^{\infty} \frac{[F_n(x) - G_m(x)]^2}{H_N(x)[1 - H_N(x)]} dH_N(x), \label{eq_AD_def} \\
 &= \frac{1}{nm} \sum_{i=1}^{N-1} \frac{(M_i N - ni)^2}{i(N - i)}, \label{eq_AD}
\end{align}
where $N = n+m$ is the combined sample size, $H_N(x) = [nF_n(x) + mG_m(x)]/N$ is the empirical distribution function for the combined sample, and $M_i$ is the number of observations less than or equal to the $i^{\rm th}$ smallest in the combined sample \citep{P1976}. 

The KS distance is simply the maximum absolute difference in the cumulative distribution functions (CDFs) and is thus most sensitive to the difference in the bulk locations of the two distributions. The AD distance on the other hand, more heavily weights the tails of the distributions and is thus more sensitive to differences at the extremes of the distributions. In practice, we find that the standard AD distance given by Equation (\ref{eq_AD}) does not sufficiently account for the differences in the sample sizes $n$ and $m$; in other words, two samples can give a very small AD distance even when $n$ and $m$ are vastly different. In the context of our models, this has the consequence that very small simulated observed catalogues (i.e. with only a handful of observed planets) can still result in small AD distances, even enough to counteract larger distances of $D_f$ for the overall rate of planets per star.\footnote{This is apparent if one considers the effect of the $\frac{nm}{N}$ term in Equation (\ref{eq_AD_def}): let $m$ be the number of observed planets in the \Kepler{} data and $n$ be the number of observed planets given by the model. If $n \simeq m$, this term is roughly $\sim m/2$; but if $n \sim 1 \ll m$, this term becomes roughly $\frac{m}{m+1} \sim 1$. Since $m$ is relatively large, the $\frac{nm}{N}$ term is thus much smaller when $n \ll m$ than when $n \simeq m$.} This is clearly undesirable as such models are terrible fits to the \Kepler{} observations. In order to counteract this unintended consequence, we modify the AD distance to penalize such models (i.e. those that result in almost no observed planets) by dividing out the $nm/N$ constant in front of the integral, giving:
\begin{align}
 \mathcal{D}_{\rm AD'} &= \int_{-\infty}^{\infty} \frac{[F_n(x) - G_m(x)]^2}{H_N(x)[1 - H_N(x)]} dH_N(x) \\
 &= \frac{N}{(nm)^2} \sum_{i=1}^{N-1} \frac{(M_i N - ni)^2}{i(N - i)}. \label{eq_AD_mod}
\end{align}
For the remainder of the paper, we will refer to our modified AD distance given by Equation (\ref{eq_AD_mod}) as simply ``the AD distance'' unless otherwise noted. We refer to the total weighted distance functions involving the KS or rescaled AD distance terms as $\mathcal{D}_{W,\rm KS}$ and $\mathcal{D}_{W,\rm AD'}$, respectively.

In order to compute the weights $w_{i'}$ for the individual distance terms, we first generate a simulated observed catalogue using the clustered periods and sizes model which serves as a reference catalogue, and then repeatedly simulate catalogues using the same model (i.e. with identical model parameters) which would be a ``perfect'' model. The RMS ($\hat{\sigma}(\mathcal{D})$) of each individual distance term and the weights (the inverse of the RMS, $w = 1/\hat{\sigma}$) computed this way are listed in Table \ref{tab:weights}, while the model parameters used to generate this reference catalogue are given in Table \ref{tab:param_ranges}. The distances for each term are not zero because the simulations involve Monte Carlo noise and only a finite number of planets are drawn per iteration. Indeed, the true \Kepler{} catalogue of observed planets is finite in size. Thus, we simulate a reference catalogue that contains a similar number of observed planets given the same number of target stars as the \Kepler{} mission. However, for the purposes of computing the weights and optimizing the distance function to find the best-fitting model parameters (see \S\ref{Optimization}), we wish to reduce stochastic noise in order to improve our power to distinguish between different models. Therefore, we simulate catalogues from the ``perfect'' model with five times as many stars to give observed catalogues that are five times as large as that of our \Kepler{} sample, balancing the desire to reduce stochastic noise and computational time.  We compute the weights with 1000 repeated simulations from the ``perfect'' model.

By summing the distances weighted by their variations for a ``perfect'' model, the total weighted distance emphasizes the distances of observables that are well characterized and prevents any single distance term with high stochastic noise from dominating the total distance function. Also, this implies that each individual distance term included in Equation (\ref{eq_dist}) typically contributes a weighted value of roughly 1 to the total distance in the case of using the perfect model.

\begin{table}
\centering
\caption{Table of weights for the individual distance terms as computed from a reference clustered periods and sizes model. The weights $w$ are computed from the root mean squares of the distances, $\hat{\sigma}(\mathcal{D})$, and are shown here as rounded whole numbers simply for guidance purposes.}
\begin{tabular}{ l | c | c | c c |}
 \hline
 \hline
 Distance term & \multicolumn{2}{c|}{$\hat{\sigma}(\mathcal{D})$} & \multicolumn{2}{c|}{$w = 1/\hat{\sigma}$} \\
 \hline
 $D_f$ & \multicolumn{2}{c|}{0.000409} & \multicolumn{2}{c|}{2443} \\
 $D_{\rm mult}$ & \multicolumn{2}{c|}{0.00123} & \multicolumn{2}{c|}{816} \\
 \hline
  & \multicolumn{2}{c|}{KS} & \multicolumn{2}{c|}{AD'} \\
 \hline
  & $\hat{\sigma}(\mathcal{D})$ & $w = 1/\hat{\sigma}$ & $\hat{\sigma}(\mathcal{D})$ & $w = 1/\hat{\sigma}$ \\
 $\mathcal{D}_1$ (for $\{P\}$) & 0.0173 & 58 & 0.000448 & 2230 \\
 $\mathcal{D}_2$ (for $\{\mathcal{P}\}$) & 0.0288 & 35 & 0.00113 & 882 \\
 $\mathcal{D}_3$ (for $\{\delta\}$) & 0.0179 & 56 & 0.000480 & 2085 \\
 $\mathcal{D}_4$ (for $\{\delta_{i+1}/\delta_i\}$) & 0.0299 & 33 & 0.000911 & 1098 \\
 $\mathcal{D}_5$ (for $\{t_{\rm dur}\}$) & 0.0213 & 47 & 0.000524 & 1907 \\
 $\mathcal{D}_6$ (for $\{\xi_{\rm res}\}$) & 0.0678 & 15 & 0.00652 & 153 \\
 $\mathcal{D}_7$ (for $\{\xi_{\rm non-res}\}$) & 0.0328 & 30 & 0.00121 & 827 \\
 \hline
 \hline
\end{tabular}
\label{tab:weights}
\end{table}

\subsubsection{The \Kepler{} catalogue: stellar and planetary properties} \label{Data}

\begin{figure}
\begin{tabular}{c}
 \includegraphics[scale=0.425,trim={0 0.5cm 0 0.3cm},clip]{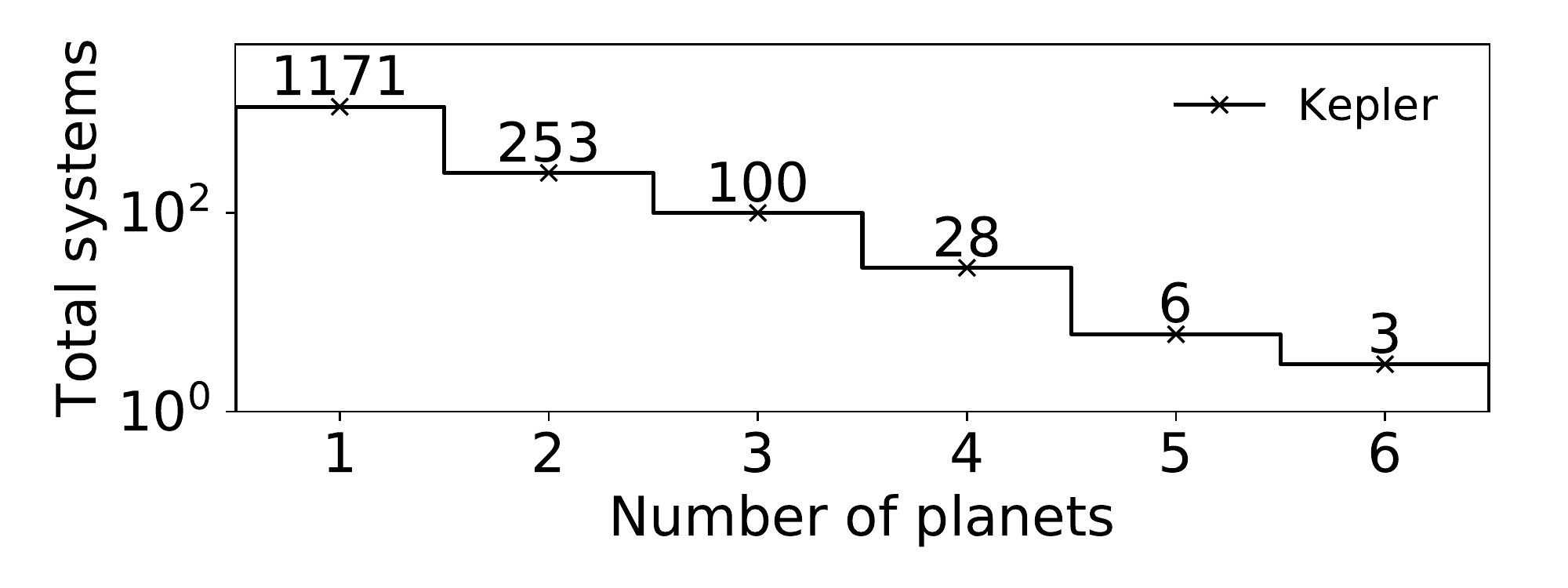} \\
 \includegraphics[scale=0.425,trim={0 0.5cm 0 0.3cm},clip]{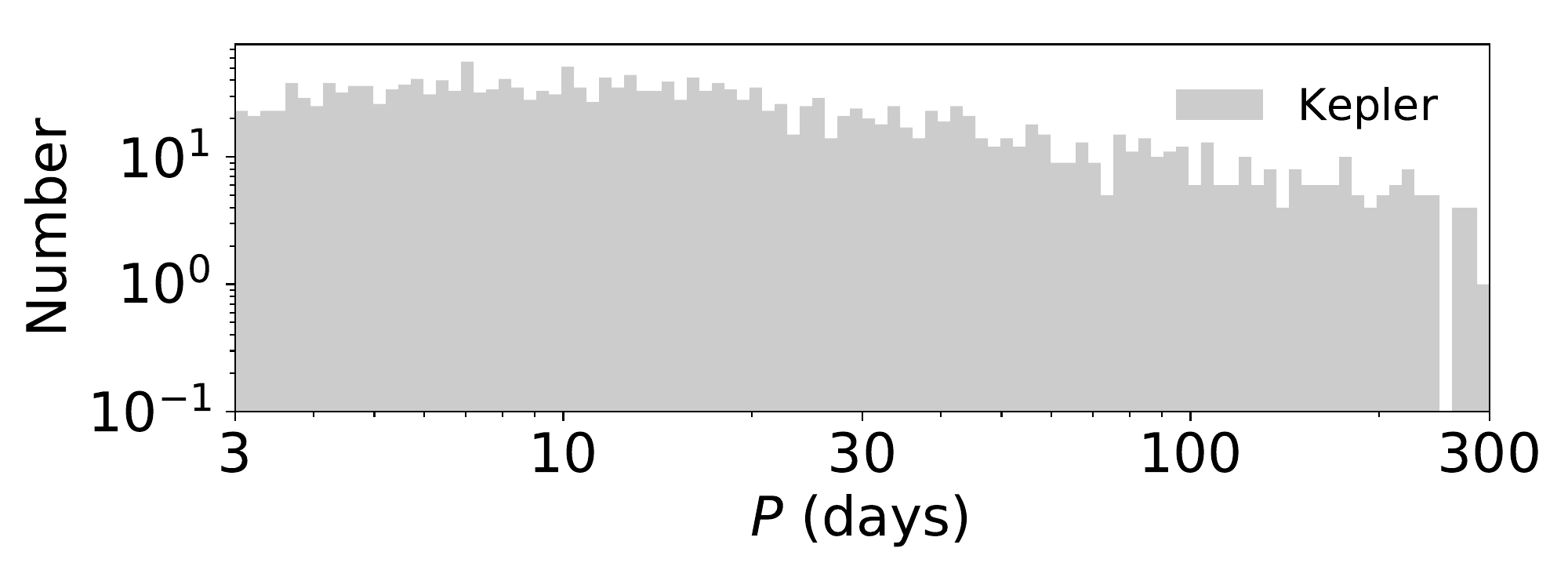}\\
 \includegraphics[scale=0.425,trim={0 0.5cm 0 0.3cm},clip]{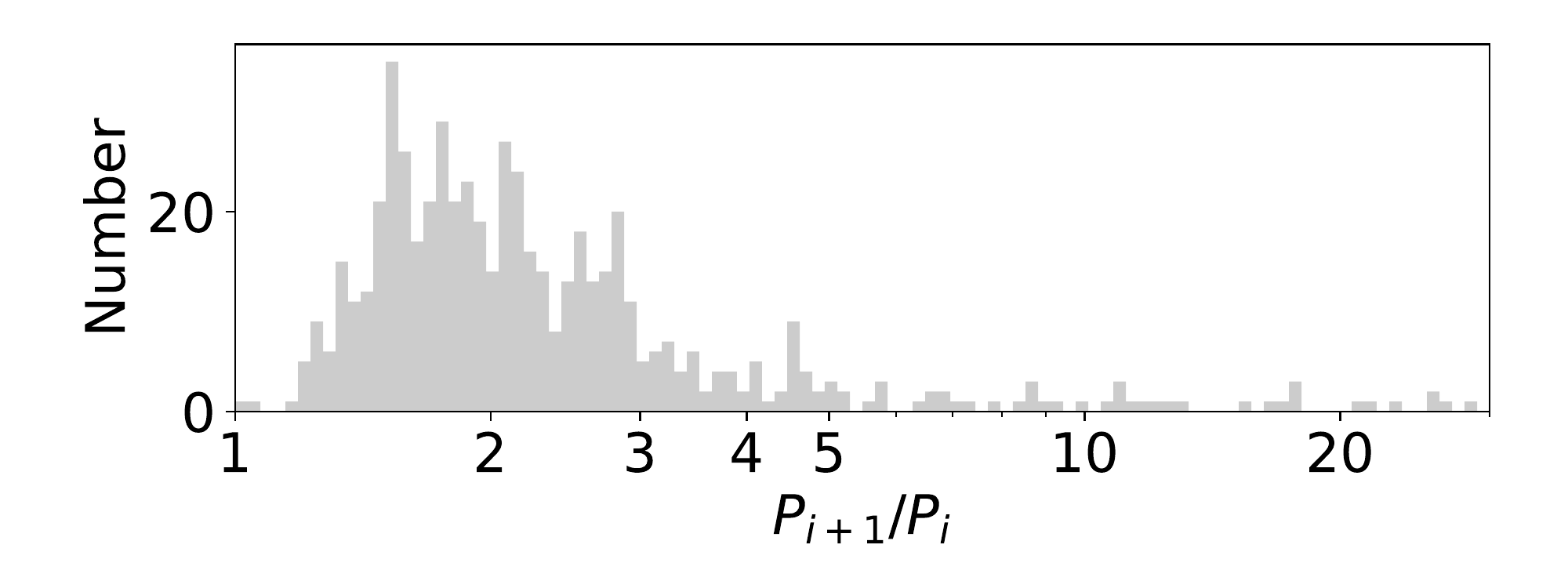}\\
 \includegraphics[scale=0.425,trim={0 0.5cm 0 0.3cm},clip]{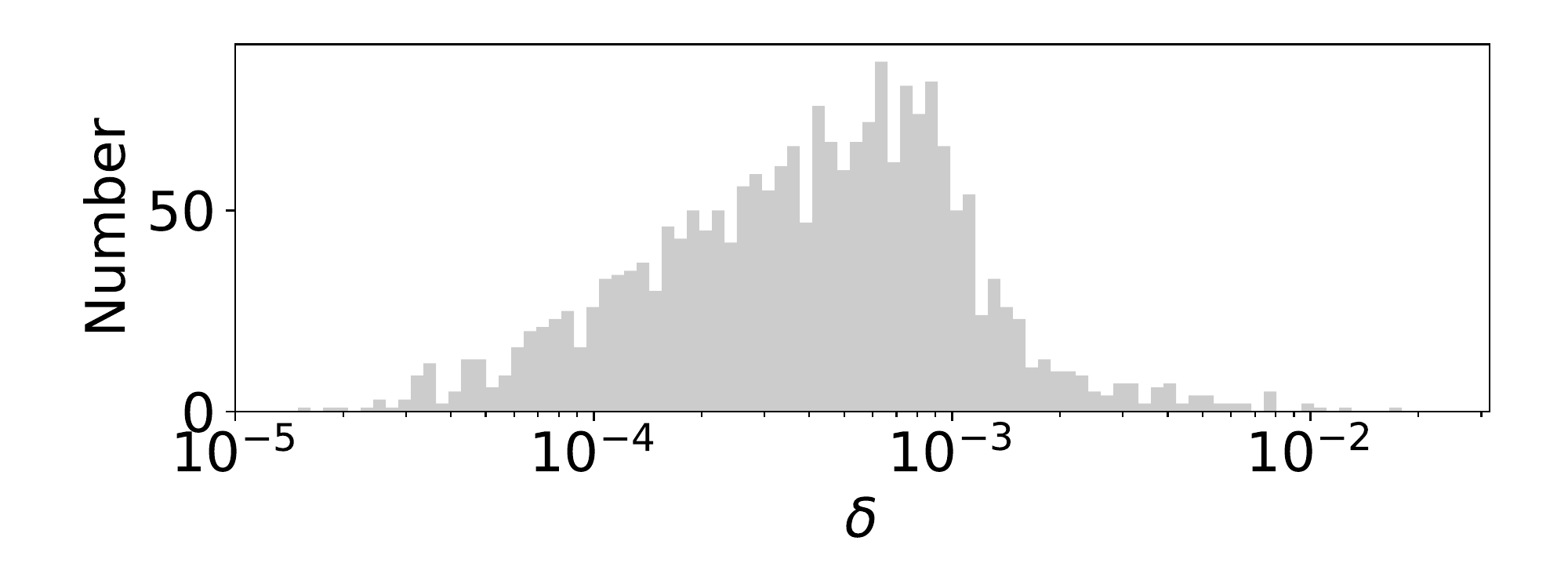}\\
 \includegraphics[scale=0.425,trim={0 0.5cm 0 0.3cm},clip]{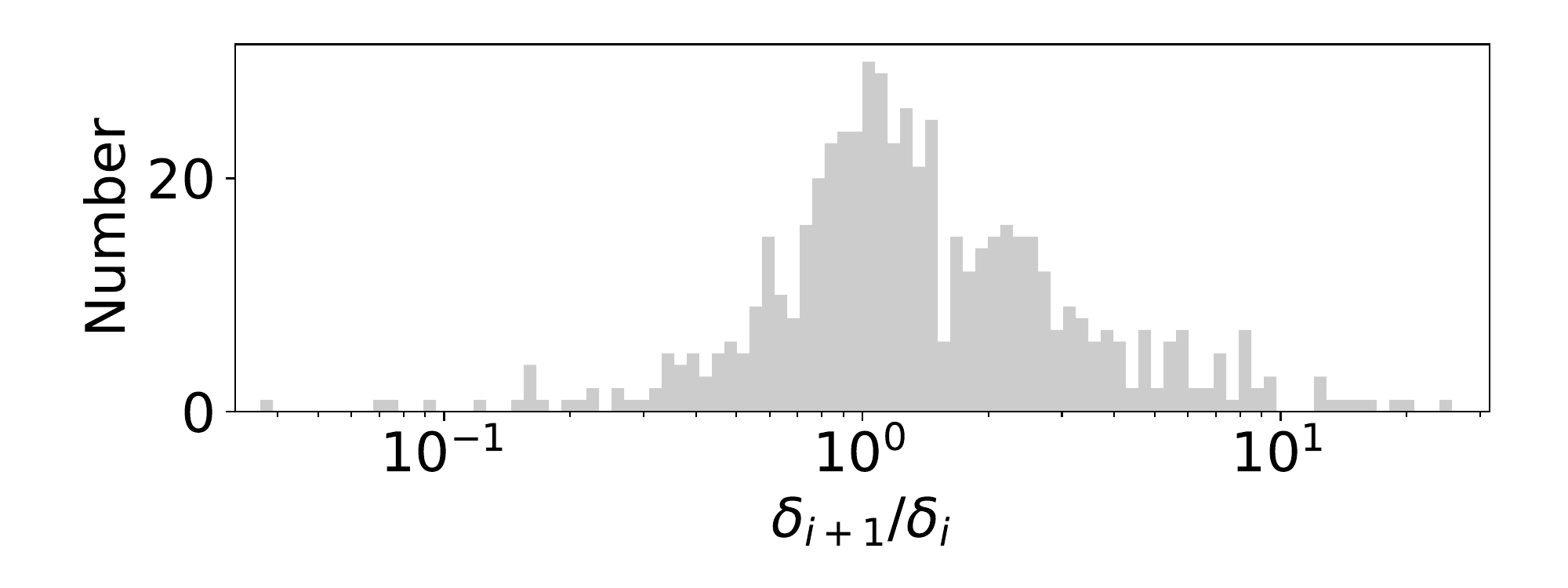}\\
 \includegraphics[scale=0.425,trim={0 0.5cm 0 0.3cm},clip]{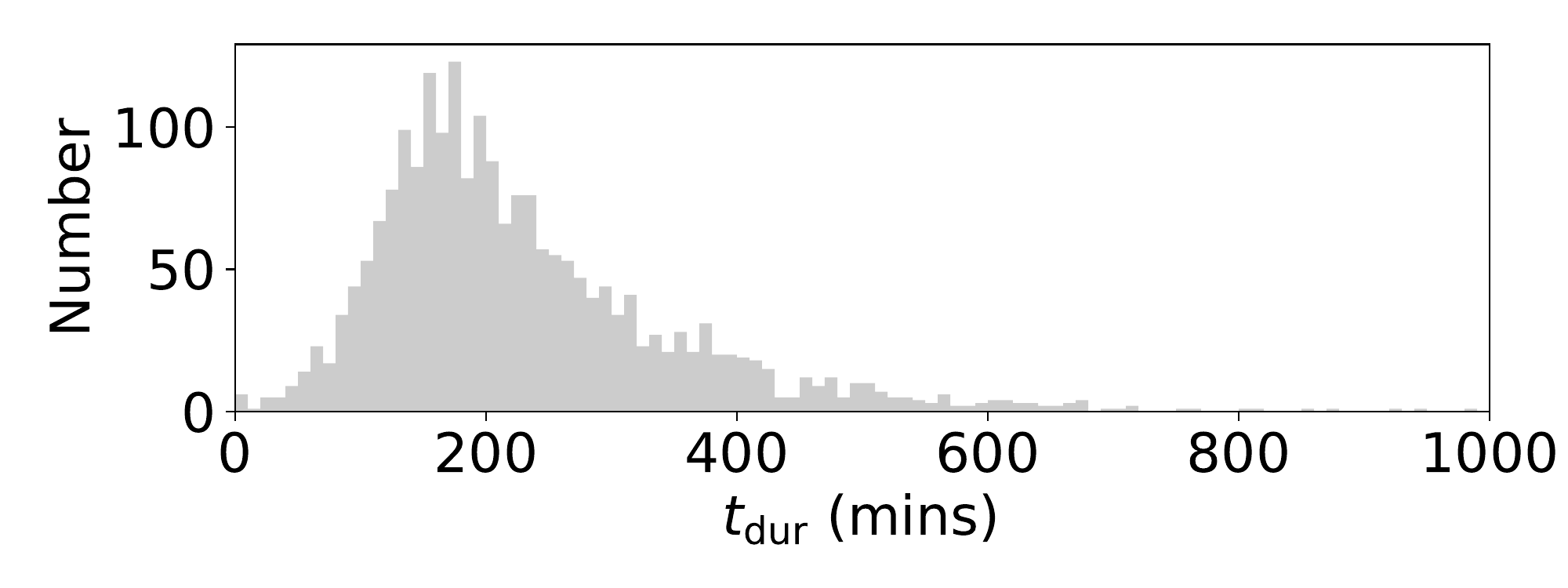}\\
 \includegraphics[scale=0.425,trim={0 0.5cm 0 0.3cm},clip]{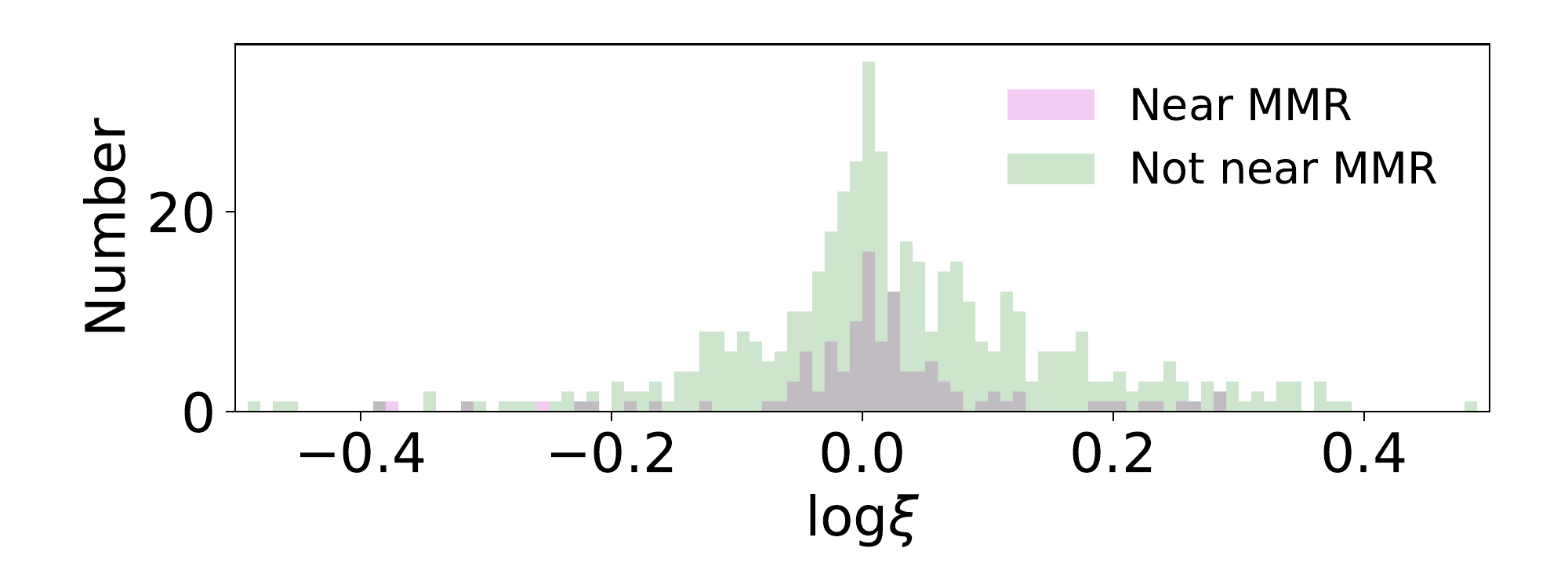}\\
\end{tabular}
\caption{The \Kepler{} Q1-Q17 DR25 population of KOIs vetted as planet candidates by the robovetter, given our cuts as described in \S\ref{Data}. From top to bottom: histograms of the observed planet multiplicities, periods ($P$), period ratios ($\mathcal{P}$), transit depths ($\delta$), transit depth ratios ($\delta_{i+1}/\delta_i$), transit durations ($t_{\rm dur}$), and period-normalized transit duration ratios ($\xi$). We separate the $\xi$ distribution into two, for planets near- and not-near-MMRs as defined in \S\ref{Incl} (most planets are not near an MMR). The period, period ratio, transit depth, and transit depth ratio bins are uniformly spaced in log, while the bins in the other histograms are linearly spaced.}
\label{fig:Kepler_DR25}
\end{figure}

{\em Stellar Catalogue:} 
We couple the \Kepler{} DR25 stellar properties catalogue with the results of the second \textit{Gaia} data release (DR2) \citep{A2018, Gaia2018} in order to take advantage of its significantly improved stellar parameters for a large fraction of the \Kepler{} target stars. These improvements (of primary interest, the stellar radii) result from the refined parallax and thus distance measurements of the Gaia mission. 
Furthermore, the Gaia DR2 parallaxes allow for a cleaner sample of main--sequence target stars thanks to a more precise positioning of the targets in color-luminosity space.  Additionally, the astrometric information allows for identification of targets likely consisting of multiple stars with comparable luminosity.

We use the same \Kepler{} target list as in \citet{H2019}, who identified a clean sample of FGK main-sequence (M-S) stars for occurrence rate studies. 
This involved performing a series of cuts on the \Kepler{} DR25 stellar table based on measurements reported in the Gaia DR2 and updating the stellar radii with values reported in the Gaia DR2.
For full details, see \S3.1 of \citet{H2019}.

In order to minimize sensitivity to uncertainties in stellar radii, impact parameters, and the limb darkening model, we have chosen a distance function based on the distribution of transit depths instead of the distribution of planet radii or planet--star radius ratios, since the measured transit depth does not depend on our knowledge of the stellar radius (like the planet radius) and is better modelled as a Gaussian distribution than the planet--star radius ratio (due to effects of limb darkening and covariance with impact parameter). Nevertheless, the uncertainties in stellar radii still affect our simulations via the transit depths of the simulated planets.  

{\em Planet Catalogue:} 
We start from the \Kepler{} Data Release 25 (DR25)
\citep{T2018} \Kepler{} Objects of Interest (KOI) catalogue as the basis for our study. This table is most suitable for a population study of the \Kepler{} exoplanets because it involves uniform vetting of the Q1-Q17 light curves obtained by \Kepler{}, and thus does not involve human biases across individual systems. It is derived from processing using the SOC pipeline release 9.3, and involves fully automated dispositioning of the Threshold Crossing Events (TCEs) using the \textit{Kepler Robovetter} \citep{T2015, T2016}. Specifically, the automated procedure involves determining whether the TCEs are transit-like, and if so, tests whether there is any evidence of an eclipsing binary, shift in the in-transit centroid position, or contamination from another source \citep{M2015, C2016}. TCEs that pass all the tests are dispositioned as planetary candidates, and their planetary and orbital parameters are computed using a Markov Chain Monte Carlo (MCMC) fitting algorithm \citep{R2015}.

Starting from the \Kepler{} DR25 KOI catalogue, we remove planet candidates around targets that were excluded from the stellar catalogue, as described above.  
Next, we keep only KOIs designated as planet candidates by the \textit{Kepler Robovetter}. 
Then, we replace the transit depths and durations in the \Kepler{} DR25 catalogue (which were maximum likelihood estimators) with the median values from the MCMC-based posterior samples described in \citet{R2015}. 
We update the planet radii based on the observed transit depths, the updated stellar radii from \textit{Gaia} DR2, and the limb darkening model from the DR25 stellar catalogue.  
Finally, we limit the planetary catalogue to include planet candidates with periods between $P_{\rm min}=3$ d and $P_{\rm max}=300$ d (see \S\ref{Methods}) and with updated planet radii between $R_{p,\rm min} = 0.5 R_\oplus$ and $R_{p,\rm max} = 10 R_\oplus$. 

These cuts result in a total of 79 935 \Kepler{} targets (hereafter denoted by $N_{\rm stars,Kep}$), with 2137 total planet candidates in 1561 systems (with periods between 3 and 300 d and planet radii between 0.5 and 10 $R_\oplus$). Of these, a total of 390 multiplanet systems with 966 planets are included, with the remaining 1171 planets being in single systems. The resulting population of KOIs is shown in Figure \ref{fig:Kepler_DR25}, where we plot histograms of the number of planets per system ($N_m$, planet multiplicity), periods ($P$), period ratios of apparently adjacent planet pairs ($P_{i+1}/P_i$), transit durations ($t_{\rm dur}$), period-normalized transit duration ratios ($\xi$), transit depths ($\delta$), and transit depth ratios ($\delta_{i+1}/\delta_i$).
These distributions serve as the target observed distributions for our models.
In particular, the distributions of the period ratios, transit duration ratios, and transit depth ratios are especially insightful to model because they probe the architectures of planetary systems, yet are insensitive to the stellar parameters.

\subsection{Optimization of the distance function} \label{Optimization}
In order to compare our forward models to the \Kepler{} observations, we need to find model parameters that result in simulated catalogues that are similar to the \Kepler{} DR25 catalogue.
Given the complexity and computational expense of the model, we take a multistage approach.  
First, we use an optimizer to identify good regions of parameter space.  
Results from the optimizer are then used to train a Gaussian Process emulator (described in \S\ref{GP}). Finally, we draw samples from prior distributions for model parameters and use rejection sampling to construct the ABC posterior for inference.

Here we describe the optimization stage that seeks the set of model parameters that minimize the distance function (for given a model, observed data set, and distance function).  This is a challenging problem for several reasons.  
First, evaluating the distance function is computationally expensive, primarily due to the catalogue simulation. 
Secondly, the parameter space is large. Our models involve several free parameters: 8, 10, and 11 for the non-clustered, clustered periods, and clustered periods and sizes models, respectively (even after fixing the break radius $R_{p,\rm break}$ and minimum separation for stability $\Delta_c$).  There can be correlations or potentially complicated interplay between model parameters.
The third and most challenging factor is that the forward model is stochastic due to sampling variance and the finite number of targets. Even for fixed model parameters, the computed distance varies from one realization to the next due to Monte Carlo randomness in drawing target properties, physical properties of planetary systems, and simulated measurement noise. This means that traditional optimization algorithms that assume a deterministic function are not appropriate for our problem.
In theory, one could reduce the variance in the distances drawn from our forward model by simulating significantly more targets than \Kepler{} observed. While this could be a useful (but expensive) way to find the ``best-fitting'' model parameters, it is not appropriate for accurately characterizing the uncertainties in the model parameters. The summary statistics for the \Kepler{} catalogue include features that might be real or merely the result of small number statistics. In ABC, our forward model should also have variance due to the finite number of targets observed, in order for the ABC posterior to properly weight model parameters accounting for the extent of variations due to the finite number of targets. Therefore, we use the same number of targets as in the \Kepler{} catalogue during both the optimization stage (this section) and the emulator stage (as described in the next section).

We use an adaptive Differential Evolution optimizer function with radius limited sampling, implemented by ``BlackBoxOptim'' (\url{https://github.com/robertfeldt/BlackBoxOptim.jl}). This package includes a general purpose optimization function ``bboptimize()'' which provides various algorithms, some of which are designed to deal with stochastic noise. For our purposes, we use the optimizer ``adaptive\_de\_rand\_1\_bin\_radiuslimited''. This optimizer uses a population-based algorithm that iteratively searches a specified region in parameter space in order to try and minimize a target fitness function without assuming that the function is differentiable. 
Thus it is well suited to high dimensional problems with stochastic noise. 
We use the total weighted distance given by Equation (\ref{eq_dist}) as the target fitness function, and leave most of the key model parameters of each model as free parameters with an allowed search range (minimum, maximum) for each parameter. 
Table \ref{tab:param_ranges} lists the search ranges we used for each free parameter, in each model.

We repeat 50 runs of the optimizer on each of our models, each with a different set of initial values for the free parameters drawn randomly (uniformly, while for $\lambda_c$ and $\lambda_p$, uniformly in log) within each of the search ranges for each parameter. For each run, we set the ``population size'' parameter equal to $4n_{\rm params}$, where $n_{\rm params}$ is the number of free parameters in the model. We simulate $N_{\rm stars,sim} = 79 935$ targets for each generation of the model. In order to enforce the criteria $\sigma_{i,\rm low} \leq \sigma_{i,\rm high}$, we transform these two parameters into two dummy variables $r_1$, $r_2$ using a mapping from the unit square to a triangle \citep{OFCD2002}.\footnote{\citet{OFCD2002} prescribe the equation $P = (1-\sqrt{r_1})A + \sqrt{r_1}(1-r_2)B + {r_2}\sqrt{r_1}C$ in their Equation (1) of Section 4.2 to uniformly map the unit square to any arbitrary triangle of vertices $(A,B,C)$, i.e. by drawing $r_1$ and $r_2$ randomly between 0 and 1. We adopt this transformation and set the vertices to $A = (0^\circ, 0^\circ)$, $B = (90^\circ, 90^\circ)$, and $C = (90^\circ, 0^\circ)$ for $(\sigma_{i,\rm high}, \sigma_{i,\rm low})$.} We also set a condition to avoid simulating the model if $\ln{(\lambda_c)} + \ln{(\lambda_p)} > 2.5$, in order to avoid wasting time on models with too many planets.

For computational expediency, we stop the optimization process after 5000 model iterations have been computed for each of the 50 runs.
We observe from preliminary runs that the best models found during this stage do not improve appreciably with significantly more iterations (e.g. 10,000 iterations). We verify that the local minima found by each of the runs are in a similar region of parameter space.

\begin{table}
\centering
\caption{Table of the free parameters and their search ranges (min, max) explored, of each model. The last column lists the parameter values of the clustered periods and sizes model used to generate the reference catalogue as described in \S\ref{Distance}. We set $R_{p,\rm break} = 3 R_\oplus$ and $\Delta_c = 8$.}
\begin{tabular}{ l | c | c | c l c |}
 \hline
 \hline
 Parameter & Non- & Clustered & Clustered periods & Ref. \\
  & clustered & periods & and sizes & catalogue \\
 \hline
 $f_{\sigma_{i,\rm high}}$ & $(0, 1)$ & $(0, 1)$ & $(0, 1)$ & 0.5 \\
 $\lambda_c$ & - & $(0.2, 5)$ & $(0.2, 5)$ & 1.6 \\
 $\lambda_p$ & $(1, 8)$ & $(0.5, 10)$ & $(0.5, 10)$ & 1.6 \\
 $\alpha_P$ & $(-2, 2)$ & $(-2, 2)$ & $(-2, 2)$ & $0$ \\
 $\alpha_{R1}$ & $(-4, 2)$ & $(-4, 2)$ & $(-4, 2)$ & $-1$ \\
 $\alpha_{R2}$ & $(-6, 0)$ & $(-6, 0)$ & $(-6, 0)$ & $-4$ \\
 $\sigma_e$ & $(0, 0.1)$ & $(0, 0.1)$ & $(0, 0.1)$ & 0.01 \\
 $\sigma_{i,\rm high}$ ($^\circ$) & $(0, 90)$ & $(0, 90)$ & $(0, 90)$ & 10 \\
 $\sigma_{i,\rm low}$ ($^\circ$) & $(0, \sigma_{i,\rm high})$ & $(0, \sigma_{i,\rm high})$ & $(0, \sigma_{i,\rm high})$ & 1 \\
 $\sigma_R$ & - & - & $(0, 0.5)$ & 0.25 \\
 $\sigma_P$ & - & $(0, 0.3)$ & $(0, 0.3)$ & 0.15 \\
 \hline
 \hline
\end{tabular}
\label{tab:param_ranges}
\end{table}

\subsection{Exploring the parameter space with a Gaussian process emulator} \label{GP}

Computing the full forward model as detailed in \S\ref{Procedure}-\S\ref{Obs_pipeline} is computationally expensive. 
Generating a \textit{physical catalogue} with $N_{\rm stars,sim} = N_{\rm stars,Kep} = 79 935$ typically takes $\sim$10--20 s for reasonable model parameters, although this is highly dependent on the mean rates of clusters and planets per cluster $\lambda_c$, $\lambda_p$, and the cluster width in log-period per planet $\sigma_P$ due to repeated draws for stability.
(The procedure to simulate an \textit{observed catalogue} from a pre-existing \textit{physical catalogue} is faster, taking just a few seconds.) 
Thus, it would be prohibitively time-consuming to simulate the full model for millions of iterations. 
Fortunately, for the purpose of finding the region(s) of parameter space that result in observed catalogues similar to the \Kepler{} catalogue, we only need to store the total weighted distance for each proposed set of model parameters and do not necessarily need to save all the information in the catalogues.  
Based on our initial exploratory analyses, the total weighted distance has a single dominant mode for each physical model, which is identified by the optimizer from \S\ref{Optimization}.  
In the vicinity of that mode, the mean of distance function varies smoothly, but with considerable variance due to Monte Carlo noise from the finite number of stars.
Thus, we can dramatically accelerate the exploration of parameter space by approximating the total weighted distance of our full model using a fast emulator.  
We adopt the machinery of Gaussian processes (GPs) to train an emulator for our distance function and use the GP to explore the model parameter space in a computationally feasible manner (see the textbook by \citealt{RW2006} for an extensive guide to and discussion of GPs, and \citealt{O2004} for an introduction to the GP emulator approach).

A Gaussian process emulator is a statistical model that aims to mimic a more complicated and expensive function by ``emulating'' the outputs of the expensive function given the same inputs. 
A covariance function specifies the correlation between draws from the GP for any pair of input values.  
For any set of inputs (i.e., model parameters), the GP emulator returns a Gaussian distribution for its prediction of the model output (i.e., distance) that depends on the observed values of the function at a set of training points.  
While the detailed outputs of our physical (clustered and non-clustered) models are complex, we only require the GP emulator to provide a good approximation to the total weighted distance in a region of parameter space that results in simulated catalogues that are a good match to the \Kepler{} data. 
For each set of model parameters, the GP emulator returns a distribution, which approximates the mean and variance of the distribution of distances that would be returned by the full model (including the effects of the finite number of targets).
This allows us to explore the parameter space very quickly in order to efficiently estimate how often realizations with a given set of model parameters would result in a weighted distance less than the maximum acceptable distance for our ABC posterior sample.

Here we provide an overview of the GP emulator before providing more specific details below.  
For the remainder of this paper, we let $\bm{\theta}$ denote the free parameters of our physical models (e.g. $\lambda_p$, $\alpha_P$, etc. as listed in Table \ref{tab:param_ranges}, of our non-clustered and clustered models) and $\bm{\phi}$ denote the (hyper) parameters of the GP emulator. 
We have a distance function $\mathcal{D}_W(\bm{\theta})$ (this is either $\mathcal{D}_{W,\rm KS}(\bm{\theta})$ or $\mathcal{D}_{W,\rm AD'}(\bm{\theta})$) that we evaluated at a large number of points $\bm\theta$ during the optimization stage. 
As discussed below, we use a subset of points that yielded low distances during the optimization stage as training points for the GP emulator (see \S\ref{secTrainingPoints}).
For a given model and set of training points, we find the ``best-fitting'' values of the hyperparameters $\bm\phi_{bf}$ which maximize the log likelihood for the GP.
Then, we use the emulator with $\bm\phi_{bf}$ to predict the distance $\mathcal{D}_W(\bm{\theta})$ for a much larger number of points in the model parameter space.
We draw trial values of $\bm{\theta}$ from a prior and reject draws that result in a predicted distance greater than a distance threshold. 
The distribution of the accepted points provides a sample from the ABC posterior (see \S\ref{CredibleRegions}).
We compute credible regions for each model parameter $\theta$ from the quantiles of the accepted points.

\subsubsection{Choice of mean and covariance function}
\label{secCovarFunction}

GPs are especially useful in our context due to their flexibility in modelling stochastic processes with intractable functional forms. 
This is exactly the case for our distance function.

Mathematically, a GP is described by a prior mean function $m(\bm{x})$ and a covariance (i.e. kernel) function $k(\bm{x},\bm{x'};\bm{\phi})$:
\begin{align}
 f(\bm{x}) \sim \mathcal{GP}\big(m(\bm{x}), k(\bm{x},\bm{x'};\bm{\phi})\big), \label{eq_GP}
\end{align}
where $f(\bm{x})$ is the function we wish to model ($f(\bm{x}) = \mathcal{D}_W(\bm{\theta})$ for our purposes), and $\bm{\phi}$ are the hyperparameters of the kernel.  

The prior mean function can be used to model an underlying deterministic process if one is known (such as the periodic motion of a star in a series of radial velocity measurements, e.g. in \citealt{Rj2015}). For our problem, an underlying functional dependence (i.e. $\mathcal{D}_W$ as a function of the model parameters $\bm{\theta}$) is not known; thus we use a constant prior mean function.  
When evaluating the emulator near training points, the predicted values will be strongly affected by the training points and only minimally affected by the prior mean.  The choice of our mean function will be important when evaluating the emulator far from training points.

In an ideal world, we would supply enough training points to adequately characterize the model behavior over the full parameter space, so the emulator would return accurate predictions at any given point.  However, this is not practical for the entirety of our 8--11 dimensional parameter space, as the number of training points is limited to just a few thousand, due to the computational cost of training the GP emulator.
Therefore, we select training points to be in the vicinity of the best-fitting region as found during optimization. 
We verified that the distribution of distances returned by the GP emulator is accurate in the region of interest.
In order to ensure that the GP emulator consistently returns values well above the distance threshold in other regions, we set the prior GP mean to a large value, i.e., near the high end of the distance for the training points (e.g., $m(\bm{x}) = 75$ for emulating the distance function involving KS distance terms, $\mathcal{D}_{W,\rm KS}$, and $m(\bm{x}) = 250$ for emulating the distance function involving AD distance terms, $\mathcal{D}_{W,\rm AD'}$).
Since the prior GP mean is significantly larger than the distance threshold to be used by ABC, the emulator will almost certainly return emulated distances well above distance threshold, when it is given parameter values far from the training points.
For the emulator to have a reasonable chance of returning a distance that would be accepted, the model parameters must be near a sizeable number of training points that cause the mean of the prediction to drop well below the prior mean.

For the covariance function, we choose the squared exponential kernel with a separate length scale, $\lambda_i$, for each dimension (i.e. model parameter $\theta_i$, with $i = 1,2,...,d$ where we let $d$ denote the number of dimensions/model parameters),
\begin{align}
 k(\bm{x},\bm{x'};\bm{\phi}) = \sigma_f^2 {\rm exp} \Bigg[-\frac{1}{2} \sum_i \frac{(x_i - {x_i}')^2}{\lambda_i^2} \Bigg], \label{eq_kernel}
\end{align}
where $\bm{\phi} = (\sigma_f, \lambda_1, \lambda_2, ..., \lambda_d)$ are the hyperparameters of the kernel. The hyperparameter $\sigma_f$ controls the strength of correlation between points, and also serves as the standard deviation of the Gaussian prior (i.e. far away from any training data). Intuitively, the length scales $\lambda_i$ govern how far points can be from one another, in each dimension, before they become uncorrelated with each other.

We find the best values for the hyperparameters $\bm{\phi}$ by attempting to maximizing the log marginal likelihood:
\begin{align}
 \log p(\bm{y}|\bm{X},\bm{\phi}) = -\frac{1}{2}\bm{y}^T k_y^{-1}\bm{y} - \frac{1}{2}\log{| k_y |} - \frac{n}{2}\log{2\pi}, \label{eq_log_likelihood}
\end{align}
where $\bm{y} = \bm{\mathcal{D}_W}(\bm{\theta})$ are the function values at the training points, $k_y = k(\bm{\theta},\bm{\theta'};\bm{\phi}) + \sigma_n^2 I$ is the covariance matrix for $\bm{y}$ (and $\sigma_n = \hat{\sigma}(\mathcal{D}_W)$ are the estimated uncertainties at the training points), and $n$ is the total number of training points ($\bm{X} = \bm{\theta}$). The choice of training points is described in \S\ref{secTrainingPoints}.

In practice, optimizing a 8--11D function is computationally expensive. Rather than simultaneously optimizing all of the hyperparameters, we set $\sigma_f = 1$ and fixed values for the relative length scales ${\lambda_i}'$ (listed in Table \ref{tab:GP_hparams}) informed by inspection of the distribution of training points. Then, we maximize the log-marginal likelihood by varying an overall length scale factor $\lambda_{\rm global}$, where $\lambda_i = \lambda_{\rm global}{\lambda_i}'$. Note that for the purposes of training the emulator for our clustered models, we also transform the parameters $\lambda_c$ and $\lambda_p$ into a sum and difference of their log-values, $\ln{(\lambda_c \lambda_p)}$ and $\ln{(\frac{\lambda_p}{\lambda_c})}$, as the region of best values for these transformed parameters are more Gaussian than that of the rates of clusters and planets per cluster parameters themselves. 

\begin{table}
\centering
\caption{Table of length scale hyperparameters $\lambda_i$ and bounds for computing credible regions for the parameters of each model. The same sets of hyperparameters and box bounds were used for both the analyses involving the KS and the AD distances. Trial sets of model parameters were drawn uniformly in each bounded interval.}
\begin{threeparttable}
\makebox[\linewidth]{%
\begin{tabular}{l|cc|cc|cc}
 \hline
 \hline
 Parameter & \multicolumn{2}{c|}{Non-clustered} & \multicolumn{2}{c|}{Clustered periods} & \multicolumn{2}{c}{Clustered periods} \\
 & \multicolumn{2}{c|}{} & \multicolumn{2}{c|}{} & \multicolumn{2}{c}{and sizes} \\
 \hline
 & $\lambda_i$ & Bounds & $\lambda_i$ & Bounds & $\lambda_i$ & Bounds \\
 $f_{\sigma_{i,\rm high}}$ & 0.05 & $(0, 0.15)$ & 0.2 & $(0.1, 0.7)$ & 0.2 & $(0.1, 0.7)$ \\
 $\ln{(\lambda_c \lambda_p)}$* & 0.03 & $(0.7, 1.9)$ & 0.6 & $(0, 2)$ & 0.6 & $(0, 2.2)$ \\
 $\ln{(\frac{\lambda_p}{\lambda_c})}$ & - & - & 1 & $(0, 3)$ & 1 & $(-1, 3)$ \\
 $\alpha_P$ & 0.2 & $(-0.5, 0.3)$ & 0.8 & $(-1, 1.5)$ & 1 & $(-1, 2)$ \\
 $\alpha_{R1}$ & 0.5 & $(-2, 0)$ & 1 & $(-2, 1)$ & 0.8 & $(-3, 1)$ \\
 $\alpha_{R2}$ & 1 & $(-6, -4)$ & 1 & $(-6, -3)$ & 1.5 & $(-6, -2)$ \\
 $\sigma_e$ & 0.01 & $(0, 0.03)$ & 0.02 & $(0, 0.04)$ & 0.02 & $(0, 0.05)$ \\
 $\sigma_{i,\rm high}$ ($^\circ$) & 30 & $(0, 90)$ & 30 & $(0, 90)$ & 30 & $(0, 90)$ \\
 $\sigma_{i,\rm low}$ ($^\circ$) & 0.25 & $(0, 0.7)$ & 1 & $(0, 2.5)$ & 1 & $(0, 3)$ \\
 $\sigma_R$ & - & - & - & - & 0.15 & $(0.1, 0.5)$ \\
 $\sigma_P$ & - & - & 0.1 & $(0.1, 0.3)$ & 0.1 & $(0.1, 0.3)$ \\
 \hline
 \hline
\end{tabular}}
\begin{tablenotes}
 \item \textit{Note.} *This is $\ln({\lambda_p})$ for the non-clustered model.
\end{tablenotes}
\end{threeparttable}
\label{tab:GP_hparams}
\end{table}

\subsubsection{Training points}
\label{secTrainingPoints}
For each model, we choose a set of training points $\{\bm{\theta}\}$ for the GP emulator by taking a subset of the model evaluations from the optimization runs as described in \S\ref{Optimization}. Since we ran 50 individual optimizers with 5000 total model evaluations per run, we have a pool of $2.5\times10^5$ model evaluations scattered around the $d$-dimensional parameter space. We rank order these points by $\mathcal{D}_W$ and choose the top $10^5$ points, keeping every $10^{\rm th}$ point for a total of $10^4$ points. The choice of keeping every $10^{\rm th}$ point instead of all points is due to (1) the computational limits in calculating and inverting the kernel matrix $k(\bm{\theta},\bm{\theta'};\bm{\phi})$, which scales as $n^3$ and thus prevents us from using more than a few thousand training points, and (2) the desire to include some points far enough away from the minimum so that the GP emulator makes reasonable predictions for a wider region of parameter space.

The combination of stochastic noise in $\mathcal{D}_W$ due to the Monte Carlo noise of our simulations and keeping the best rank-ordered points would introduce a bias towards smaller than average distances at these points.  In order to avoid this bias, we recompute the values $\mathcal{D}_W$ by simulating a new catalogue with the full SysSim model at each of these points. We train the GP emulator using updated values of $\mathcal{D}_W$ for a random subset of 2000 points of the $10^4$ available points.

\subsubsection{Computing credible regions for model parameters}
\label{CredibleRegions}

The prior mean function, kernel function, and training points fully define a GP model. For each model and distance function ($\mathcal{D}_{W,\rm KS}$ or $\mathcal{D}_{W,\rm AD'}$) combination\footnote{We do not train an emulator or compute credible regions for the non-clustered model using the AD distances, because upon inspection of the observed catalogues generated using the best model parameters resulting from the optimization scheme in Section \S\ref{Optimization}, we find that the overall rate of planets is such a poor fit with this distance function and model that it is not worthy of more detailed investigation.}
we train a GP emulator and use it to predict the distance function at a large number of points $\bm\theta$ in the $d$-dimensional model parameter space. 
In order to improve computational efficiency, we draw samples from a reduced range for each parameter (based on inspecting the results of the optimization stage). 
Effectively, we assume a uniform prior for each model parameter by drawing points uniformly in the $d$-dimensional box, which the bounds for each parameter are specified in Table \ref{tab:GP_hparams}. 

For each draw of model parameters from the prior distribution, we draw an emulated distance from the GP emulator. Finally, we only accept draws if the emulated distance is less than a distance threshold.  We repeat this procedure until we have accumulated $10^4$ draws from the ABC posterior.

We adopt distance thresholds of $\mathcal{D}_{W,\rm KS} = 35$ and $\mathcal{D}_{W,\rm AD'} = 140$ for both the clustered models, and a distance threshold of $\mathcal{D}_{W,\rm KS} = 55$ for the non-clustered model. 
The distance thresholds are less than the medians of the training points, so that the mean and variance of the GP emulators are constrained both in and around the perimeter of the regions of parameter space that are plausibly good fits. 
Given the stochastic nature of our forward model, even a perfect model would typically return a distance of $\sim 9 \times \sqrt{5}\simeq 20$, i.e., the number of component distance functions in our total weighted distance times the square root of the ratio of the number of target stars used when generating weights to the number of target stars used during the inference step). 
The best distances found during the optimization stage ($\sim 20$ for the clustered models and $\sim 45$ for the non-clustered model, when using the KS distance) set the smallest distance thresholds we could have chosen.  
The use of larger distance thresholds, means that credible regions based on our ABC posterior samples will be somewhat larger than the ideal posterior if we had been able to use a smaller distance threshold.  
In practice, the size of the ABC posterior credible intervals appear to be shrinking only slowly as we decrease the distance threshold further.  
Since the credible regions can only shrink with a smaller distance threshold, the credible regions that we report are conservative, i.e., larger than if we would have obtained with infinite computational resources.

\begin{figure*}
\includegraphics[scale=0.28,trim={0 0 0 0},clip]{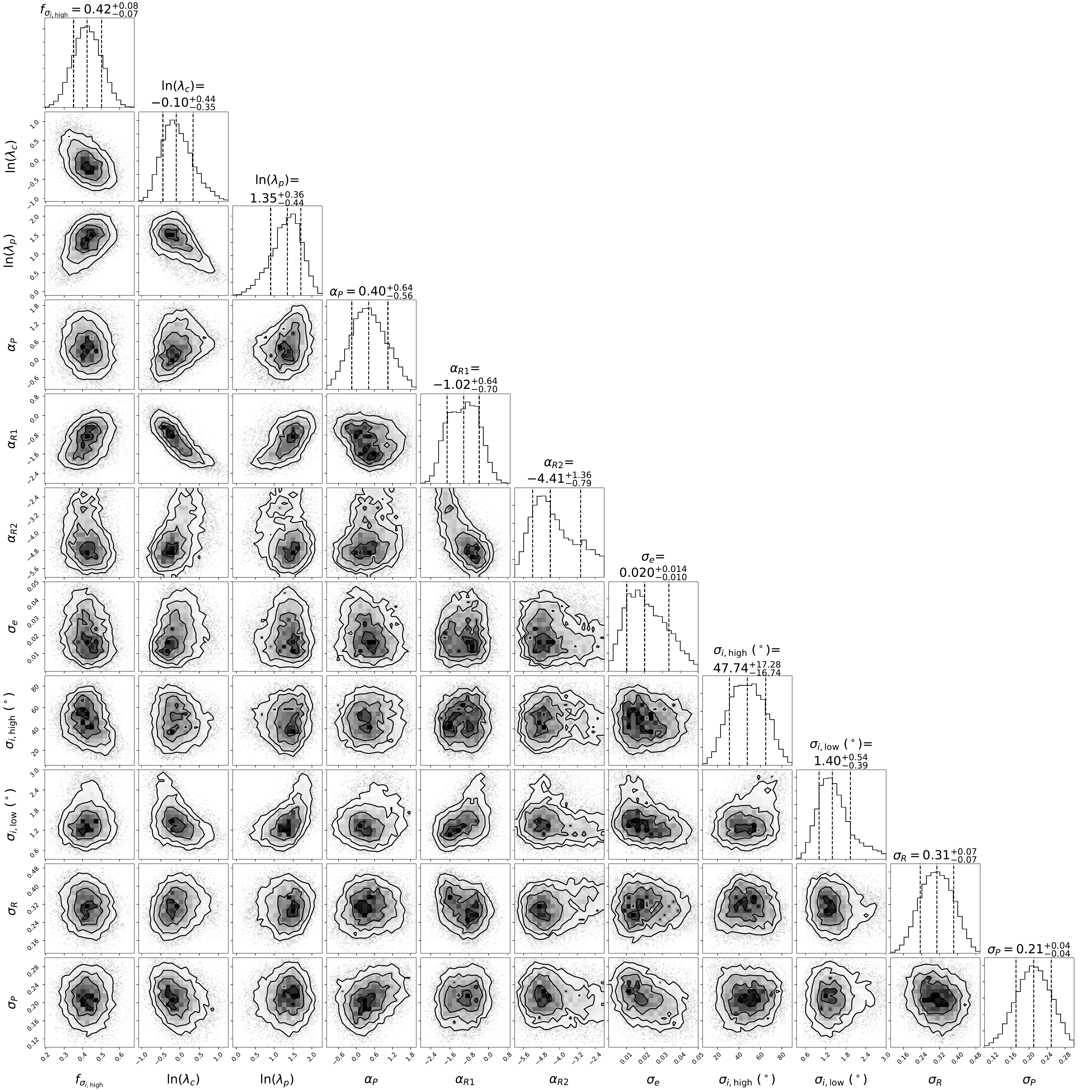}
\caption{Marginal ABC posterior distributions of the model parameters for the clustered periods and sizes model, showing the projections of the points that pass a distance threshold of $\mathcal{D}_{W,\rm KS} = 35$ as drawn from the GP emulator. The prior mean function was set to a constant value of 75.}
\label{fig:clustered_P_R_corner_KS}
\end{figure*}

In Figure \ref{fig:clustered_P_R_corner_KS}, we display a ``corner'' plot (plotted using \texttt{corner.py}; \citealt{Fm2016}) showing the ABC posterior based on a sample of  $10^4$ for the clustered periods and sizes model with the KS distance function. Similar figures for the non-clustered, clustered periods, and alternative MMR inclinations models are available in supplementary online material.   
Similarly, supplemental Figure \ref{fig:clustered_P_R_corner_AD} shows the analogous plot using the AD distance function.
We interpret and discuss the meaning of these results in detail in the next section.

\section{Results} \label{Model_params}

\begin{table*}
\centering
\caption{Table of the best-fitting values for the free parameters of each model, with $1\sigma$ credible intervals. The last two columns (under the label ``Alt. MMR inclinations'') show the results for the clustered periods and sizes model without the lowering of mutual inclinations for near-MMR planets. The rows with $\lambda_c$ and $\lambda_p$ are equivalent to those with $\ln({\lambda_c})$ and $\ln({\lambda_p})$, respectively, and are shown for interpretability. We set $R_{p,\rm break} = 3 R_\oplus$ and $\Delta_c = 8$.}
\begin{tabular}{ l | cc | ccc | ccc | cc }
 \hline
 \hline
 Parameter & \multicolumn{2}{c|}{Non-clustered} & \multicolumn{3}{c|}{Clustered periods} & \multicolumn{3}{c}{Clustered periods and sizes} & \multicolumn{2}{c}{Alt. MMR inclinations} \\
 \hline
 & Fig. \ref{fig:non_clustered_model}, \ref{fig:models_underlying} & Best-fit KS & Fig. \ref{fig:clustered_P_model}, \ref{fig:models_underlying} & Best-fit KS & Best-fit AD & Fig. \ref{fig:clustered_P_R_model}, \ref{fig:models_underlying} & Best-fit KS & Best-fit AD & Best-fit KS & Best-fit AD \\[5pt]
 $f_{\sigma_{i,\rm high}}$ & 0.03 & $0.03_{-0.02}^{+0.02}$ & 0.4 & $0.42_{-0.10}^{+0.10}$ & $0.41_{-0.13}^{+0.11}$ & 0.4 & $0.42_{-0.07}^{+0.08}$ & $0.40_{-0.12}^{+0.11}$ & $0.32_{-0.06}^{+0.06}$ & $0.31_{-0.10}^{+0.10}$ \\[5pt]
 $\ln{(\lambda_c)}$ & - & - & $-0.51$ & $-0.39_{-0.31}^{+0.32}$ & $-0.28_{-0.27}^{+0.32}$ & $-0.22$ & $-0.10_{-0.35}^{+0.44}$ & $0.24_{-0.40}^{+0.33}$ & $0.04_{-0.45}^{+0.46}$ & $0.35_{-0.41}^{+0.32}$ \\[5pt]
 $\lambda_c$ & - & - & $0.6$ & $0.68_{-0.18}^{+0.25}$ & $0.76_{-0.18}^{+0.28}$ & $0.8$ & $0.90_{-0.26}^{+0.50}$ & $1.27_{-0.42}^{+0.50}$ & $1.04_{-0.38}^{+0.61}$ & $1.42_{-0.48}^{+0.54}$ \\[5pt]
 $\ln{(\lambda_p)}$ & $1.25$ & $1.26_{-0.15}^{+0.16}$ & $1.44 $ & $1.50_{-0.50}^{+0.30}$ & $1.27_{-0.50}^{+0.38}$ & $1.31$ & $1.35_{-0.44}^{+0.36}$ & $0.73_{-0.56}^{+0.60}$ & $1.05_{-0.34}^{+0.39}$ & $0.56_{-0.43}^{+0.50}$ \\[5pt]
 $\lambda_p$ & $3.5$ & $3.53_{-0.50}^{+0.61}$ & $4.2$ & $4.48_{-1.76}^{+1.57}$ & $3.56_{-1.40}^{+1.65}$ & $3.7$ & $3.86_{-1.38}^{+1.67}$ & $2.08_{-0.89}^{+1.70}$ & $2.86_{-0.82}^{+1.36}$ & $1.75_{-0.61}^{+1.14}$ \\[5pt]
 $\alpha_P$ & $-0.1$ & $-0.13_{-0.07}^{+0.10}$ & 0.1 & $0.13_{-0.48}^{+0.73}$ & $0.01_{-0.29}^{+0.71}$ & 0.4 & $0.40_{-0.56}^{+0.64}$ & $0.07_{-0.45}^{+0.66}$ & $0.19_{-0.57}^{+0.69}$ & $-0.05_{-0.33}^{+0.37}$ \\[5pt]
 $\alpha_{R1}$ & $-1.0$ & $-1.08_{-0.20}^{+0.31}$ & $-0.8$ & $-0.75_{-0.46}^{+0.46}$ & $-0.85_{-0.26}^{+0.26}$ & $-1.0$ & $-1.02_{-0.70}^{+0.64}$ & $-1.27_{-0.25}^{+0.26}$ & $-1.28_{-0.57}^{+0.72}$ & $-1.32_{-0.29}^{+0.30}$ \\[5pt]
 $\alpha_{R2}$ & $-5.3$ & $-5.30_{-0.40}^{+0.53}$ & $-4.9$ & $-4.91_{-0.62}^{+0.96}$ & $-5.49_{-0.33}^{+0.42}$ & $-4.4$ & $-4.41_{-0.79}^{+1.36}$ & $-5.08_{-0.54}^{+0.71}$ & $-4.50_{-0.70}^{+1.03}$ & $-4.81_{-0.71}^{+0.79}$ \\[5pt]
 $\sigma_e$ & 0.003 & $0.003_{-0.002}^{+0.003}$ & 0.01 & $0.014_{-0.007}^{+0.009}$ & $0.010_{-0.006}^{+0.008}$ & 0.02 & $0.020_{-0.010}^{+0.014}$ & $0.014_{-0.008}^{+0.010}$ & $0.019_{-0.009}^{+0.011}$ & $0.014_{-0.008}^{+0.010}$ \\[5pt]
 $\sigma_{i,\rm high}$ ($^\circ$) & 60 & $62_{-50}^{+16}$ & 50 & $49_{-21}^{+22}$ & $49_{-25}^{+25}$ & 50 & $48_{-17}^{+17}$ & $49_{-25}^{+23}$ & $49_{-18}^{+19}$ & $42_{-20}^{+26}$ \\[5pt]
 $\sigma_{i,\rm low}$ ($^\circ$) & 0.3 & $0.31_{-0.10}^{+0.10}$ & 1.1 & $1.13_{-0.33}^{+0.38}$ & $1.16_{-0.26}^{+0.31}$ & 1.4 & $1.40_{-0.39}^{+0.54}$ & $1.29_{-0.32}^{+0.35}$ & $1.38_{-0.38}^{+0.43}$ & $1.29_{-0.31}^{+0.34}$ \\[5pt]
 $\sigma_R$ & - & - & - & - & - & 0.3 & $0.31_{-0.07}^{+0.07}$ & $0.32_{-0.07}^{+0.07}$ & $0.30_{-0.07}^{+0.07}$ & $0.30_{-0.06}^{+0.06}$ \\[5pt]
 $\sigma_P$ & - & - & 0.2 & $0.20_{-0.04}^{+0.03}$ & $0.17_{-0.03}^{+0.04}$ & 0.2 & $0.21_{-0.04}^{+0.04}$ & $0.20_{-0.04}^{+0.04}$ & $0.21_{-0.04}^{+0.04}$ & $0.18_{-0.03}^{+0.04}$ \\
 \hline
 \hline
\end{tabular}
\label{tab:param_fits}
\end{table*}

In Table \ref{tab:param_fits}, we report the best-fitting values of the free model parameters for each of the models.
We discuss the results and the effect of each of the free model parameters on the simulated observed population in this section.

\subsection{Comparison of clustered and non-clustered models} \label{secCompClusteredNonClustered}

First, we briefly summarize how our models compare with each other, before reporting the results for each parameter of each model in detail. 
Our clustered models are clearly preferred over the non-clustered model, as evidenced by the significantly smaller best-fitting weighted distances as described in \S\ref{GP}.  
The improvement in the clustered models can be traced to improvements in the component distances for the multiplicities, period ratios, transit duration ratios for planet pairs not near resonance, and to a lesser extent periods and transit durations.  
Going from clustered periods to clustered periods and radii results in some component distance improving (based on period ratios and radius ratios), but the change in total distance is more modest.

All three models are able to match the overall observed rate of planets (via our analysis involving KS distances; the analysis involving AD distances fails for the non-clustered model). 
The period and radius (power/broken-power law) distributions are broadly consistent across all the models and distances, suggesting a shallowly increasing number of planets in log-period, a roughly flat distribution of planet sizes below the break radius, and a sharply decreasing occurrence of planets above the break radius. 
Both clustered models prefer: (1) a sizable fraction ($\sim 40\%$) of highly mutually inclined systems and (2) the remaining majority of systems with small mutual inclinations of $\sim 1^\circ$.  
These findings are consistent across analyses using either the KS and AD distances, corroborating previous studies (e.g., \citealt{Mu2018}). 
All three models imply that most planets have small orbital eccentricities. 
Finally, the clustered models suggest that both periods and planet sizes are highly clustered, with consistent results for the cluster widths in log-period and radius using both KS and AD distances.

Our results are also relatively unchanged when considering the alternative MMR inclinations model, i.e. our fully clustered model without the reduction of mutual inclinations for the near-MMR planets. All the parameters with the exception of $f_{\sigma_{i,\rm high}}$ remain consistent within statistical uncertainties. While $f_{\sigma_{i,\rm high}}$ is somewhat reduced ($\sim 0.32$), the true fraction of highly mutually inclined planets is similar, since in our standard clustered models the $f_{\sigma_{i,\rm high}} \sim 40\%$ of highly mutually inclined systems includes planets near an MMR that have low mutual inclinations. Overall, the rate of planets per star, the period and radius distributions, the eccentricity and mutual inclination scales, and the period and radius cluster widths all remain the same.

\subsection{Rates of clusters per system and planets per cluster \\ ($\lambda_c$, $\lambda_p$)} \label{Params_rates}

In our two clustered models, the parameters $\lambda_c$ and $\lambda_p$ parameterize the mean numbers of clusters and planets per cluster, respectively, before rejection-sampling and any truncation. The true mean rate of clusters in our simulations is somewhat less than $\lambda_c$, while the true mean rate of planets per cluster is typically greater than $\lambda_p$ for small values, and always greater than one, due to the draws from the zero-truncated Poisson distribution. These parameters largely control the overall rates of planets and how they are distributed within and between clusters.

It is clear that with increasing $\lambda_c$ and $\lambda_p$, the overall number of observed planets increases, due to there being intrinsically more planets around each star. This effect is perhaps slightly more sensitive to $\lambda_c$, likely because increasing $\lambda_p$ is more likely to result in rejected systems which slows the increase in the number of planets with increasing $\lambda_p$ (since it is harder to fit more planets into the same cluster than to increase the number of clusters, due to stability). The product of these two parameters is most directly constrained by the total rate of planets per star (i.e. the distance term $D_f$ in equation \ref{eq_dist}).

In our clustered periods and sizes model, we find that $\lambda_c \simeq 0.90_{-0.26}^{+0.50}$ indicating that $\sim 80\%$ of systems with planets have one cluster. The number of planets per cluster peaks at four with $\lambda_p \simeq 3.86_{-1.38}^{+1.67}$. The combination of these gives a typical $\lambda_c \lambda_p \sim 3.5$ planets per star. As seen in Table \ref{tab:param_fits}, using the AD distances leads to more clusters per star (1.27) and fewer planets per cluster (2.08), though the 1$\sigma$ credible regions overlap. The overall rate of planets per star ($\sim 2.6$) is similar. 

For the clustered periods model, we find similar values for the rates of clusters and planets per cluster using both KS and AD distances: $\lambda_c \simeq 0.68_{-0.18}^{+0.25}$ and $\lambda_p \simeq 4.48_{-1.76}^{+1.57}$ using KS distances. Again, while the KS analysis results in slightly more planets per cluster and fewer clusters than the AD analysis, the difference is not significant given the credible regions. These results are very similar to those of the clustered periods and sizes model using KS distances.

For our non-clustered model, the number of planets per system $N_p$ is described by a single parameter, $N_p \sim {\rm Poisson}(\lambda_p)$. We find that using KS distances, $\lambda_p \simeq 3.53_{-0.50}^{+0.61}$, implying a median value of $\sim 3.5$ for the mean number of planets per system which is comparable to the value in our best-fitting clustered models. As discussed before, we do not recover meaningful results for this model using AD distances.

Returning to our fully clustered model, the values of $\lambda_c$ and $\lambda_p$ are somewhat shifted in our alternative MMR inclinations model. There are slightly more clusters per system and fewer planets within each cluster, using both KS and AD distances, although these differences are within the statistical uncertainties. The overall rate of planets per system effectively remains the same.

We conclude that the \Kepler{} observations robustly constrain the mean number of planets per system and the results are insensitive to the choice of forward model or distance function.
While there is more variance across models and distance functions for the inferred ratio of planets per cluster to clusters per system, the differences are less than the statistical uncertainties for three of the four cases considered (i.e., clustered periods and fully clustered models, each with KS and AD distances).

\subsection{Period distribution ($\alpha_P$)} \label{Params_periods}

For all our models, the overall period distribution is described by a single power law between 3 and 300 d with index $\alpha_P$. We allowed $\alpha_P$ to vary between $-2$ and 2 during the optimization process (note that $\alpha_P = -1$ corresponds to a flat distribution in log-period).
All three of our models result in an increasing occurrence of planets with log-period, strongly disfavouring a flat distribution ($\alpha_P = -1$).
The estimates of $\alpha_P$ are consistent within statistical uncertainties across all of our clustered models.  
However, assuming a non-clustered model results in both a shallower slope and significantly smaller uncertainties on $\alpha_P$.  

The clustered periods and sizes model results in $\alpha_P \simeq 0.40_{-0.56}^{+0.64}$ using KS distances, and favours a shallower slope when using AD distances, $\alpha_P \simeq 0.07_{-0.45}^{+0.66}$. 
The results for the clustered periods models are very similar for the KS and AD analyses, giving $\alpha_P \simeq 0.13_{-0.48}^{+0.73}$ and $\alpha_P \simeq 0.01_{-0.29}^{+0.71}$, respectively. The constraint on $\alpha_P$ appears to be much tighter for the non-clustered model, with $\alpha_P \simeq -0.13_{-0.07}^{+0.10}$. 
We conclude that one should be cautious about overinterpreting measurements of $\alpha_P$ that have assumed orbital periods are drawn independently for planets within the same system. 

\subsection{Radius distribution ($\alpha_{R1}$, $\alpha_{R2}$)} \label{Params_radii}
 
We allowed $\alpha_{R1}$ and $\alpha_{R2}$, the power law indices for the broken power-law as given in Equation (\ref{eq_brokenR}), to vary from $-4$ to 2 and $-6$ to 0, respectively, for all our models (Table \ref{tab:param_ranges}). These two parameters along with $R_{p,\rm break}$ control the overall distribution of planetary radii; we fix $R_{p,\rm break} = 3 R_\oplus$ here, so $\alpha_{R1}$ controls the distribution of Earth-sized planets between 0.5 and 3$R_\oplus$ and $\alpha_{R2}$ controls the distribution of larger planets between 3 and 10$R_\oplus$. The continuity of $p(R_p)$ induces a correlation between $\alpha_{R1}$ and $\alpha_{R2}$.

In our distance function, the most direct constraint on the radius distribution comes from fitting the transit depth and transit depth ratio distributions. The sizes of planets also affect the observed period ratio distribution, since small planets can be packed closer to each other but may be harder to detect. Formally, the radius distribution must be inferred simultaneously with the period distribution and other model parameters \citep{Y2011}. Nevertheless, it appears that one could infer the overall radius distribution reasonably well even without simultaneously modelling the planetary architectures.

There is a clear need for a broken power law for the radius distribution, given the joint posterior for $\alpha_{R1}$ and $\alpha_{R2}$ strongly excludes equal values. Moreover, the radius distribution falls much more quickly above $R_{p,\rm break}$ than below it; $\alpha_{R2} \simeq -4.41_{-0.79}^{+1.36}$ while $\alpha_{R1} \simeq -1.02_{-0.70}^{+0.64}$ for our clustered periods and sizes model with KS distances (and slightly steeper values using AD distances). Similarly, the clustered periods and non-clustered models also prefer $\alpha_{R2} \lesssim -5$ and $\alpha_{R1} \simeq -1$. This is expected given that the \Kepler{} population does not have a high occurrence of larger, Jupiter-sized planets. 
The differences in the constraints on $\alpha_{R1}$ and $\alpha_{R2}$ across models and distances functions are smaller than the statistical uncertainties.  
The 68.3\% credible interval for $\alpha_{R1}$ includes a flat distribution of $\alpha_{R1} = -1$ for almost all our models and distance functions, consistent with previous results \citep[e.g.,][]{H2012, PMH2013b, Mu2018}.
While each of our analyses yields similar constraints on $\alpha_{R1}$, we caution that our radius model cannot include a local minima, i.e., the radius valley in the underlying distribution \citep{F2017,vE2017,H2019}.  Therefore, we leave the generalization of our model to allow for a more flexible radius distribution for future studies.

\subsection{The eccentricity distribution ($\sigma_e$)} \label{Params_ecc}

The orbital eccentricity distribution primarily affects the transit duration and duration ratio distributions, with a slight influence on the period ratio distribution due to the stability criteria. In all our models, the eccentricities are very low; for the clustered models, both KS and AD analyses result in $\sigma_e \simeq 0.01$--0.02 for the Rayleigh distribution.

The eccentricity scale preferred by our non-clustered model is even smaller, centred around $\sigma_e \sim 0.003$. We speculate that this may be caused by the poor fit to the period ratio distribution (see \S\ref{Non_clustered_model}): this model is unable to produce enough observed planets with small period ratios, and thus desires near circular orbits in order to space planets as close as possible while remaining stable. As such, the fits to the transit duration and $\xi$ distributions are also worse in the non-clustered model (see supplemental Figure \ref{fig:distances}).

Our results are consistent with those of several previous studies based on properties of systems with multiple transiting planets.
\citet{F2014} used an early sample of 899 planets in multitransiting \Kepler{} systems and fit to the $\xi$ distribution to constrain both eccentricity and mutual inclination dispersions.  
They found a best fit of $\sigma_e \simeq 0.032$ (also assuming a Rayleigh distribution of $e$), but caution that a wide range of $\sigma_e$ were plausible since the $\xi$ distribution is not as sensitive to $e$ as it is to mutual inclinations. 
\citet{WL2013}, who found $e \sim 0.01$, and \citet{HL2014}, who reported an rms of $\sigma_e \simeq 0.018_{-0.004}^{+0.005}$, both performed analyses of TTVs to arrive at these results. \citet{vEA2015} assumed a Rayleigh distribution and arrived at a slightly larger value of $\sigma_e = 0.049 \pm 0.013$. More recently, \citet{W2019} examined the dynamical stability of multiplanet systems and the period ratio distribution, and found an upper limit of $\sigma_e = 0.03$ (also using the Rayleigh distribution).

Previous studies that included systems with only a single detected transiting planet have led to more varied conclusions. For example, \citet{M2011} assumed a Rayleigh distribution of eccentricities and performed an analysis comparing normalized transit durations (durations assuming eccentric orbits divided by durations for circular orbits) with a small subset (104 planets) of the initial \Kepler{} discoveries.  They found a larger mean eccentricity in the range of 0.1--0.25 for their population, but cautioned that the measurement uncertainty and potential systematic effects in the host stellar densities could bias the derived eccentricities.  This concern has now been mitigated thanks to improved stellar properties and the ability to select a cleaner sample of host stars thanks to Gaia DR2.  
\citet{SDC2015} investigated the eccentricity distribution of a small sample of mostly single, large, short-period planets using both transit and occultation ($h = e\cos{\omega}$ and $k = e\sin{\omega}$) measurements.  When they assumed a one-component Gaussian distribution for $h$ and $k$ (equivalent to a Rayleigh distribution for $e$), they found a value of $\sigma_e = 0.081_{-0.003}^{+0.014}$.  However, they also show that adopting a two-component Gaussian (very similar to a mixture of two Rayleigh distributions) is favoured.  That model results in about 90\% of planets with $\sigma_e = 0.01_{-0.002}^{+0.014}$ and 10\% with $\sigma_e = 0.22_{-0.026}^{+0.1}$.

Some studies have found evidence suggesting that the eccentricity distribution may differ between systems with a single transiting planet and multiple transiting planets.  
\citet{M2011} found a difference in the transit duration distribution (normalized by the duration estimated for a transit over the diameter of the star with a circular orbit of the same period) that was statistically significant, but cautioned about the potential impact of uncertainties in the stellar parameters.  
\citet{X2016} also found a difference in the eccentricity distribution for \Kepler{} single transiting planets ($\bar{e} \approx 0.3$) and systems with multiple detected transiting planets ($\bar{e} = 0.04_{-0.04}^{+0.03}$), based on a larger catalogue and stellar properties updated by LAMOST. 
A similar finding is reported by \citet{vE2019}, who find $\sigma_e = 0.32 \pm 0.06$ for singles and $\sigma_e = 0.083_{-0.020}^{+0.015}$ for multis, using a smaller sample of planets but with much more precisely measured stellar properties, thanks to asteroseismology.
\citet{M2019} updated stellar properties based on \textit{Gaia} and spectra from the California--Kepler Survey for a large sample of planets.  They found that singles have higher mean eccentricities, with $\bar{e} = 0.167_{-0.008}^{+0.013}$ for singles and $\bar{e} = 0.036 \pm 0.012$ for multis.  
\citet{M2019} proceeded to fit a mixture model for the single transiting planet systems,
that included $\simeq 69\%$ of singles being drawn from a low-eccentricity population with $\sigma_{e,\mathrm{low}}<0.05$ and the remainder coming from a high-eccentricity population with $\sigma_{e,\mathrm{high}}>0.3$.  
Formally, our low-inclination population contributes a small number of systems with a single transiting planet detected.  However, we find that this fraction is so small that the transit duration distribution of singles will remain a robust probe of the properties of the eccentricity distribution of the highly excited population.  
Several of these studies also find correlations of eccentricities with planet radii (e.g., \citealt{HL2014, SDC2015,DLC2016}) and stellar metallicity (e.g., \citealt{SDC2015, M2019}).

In this study, we adopted a single Rayleigh distribution of $e$ for simplicity and to avoid increasing the number of model parameters.  Thus, at some level, our results are likely averaging over a small population with larger eccentricities and/or any correlations between eccentricity and planetary or stellar properties.  However, we expect this to have a relatively small effect on our results for the eccentricity distribution, since the $\xi$ distribution is only observable for systems with multiple transiting planets.  While the transit duration distribution is affected by planets regardless of multiplicity, this provides a weaker constraint than the $\xi$ distribution as almost half of the known transiting planets are in systems with multiple transiting planets.  We encourage future work to explore the impact of allowing the high-inclination population to have a broader distribution of eccentricities and potentially correlations with stellar properties.

\subsection{The mutual inclination distribution ($f_{\sigma_{i,\rm high}}$, $\sigma_{i,\rm high}$, $\sigma_{i,\rm low}$)} \label{Params_incl}

The mutual inclination angles between planets have pronounced effects on the multiplicity distribution and the period-normalized transit duration ratio, as discussed in \S\ref{Summary_stats}. As listed in Table \ref{tab:param_ranges}, we allow the fraction of systems with broad mutual inclinations $f_{\sigma_{i,\rm high}}$ to vary in the entire range between 0 and 1, and the mutual inclination scales $\sigma_{i,\rm high}$ and $\sigma_{i,\rm low}$ for the Rayleigh distributions to vary between $0^\circ$ (coplanar orbits) and $90^\circ$ (effectively isotropic), with the constraint that $\sigma_{i,\rm low} \leq \sigma_{i,\rm high}$. As discussed in \S\ref{Incl}, $\sigma_{i,\rm low}$ also governs the mutual inclinations for planets near MMRs.

For both of our clustered models, the fraction of systems with $\sigma_{i,\rm high}$ is close to 40\%; $f_{\sigma_{i,\rm high}} \simeq 0.42_{-0.07}^{+0.08}$ ($0.40_{-0.12}^{+0.11}$) using KS (AD) distances for the clustered periods and sizes model. Results with the clustered periods model are very similar. The broad mutual inclination scale $\sigma_{i,\rm high}$ is clearly greater than $\sim 10^\circ$, but poorly unconstrained at significantly larger inclinations, since these systems contribute almost solely to the number of observed single-transiting systems.

Our result of $f_{\sigma_{i,\rm high}} \sim 0.4$ is very similar to the value of $f_{\rm iso} \simeq 0.38$ reported in \citet{Mu2018} and consistent with previous studies on the \Kepler{} dichotomy (e.g. \citealt{BJ2016} for M dwarfs). 
For the rest of the systems, the mutual inclinations between planets are only a few degrees, $\sigma_{i,\rm low} \simeq 1.40_{-0.39}^{+0.54}$ ($1.29_{-0.32}^{+0.35}$) degrees using KS (AD) distances.
Additionally, $\simeq 30\%$ of the planets from the $\simeq 40\%$ of systems labelled high mutual inclination, were actually assigned a low inclination to the system's reference plane, due to being near a first-order MMR with an adjacent planet.
While $\sigma_{i,\rm low}$ is small, strictly coplanar orbits are strongly prohibited primarily due to the $\log{\xi}$ distribution.

We note that in the clustered periods and sizes model, if we do not include the lowering of mutual inclinations for planets near an MMR, $f_{\sigma_{i,\rm high}}$ decreases to $\sim 0.32$. This is expected because more systems need to host multiple planets with relatively small mutual inclinations in order to make up for the lost contribution to the multitransiting (2+ observed planets) systems from the near-MMR planets. Thus, this value is comparable to the true fraction of systems with highly mutually inclined planets in our usual clustered models. Notably, the values of $\sigma_{i,\rm high}$ and $\sigma_{i,\rm low}$ do not change in this model, highlighting their robustness to our treatment of the near-MMR planets.

The non-clustered model is unable to adequately match the observed multiplicity distribution as shown in Figure \ref{fig:non_clustered_model} and discussed later in \S\ref{Non_clustered_model}.  
This leads the non-clustered model to favour very small values of $f_{\sigma_{i,\rm high}}$ and $\sigma_{i,\rm low}$.  Since only a few percent of systems 
have high inclinations, the value of $\sigma_{i,\rm high}$ is only weakly constrained.  This demonstrates the importance of adopting a clustered model, not just for inferring the distribution of orbital periods, but also for inference about the distribution of mutual inclinations. One reason for this is that the correlation between the true multiplicity and inclination distribution noticed by earlier authors \citep[e.g.,][]{Li2011b,TD2012} is partially broken when including the period ratio distribution, which sets the typical spacing of planets and affects the observed multiplicity in a non-clustered model.

Our results for the mutual inclination distribution are in strong agreement with many previous studies. The first study on the mutual inclinations of the \Kepler{} exoplanets was done by \citet{Li2011b}, who also simulated planetary systems in order to match the observed multiplicities of the first few months of \Kepler{} data. They found an excess of single transiting systems, and by fitting the multis alone with a mixture of 3- and 4-planet systems found a best fit of $\sigma_i = 2^\circ$ using a Rayleigh distribution of mutual inclinations. \citet{TD2012} combined both \Kepler{} and RV survey results to estimate mean inclinations in the range 0--$5^\circ$; a similar result of $\leq 5^\circ$ was reported by \citet{J2012}, who tried to match the number of single, double, and triple transiting systems by assuming all systems are intrinsic triples. \citet{F2012} also used \Kepler{} and RV (HARPS) data to constrain $\sigma_i \leq 1^\circ$ assuming a Rayleigh distribution, although they included only planets larger than $2 R_\oplus$ and noted that larger inclinations of $\sim 5^\circ$ are possible if the mass--radius relationship is more extreme. \citet{FM2012} explored combinations of bounded-uniform and zero-truncated Poisson distributions of multiplicities, with Rayleigh and Rayleigh of Rayleigh distributions of mutual inclinations, in order to fit the observed multiplicity and $\xi$ distributions. They found the best fits with the bounded-uniform and either Rayleigh or Rayleigh of Rayleigh distributions with $\sigma = 1^\circ$ (or $\sigma_\sigma = 1^\circ$). \citet{F2014} also used the observed multiplicity and $\xi$ distributions to constrain mutual inclinations in the range 1--$2.2^\circ$. A similar range of 0.3--$2.2^\circ$ (for mean mutual inclinations) was reported by \citet{X2016}. Thus, our results from the clustered models corroborate these previous studies and also serve as the tightest constraints on the mutual inclination distribution to date, with consistent results via two independent analyses involving KS and AD distances.

\citet{Z2018} took a different approach to modelling the mutual inclination distribution compared to these previous works, by parametrizing the mutual inclination dispersion as a function of planet multiplicity $k$, with $\sigma_{i,k} \propto k^\alpha$. By matching to both the observed transiting multiplicity and TTV-inferred multiplicity distributions, they found a steep inverse relation, $-4 \lesssim \alpha < -2$, and a normalization of $\sigma_{i,5} = 0.8^\circ$. The results of our clustered models, considering only the low mutual inclination population, are generally consistent with their mutual inclination dispersion for high-multiplicity systems.

\subsection{The period and radius clustering of planets ($\sigma_P$, $\sigma_R$)} \label{Params_clustering}

The parameters $\sigma_P$ and $\sigma_R$ refer to the scale factors for the cluster period and radius distributions, respectively. While $\sigma_P$ is effectively the width of the cluster in log-period, per planet in the cluster, $\sigma_R$ is the fixed width of the cluster in log-radius regardless of the number of planets (recall Equations \ref{eq_P_clusP} and \ref{eq_R_clusR}). These parameters are largely constrained by the period ratio and transit depth (i.e. radius) ratio distributions, respectively.

We find that $\sigma_P \simeq 0.20_{-0.04}^{+0.03}$ ($0.17_{-0.03}^{+0.04}$) using KS (AD) distances for our model with just clustered periods and very similar values for the model that includes clustering in radii, where $\sigma_R \simeq 0.31_{-0.07}^{+0.07}$ using KS ($0.32_{-0.07}^{+0.07}$ using AD) distances. We find that extremely small values of $\sigma_P \lesssim 0.05$ are highly excluded, likely due to the effects of our stability criteria which causes some clusters to be discarded after many attempts to fit planets in and thus significantly decreasing the overall rate of observed planets.

These values suggest that the periods and radii of planets are highly clustered, as smaller values imply a greater degree of clustering (i.e. smaller cluster widths). For example, if we assume $\sigma_P \simeq 0.2$ and consider a four-planet cluster centred around $P_c = 20$ d, this implies that $\simeq 68\%$ of draws ($\sim 3$ planets) to populate this cluster will have periods between 9 and 45 d. Similarly, $\simeq 68\%$ of planets drawn for a cluster of super-Earth sized planets around $2 R_\oplus$ will have radii between $1.48$ and $2.7 R_\oplus$, assuming $\sigma_R \simeq 0.3$.

\section{Discussion} \label{Discussion}

\subsection{Comparison of models for reproducing the observed \Kepler{} data} \label{Model_comparison_observed}

\begin{figure*}
\begin{tabular}{cc}
 \includegraphics[scale=0.425,trim={0 0.4cm 0 0.2cm},clip]{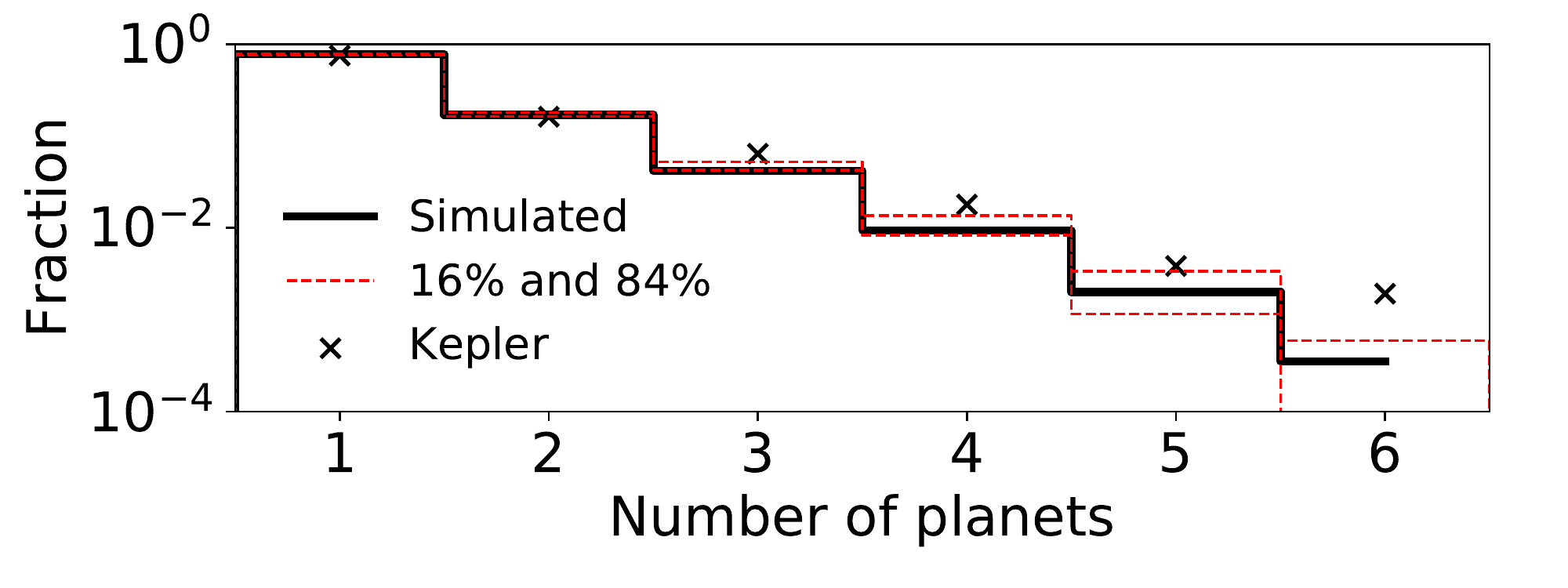} & 
 \includegraphics[scale=0.425,trim={0 0.4cm 0 0.2cm},clip]{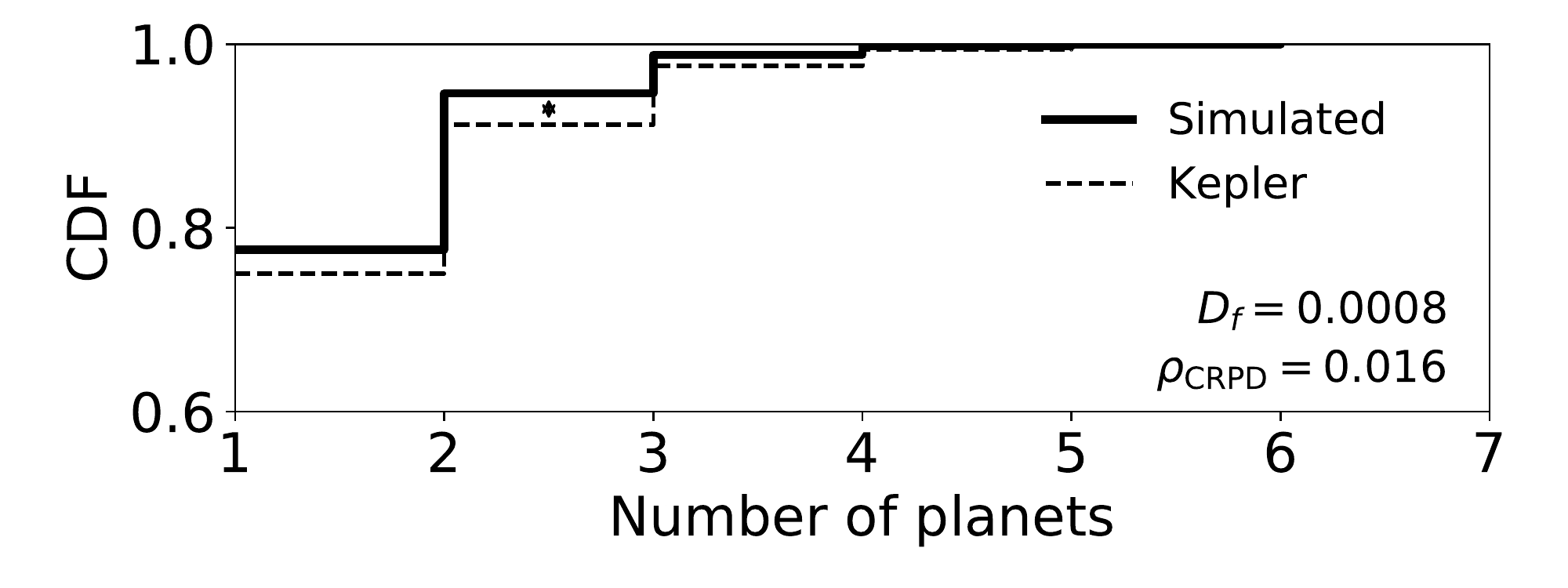} \\
 \includegraphics[scale=0.425,trim={0 0.4cm 0 0.2cm},clip]{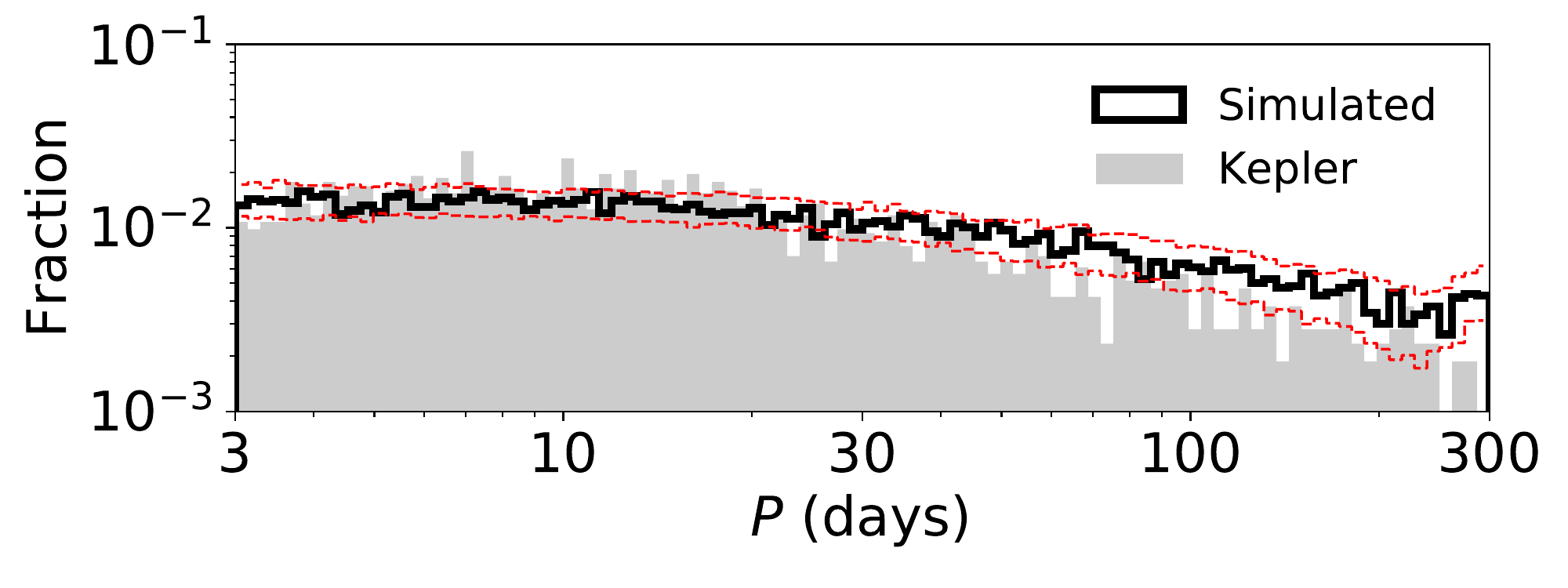} & 
 \includegraphics[scale=0.425,trim={0 0.4cm 0 0.2cm},clip]{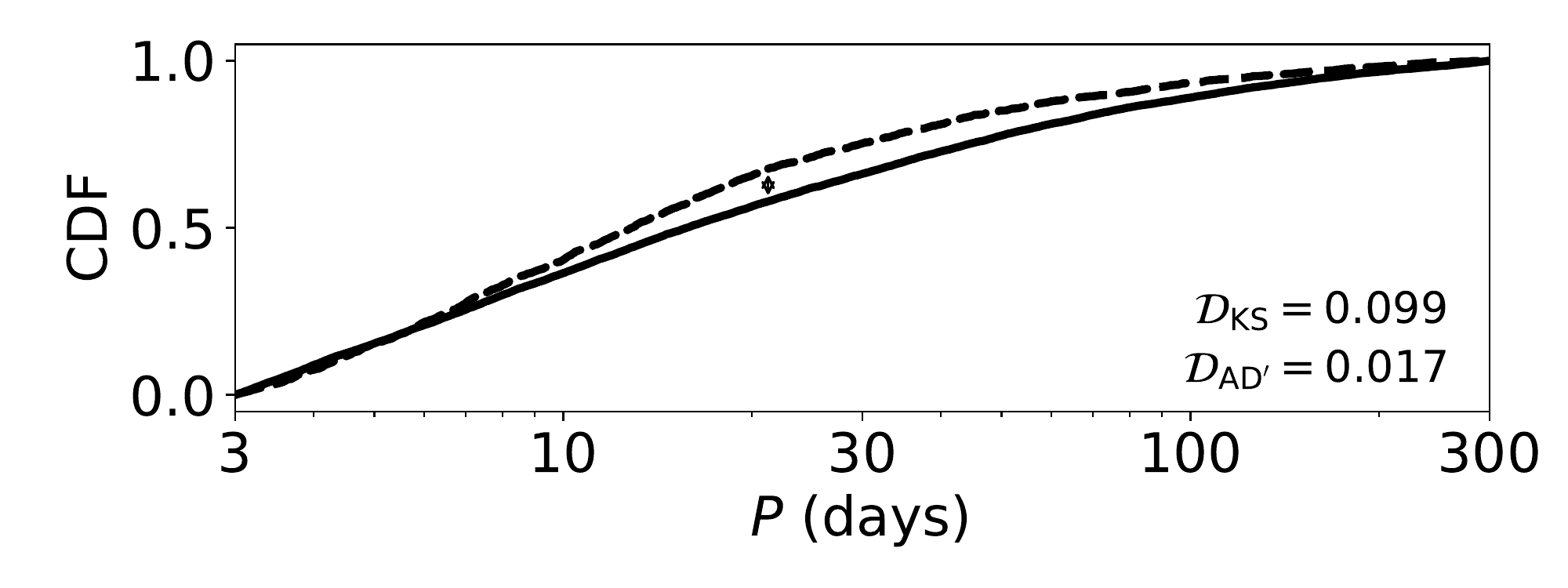} \\
 \includegraphics[scale=0.425,trim={0 0.4cm 0 0.2cm},clip]{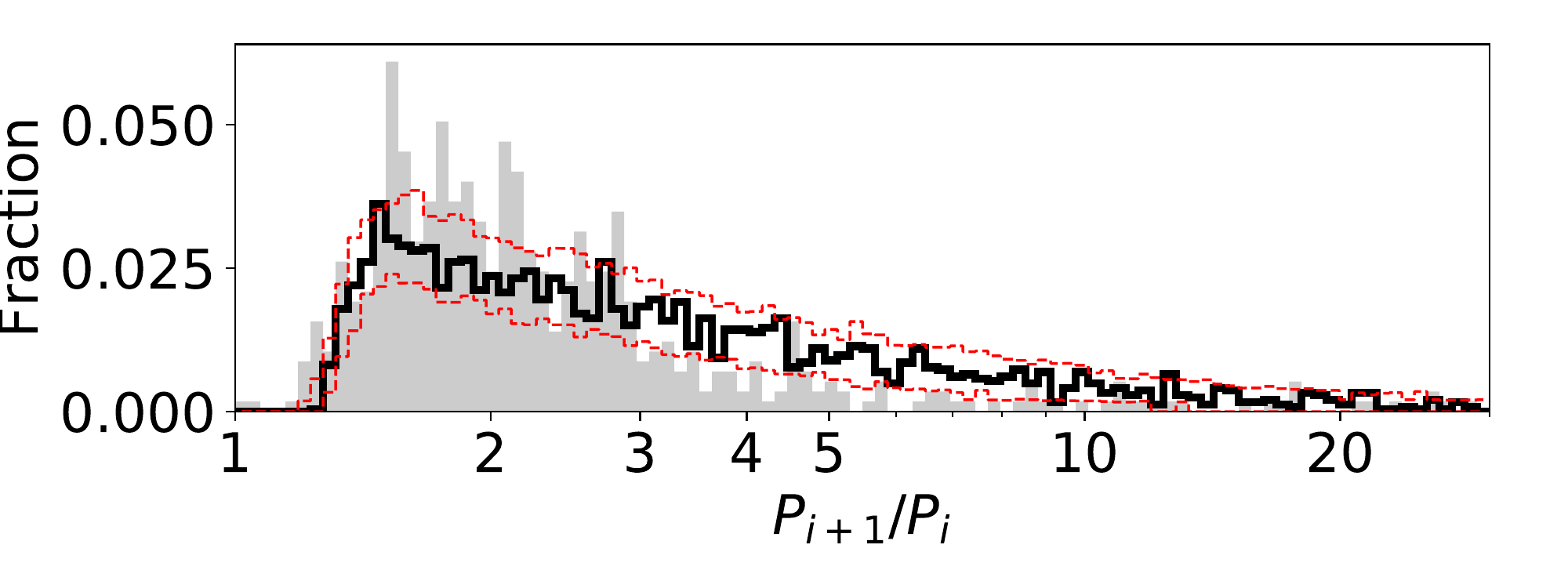} & 
 \includegraphics[scale=0.425,trim={0 0.4cm 0 0.2cm},clip]{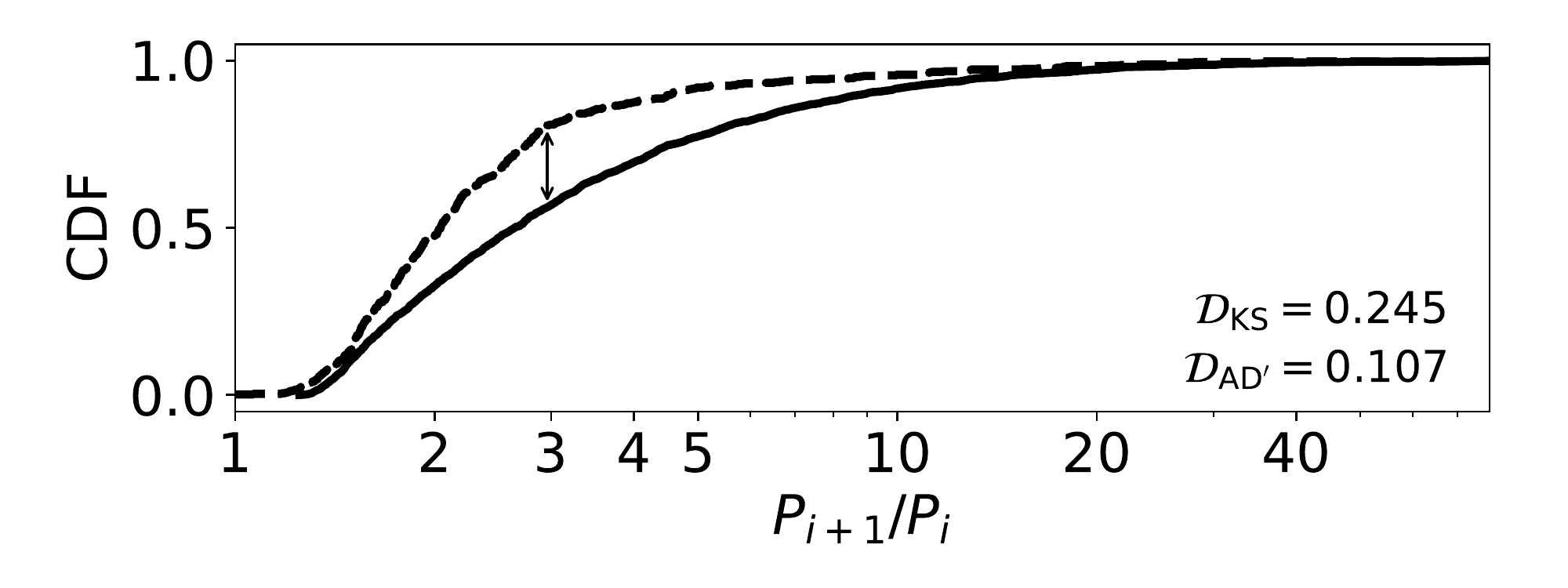} \\
 \includegraphics[scale=0.425,trim={0 0.4cm 0 0.2cm},clip]{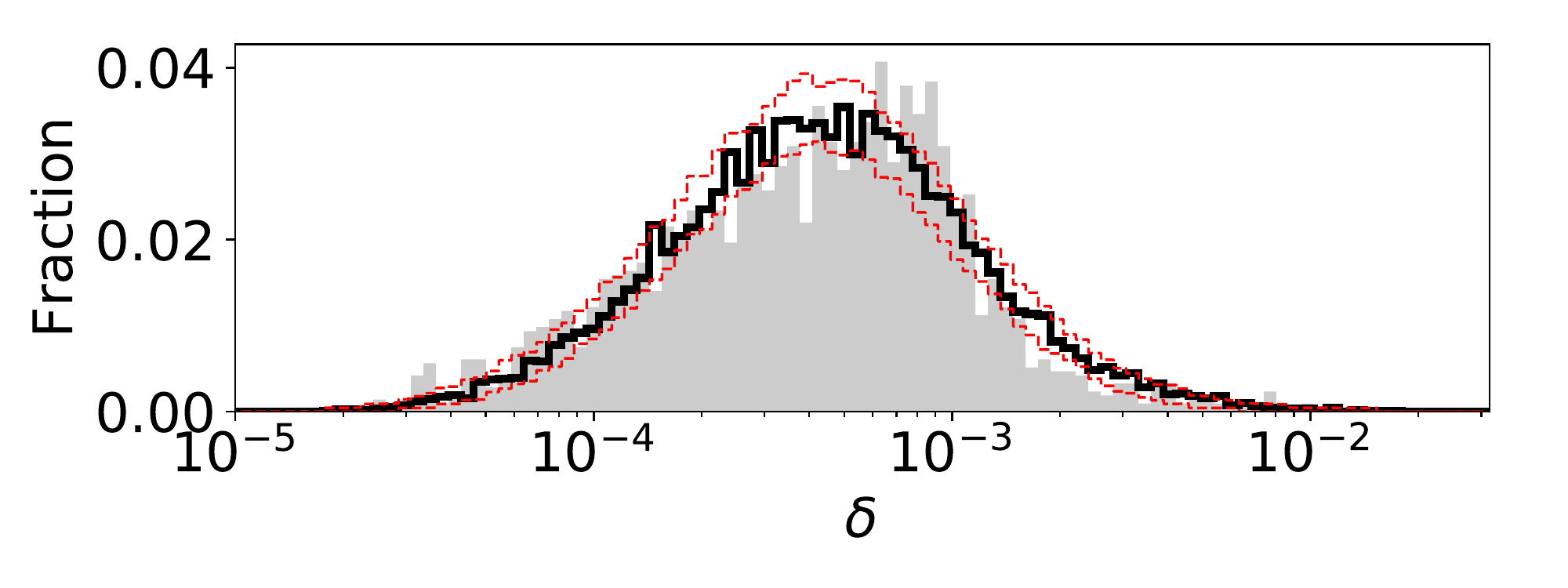} & 
 \includegraphics[scale=0.425,trim={0 0.4cm 0 0.2cm},clip]{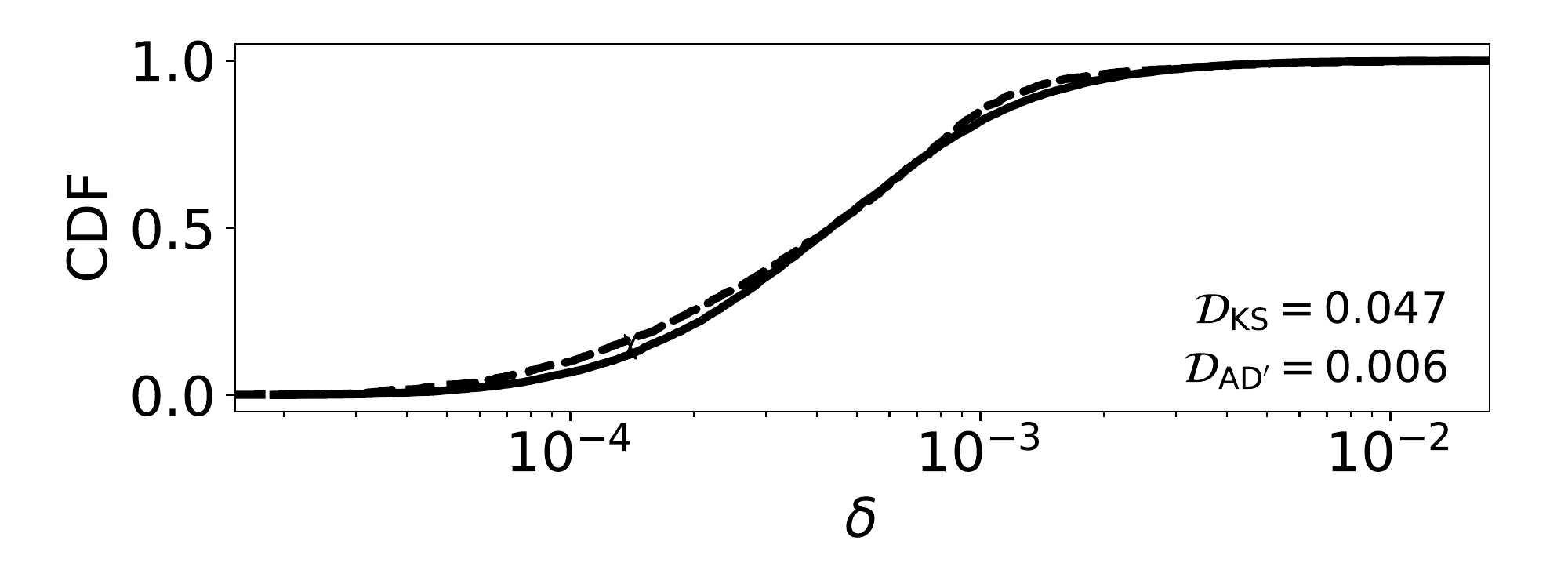} \\
 \includegraphics[scale=0.425,trim={0 0.5cm 0 0.2cm},clip]{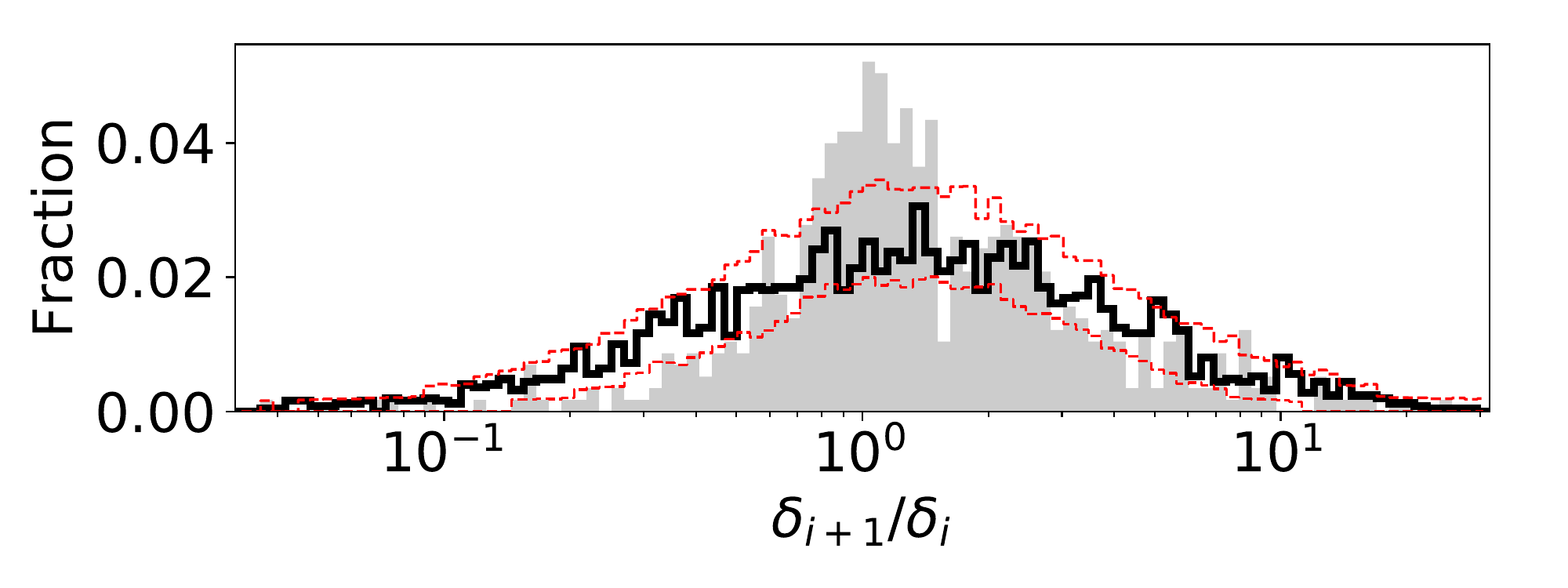} & 
 \includegraphics[scale=0.425,trim={0 0.5cm 0 0.2cm},clip]{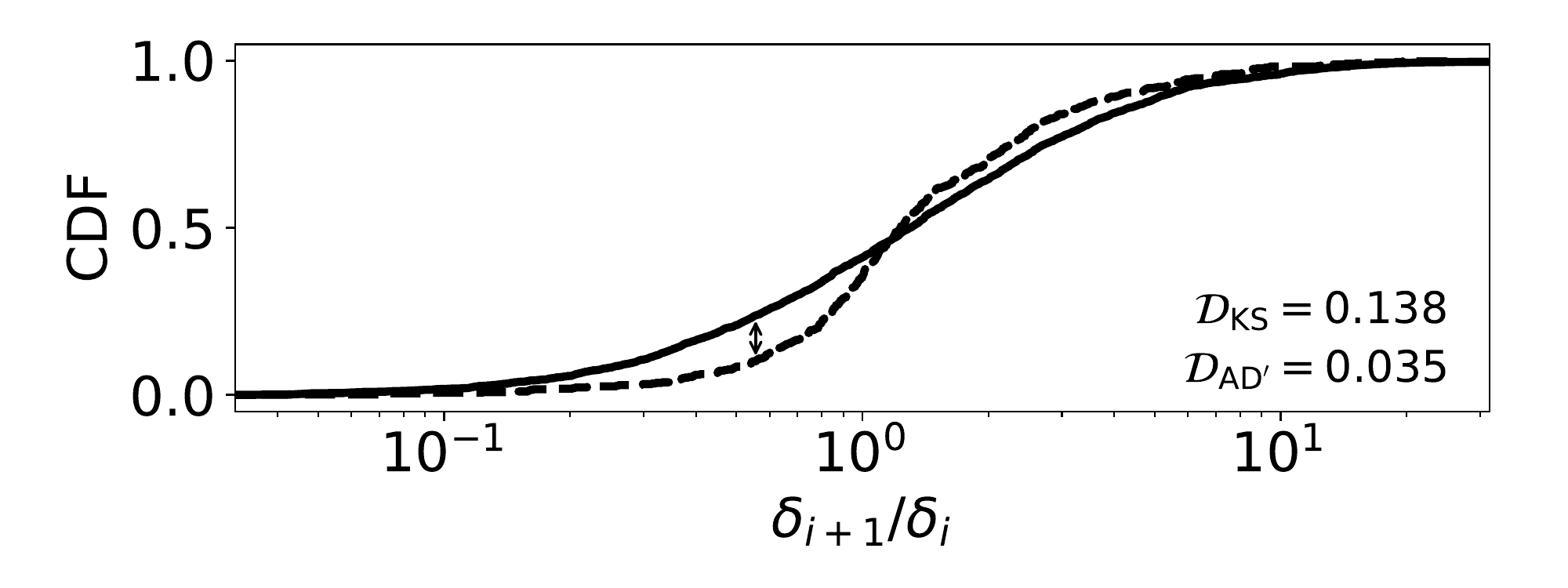} \\
 \includegraphics[scale=0.425,trim={0 0.4cm 0 0.2cm},clip]{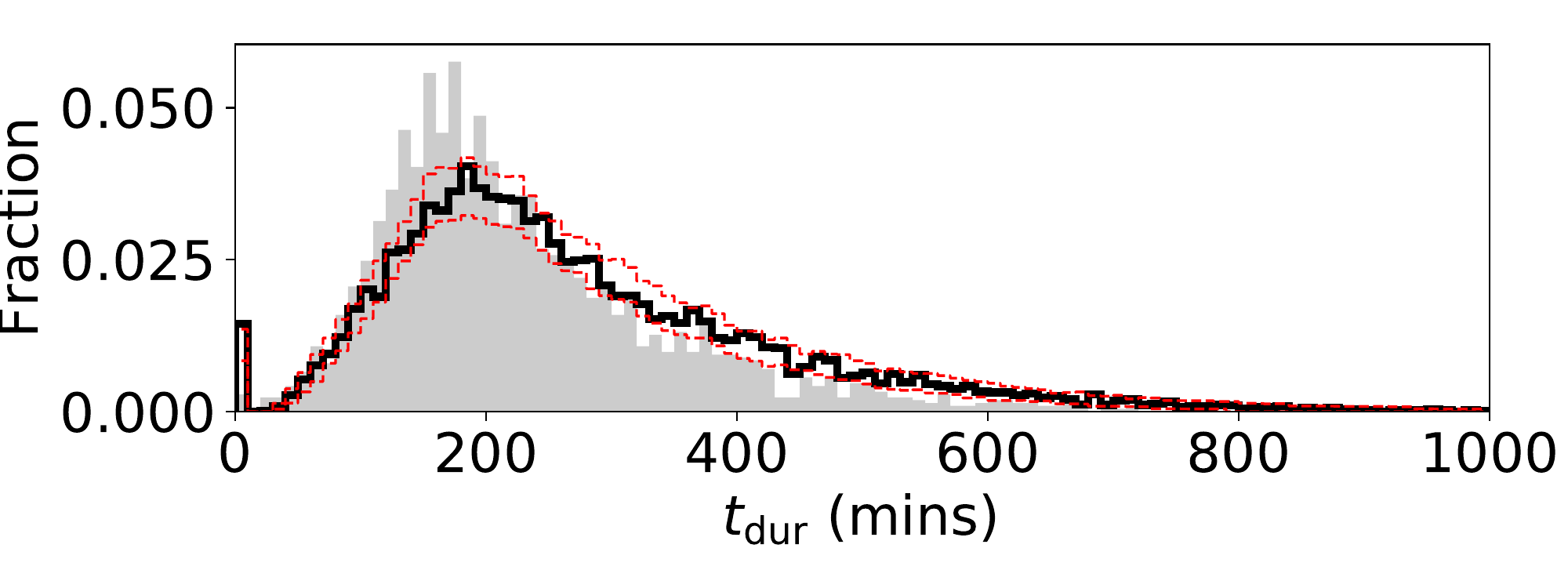} & 
 \includegraphics[scale=0.425,trim={0 0.4cm 0 0.2cm},clip]{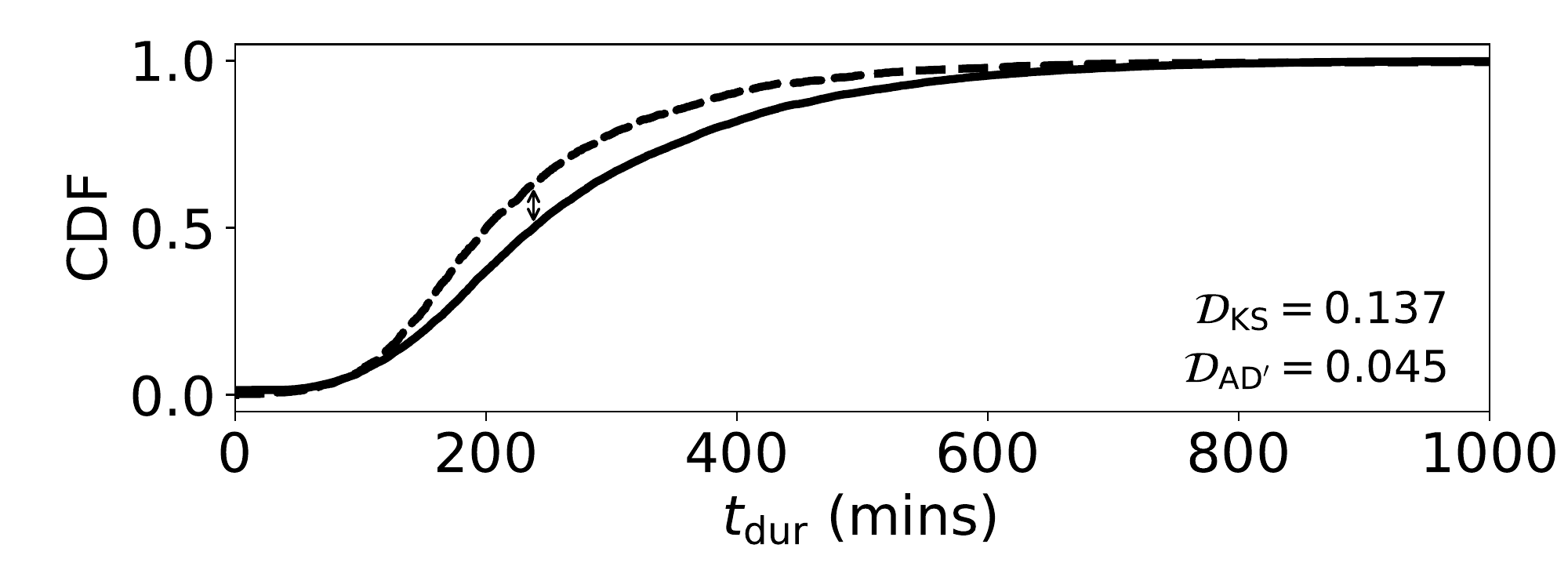} \\
 \includegraphics[scale=0.425,trim={0 0.4cm 0 0.2cm},clip]{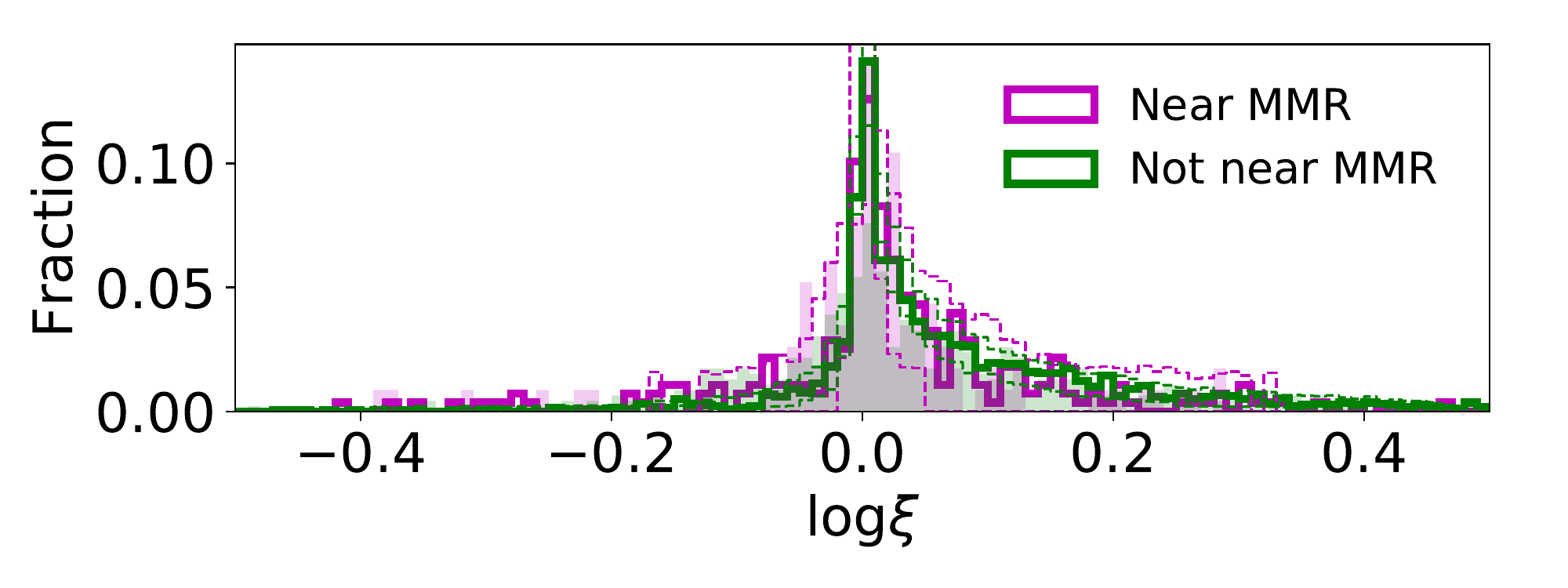} & 
 \includegraphics[scale=0.425,trim={0 0.4cm 0 0.2cm},clip]{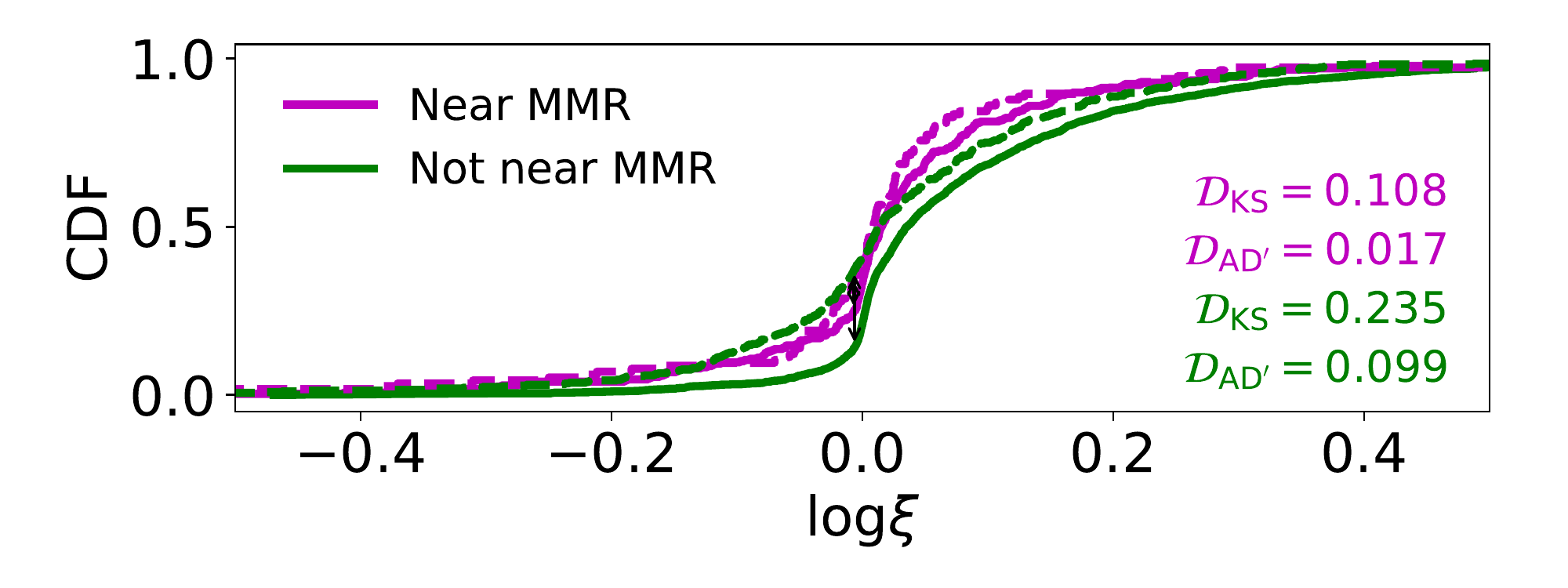} \\
\end{tabular}
\caption{\textbf{Non-clustered model:} a simulated observed catalogue of exoplanet systems generated from our non-clustered model. \textbf{Left-hand panels:} histograms of the summary statistics, with the same panels as the ones in Figure \ref{fig:Kepler_DR25}.
The black hollow histograms show one simulated catalogue (model parameters listed in Table \ref{tab:param_fits}) while the \Kepler{} DR25 exoplanets are plotted as grey shaded histograms for comparison. The red dashed histograms show the 16 and 84 percentiles of each bin based on 100 simulated catalogues with parameters drawn from our emulator with $\mathcal{D}_{W,\rm KS} < 55$.
\textbf{Right-hand panels:} the corresponding CDFs to the left-hand panels. The distances for each summary statistic are shown and labelled as arrows between the CDFs.
The observed multiplicities are poorly fit, with too few multiplanet systems. While the period distribution is reasonably reproduced, the period ratio distribution is very poorly modelled. Similarly, the transit depth ratio distribution of the model does not peak as sharply around unity as the data.
}
\label{fig:non_clustered_model}
\end{figure*}

\begin{figure*}
\begin{tabular}{cc}
 \includegraphics[scale=0.425,trim={0 0.4cm 0 0.2cm},clip]{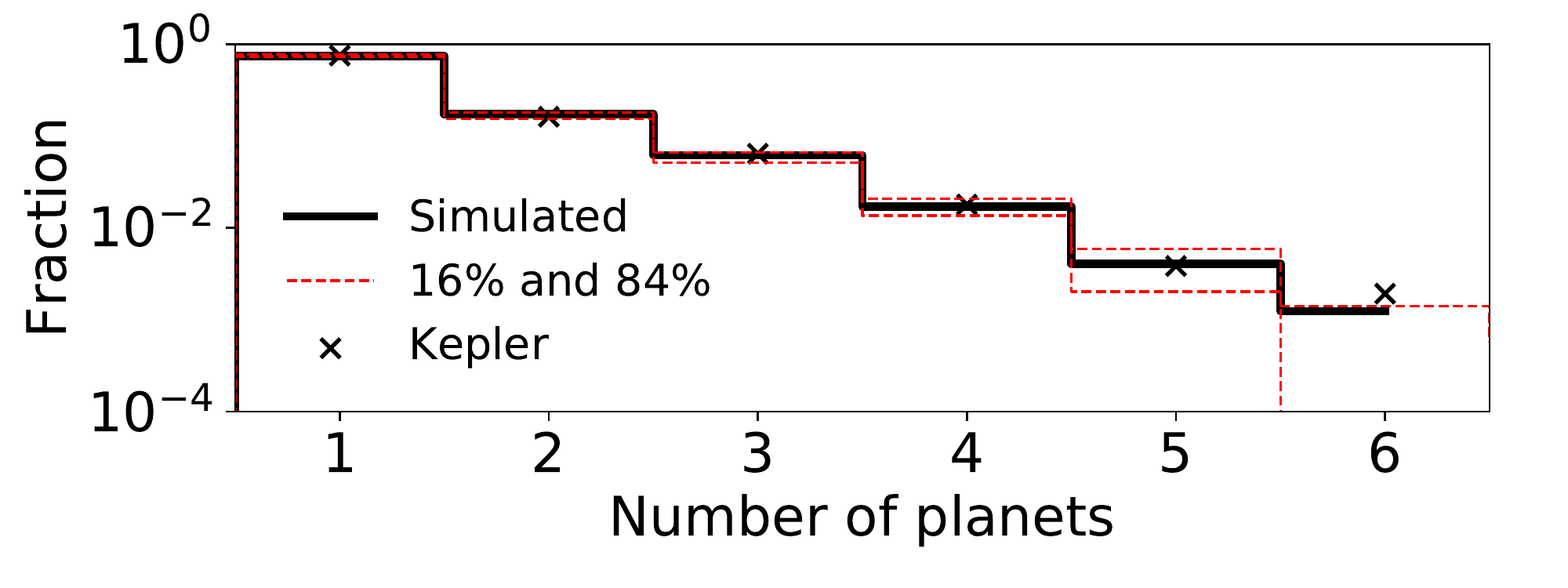} & 
 \includegraphics[scale=0.425,trim={0 0.4cm 0 0.2cm},clip]{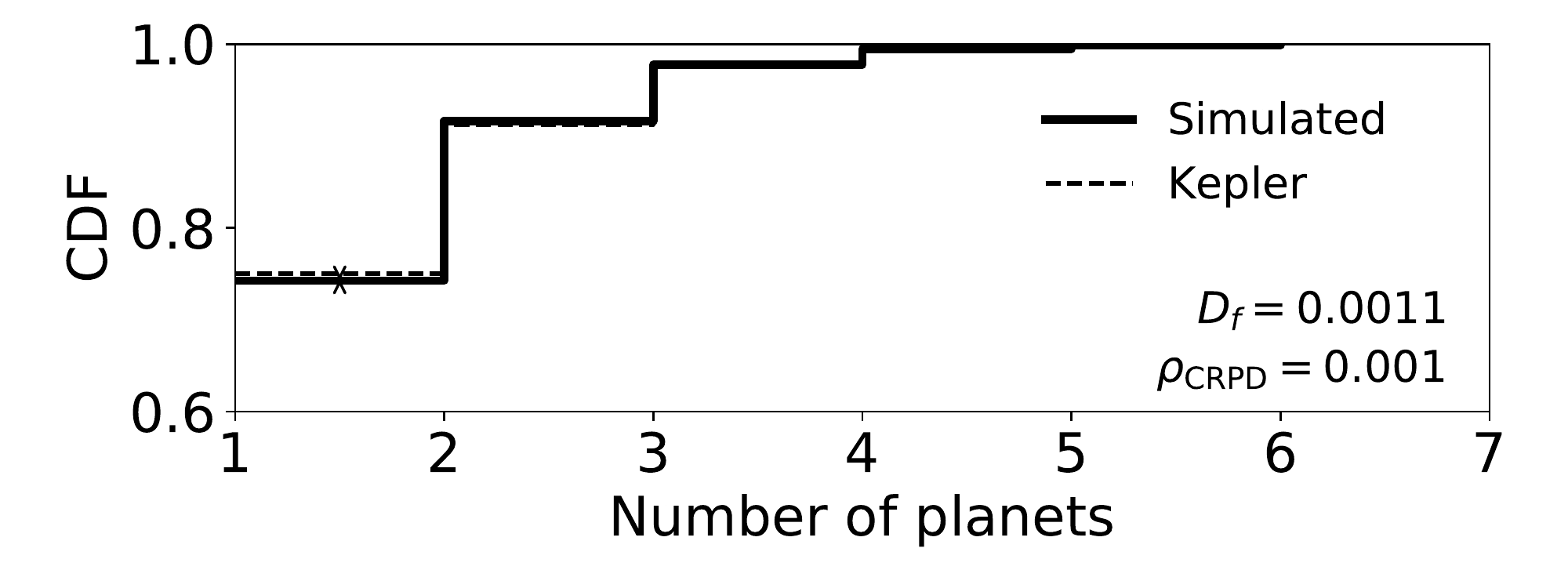} \\
 \includegraphics[scale=0.425,trim={0 0.4cm 0 0.2cm},clip]{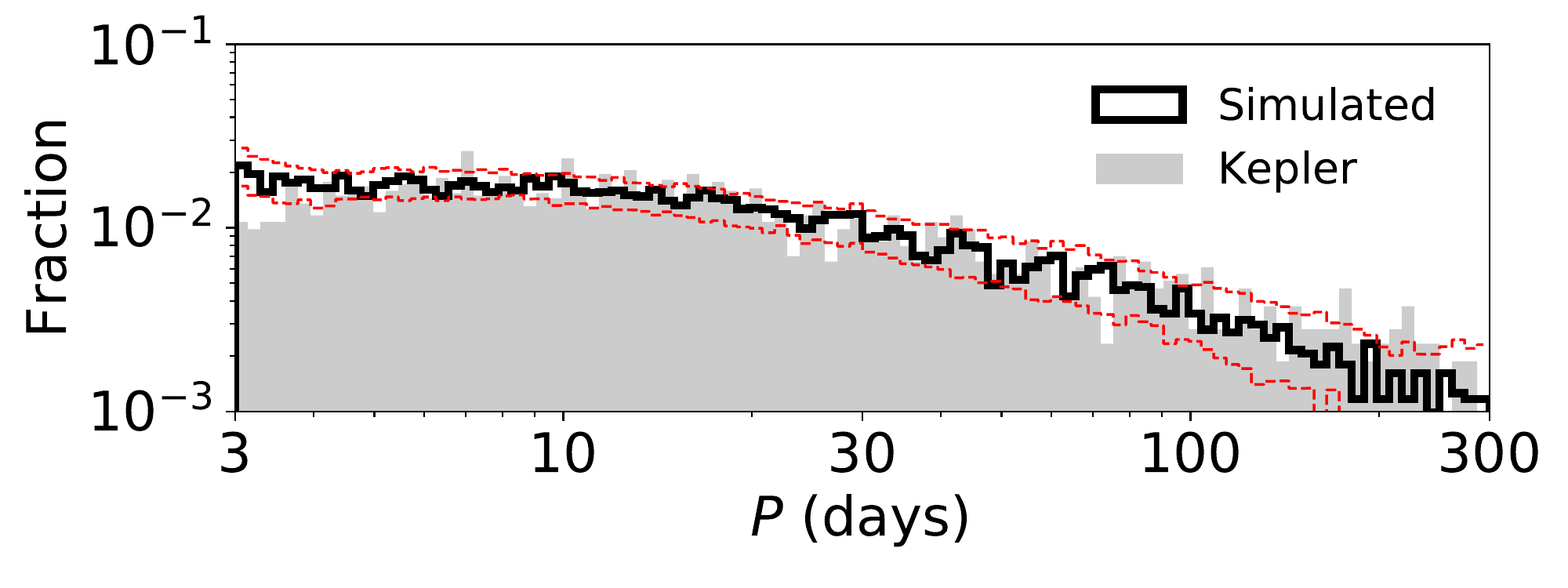} & 
 \includegraphics[scale=0.425,trim={0 0.4cm 0 0.2cm},clip]{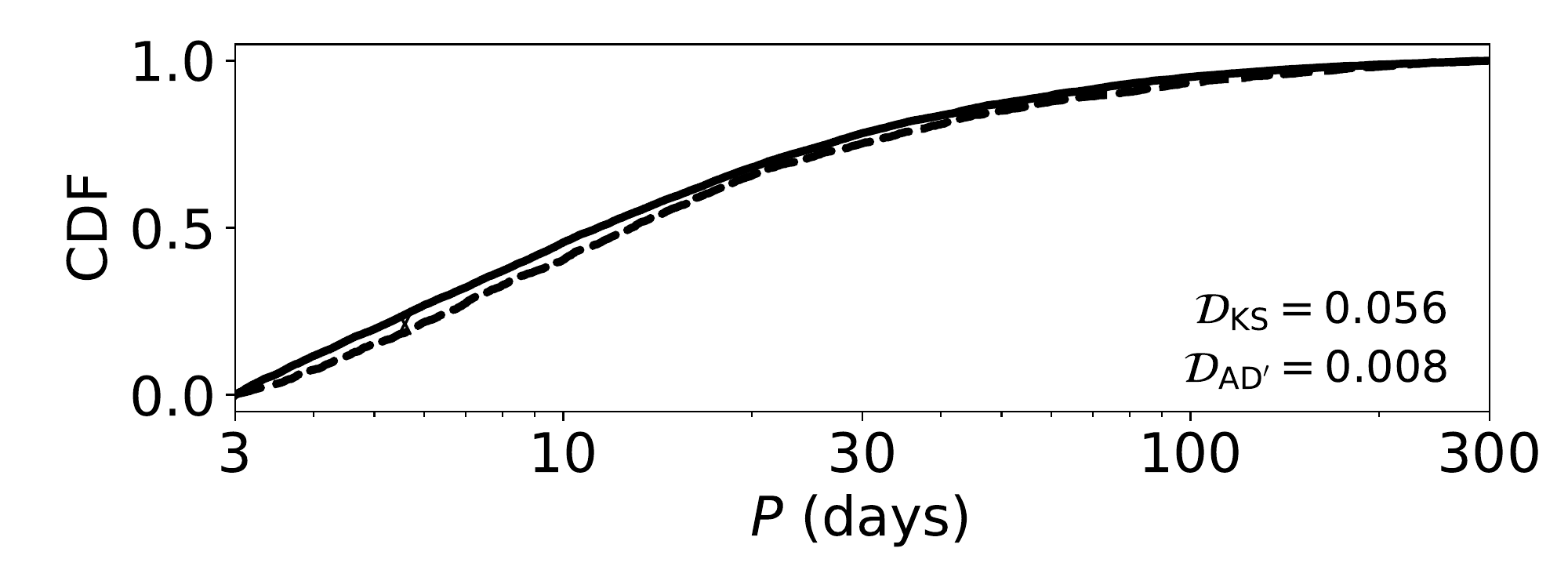} \\
 \includegraphics[scale=0.425,trim={0 0.4cm 0 0.2cm},clip]{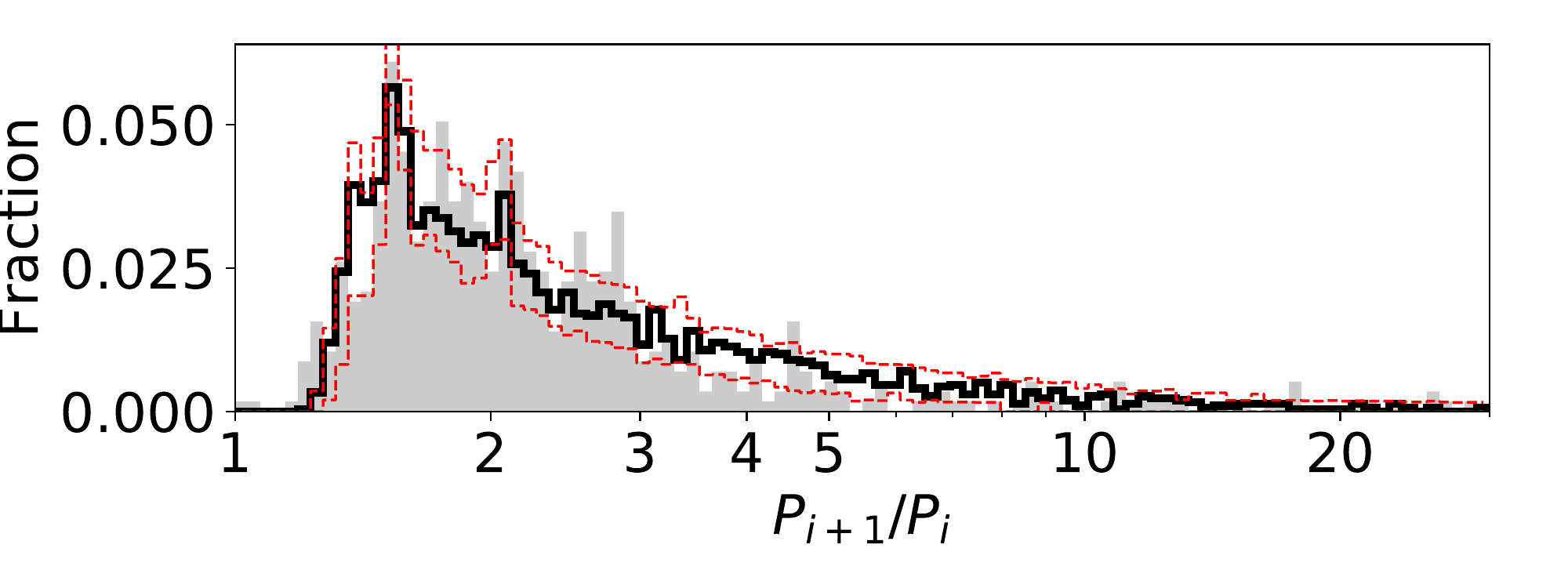} & 
 \includegraphics[scale=0.425,trim={0 0.4cm 0 0.2cm},clip]{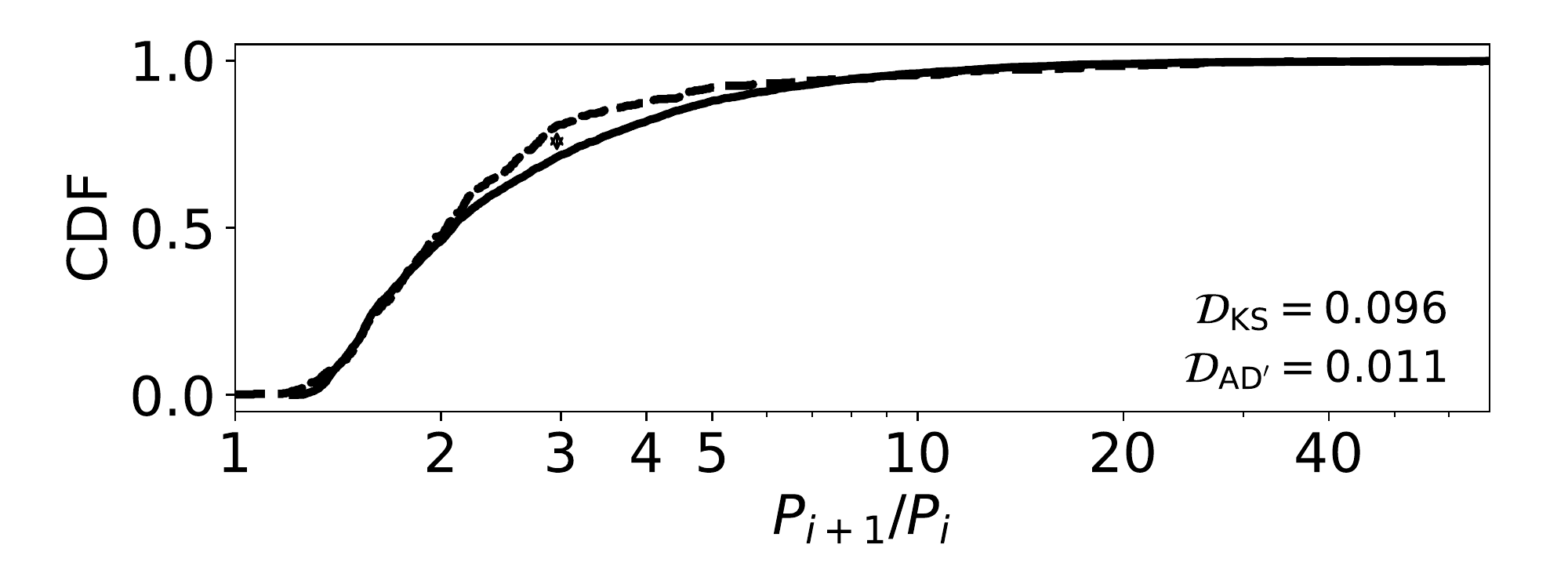} \\
 \includegraphics[scale=0.425,trim={0 0.4cm 0 0.2cm},clip]{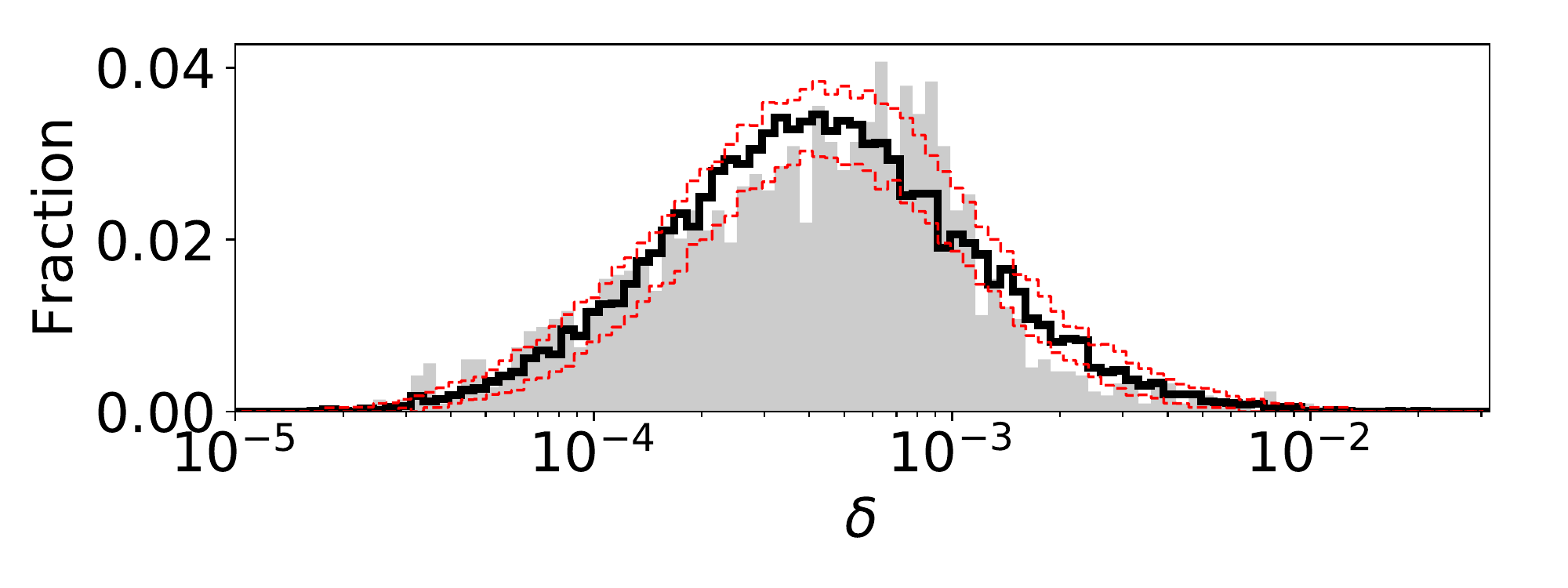} & 
 \includegraphics[scale=0.425,trim={0 0.4cm 0 0.2cm},clip]{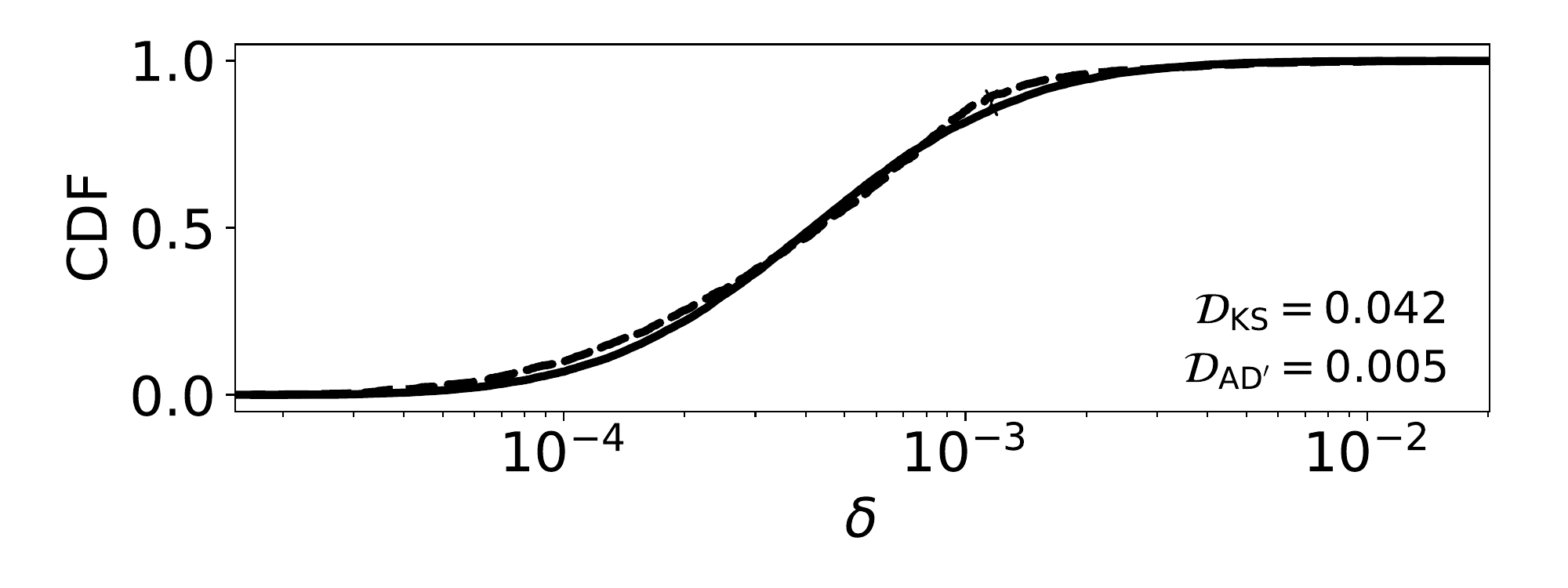} \\
 \includegraphics[scale=0.425,trim={0 0.5cm 0 0.2cm},clip]{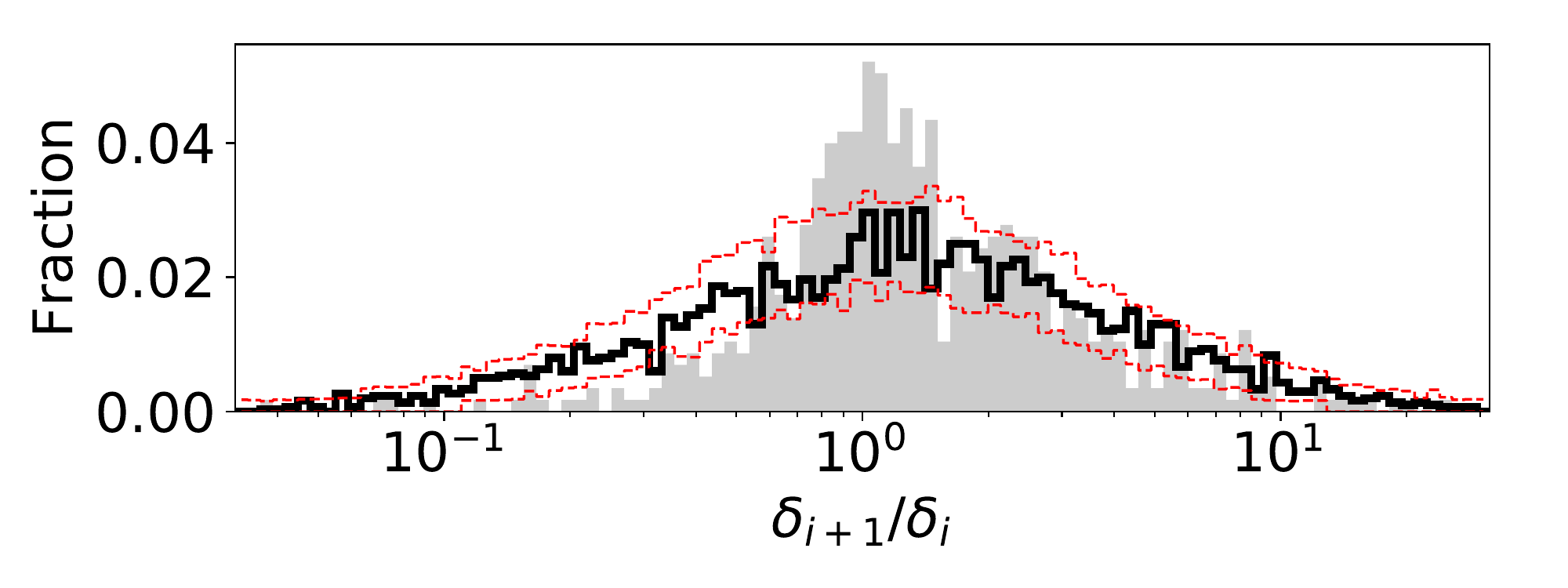} & 
 \includegraphics[scale=0.425,trim={0 0.5cm 0 0.2cm},clip]{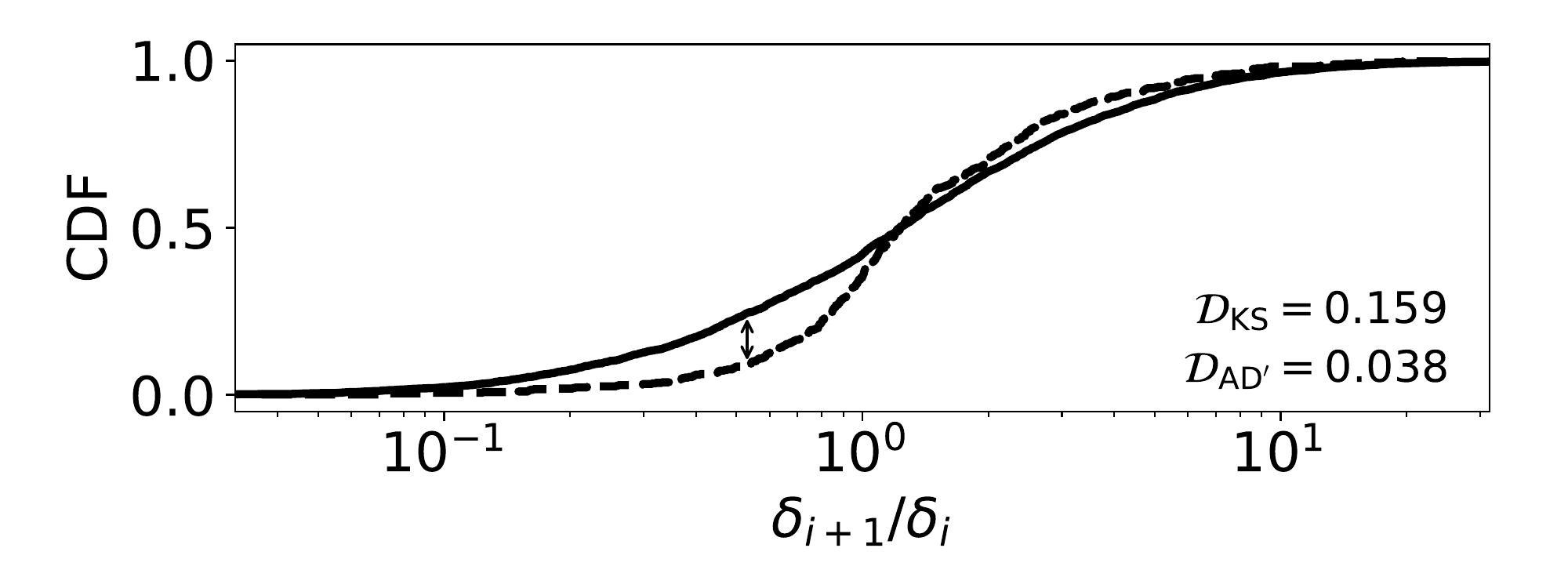} \\
 \includegraphics[scale=0.425,trim={0 0.4cm 0 0.2cm},clip]{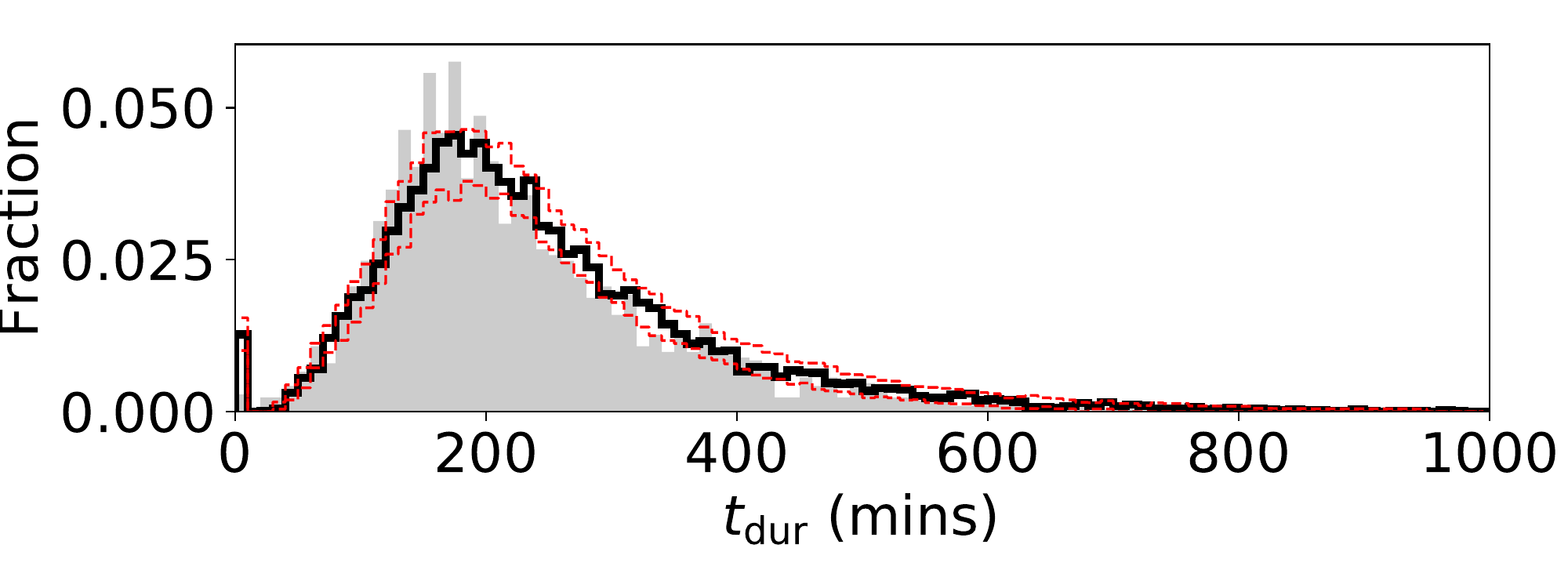} & 
 \includegraphics[scale=0.425,trim={0 0.4cm 0 0.2cm},clip]{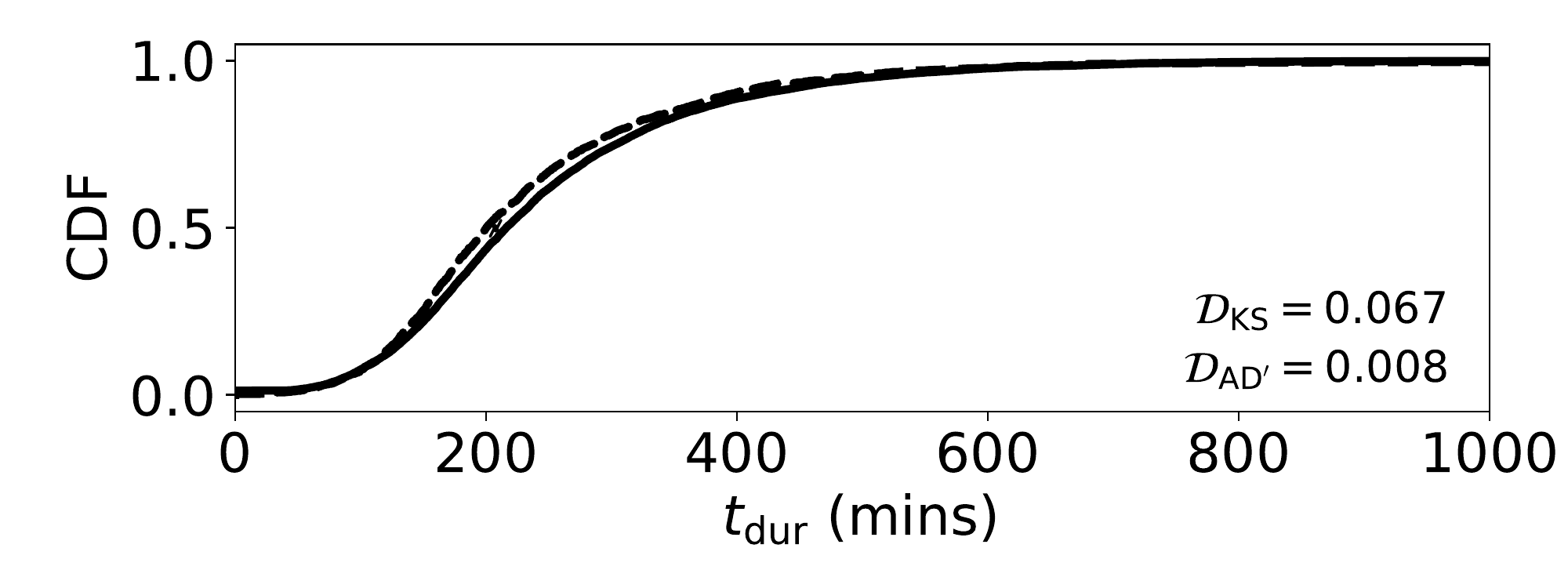} \\
 \includegraphics[scale=0.425,trim={0 0.4cm 0 0.2cm},clip]{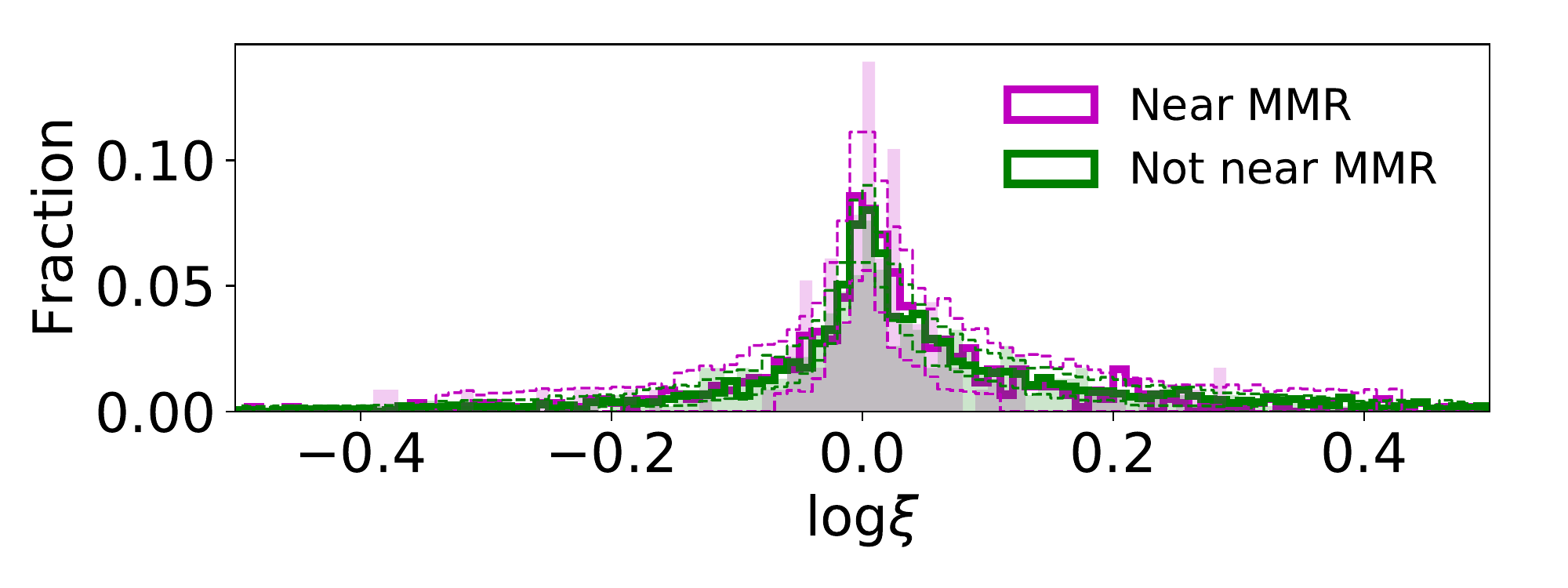} & 
 \includegraphics[scale=0.425,trim={0 0.4cm 0 0.2cm},clip]{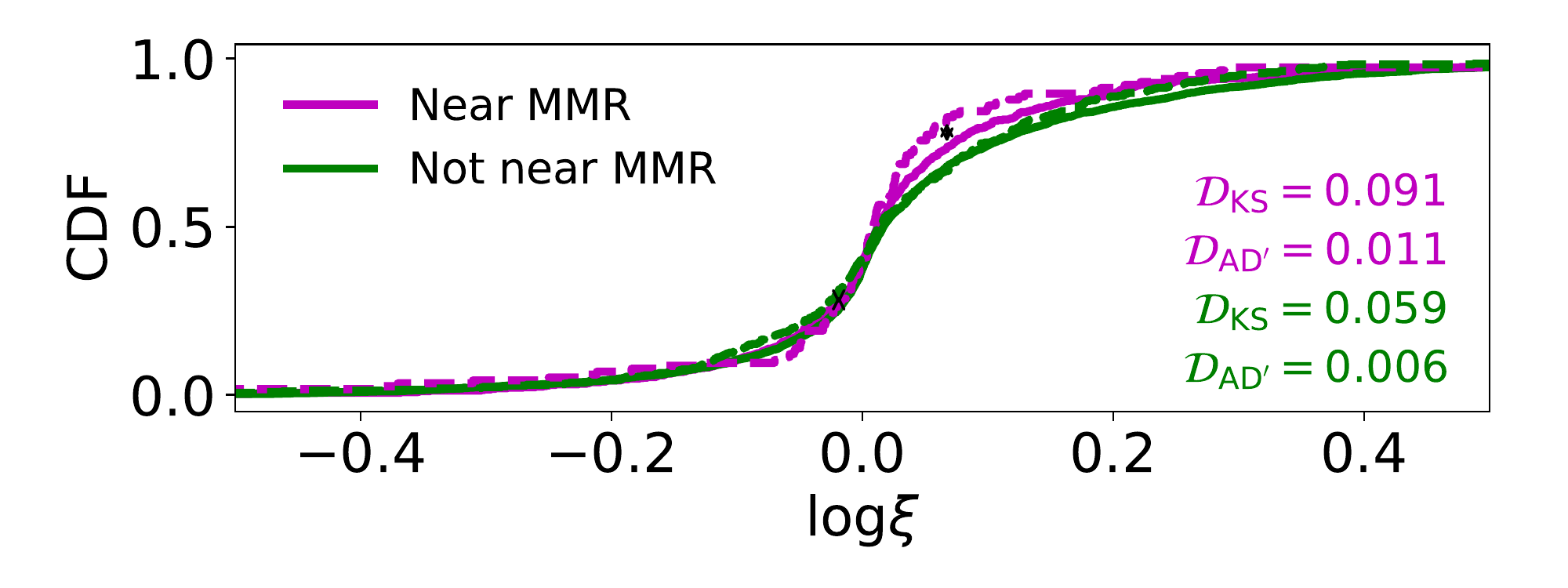} \\
\end{tabular}
\caption{\textbf{Clustered periods model:} a simulated observed population of exoplanet systems generated from our clustered periods model. The panels are the same as the ones in Figure \ref{fig:non_clustered_model}. The black hollow histograms show one simulated catalogue (model parameters listed in Table \ref{tab:param_fits}) while the \Kepler{} DR25 exoplanets are plotted as grey shaded histograms for comparison. The red dashed histograms show the 16 and 84 percentiles of each bin based on 100 simulated catalogues with parameters drawn from our emulator with $\mathcal{D}_{W,\rm KS} < 35$.
The observed multiplicity and period ratio distributions are remarkably improved compared to those of the non-clustered model. However, the transit depth ratio distribution looks similar as before, since no clustering in sizes is present in this model, and thus still poorly fits the data.
}
\label{fig:clustered_P_model}
\end{figure*}

As outlined in \S\ref{Methods}, our forward modelling procedure produces a simulated observed catalogue of transiting exoplanets, including measured planet multiplicities per system, observed properties for each planet (period, transit duration, and transit depth), and the ratios of period, transit depth, and transit duration for pairs of apparently adjacent planets. In this section, we display simulated \textit{observed catalogues} from each of our three models, non-clustered, clustered periods, and clustered periods and sizes, in Figures \ref{fig:non_clustered_model}--\ref{fig:clustered_P_R_model} respectively. For each model, we generate one catalogue with $N_{\rm stars,sim} = 5N_{\rm stars,Kep} = 399 675$ in order to reduce stochastic noise and provide smoother marginal distributions for comparison. The model parameters $\bm{\theta}$ used to generate these catalogues are listed in Table \ref{tab:param_fits}.  In order to compute the 16 and 84 percentiles for each bin, we also generate 100 additional catalogues with model parameters drawn from our ABC posterior and that pass our (KS) distance thresholds after generating an observed catalogue using the full forward model. Appendix Figure \ref{fig:distances} shows how each of these catalogues compare to the \Kepler{} DR25 catalogue in terms of the total and individual (weighted and unweighted) distance terms. We discuss how each of our models compare to the \Kepler{} data below.

\subsubsection{Non-clustered model} \label{Non_clustered_model}

In Figure \ref{fig:non_clustered_model}, we show the results of our non-clustered model (the model parameters for the population plotted in black are listed in Table \ref{tab:param_fits}). The panels from top to bottom display the marginal distribution of each summary statistic while the left-hand and right-hand sides show the empirical PDFs and CDFs, respectively.  
The \Kepler{} DR25 population (as shown in Figure \ref{fig:Kepler_DR25}) is overplotted as grey histograms for comparison.  
The lower right corners of each CDF panel display the relevant distance terms between the model and the \Kepler{} population.

The non-clustered model is able to produce a population of simulated observed exoplanets that resembles the \Kepler{} DR25 population in some regards.
The overall rate of observed planets per target can be matched almost exactly (the $D_f$ term is almost zero).
The bulk of the period distribution appears well modelled by a simple power law with $\alpha_P \sim -0.1$ (i.e. shallowly increasing in log-period, see \S\ref{Model_comparison_underlying} for the underlying distribution), although there appears to be a slight excess of planets at long periods.
The transit depth distribution is reasonably well modelled with planet radii drawn independently from a broken power law, although the observed distribution of (log) $\delta$  is slightly left-skewed (i.e., mode is at larger value than the median) whereas our model produces a symmetric distribution.
The transit duration distribution is also reasonably well modelled, although we produce an excess of grazing transits with zero duration. The distribution is also somewhat shifted to longer durations due to the extremely small eccentricities. The $\xi$ distribution of our simulated catalogue appears to be slightly more skewed towards values $\log{\xi} > 0$, suggesting that the mutual inclinations of the \Kepler{} planets are larger than $\sigma_{i,\rm low} \sim 0.3^\circ$.

However, there are several major, clear differences between the non-clustered model and the \Kepler{} planet catalogues. First, the planet multiplicity distribution is poorly modelled and produces far too few multiplanet systems.  This shortcoming of the model is persistent for many realizations near the best-fitting parameters.  
Secondly, the period ratio distribution is clearly not well captured by this model, motivating our use of clustered models.  
While most of the \Kepler{} adjacent planet pairs have small period ratios and the distribution falls rapidly above $\mathcal{P} \sim 3$, the non-clustered model produces a much more gradual decline in the tail of the distribution. 
A stability criteria of $\Delta_c = 8$ appears to sculpt the inner edge of the period ratio distribution in a similar manner as seen in the data. 
Finally, the transit depth ratio distribution is inadequately described by our non-clustered model; the model produces a wide, symmetric distribution around unity whereas the observed \Kepler{} radii ratios are highly peaked around unity. 
In summary, the non-clustered model described in the previous section performed well in some respects, but there are at least three obvious differences between the simulated observed and \Kepler{} DR25 populations: the observed multiplicity, period ratio, and transit depth ratio distributions.
These shortcomings motivate our clustered models.

\subsubsection{Clustered periods model} \label{Clustered_P}

Before we consider our full clustered periods and sizes model, we first explore the effect of adding just a clustering of orbital periods, i.e., our clustered periods model (the planetary radii are drawn in the same manner as in our non-clustered model). In Figure \ref{fig:clustered_P_model}, we show the results of this model with the \Kepler{} DR25 population (again, the model parameters used are listed in Table \ref{tab:param_fits}).

Our clustered periods model is a significantly better description of the \Kepler{} data than our non-clustered model. Notably, the observed multiplicity and period ratio distributions bear a much closer resemblance to the data, and have significantly reduced distances (CRPD distance for the observed multiplicity counts; KS and AD distances for the period ratio distribution). This is perhaps not surprising given that the main difference between this model and the previous model is the introduction of period clusters. The clustering in periods allows for some systems to contain planets that are closely packed more often than in the case of independently drawn periods, while also allowing for a more gradually falling tail to large orbital period ratios. The clustering in periods also provide a more flexible model for the numbers of planets per system since there are two parameters ($\lambda_c$ and $\lambda_p$) instead of just one ($\lambda_p$ of the non-clustered model).
The fits to the period and transit duration distributions are also slightly improved, while the transit duration ratio distribution (for planets not near resonances; $\{\xi_{\rm non-res}\}$) is significantly improved, in both KS and AD distances. This suggests that the distribution of orbital eccentricities is not as low as what is implied by the non-clustered model, and is instead well described by a Rayleigh distribution with $\sigma_e \simeq 0.01$. These results also imply that the mutual inclination distribution is well described by $\sigma_{i,\rm low} \simeq 1.1^\circ$ for $\sim 60\%$ of systems and planets near resonances. This also appears to produce noticeable peaks near the first-order MMRs in the period ratio distribution (most apparent near $\mathcal{P} \simeq 1.5$ and to a smaller extent at $\mathcal{P} \simeq 2$), although we only observe a marginal improvement to the KS and AD distances for the $\xi_{\rm res}$ distribution. There is no significant change in the fit to the transit depth distribution. The transit depth ratio distribution, on the other hand, is modelled just as poorly as and perhaps even worse than before. These results show that clustered periods alone cannot reproduce the highly peaked nature of the planet radii ratios observed in the \Kepler{} multiplanet systems.

The clustering in orbital periods is able to substantially improve the modelling of the period ratio distribution, but the transit depth ratio distribution remains poorly modelled by both the non-clustered and clustered periods models. The transit depth ratios are more highly peaked around one in the actual \Kepler{} population than in the simulated catalogues of these models (see Figures \ref{fig:non_clustered_model} and \ref{fig:clustered_P_model}), motivating our next investigation of the clustered periods and sizes model.

\begin{figure*}
\begin{tabular}{cc}
 \includegraphics[scale=0.425,trim={0 0.4cm 0 0.2cm},clip]{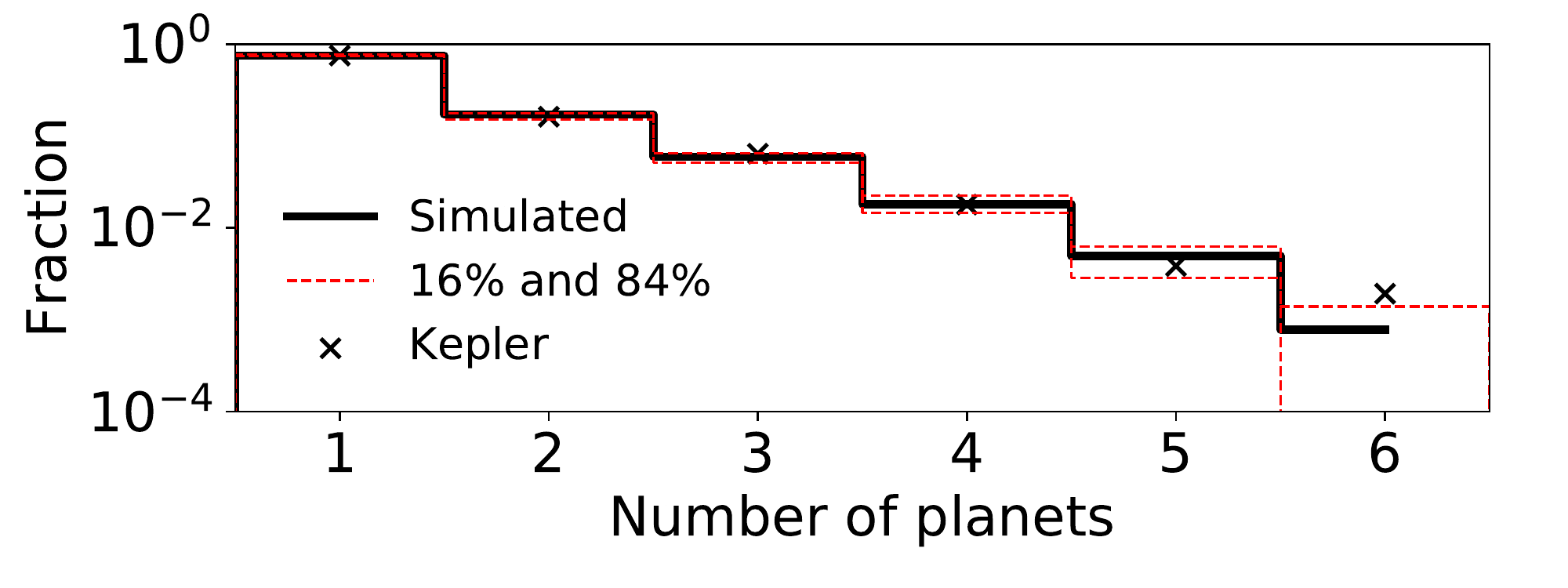} & 
 \includegraphics[scale=0.425,trim={0 0.4cm 0 0.2cm},clip]{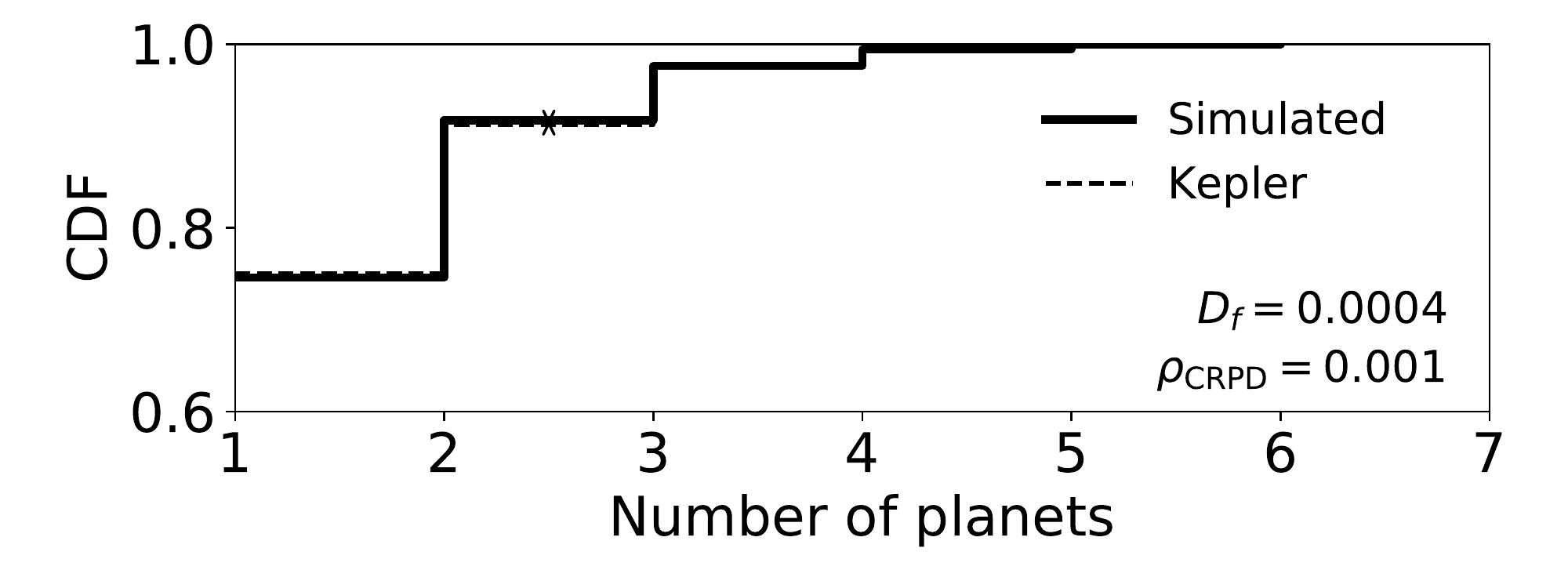} \\
 \includegraphics[scale=0.425,trim={0 0.4cm 0 0.2cm},clip]{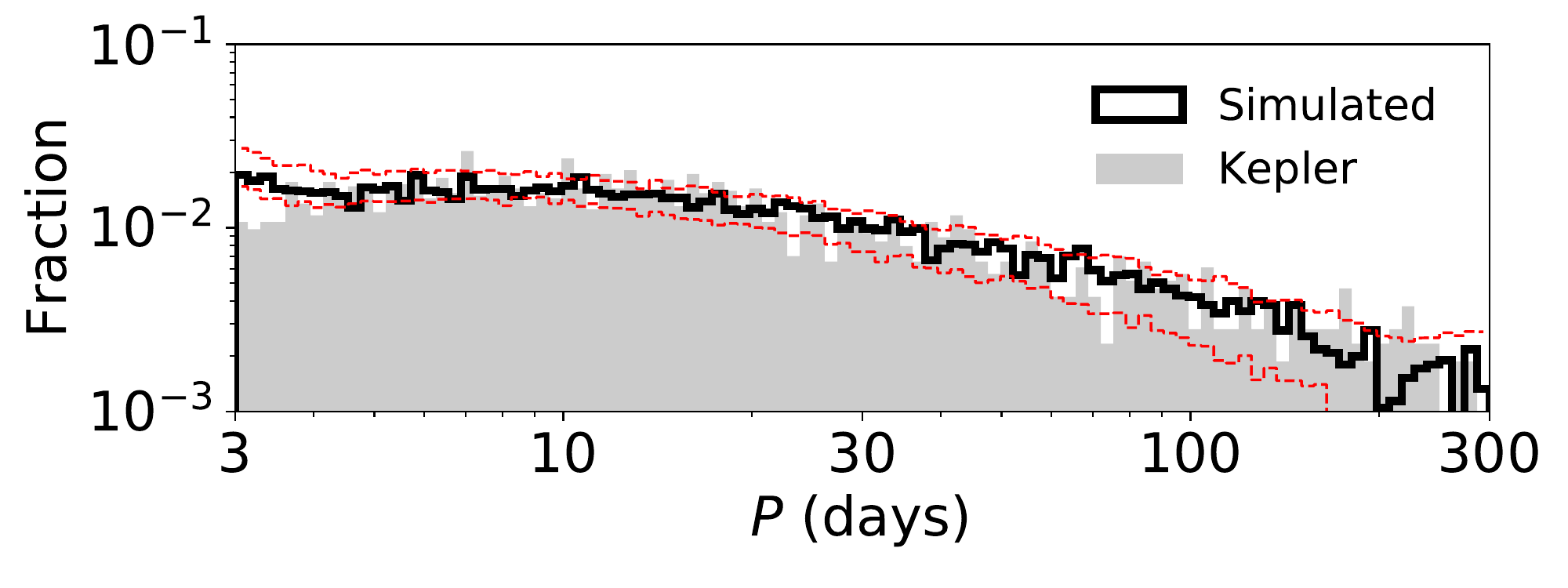} & 
 \includegraphics[scale=0.425,trim={0 0.4cm 0 0.2cm},clip]{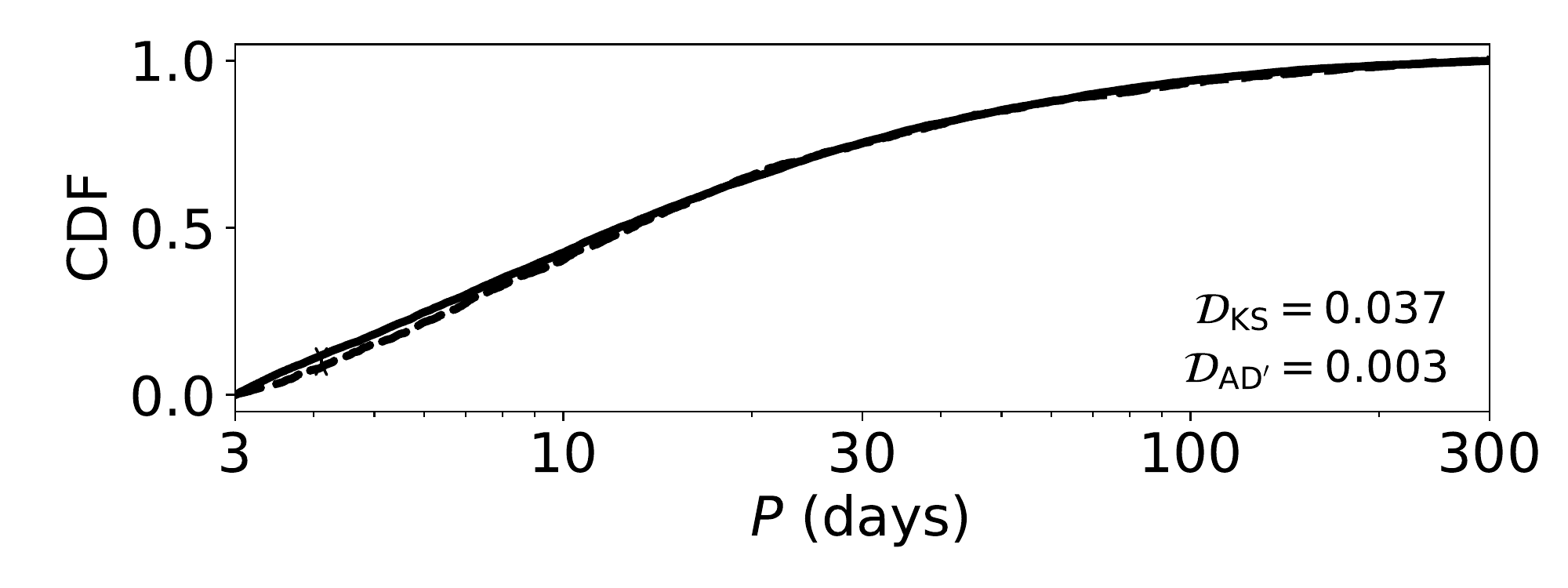} \\
 \includegraphics[scale=0.425,trim={0 0.4cm 0 0.2cm},clip]{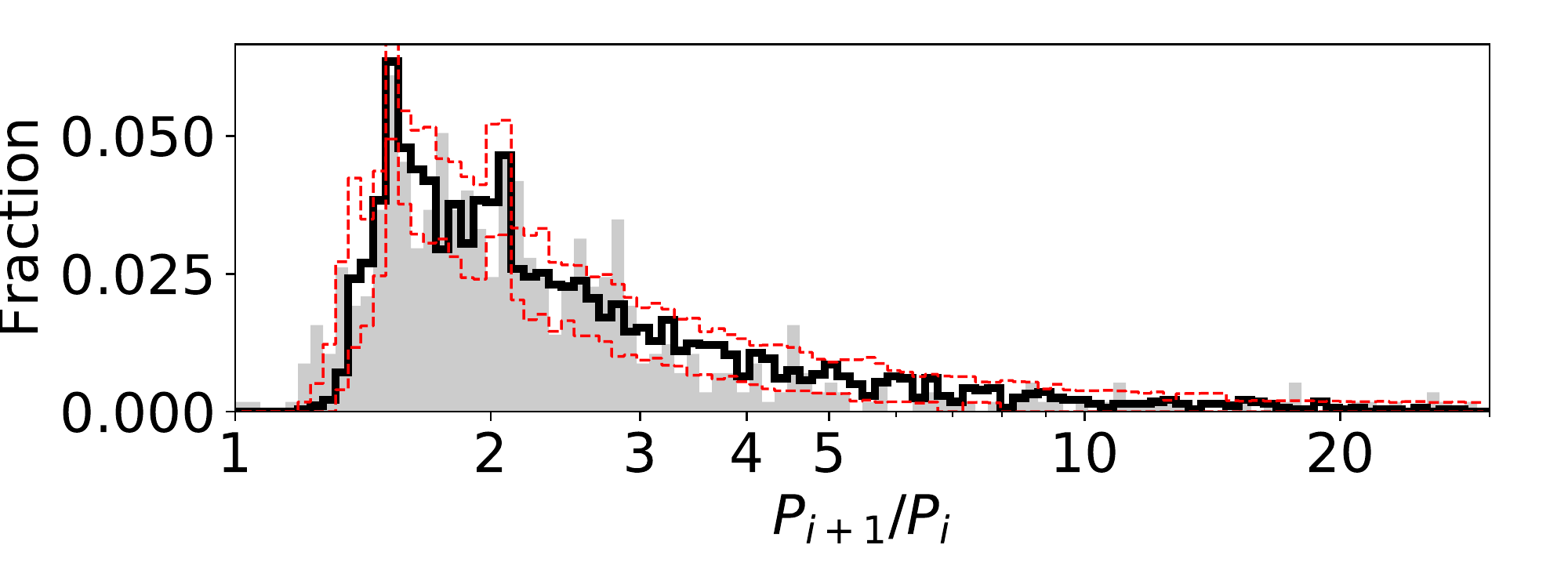} & 
 \includegraphics[scale=0.425,trim={0 0.4cm 0 0.2cm},clip]{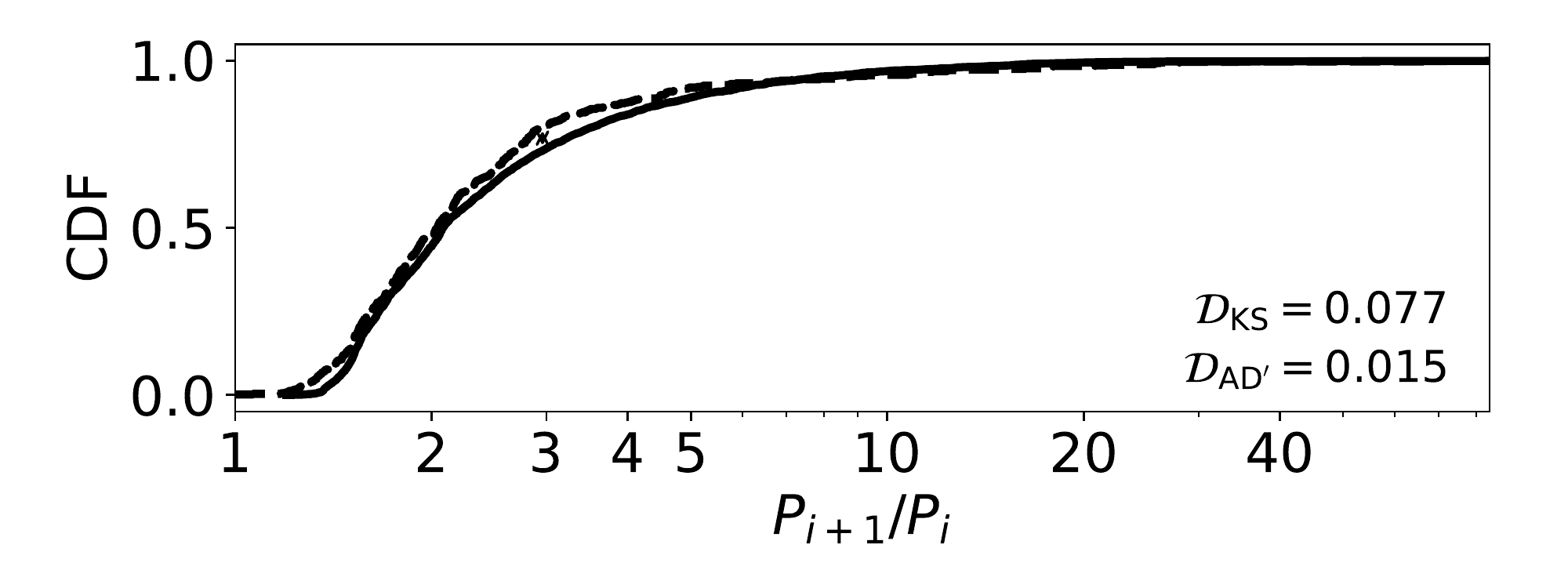} \\
 \includegraphics[scale=0.425,trim={0 0.4cm 0 0.2cm},clip]{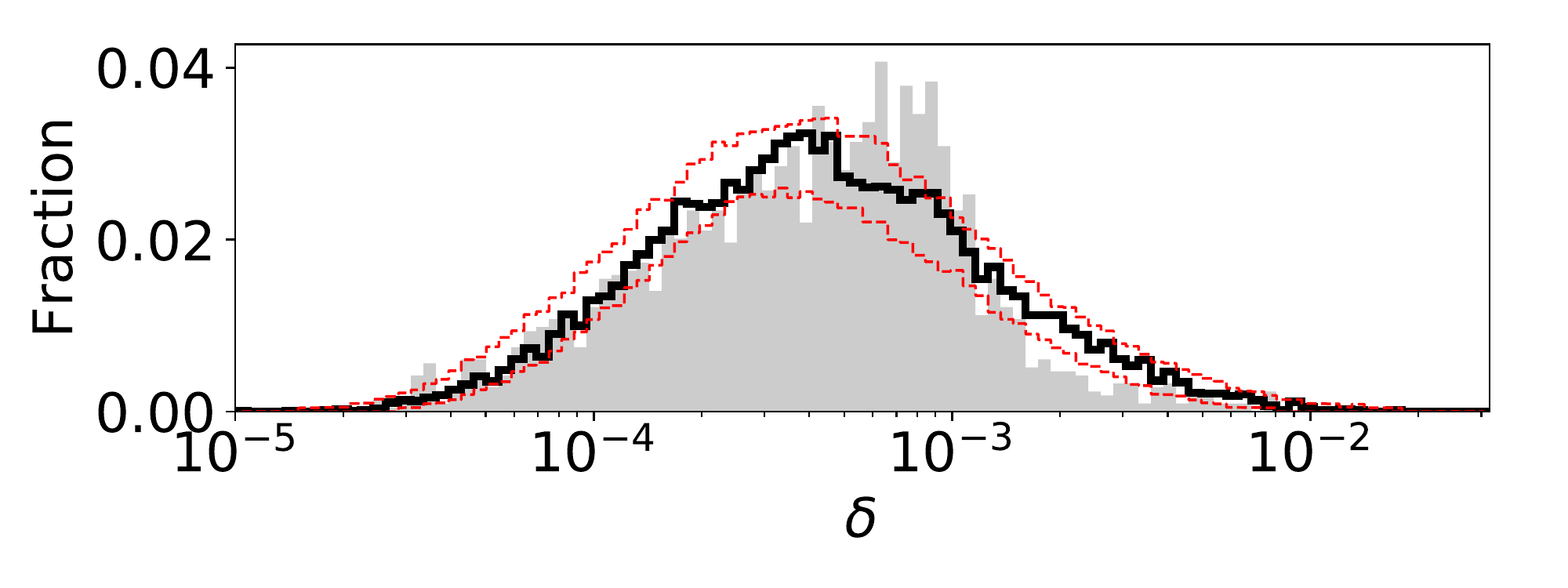} & 
 \includegraphics[scale=0.425,trim={0 0.4cm 0 0.2cm},clip]{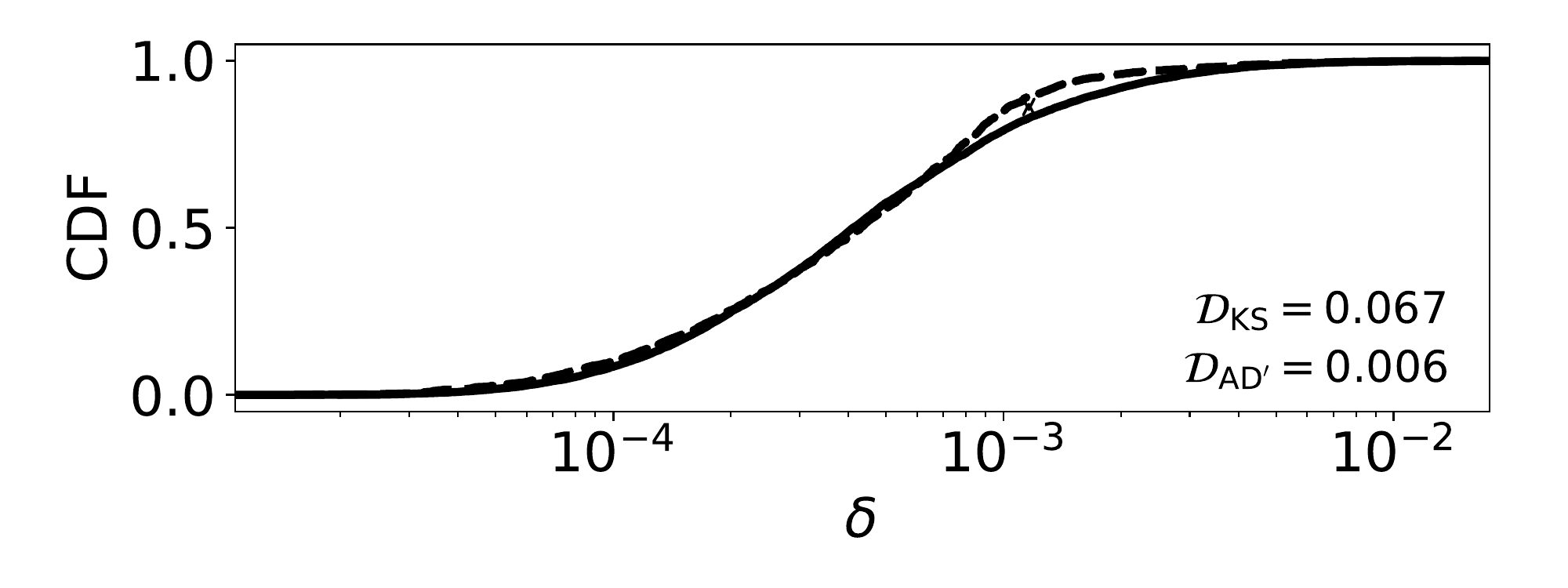} \\
 \includegraphics[scale=0.425,trim={0 0.5cm 0 0.2cm},clip]{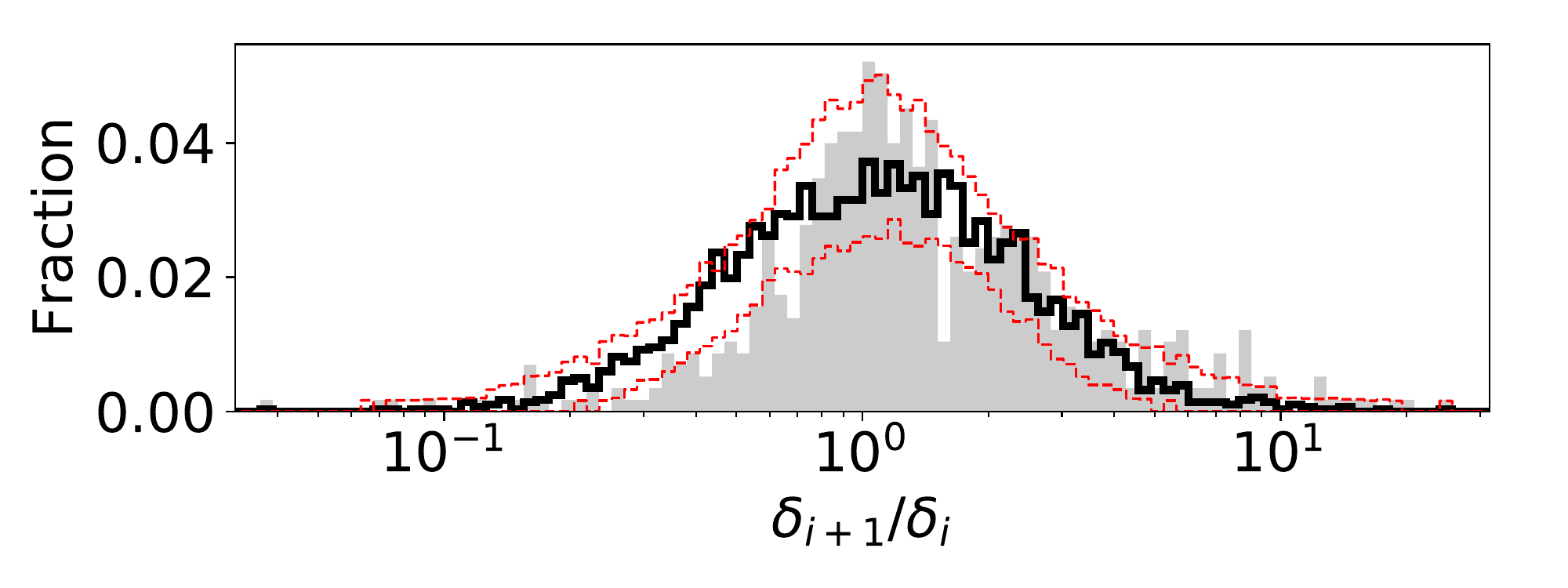} & 
 \includegraphics[scale=0.425,trim={0 0.5cm 0 0.2cm},clip]{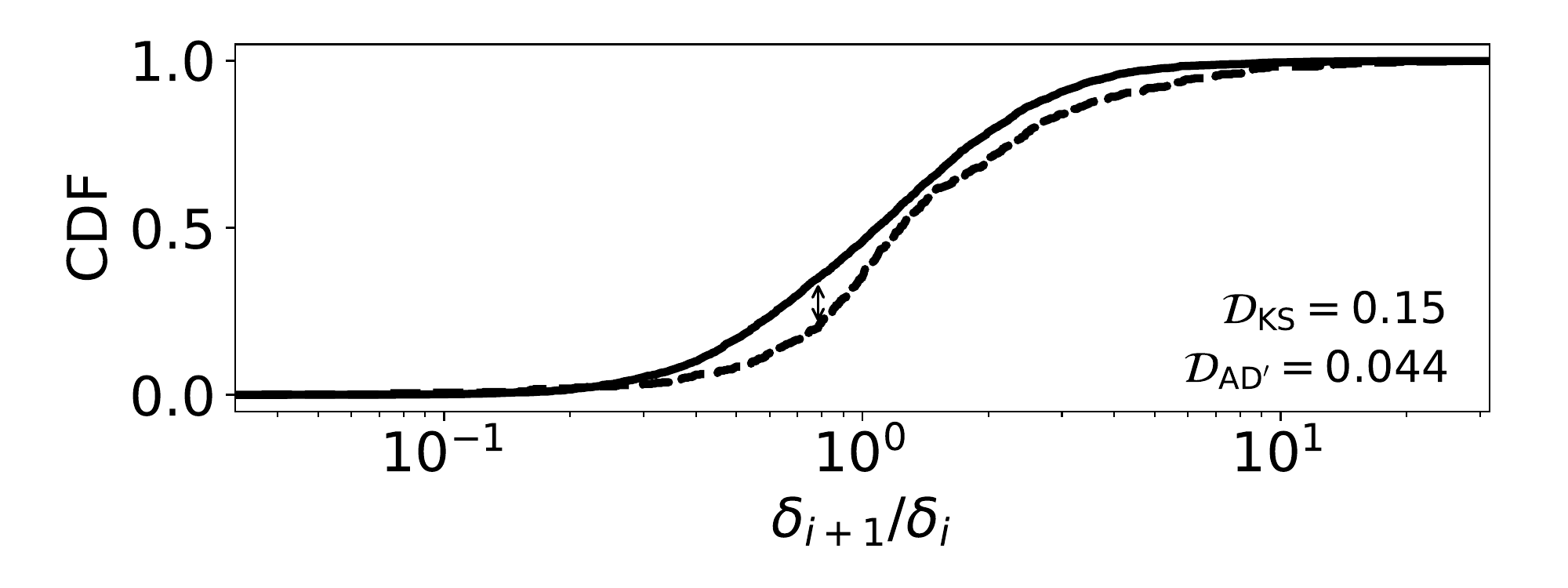} \\
 \includegraphics[scale=0.425,trim={0 0.4cm 0 0.2cm},clip]{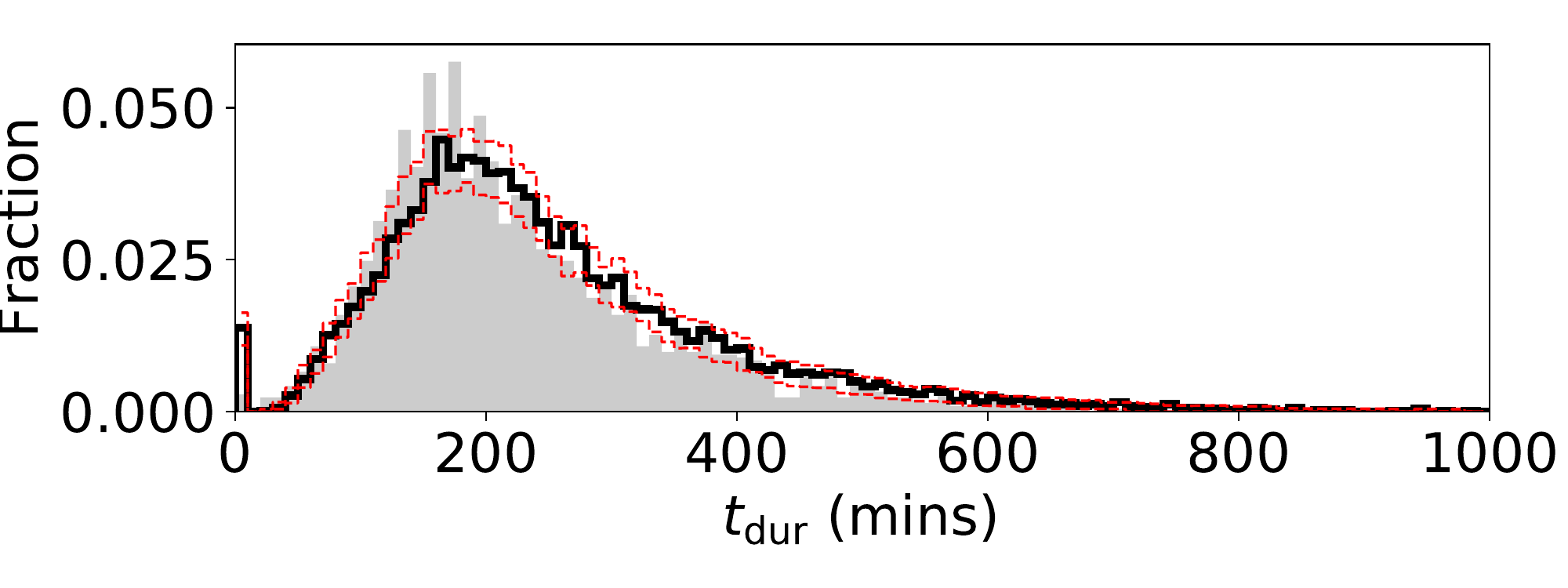} & 
 \includegraphics[scale=0.425,trim={0 0.4cm 0 0.2cm},clip]{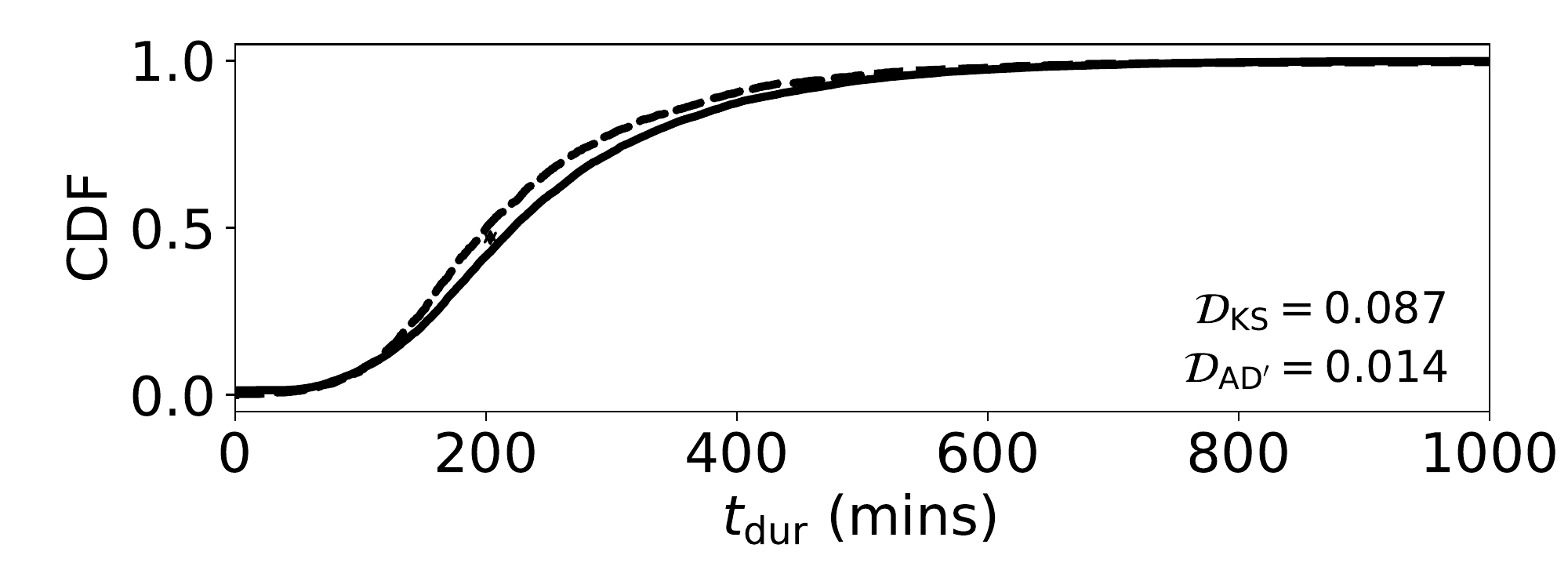} \\
 \includegraphics[scale=0.425,trim={0 0.4cm 0 0.2cm},clip]{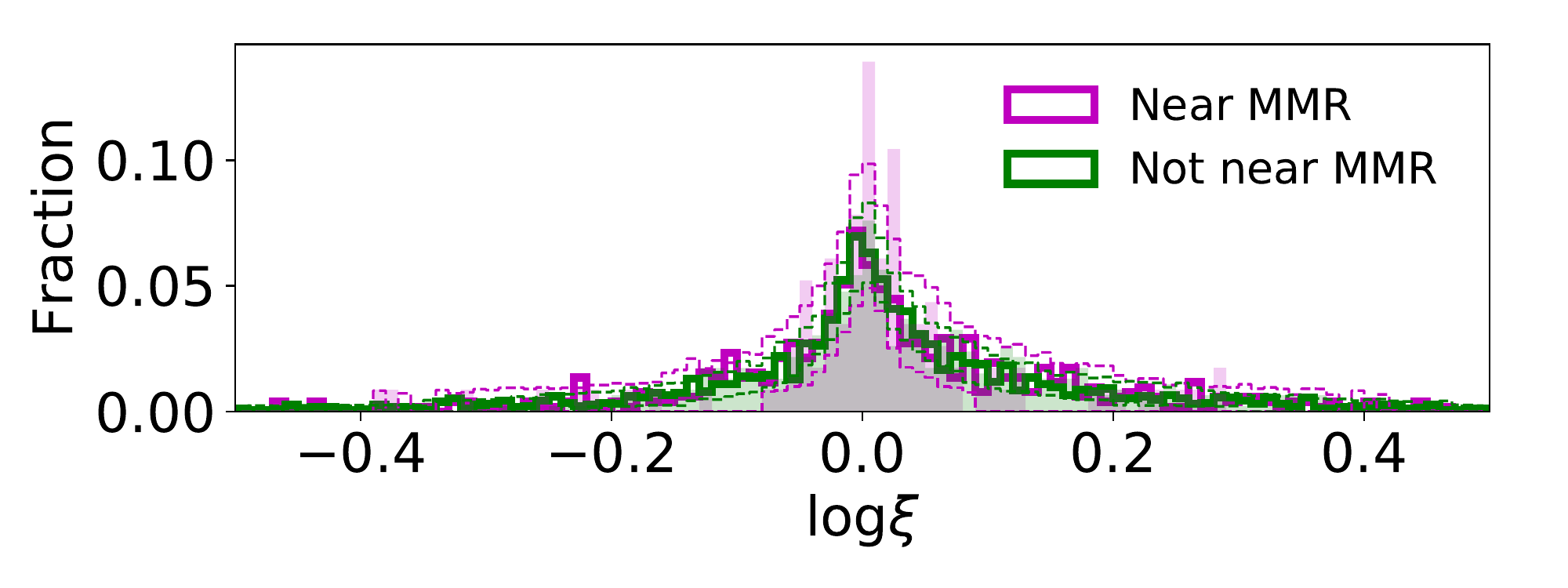} & 
 \includegraphics[scale=0.425,trim={0 0.4cm 0 0.2cm},clip]{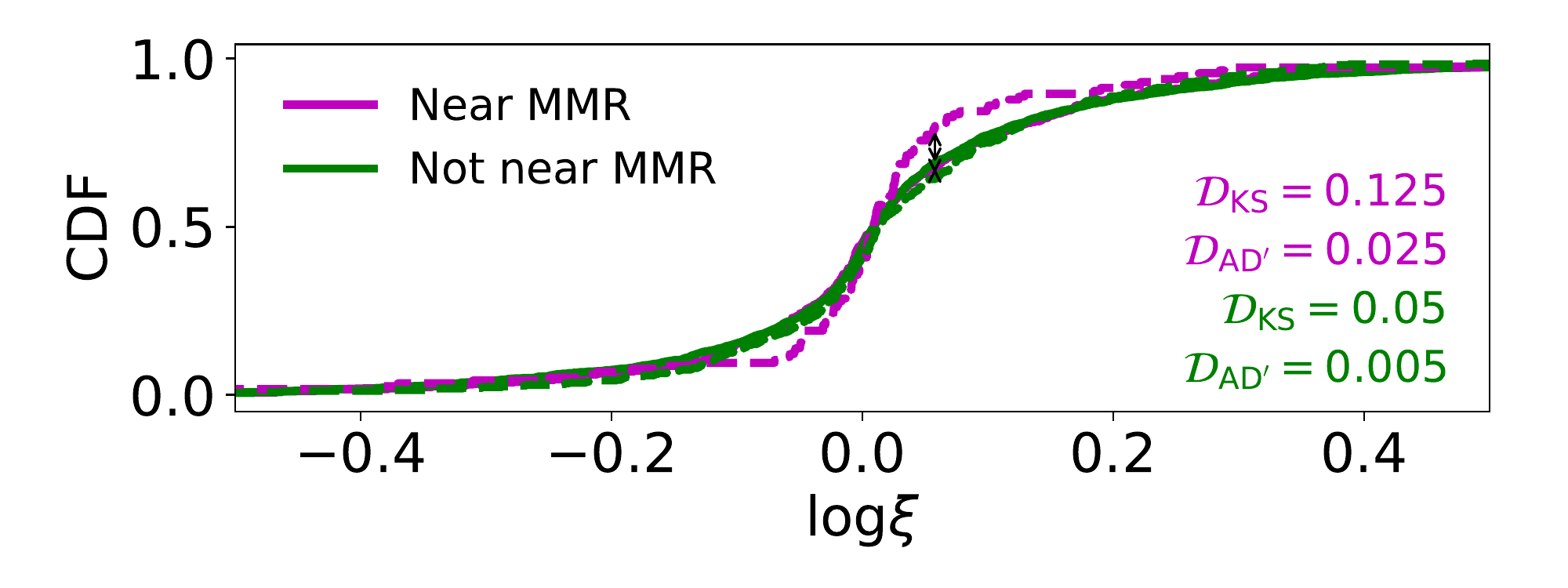} \\
\end{tabular}
\caption{\textbf{Clustered periods and sizes model:} a simulated observed population of exoplanet systems generated from our fully clustered model. The panels are the same as the ones in Figures \ref{fig:non_clustered_model} and \ref{fig:clustered_P_model}. The black hollow histograms show one simulated catalogue (model parameters listed in Table \ref{tab:param_fits}) while the \Kepler{} DR25 exoplanets are plotted as grey shaded histograms for comparison. The red dashed histograms show the 16 and 84 percentiles of each bin based on 100 simulated catalogues with parameters drawn from our emulator with $\mathcal{D}_{W,\rm KS} < 35$.
This model produces a great fit to the observed summary statistics of the data. While the transit depth ratio distribution is improved and peaks more closely around unity due to the clustering in planet sizes, the distances ares not reduced compared to that of the non-clustered model, suggesting that there are still differences yet to be modelled. In particular, the observed \Kepler{} transit depth ratio distribution appears slightly more asymmetric than what our models can produce.
}
\label{fig:clustered_P_R_model}
\end{figure*}

\subsubsection{Clustered periods and sizes model} \label{Clustered_P_R}
Several previous studies have found that the radii of exoplanets observed by \Kepler{} around a single target star are correlated, causing planets to be more similar in size within each system compared to between systems \citep{C2013, M2017, W2018a}. However, most recently these results have been called into question by \citet{Z2019}, who show that the clustered radii can be reproduced by a re-sampling of signal-to-noise ratios and conclude that these correlations are largely due to detection biases. Thus, we extend the clustering point process to include clustering of the planetary radii (and implicitly planet masses). This is the full implementation of our clustered periods and sizes model as detailed in \S\ref{Procedure}.

We plot the results of this model in Figure \ref{fig:clustered_P_R_model} and list the model parameters in Table \ref{tab:param_fits}. 
As is the case in our previous clustered periods model, the distances for the observed multiplicity, period, and period ratio distributions are small, and are significantly improved compared to the non-clustered model. There are no significant differences in the distances (KS and AD) for these observables between the two clustered models. A similar conclusion can be drawn for the transit durations and transit duration ratios; while both clustering in periods and clustering in both periods and sizes fit these marginal distributions equally well, both clustered models provide substantially better descriptions of $\{t_{\rm dur}\}$ and $\{\xi_{\rm non-res}\}$ and perhaps slightly better fits to $\{\xi_{\rm res}\}$ as compared to the non-clustered model.
However, the transit depth distribution appears to be modelled slightly worse than in the previous models. Also, while the transit depth ratios qualitatively appear to be better fit with this model, there is almost no improvement in the KS or AD distance.
A closer examination reveals that there is a slight offset in the CDFs, due to the observed distribution of transit depth ratios of adjacent \Kepler{} planet pairs being asymmetric (in log). 
This suggests that while models with non-clustered planet sizes fail to predict the highly peaked nature of the transit depth ratio distribution, a model with clustered sizes still does not adequately reproduce this property; there are additional features shaping the distribution of adjacent-planet radii ratios that require a more complex model to explain. In particular, modelling the \Kepler{} depth and depth ratio distributions would likely be improved by allowing for a valley in the planet radius distribution (perhaps due to photoevaporation, core heating, or some other process, see \S\ref{Introduction}) that is not included in our model.
We discuss our speculations for how future models can be generalized to better match these features in \S\ref{secFuture}.

\subsection{Differences in the underlying populations (\textit{physical catalogues}) between the models} \label{Model_comparison_underlying}

\begin{figure*}
\begin{tabular}{cc}
 \includegraphics[scale=0.425,trim={0 0.3cm 0 0.2cm},clip]{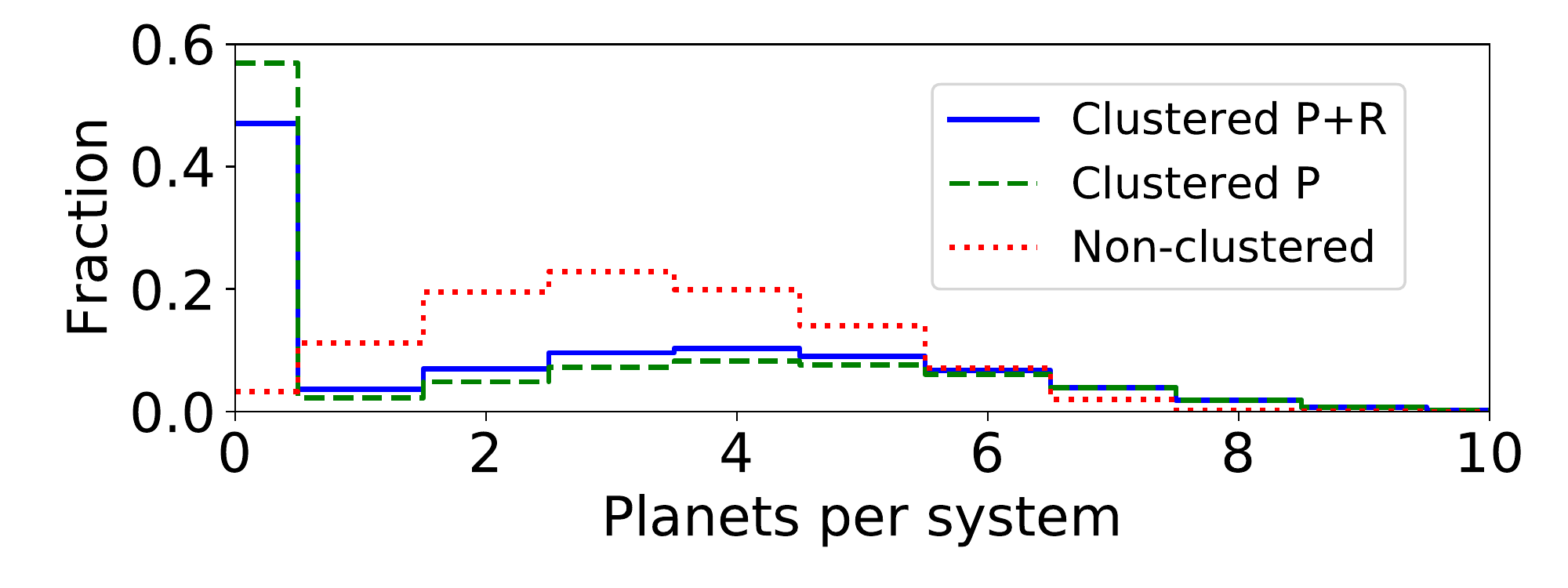} & 
 \includegraphics[scale=0.425,trim={0 0.3cm 0 0.2cm},clip]{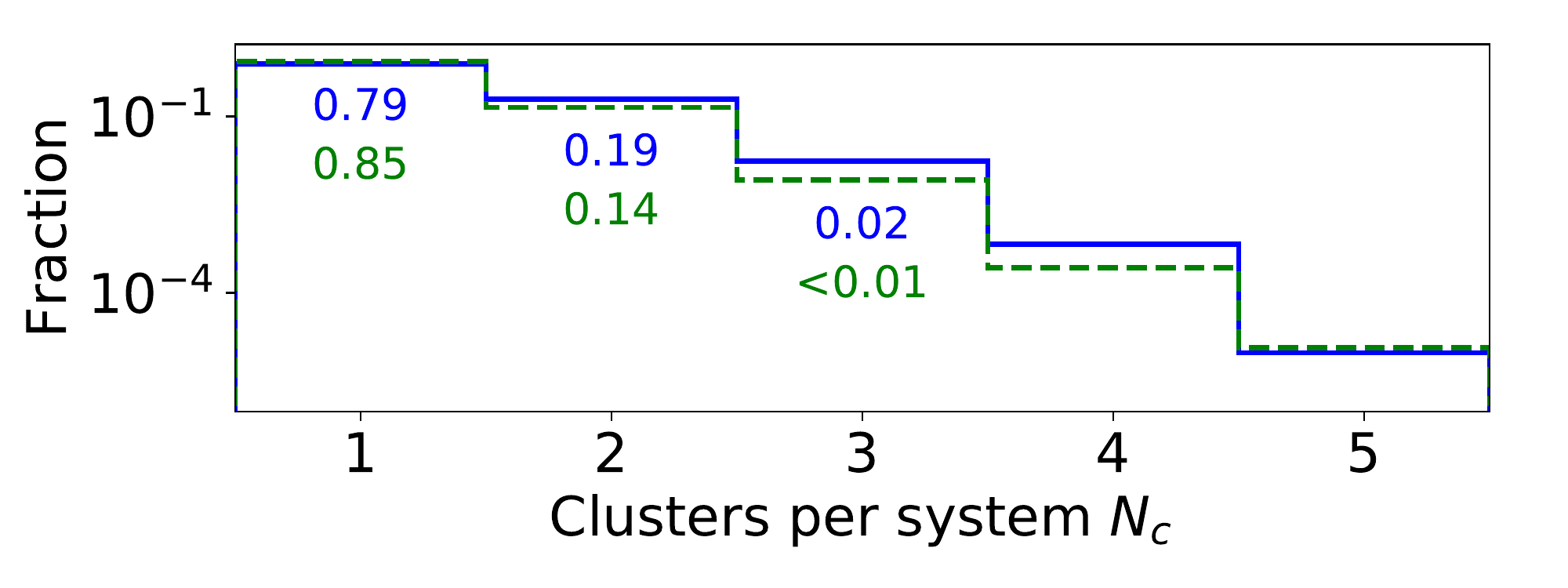} \\
 \includegraphics[scale=0.425,trim={0 0.3cm 0 0.2cm},clip]{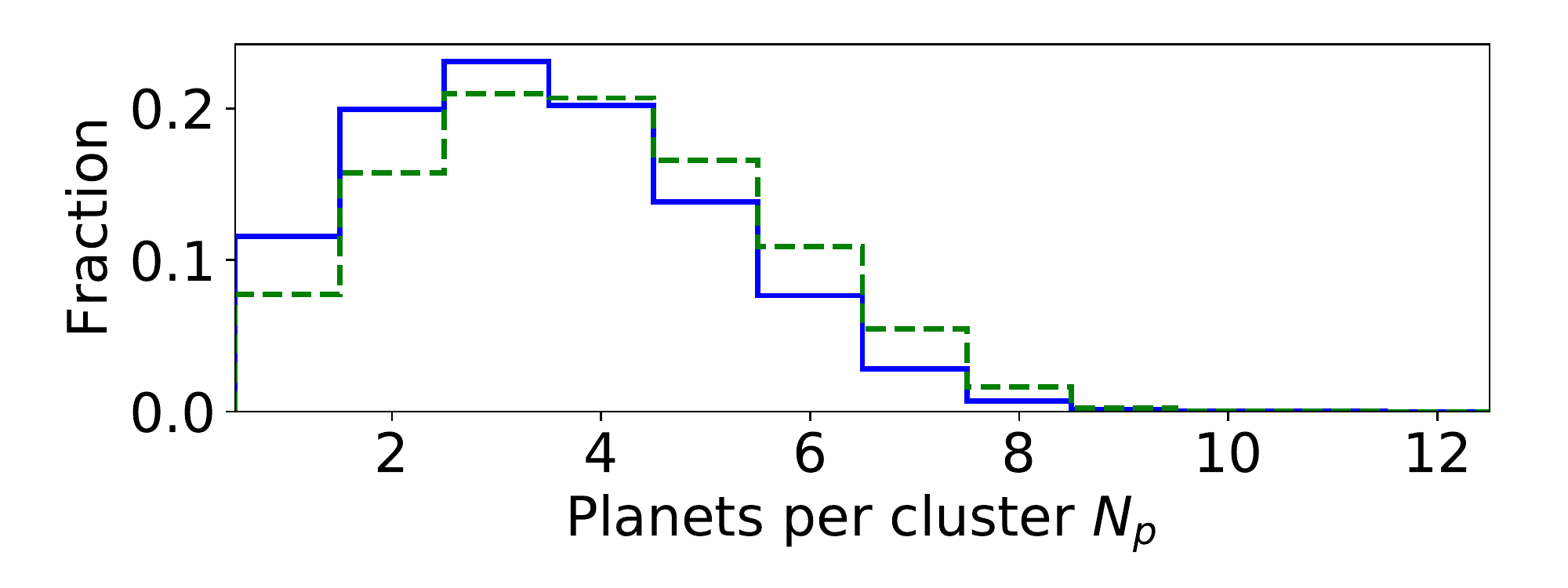} & 
 \includegraphics[scale=0.425,trim={0 0.3cm 0 0.2cm},clip]{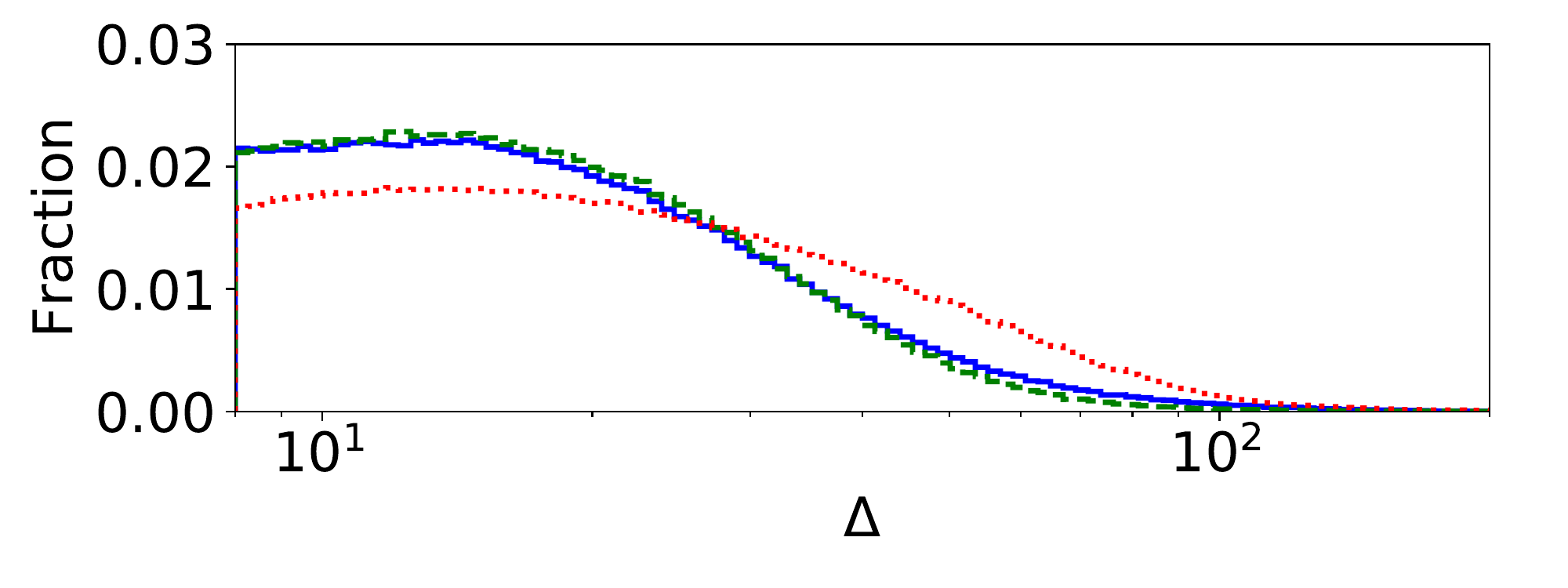} \\
 \includegraphics[scale=0.425,trim={0 0.3cm 0 0.2cm},clip]{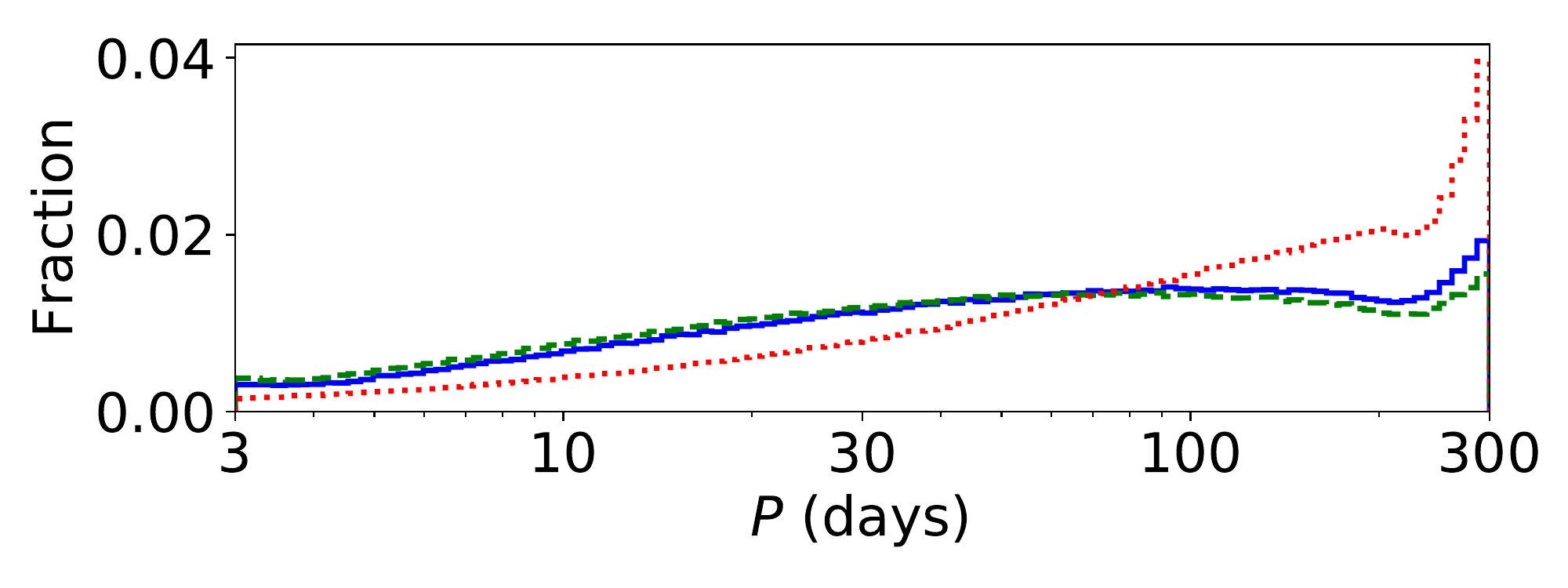} & 
 \includegraphics[scale=0.425,trim={0 0.3cm 0 0.2cm},clip]{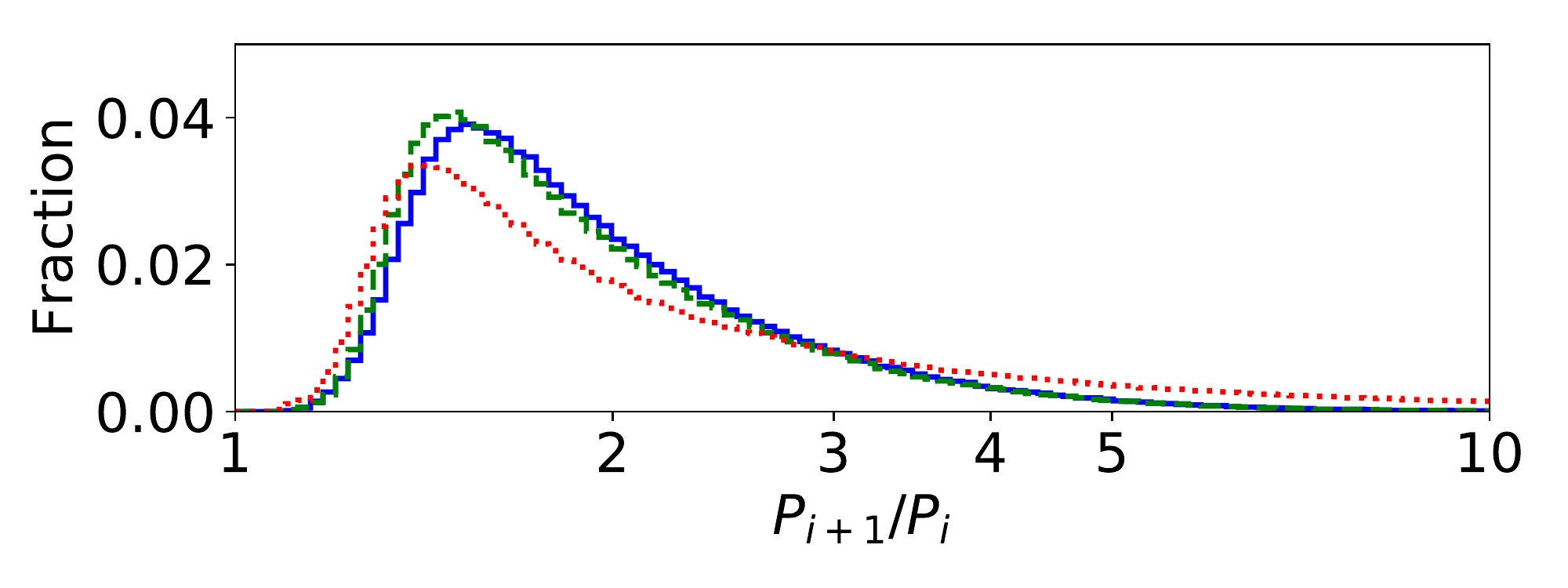} \\
 \includegraphics[scale=0.425,trim={0 0.3cm 0 0.2cm},clip]{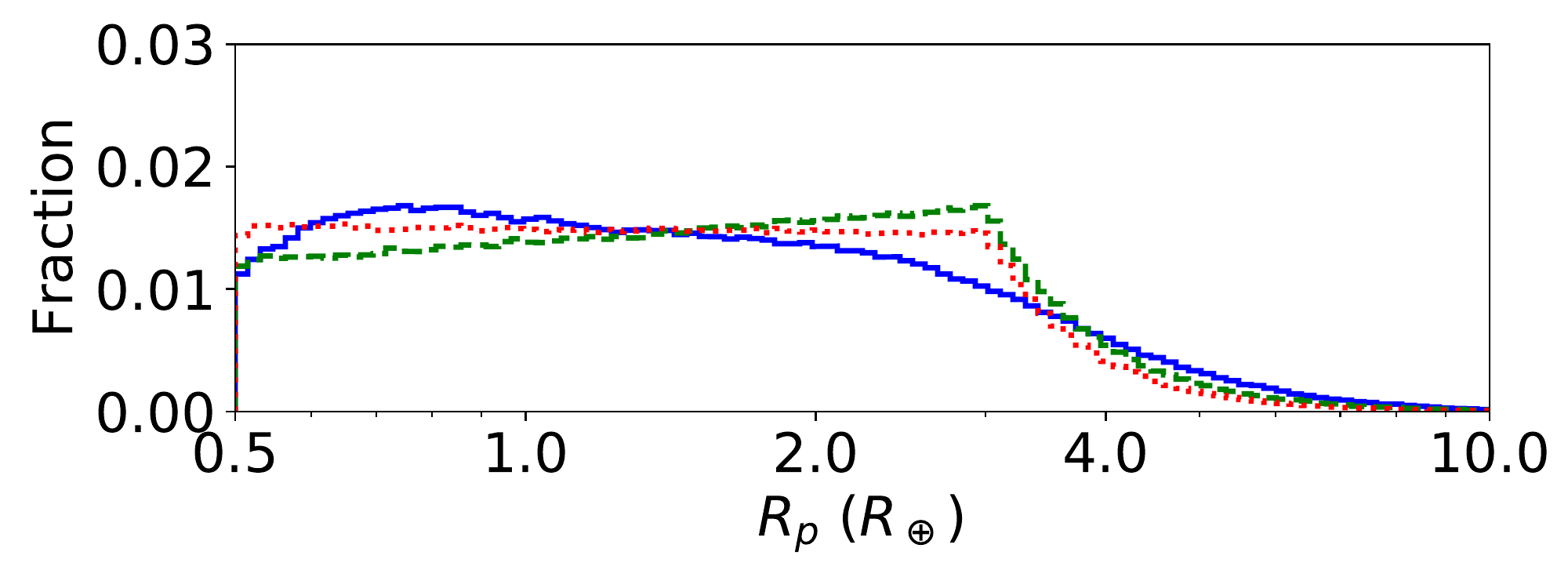} & 
 \includegraphics[scale=0.425,trim={0 0.3cm 0 0.2cm},clip]{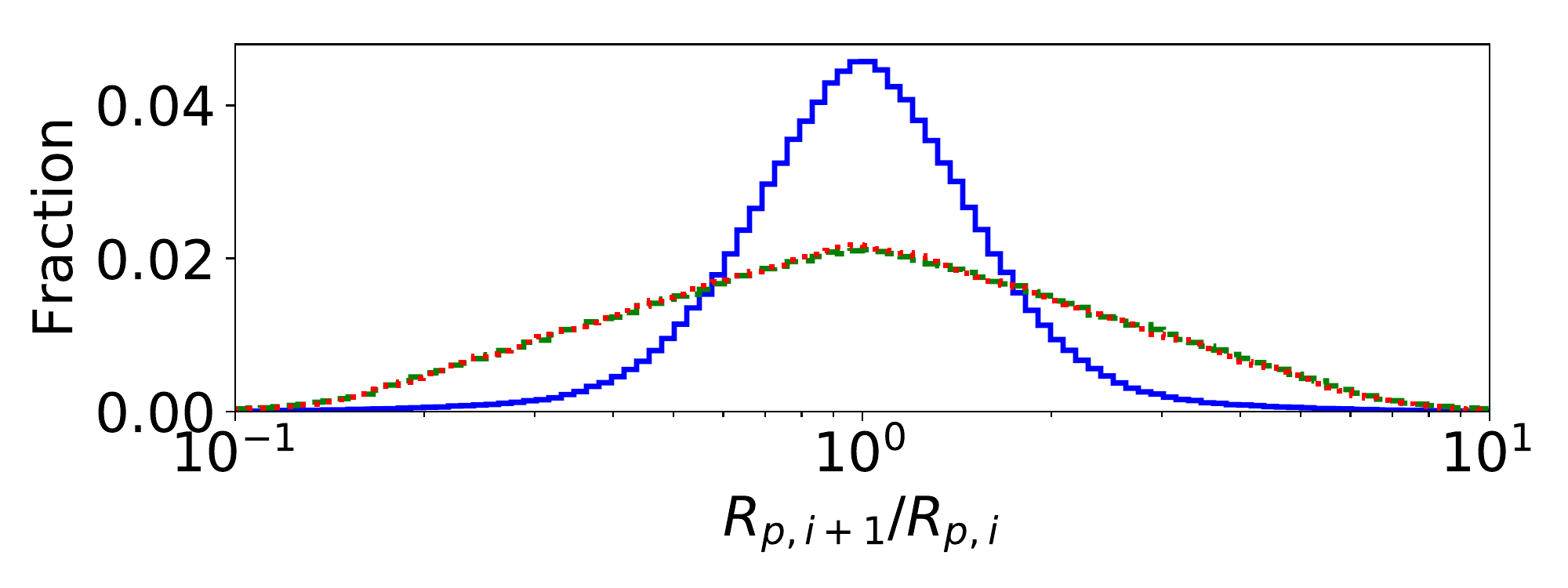} \\
\end{tabular}
\caption{Simulated \textit{physical catalogues} generated by each of our three models. These physical populations correspond to the same populations used to generate the \textit{observed catalogues} shown as black histograms in Figures \ref{fig:non_clustered_model}--\ref{fig:clustered_P_R_model}. From left to right, top to bottom: histograms of the total planet multiplicities, numbers of clusters per system ($N_c$), numbers of planets per cluster ($N_p$), separations in mutual Hill radii ($\Delta$), periods ($P$), period ratios ($\mathcal{P}$), true planet radii ($R_p$), and planet radii ratios ($R_{p,i+1}/R_{p,i}$). 
The non-clustered model (dotted red lines) produces fewer stars with no planets due to the single Poisson distribution describing the number of planets $N_p$, while the clustered models (dashed green and solid blue lines) produce many more such systems due to draws of zero-cluster systems from $N_c \sim {\rm Poisson}(\lambda_c)$. All three models produce shallowly rising period power laws, with a slight pile-up near $P_{\rm max} = 300$ d likely due to edge effects of our rejection-sampling algorithm. Given the same stability criteria of $\Delta \geq \Delta_c = 8$, the clustered models produce narrower $\Delta$ and period ratio distributions, suggesting that planetary systems are tightly spaced. All three models shown here exhibit a radius power law that is relatively flat for small sizes towards the break radius ($R_{p,\rm break} = 3 R_\oplus$) and sharply falls above it. Only the clustered periods and sizes (solid blue) model produces an intrinsic radius ratio distribution strongly peaked around unity; the radius ratio distributions of the other two models reflect what happens if planet pairs with radii drawn from a broken power law are randomly paired.}
\label{fig:models_underlying}
\end{figure*}

\begin{figure*}
\begin{tabular}{cc}
 \includegraphics[scale=0.425,trim={0 0.3cm 0 0.2cm},clip]{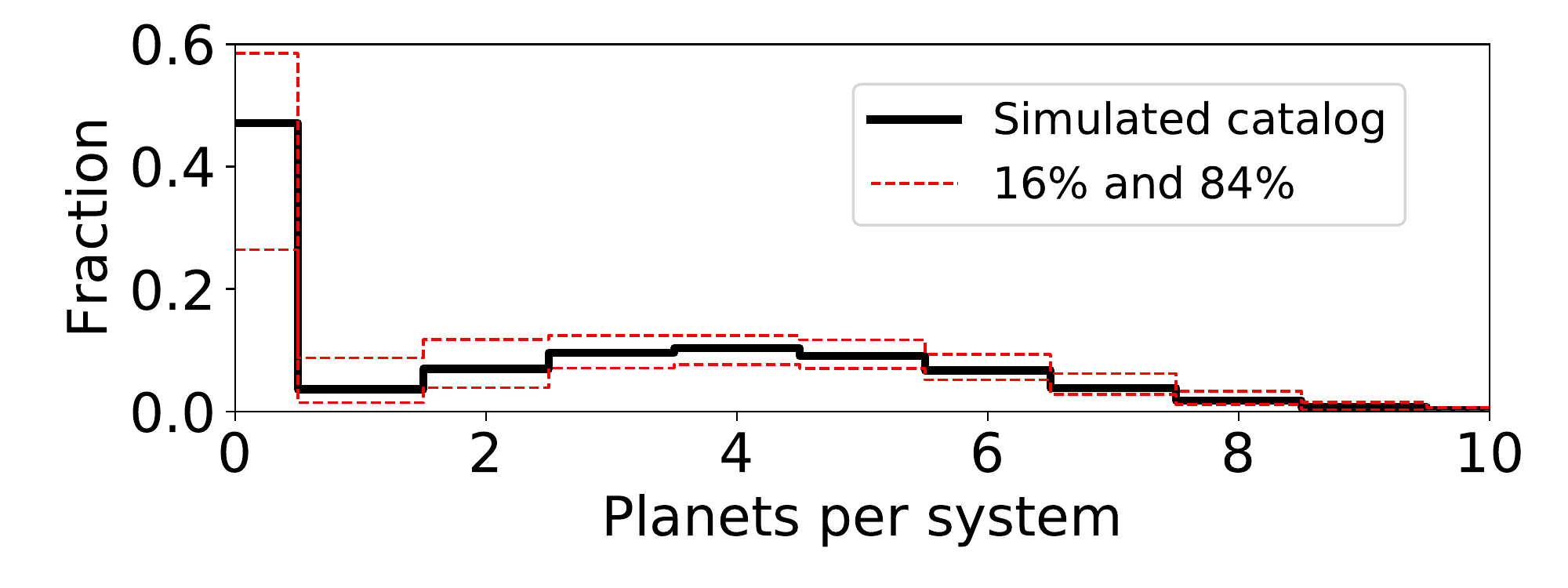} & 
 \includegraphics[scale=0.425,trim={0 0.3cm 0 0.2cm},clip]{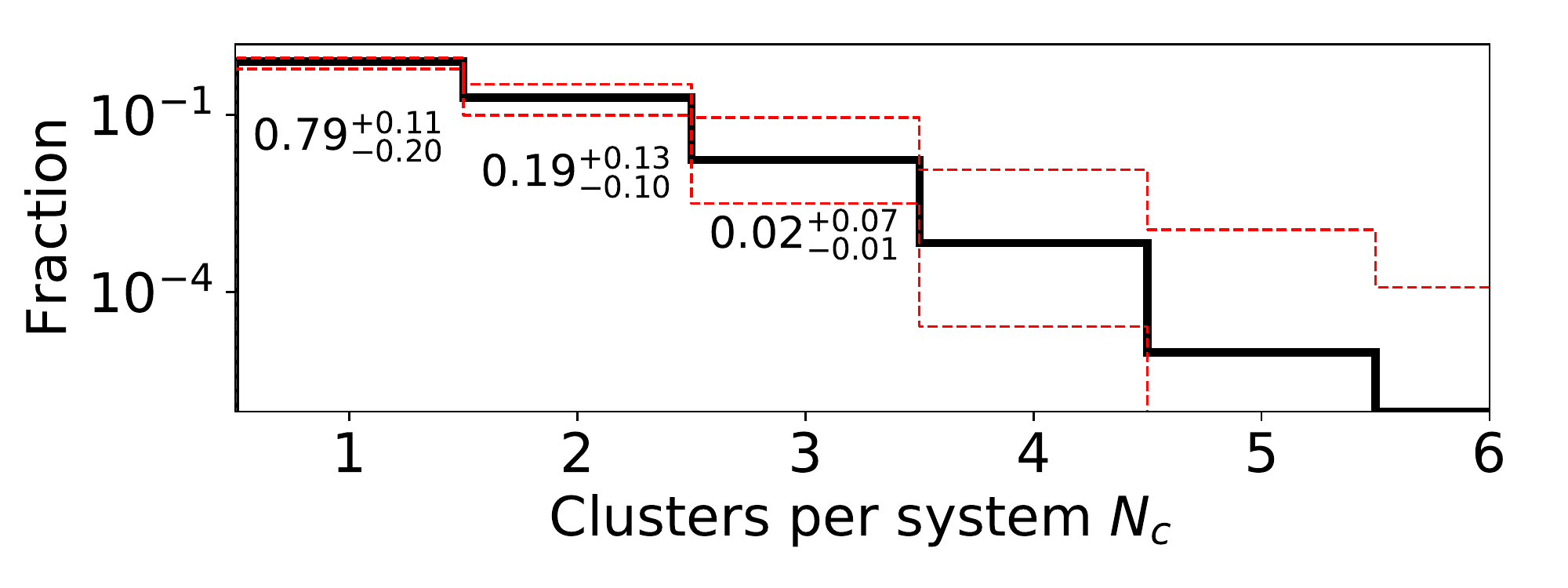} \\
 \includegraphics[scale=0.425,trim={0 0.3cm 0 0.2cm},clip]{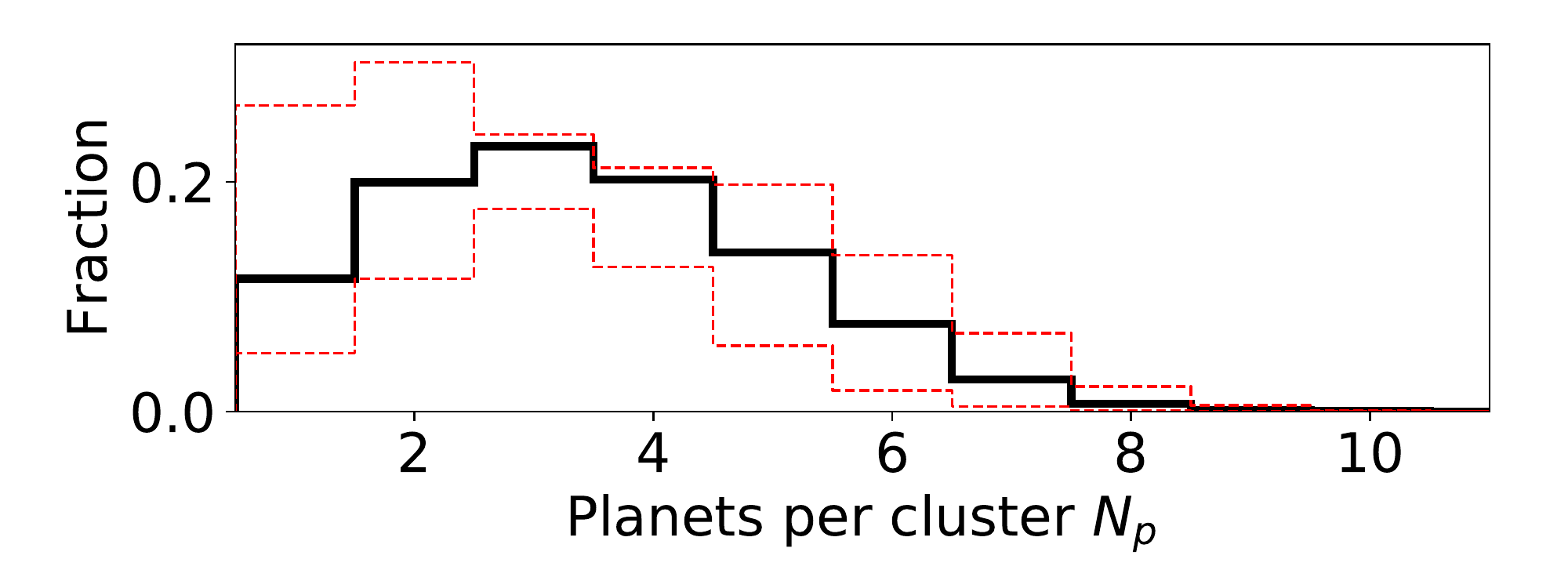} & 
 \includegraphics[scale=0.425,trim={0 0.3cm 0 0.2cm},clip]{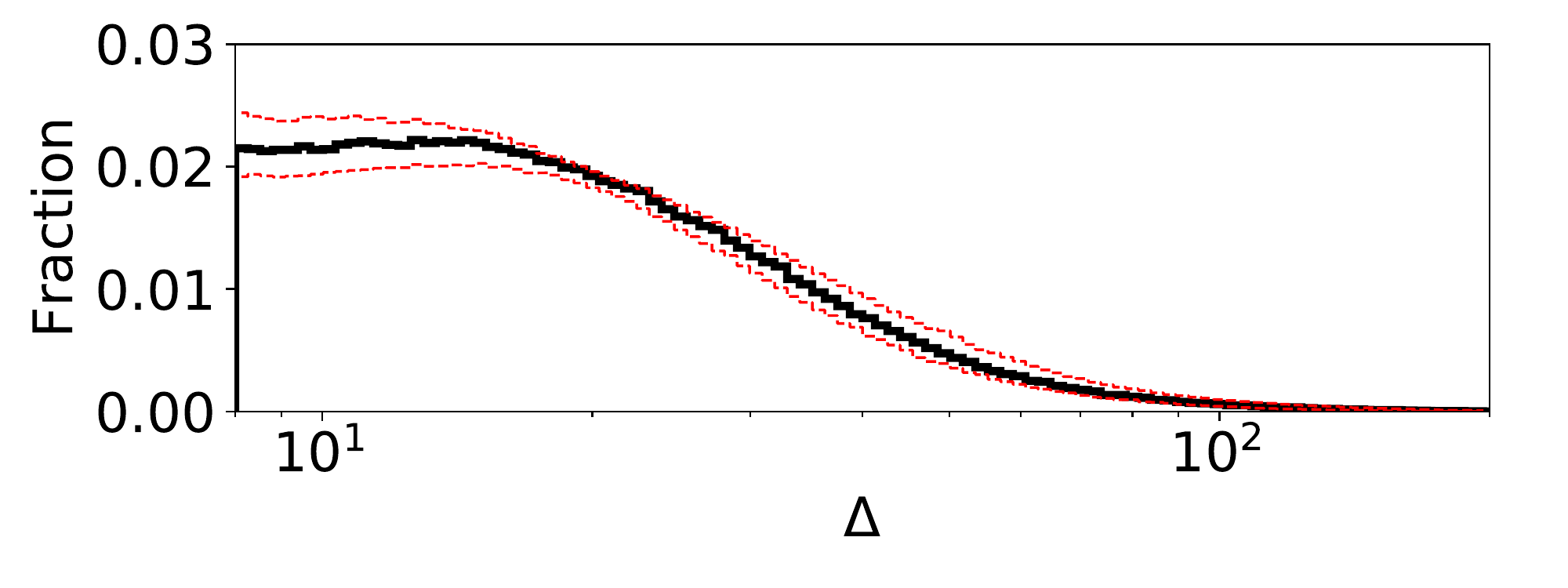} \\
 \includegraphics[scale=0.425,trim={0 0.3cm 0 0.2cm},clip]{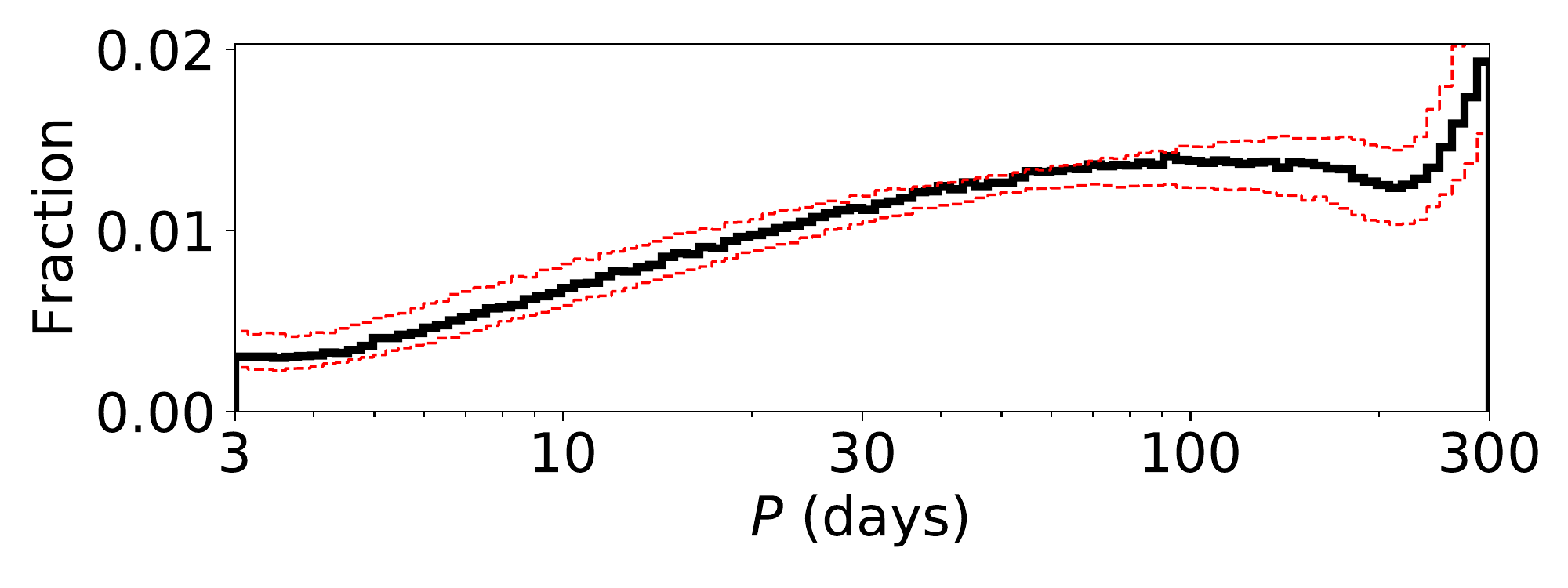} & 
 \includegraphics[scale=0.425,trim={0 0.3cm 0 0.2cm},clip]{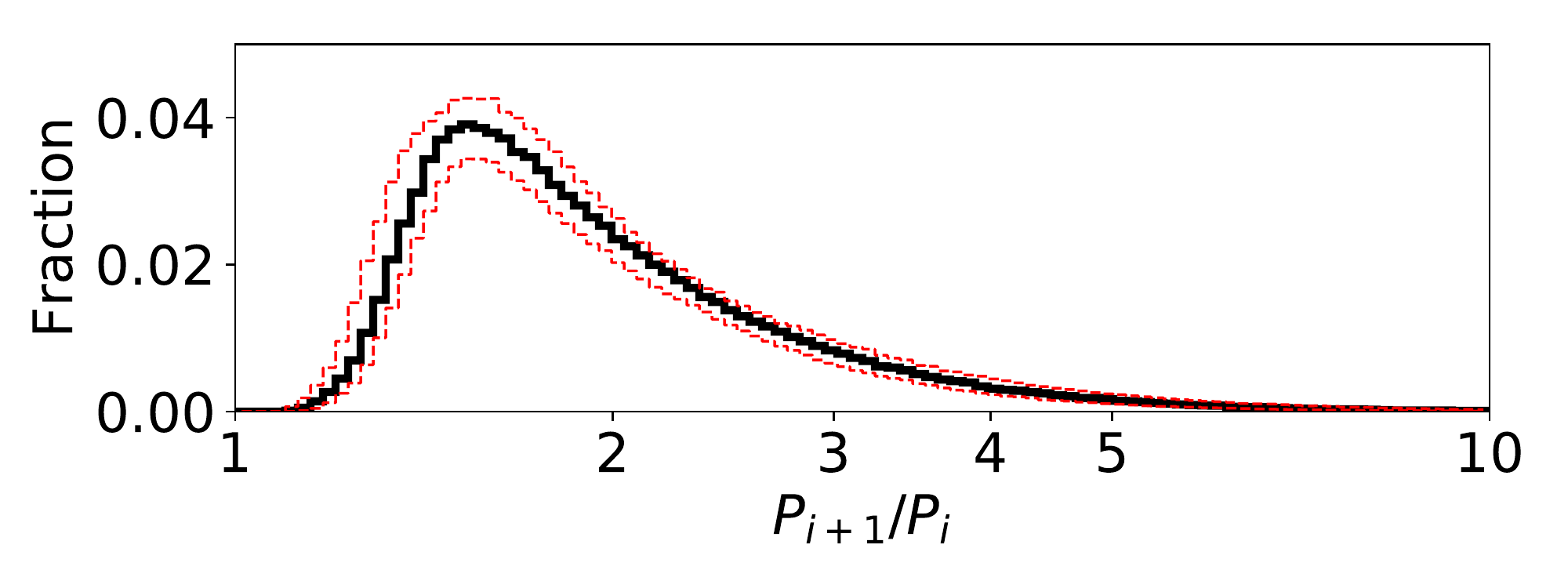} \\
 \includegraphics[scale=0.425,trim={0 0.3cm 0 0.2cm},clip]{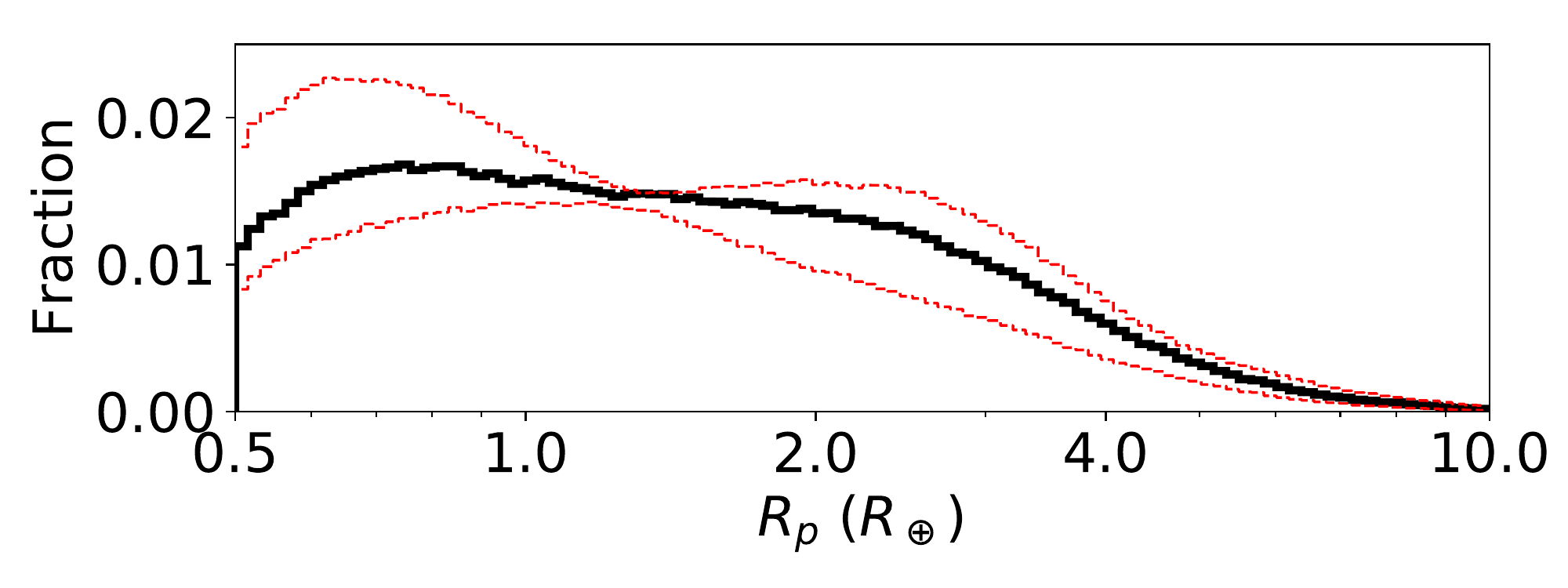} & 
 \includegraphics[scale=0.425,trim={0 0.3cm 0 0.2cm},clip]{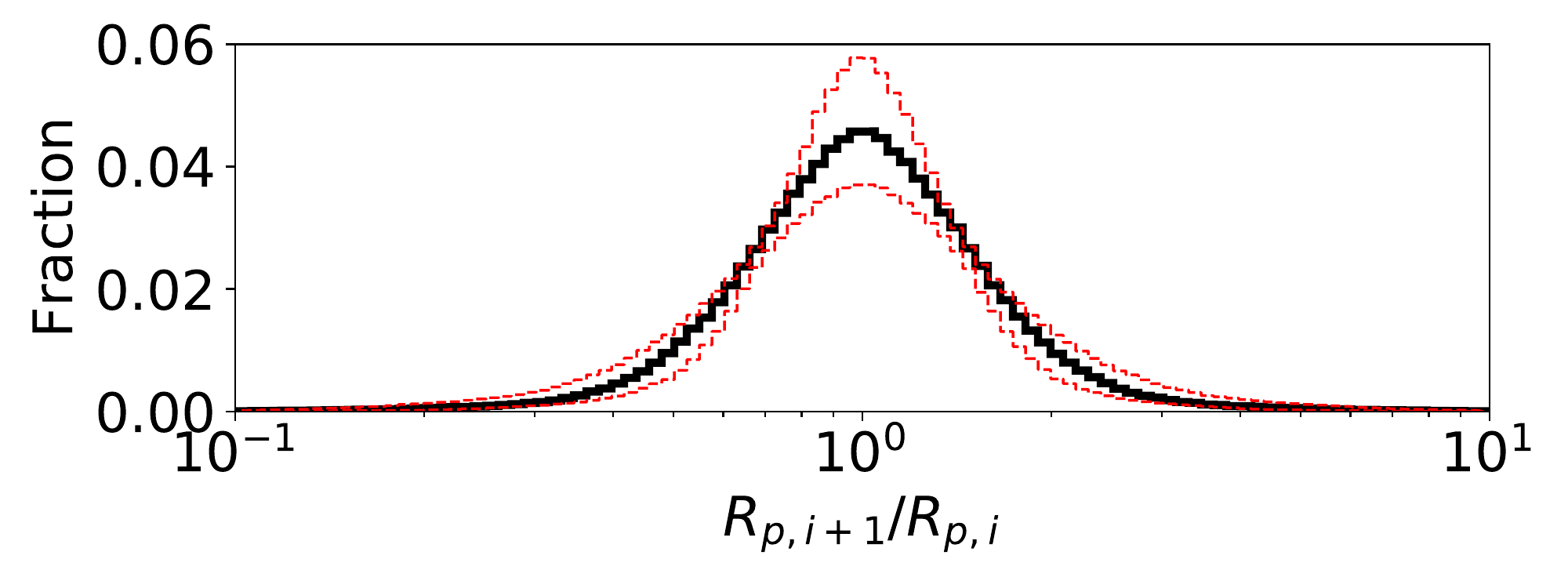} \\
\end{tabular}
\caption{Simulated \textit{physical catalogues} generated from our clustered periods and sizes model. The panels are the same as the ones in Figure \ref{fig:models_underlying}. The solid black histograms correspond to the same population used to generate the \textit{observed catalogue} in Figure \ref{fig:clustered_P_R_model}, while the dashed red histograms correspond to the 16 and 84 percentiles of each bin based on 100 simulated catalogues with parameters drawn from our emulator with $\mathcal{D}_{W,\rm KS} < 35$.
}
\label{fig:clustered_P_R_underlying}
\end{figure*}

A major benefit of forward modelling the \Kepler{} mission is that we can directly analyze the predictions of our models for the true, underlying exoplanetary systems. While the model parameters describe the underlying system properties including the rate of planets and clusters per system, the period and radius distributions, and the orbital architectures, the \textit{physical catalogues} generated by our models can be directly examined to see Monte Carlo realizations of populations of planetary systems. In Figures \ref{fig:models_underlying} and \ref{fig:clustered_P_R_underlying}, we plot the underlying distributions of planetary systems as generated by one instance of each model and 100 realizations of the clustered periods and sizes model, respectively. The \textit{physical catalogues} shown here are the same simulated populations generating the \textit{observed catalogues} as shown in Figures \ref{fig:non_clustered_model}--\ref{fig:clustered_P_R_model}. We make the following observations:

\begin{itemize}
 \item \textbf{Planets and clusters per system:} There is a huge difference in the way planets are distributed across different systems between our clustered and non-clustered models. The non-clustered model (dotted red) produces very few ($\lesssim 5\%$) systems with no planets due to the single Poisson distribution describing the number of planets, $N_p \sim {\rm Poisson}(\lambda_p)$. In this model, the most common planetary system consists of three planets (between 3 and 300 d and 0.5--10$R_\oplus$). The clustered models (dashed green and solid blue) produce many more systems with no planets due to draws of zero-cluster systems from $N_c \sim {\rm Poisson}(\lambda_c)$ (since the number of planets per cluster is a zero-truncated Poisson distribution, $N_p \sim {\rm ZTP}(\lambda_p)$ in the clustered models, and $\lambda_c$ tends to be small). For systems with planets, the multiplicity tends to be higher; the most common planetary system is a four-planet system. While there is a clear preference for clustered planetary systems, the number of actual clusters tends to be small. For planet-harbouring stars, $\sim$80\% of systems have just a single cluster consisting of three or four planets (over the range of periods considered, 3--300 d).

 Thus, assuming that planetary systems are highly clustered, our model predicts a large fraction of stars with no planets larger than 0.5 $R_\oplus$ with periods between 3 and 300 d, while the remaining systems to have many planets in this range.  We discuss the occurrence rates of planets in more detail in \S\ref{Planet_rates}.

 \item \textbf{Orbital period distribution:} All three models exclude flat or falling power laws in log-period (i.e, $\alpha_P \leq -1$). Instead, our assumed power law for the period distribution leads to a shallowly increasing planet occurrence as a function of log-period. There is a pile-up of planets near $P_{\rm max} = 300$ d, likely due to edge effects of our rejection-sampling algorithm; this pile-up is most severe in our non-clustered model. However, this has a minimal effect on our observed catalogues, as the probability of transits decreases with longer periods. Additionally, the period distributions produced by our clustered models appear to be ``rounded'' and deviate from a power law, due to the draws of period clusters.
 
 \item \textbf{Spacings of adjacent planets:} Given the same stability criteria of $\Delta \geq \Delta_c = 8$, the clustered models produce slightly narrower separation ($\Delta$) and underlying period ratio distributions, suggesting that planetary systems are more tightly spaced than one would infer from a model with periods drawn independently. All three models produce distributions of separations in mutual Hill radii that are sharply truncated at $\Delta_c$, suggesting that many period or period scale draws are attempted more than once. The distribution of $\Delta$ begins to fall out around $20$, which is similar to the findings of \citet{W2018a} although they assumed a much simpler mass--radius relationship \citep{WM2014} and only analysed the spacings of observed planets. The underlying period ratio distributions are highly peaked around $\sim 1.5$.
 
 \item \textbf{Planet radius distribution:} As listed in Table \ref{tab:param_fits}, we chose $\alpha_{R1} = -1, -0.8$, and $-1$ for these realizations of the non-clustered, clustered periods, and clustered periods and sizes models, respectively. As expected, the radius distribution is flat for small sizes up to the break radius ($R_{p,\rm break} = 3 R_\oplus$) and sharply falls above it for this non-clustered model. However, the catalogue generated from the clustered periods and sizes model exhibits a rounded distribution for small radii and the break radius is not as clear (despite also having set $\alpha_{R1} = -1$), due to the clustering in radii which tends to smooth out the distribution. The catalogue shown here from the clustered periods model has a radius distribution that slightly increases up towards the break radius, although the credible regions for $\alpha_{R1}$ are consistent with a flat distribution for all three models.
 
 \item \textbf{Planet radius ratios:} The clustered periods and sizes (solid blue) model produces a radius ratio distribution strongly peaked around unity. The effect of this peak increases for decreasing values of $\sigma_R$ due to more highly clustered planet radii. The radius ratio distributions of the other two models reflect what the distribution looks like if planet pairs with radii drawn independently from a broken power law are randomly paired.
\end{itemize}

\subsection{Implications for the fraction of stars with planets} \label{Planet_rates}

Since we also fit our simulated results to match the overall rate of observed planets per surveyed (FGK) star in the \Kepler{} mission, we can directly estimate the true fraction of stars with planets and the mean rate of planets per star, given various ranges of period and planetary radius. We use the same 100 simulated catalogues generated for Figure \ref{fig:clustered_P_R_underlying} to calculate the following planet occurrence rates (and their 16\% and 84\% quantiles) from our clustered periods and sizes model.

{\em Planets between 0.5 and 10$R_\oplus$:}
The fraction of stars with planets (FSWP) with 3 d $<P<$ 300 d and $0.5R_\oplus < R_p < 10 R_\oplus$ is $0.56_{-0.15}^{+0.18}$.  Roughly half of all FGK stars have at least one planet in this range based on our clustered periods and sizes model.  In contrast, our non-clustered model gives a remarkably different result, producing just a few percent ($\sim 3\%$) of stars with no planets (see the dotted red curve in Figure \ref{fig:models_underlying}). This emphasizes the importance of adopting a clustered model; while both our clustered and non-clustered models can fit the overall rate of planets equally well, the distribution of these planets across systems is very different.
A similar result was drawn by \citet{WSS2012}, who studied the intrinsic multiplicity and mutual inclination distributions and concluded that either planet occurrence is correlated between planets in the same system and/or some stars are significantly more planet-rich than others.
Returning to our clustered periods and sizes model, the mean number of planets per star is $2.28_{-0.53}^{+0.94}$. This rate is somewhat less than the credible region values of $\lambda_c \lambda_p$, due to the rejection-sampling procedure. Excluding the stars with no planets (in this range of periods and sizes), the mean number of such planets per system rises to $4.26_{-0.37}^{+0.52}$.

\citet{Z2018} find that the fraction of Sun-like stars with at least one ``\Kepler-like'' planet is $30\pm3\%$ based on multiplicity and rate of TTVs, although they define ``\Kepler-like'' to only include planets larger than $1R_\oplus$ (and periods less than 400 d).
In contrast, \citet{ZCH2019} find that $0.72^{+0.04}_{-0.03}$ of stars have at least one planet (with orbital periods between 0.5 and 500 d and radii between 0.5 and $16 R_\oplus$), while modelling just the observed periods, radii, and multiplicity. \citet{ZCH2019} report $8.4\pm0.31$ planets per star hosting a planetary system (but excluding stars that host a single transiting planet with $R_p\ge 6.7R_\oplus$). This implies that most stars have at least seven planets over the range they consider and only 5.5\% of planets have one to four such planets.

Our finding of $0.56_{-0.15}^{+0.18}$ FGK dwarf stars hosting at least one planet is greater than that suggested by \citet{Z2018} and less than that suggested by \citet{ZCH2019}.  While each study considers a slightly different range of orbital periods and sizes, we attribute the bulk of the differences relative to our model as due to our use of a clustered model for the distribution of orbital periods within a planetary system.  When we applied a non-clustered model, we find $\sim 97\%$ of stars host at least one planet.  Switching to a model which accounts for clustering in orbital period and planet size (or just orbital periods) dramatically increases the fraction of stars with no planets in our model.  Similarly the mean number of planets per star with at least one planet (within the range of periods and sizes considered) increases from $3.41_{-0.39}^{+0.31}$ to $4.26_{-0.37}^{+0.52}$ when switching from a non-clustered to a clustered model (see Figure \ref{fig:clustered_P_R_underlying}).  This demonstrates the importance of allowing for clustering when performing inference on the fraction of stars with planets or mean number of planet per stars with at least one planet.

Another important way in which our model differs from those of \citet{Z2018} and \citet{ZCH2019} is that we allow for the mutual inclination distribution to have both a low and high inclination component.  Since we treat the fraction of systems from the high inclination component as a free parameter, our simulations could have resulted in $f_{\sigma_{i,\mathrm{high}}}$ consistent with zero.  Instead, values of zero were strongly excluded in both clustered models, regardless of which distance function was chosen.  Interestingly, the simulations using a non-clustered model did result in small values of $f_{\sigma_{i,\mathrm{high}}}$ and did not exclude $f_{\sigma_{i,\mathrm{high}}}=0$.
This finding also underscores the importance of allowing for clustering when performing inference on the mutual inclination distribution of planetary systems. 

{\em Earth-sized planets:}
We provide results for the rate of Earth-sized planets (here defined to be 0.75--1.25$R_\oplus$) around FGK stars, using our clustered periods and sizes model.  
The fraction of stars with at least one Earth-sized planet (between $3-300$ days) is $0.42_{-0.16}^{+0.24}$. 
The mean number of these planets per star is $0.85_{-0.30}^{+0.51}$. 
Considering systems with at least one planet (with sizes between 0.5 and 10$R_\oplus$ and periods between 3 and 300 d), the probability that a planetary system contains an Earth-sized planet (in the same period range) is $0.75 \pm 0.14$. 
If we focus on planetary systems with at least one planet (not necessarily Earth-size), then the mean number of Earth-sized planets per system is $1.62_{-0.38}^{+0.23}$ (both with periods 3--300 d). Broadly, about half of stars have an Earth-sized planet and most inner planetary systems have Earth-sized planets. Due to the increasing frequency of planets at longer periods, most of these Earth-like planets are in 100--300 d orbits.  These conclusions follow directly from combining the inferred distributions for orbital period and planet radius (that are consistent with previous results) with the inferred fraction of stars with planetary systems.

\subsection{\Kepler{} dichotomy} \label{Dichotomy}

Previous studies have shown that many simple parametrized models with a single population significantly underpredict the number of systems with only one transiting planet. Since this implies two populations, this result is known as the \Kepler{} dichotomy.  
This paper focuses on models that allow for two populations for the mutual inclinations, each with similar distributions for the remaining properties of the system (i.e., no explicit dichotomy in the intrinsic distributions of multiplicity, orbital period, planet size, or eccentricity).  
A model with a single inclination population is nested inside our model.  
Therefore, we can examine whether our posterior distribution for the fraction of stars with a high mutual inclination planetary has significant weight near zero.  
For each of the clustered models and distance functions we considered, we find that the predicted fraction of high-inclination systems ($40 \pm 10$\%) is inconsistent with zero, supporting the theorized \Kepler{} dichotomy.

Another proposed solution to the \Kepler{} dichotomy is an alternative model consisting of one population of high-multiplicity systems (i.e., very similar to our low-inclination population) and a second population of systems with a single planet (both within 3 d $<P<$ 300 d and $0.5 R_\oplus < R_p < 10 R_\oplus$)\footnote{While our range of periods and radii includes many Hot Jupiters which are thought to be relatively isolated, the fraction of stars with such systems is known to be quite low ($\sim$1\%) and does not affect our conclusions about the origin of the \Kepler{} dichotomy.}. 
As shown in \S\ref{Planet_rates}, the average number of planets per planetary system (i.e., star with at least one planet) is $\left<{\rm NPPS}\right> \sim 4.3$ and the fraction of stars with at least one planet is ${\rm FSWP} \sim 0.56$ in our fully clustered model.
Therefore, the average star coming from the high-inclination population in our model actually hosts multiple detectable transiting planets, greatly increasing the fraction of viewing angles for which the star would be perceived as hosting a single transiting planet.  

Using values from our fully clustered model, 
simply replacing all the stars with at least one planet ($\simeq 0.56$) and assigned to the high mutual inclination population ($f_{\sigma_{i,\rm high}} = 0.42$) with stars hosting a single planet, would require ${\rm FSWP} \times f_{\sigma_{i,\rm high}} \times \left<{\rm NPPS}\right> \simeq 101\%$ of stars to host a single planet in order to provide the same number of planets (or the same number of systems with a single transiting planet), leaving no stars available to host multiple planet systems.
We can reject this model for the \Kepler{} dichotomy, since the only way to fit enough systems with one transiting planet around stars that are not already spoken for is to have more than one (highly inclined) planet per star.

Before completely discarding an abundance of single-planet systems as an explanation for the \Kepler{} dichotomy, it is worth considering this argument in more detail.
In the fully clustered model, $\simeq 29\%$ of planets around stars nominally in the high-inclination population were actually assigned a low inclination (relative to the system mid-plane), due to the planet being near a first-order MMR with another planet in the same system.  
This results in a subset of the planet pairs from the high-inclination population contributing to the number of systems observed to have two transiting planets and detracting from the fraction of viewing geometries that the system would be observed as a single planet system.
We consider the possibility of replacing the planets with truly high-inclinations with a population of intrinsic single-planet systems by considering the fraction of stars hosting each of three types of planetary systems:  
(1) the fraction of stars with planetary systems (i.e. with at least one planet) initially assigned to the low-inclination population is ${\rm FSWP} \times (1 - f_{\sigma_{i,\rm high}}) \simeq 32\%$; 
(2) the fraction of stars hosting planets in a system initially assigned to the high-inclination population, but containing planets assigned a small mutual inclination due to being near a first-order MMR, is ${\rm FSWP} \times f_{\sigma_{i,\rm high}} \times 0.29 \simeq 7\%$; and 
(3) the fraction of stars required to host exactly a single planet.
The final category would need to be $101\% \times (1 - 0.29) \simeq 71\%$ of stars in order to fully replace the planets that remain assigned to high mutual inclinations.
Thus, in the fully clustered model, the sum of the three populations comes to $\simeq 110\%$ of target stars. The net effect of accounting for the reassignment of high inclination planets to the low inclinations is to reduce, but not eliminate the tension in having enough stars to accommodate the three populations. Repeating this calculation over 100 realizations of our fully clustered model results in the total number of stars required to be 110\% $\pm$ 23\%, exceeding unity 63 times. Using the clustered periods model, the sum exceeds unity for 12 of 100 realizations.

A similar conclusion is drawn by applying the same argument to our fully clustered model without the special treatment of planets near an MMR. In this model, the fraction of high mutual-inclination systems is somewhat lower ($f_{\sigma_{i,\rm high}} \sim 0.31$, as discussed in \S\ref{Params_incl}), but none of the planets in these systems are reassigned to a lower mutual inclination. The fraction of stars with planets is higher, ${\rm FSWP} \sim 0.67$. The mean number of planets per planetary system is similar, $\left<{\rm NPPS}\right> \sim 4.2$, meaning that we would require ${\rm FSWP} \times f_{\sigma_{i,\rm high}} \times \left<{\rm NPPS}\right> \simeq 87\%$ of stars to host a single planet. Again, this does not leave enough stars to host the planetary systems from the low mutual-inclination population, for which we require ${\rm FSWP} \times (1 - f_{\sigma_{i,\rm high}}) \simeq 46\%$ of stars. Repeating this calculation results in the same conclusion for 73 of 100 realizations.

Therefore, while explaining the \Kepler{} dichotomy with a population of single planet systems cannot be strictly excluded by our work, it is disfavoured and strongly constrained. Even in the case where we cannot formally exclude it, the fraction of stars with planets is quite high, requiring extremely efficient planet formation. 
Since our model does not include planets with $P<3$ d and/or $R_p<0.5R_\oplus$, including stars with such planets could further increase the number of stars required to host multiple planet systems.  
Thus, explaining the \Kepler{} dichotomy with a population of excess singles leaves very little, if any, room for stars harbouring no planets.

Similarly, we could consider an alternative model in which the excess of stars observed as a single transit system is due to a population of highly excited two-planet systems. In this scenario, the fraction of stars hosting two highly inclined planets would need to be $\simeq 36\%$, similar to the fraction of stars with planetary systems drawn from the low mutual inclination population.
While this rate is significantly higher than the ${\rm FSWP} \times f_{\sigma_{i,\rm high}} \simeq 24\%$ of stars in our standard fully clustered model, there are sufficient stars that the \Kepler{} dichotomy could be explained by population of highly inclined systems with two (or more) planets.
However, in order to have enough stars for such model, we had to relax the assumption that the number of cluster per stars is drawn from a Poisson distribution.  Instead, we must allow for the number of zero planet systems to be decoupled from $\lambda_c$, the rate of clusters per star, which is inferred from the observed multiple planet systems.
Using our clustered periods model leads to similar conclusions.
Of course, these conclusions must be interpreted in the context of the parametrization of our model, e.g., the use of Poisson and Rayleigh distributions. 
Altogether, our results suggest that the \Kepler{} dichotomy is most easily explained by two populations with different inclination distributions, as opposed to different multiplicity distributions. 

\citet{ZCH2019} proposed that the \Kepler{} dichotomy could be explained by a combination of geometric transit probability, a distribution of mutual inclinations, and a detection efficiency model that accounts for the order of planet detection by the \Kepler{} pipeline.  
For their star and planet sample, they report that the observed ratio of double to single transit systems is 4\% larger than that predicted by their model when ignoring how the detection efficiency depends on the order of detection.  When using their model for how the detection efficiency differs for subsequent planets, this difference is reduced to 2\%.
However, \citet{ZCH2019} did not attempt to make use of information contained in the distributions of orbital period ratios, transit depth ratios, or transit duration ratios of planets in a single planetary system, as was done in this study. 
We find that these provide valuable information for characterizing the distribution of planetary architectures and evidence for the \Kepler{} dichotomy.

It is also useful to compare the planet detection efficiency models of the two studies. \citet{ZCH2019} fit separate detection efficiency curves for the first planet (technically the first threshold crossing event, TCE) to be detected by the \Kepler{} pipeline and for subsequent planets (or other TCEs).  For planets with orbital periods less than 200 d (i.e., the vast majority of detections in our sample), they find that the expected MES needed for a 50\% probability of detecting the planet increases from $\simeq 8.3$  (for the first planet) to $\simeq 9.0$ (for the second planet).  The detection efficiency rises rapidly with expected MES, so the more important effect is that their model for the detection probability of a transiting planet with high signal-to-noise asymptotes to 0.982 (for the first planet), but only to 0.928 for subsequent planets in the same system.

In contrast, our study used a detailed planet detection and vetting efficiency model based on \citet{H2019}.  It was also derived by fitting a model to the results of the \citet{C2017} pixel-level transit injection study.  One key difference is that it includes both the probability that a planet was detected by the pipeline and that it was labelled as a planet by the robovetter.  
A second important difference is that the detection model of \citet{H2019} includes an explicit dependence on the number of transits observed (and an implicit dependence on the orbital period).  
Most detected planets have 38 or more observed transits, in which case the detection efficiency approaches 0.945 for high SNR.  For planets with 5--36 transits, our model asymptotes at 0.834 to 0.890 for high SNR, depending on the number of transits observed. Therefore, it appears that accounting for dependence on the number of transits (or orbital period) is more important than the order of detection. This strengthens our findings that the \Kepler{} data provide evidence for a high-inclination population of planetary systems to explain the observed \Kepler{} dichotomy. 

\subsection{Features in period ratio distribution} \label{secPratios}

{\em Planets near a mean-motion resonance (MMR):}
We find that our clustered models can produce the spikes in the period ratio distribution near first-order MMRs similar to those observed by \Kepler{} by assigning planets from the high inclination population to the low inclination population when near an MMR.\footnote{As expected, we do not get any statistically significant spikes if we do not assign planets near an MMR to have low mutual inclinations.}
Since we do not explicitly consider models with an explicit excess near MMRs, we do not exclude that possibility.  
Nevertheless, our results do provide insight into how significant an excess would be needed to create such spikes.

While $f_{\sigma_{i,\rm high}}$ describes the fraction of systems initially assigned to the high-inclination population, a fraction of the planets in these systems are still assigned a low mutual inclinations if they are near a (first-order) MMR. 
Since the distribution of period ratios for all planets is smooth (see Figures \ref{fig:models_underlying} and \ref{fig:clustered_P_R_underlying}), the spikes in the observed period ratio distribution come solely from this population. 
In our fully clustered model, we find that the fraction of all planets near an MMR (as defined in \S\ref{Incl}) with another planet is $f_{\rm mmr} = 0.29 \pm 0.04$. 
Thus, reproducing the spikes near MMRs in the observed period ratio distribution (without invoking a change in the distribution of mutual inclinations) would require a population of ${\rm FSWP} \times f_{\rm mmr} \times f_{\sigma_{i,\rm high}} \times \left<{\rm NPPS}\right> = 0.29_{-0.09}^{+0.11}$ planets per star that are near a first-order MMR.
This can be compared to a rate of ${\rm FSWP} \times (1-f_{\sigma_{i,\rm high}}) \times \left<{\rm NPPS}\right> \simeq 1.31_{-0.35}^{+0.55}$ planets per star that are responsible for the ``background'' population of multiple planet systems, excluding spikes in the period ratio distribution.
We caution that these results are based on our distance function that includes a term for the overall observed period ratio distribution, but does not know about the dynamical significance of MMRs.  
Using a distance function that explicitly considers the behavior of the period-ratio distribution near MMRs would be expected to provide a stronger constraint.  
Conversely, at least part of the observed spikes in the period ratio distribution could be due to shifting the period ratios of planet pairs from slightly less to slightly more than the value at resonance.

The period ratio distribution also provides additional constraints on proposals to explain the \Kepler{} dichotomy.  
Our parametrization of the mutual inclination distribution involves contributions to the period ratio distribution from both high and low mutual inclination populations. 
While a four-planet system has three pairs of adjacent planets, a two-planet system has only one planet pair. 
Therefore, reducing the mean multiplicity of the high-inclination population from greater than four to two (see \S\ref{Dichotomy}) would reduce the amplitude of spikes in the period ratio distribution by more than a factor of three. 
If the high-inclination population were to have a significantly lower mean number of planets per planetary system than the low-inclination population, then explaining the near-resonant peaks in the distribution of period ratios would likely require invoking an actual excess of planetary pairs near resonance.

\begin{figure}
\includegraphics[scale=0.42,trim={0 0 0 0},clip]{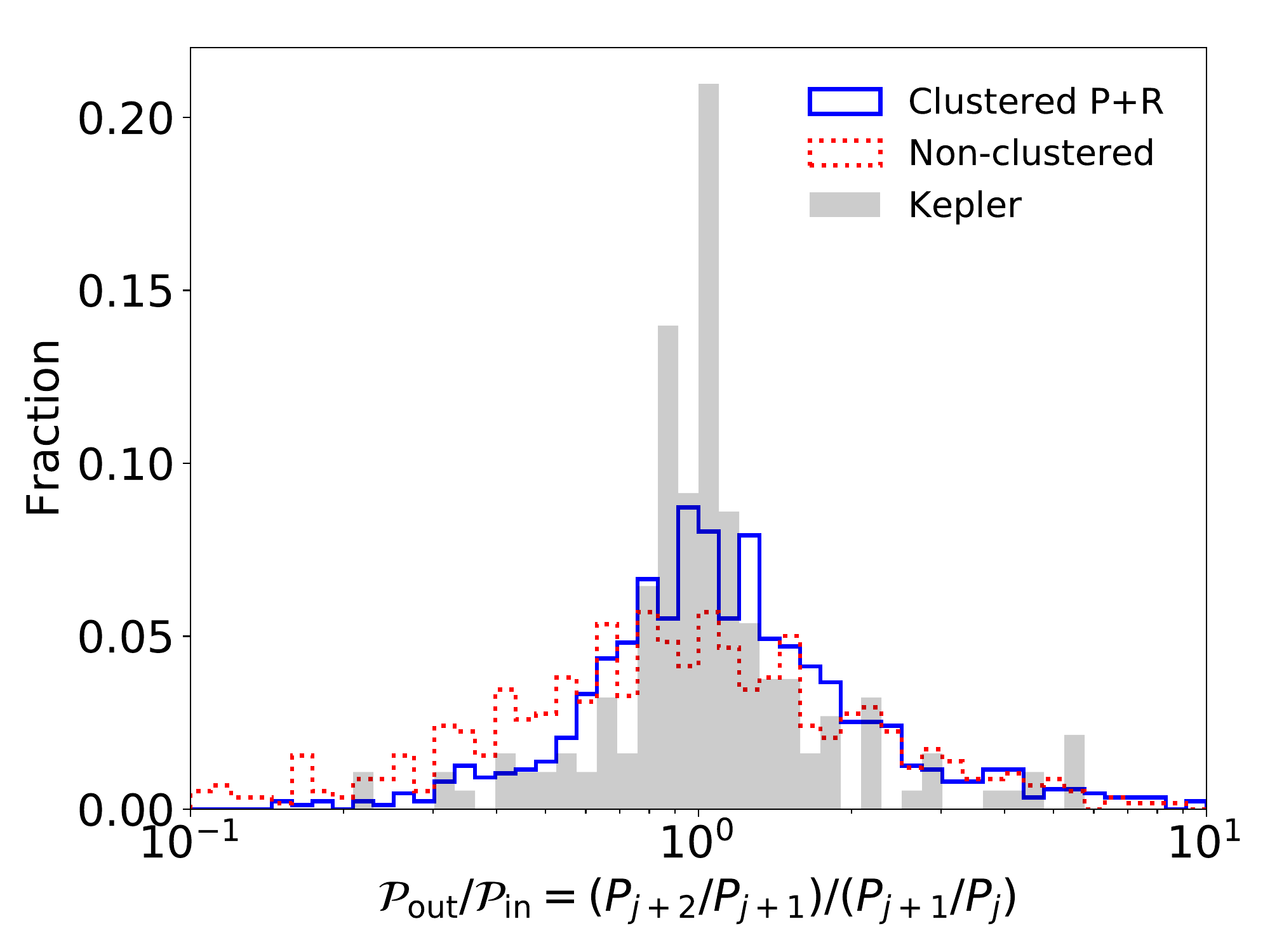}
\caption{The distributions of ratios of period ratios, for adjacent observed planets in three-planet chains. The grey shaded histogram shows the \Kepler{} distribution, while the solid blue and dotted red lines show those of the clustered periods and sizes and non-clustered models, respectively. None of our models include any explicit correlation between period ratios, but our clustered model appears to be more peaked than our non-clustered model due to the clustering in periods. Still, our clustered models are not as peaked as the data, suggesting that there is a significant correlation in the period ratios of observed planet pairs within the same system.}
\label{fig:pratio_ratio}
\end{figure}

{\em Are planets evenly spaced?}
\citet{W2018a} and \citet{Z2019} explored the period ratios of inner--outer pairs for observed three-planet chains.
\citet{W2018a} found a strong correlation between $\mathcal{P}_{\rm in} = P_{j+1}/P_{j}$ and $\mathcal{P}_{\rm out} = P_{j+2}/P_{j+1}$ (where planets are sorted so $P_j$'s are increasing) for small period ratios and concluded that planets in 3+ systems tend to be uniformly spaced.
However, \citet{Z2019} suggested that this correlation is largely due to a combination of detection biases and imposing an arbitrary upper limit on the period ratio ($\mathcal{P} < 4$ as used in \citealt{W2018a}). 
We test these assertions by also considering $\mathcal{P}_{\rm in}$ and $\mathcal{P}_{\rm out}$ as well as the their ratio ($\mathcal{P}_{\rm out}/\mathcal{P}_{\rm in}$). 
Our model does not explicitly enforce any correlation between $\mathcal{P}_{\rm in}$ and $\mathcal{P}_{\rm out}$ either in the generation of systems or in the distance function.  
Any observed correlation will be due to a combination of observational selection effects and the ``natural'' peak in  $\mathcal{P}$ given the process for drawing orbital periods.

In Figure \ref{fig:pratio_ratio}, we plot the observed distribution of the ratio of period ratios, $\mathcal{P}_{\rm out}/\mathcal{P}_{\rm in}$, for our fully clustered and non-clustered models along with that of the \Kepler{} data.
The intrinsic distributions of $\mathcal{P}_{\rm out}/\mathcal{P}_{\rm in}$ (not shown) are peaked around one in each of our models.
However, the observed distribution of $\mathcal{P}_{\rm out}/\mathcal{P}_{\rm in}$ is much more highly peaked than that observed in the non-clustered model (dotted red).
In contrast, the clustered models produce a peak in $\mathcal{P}_{\rm out}/\mathcal{P}_{\rm in}$ around unity, but it is slightly wider than the peak in the observed distribution of $\mathcal{P}_{\rm out}/\mathcal{P}_{\rm in}$ (solid blue).
Furthermore, the \Kepler{} distribution of the ratios of period ratios (shaded grey) is even more peaked than the predictions of our clustered models.
We conclude that there is a significant correlation in the period ratios of neighbouring planet pairs in actual planetary systems. 
This strengthens the conclusion of \citet{W2018a}.

\subsection{Future research}
\label{secFuture}

Future studies can build on our model and codes to incorporate additional physical processes which are not included in our current models into our forward model.
One could consider alternative parametric distributions for the distributions of intrinsic multiplicity, periods, radii, eccentricities, and mutual inclinations.
For example, one possible extension would be to incorporate a more sophisticated model for the distribution of planet radii (e.g., incorporating period dependence and/or allowing for a local minimum in the frequency as a function of planet size). \citet{ZCH2019} showed that adding a radius-valley resulted in only a modest effect on the inferred parameters for the radius and period distribution.  Nevertheless, allowing for a more flexible model for the radius-period distribution would likely lead to improving the goodness-of-fit, particularly for the somewhat poorly fit distributions of transit depth  and transit depth ratios.  

Another possible extension would be to allow for explicitly modelling a population of systems with resonant or near-resonant planet pairs or chains.  The results presented here show that the \Kepler{} observations (particularly the period ratio and transit duration ratio distributions) can be explained by a model with no excess of planets near MMRs relative to the background period ratio distribution.  However, this finding does not preclude the possibility that there still might be a small excess of near-resonant planetary systems also contributing to the observed spikes in the period ratio distribution (see \S\ref{secPratios}).  Future studies could explore models with parameters for the amplitude and/or width of such spikes in the period ratio distribution.  Such additions would likely benefit from adding new terms to the distance function that explicitly compare behavior near MMRs. Improvements to the detection efficiency model that take into account the presence of transit timing variations (TTVs) may also benefit from these efforts.

This paper adopted a distance function that takes into account each of the observed distributions of the \Kepler{} population (as described in \S\ref{Summary_stats} and shown in Figure \ref{fig:Kepler_DR25}), along with the overall rate of planets per observed star $f_{\rm Kepler}$. 
Our summary statistics are based on the marginal distributions for each observable.  
Therefore, it is possible that we might find models that reproduce our chosen summary statistics well, but differ in its prediction for the observed \Kepler{} catalogue in other ways (e.g., a correlation between orbital period and planet radius, or a correlation with planet occurrence and host star metallicity).
As shown in \S\ref{Model_params}, our choice of summary statistics and distance function already provide strong constraints on many physically interesting model parameters.
Future research adding additional summary statistics and component distances could shrink the ABC posterior further and help constrain any additional model parameters.  
For example, future research could incorporate the detection of TTVs or transit duration variations (TDVs), so as to provide stronger constraints on the abundance and properties of non-transiting planets. 

Our forward model makes use of the \Kepler{} DR25 completeness and reliability products \citep{BC2017a, BC2017b, BC2017c, C2017, T2018} to account for the vast majority of factors that influence the detectability of planets.  
One aspect in which our model of the detection efficiency could be improved relates to how the \Kepler{} pipeline's detection efficiency is affected by  multiple transiting planets.
The pipeline first detects the planet with the largest multiple event statistics (MES, analogous to combined transit signal-to-noise), masks out observations near when that planet transits, and re-searches the light curve for additional planets.
Typically, only $\sim 0.4\%$ of the light curve is lost for each additional planet \citep{SJF2017,ZCH2019}. Thus, masking out observations near transits of a more easily detected planet is expected to have only a modest effect on the integrated transit signal-to-noise and the probability of observing at least three transits. Indeed, \citet{ZCH2019} showed that the multiplicity distribution is only slightly changed when accounting for the lower detection efficiency of subsequent planets.

\section{Conclusions} \label{Conclusions}

We have developed a framework for generating populations of planetary systems and simulating observed catalogues of exoplanets under the conditions of a \Kepler{}-like mission. We compare three physically motivated models: the non-clustered model, the clustered periods model, and the clustered periods and sizes model, to the observed \Kepler{} Q1-Q17 DR25 population of uniformly vetted planet candidates. Our analysis is limited to planets with orbital periods between 3 and 300 d and radii between 0.5 and 10 $R_\oplus$, which form the bulk of the DR25 catalogue (2137 planets in our subset). 
While most of our findings are consistent with previous works \citep{Li2011b, FM2012, TD2012, F2014, Mu2018}, this work improves on previous studies by incorporating improved and more homogeneous planet and star catalogues, an improved model for \Kepler{} detection efficiency, and by providing a forward model that simultaneously fits all of the marginal distributions described in \S\ref{Summary_stats}. 

Our models are highly flexible and allow for the relatively fast generation ($\sim$10 s for a \textit{physical} and \textit{observed catalogue} with the 79 935 target stars used in this study) of simulated observed catalogues that are similar to the \Kepler{} exoplanet catalogue in terms of many observables.
We define a set of observable summary statistics and a distance function for comparing simulated models with the \Kepler{} data. We identify model parameters that result in simulated observed catalogues that closely match the \Kepler{} catalogue in terms of the observed planet multiplicity, period, period ratio, transit duration, period-normalized transit duration ratio ($\xi$), transit depth, and transit depth ratio distributions. We train a Gaussian process (GP) model for each of our physical models in order to build an emulator for rapidly predicting the distance as a function of the model parameters. Using the emulator, we draw from the ABC posterior and provide credible intervals for each of the model parameters as constrained by the data.

We find that a non-clustered model with a Poisson distribution for the true number of planets per system, single and broken power laws for the periods and planetary radii, and a simple stability criteria, can reasonably fit most of the key observable properties of the \Kepler{} population, including the overall rate of observed planets, as well as the period, transit depth, and transit duration distributions, although there are clear differences.
However, the number of observed multiplanet systems, the period ratio distribution, and the planet radius ratio distributions are poorly modelled by the non-clustered model. In contrast, clustered models (of periods) can produce observed planet multiplicity and period ratio distributions that are significantly better fits to those observed by \Kepler{}. The transit durations and duration ratios are also improved to an appreciable extent. Our fully clustered model, with clustered periods and sizes, performs similarly well and is the best description of the marginal distributions of the \Kepler{} data to date.  While this model allows for more similar sizes of planets within the same system, the fit to the observed radius ratio distribution is not significantly improved in terms of the KS or AD distances, suggesting that additional features in this distribution are still inadequately reproduced by any of our current models. Most notably, the observed radius ratio distribution appears to be quite asymmetric, which is likely a signature of stripped planetary atmospheres due to photoevaporation \citep{LFM2012, OW2017} or core-powered mass-loss \citep{GS2018}. \citet{W2018a} find a very similar effect of clustering and asymmetry in the radius ratio distribution and also a slight positive correlation between radius ratio and the difference in effective temperatures between adjacent planets.
Thus, we show with forward modelling that while the observed planet radii ratio distribution of the \Kepler{} data require more than a simple clustering in intrinsic sizes to explain, the data cannot be reproduced in a non-clustered model by detection biases alone, in contrast to the results of \citet{Z2019}.

Several investigations on the numbers of \Kepler{} single and multitransiting systems, i.e. the observed multiplicity distribution, suggest that there is an apparent excess of observed singles and that this indicates the existence of more than one underlying population of planetary systems, or a so-called ``\Kepler{} dichotomy'' (e.g., \citealt{Li2011b, J2012, HM2013, BJ2016}). While there is evidence for the excess of stars with a single transiting planet (relative to the predictions based on abundance of systems with multiple transiting planets), other studies (e.g., \citealt{W2018b}) show that the stellar properties of the singles and multis are indistinguishable, suggesting a common origin. We account for the dichotomy by including two populations of planetary systems with separate mutual inclination dispersions (${\rm Rayleigh}(\sigma_{i,\rm high})$ for a fraction $f_{\sigma_{i,\rm high}}$ of all systems and ${\rm Rayleigh}(\sigma_{i,\rm low})$ for the remaining systems) for each of our models, and find that this is necessary to fit the multiplicity distribution. In our non-clustered model, the occurrence of multitransiting systems is significantly underproduced, due to a tendency for $f_{\sigma_{i,\rm high}}$ to be very low ($\lesssim 3\%$). In our clustered models, the multiplicity distribution is extremely well reproduced, with $1 - f_{\sigma_{i,\rm high}} \simeq 60\%$ of systems having small mutual inclinations of $\sigma_{i,\rm low} \sim 1.4^\circ$ and the remaining $\sim 40\%$ of systems with broad mutual inclinations required to account for the excess observed singles. The $\log{\xi}$ distribution for planets not near any apparent MMRs with other observed planets is also well reproduced with these mutual inclinations and a Rayleigh distribution of eccentricities, $\sigma_e \sim 0.02$.
Based on models that allow for clustering in period and radius, it is unlikely that the excess of single transiting planet systems could be explained solely by a large population of intrinsically single-planet systems.  Thus, our results provide new evidence in favour of the \Kepler{} dichotomy being due to a population of planetary systems with a broader distribution of mutual inclinations than characteristic of the observed \Kepler{} multiple planet systems.

Previous studies also show that most observed \Kepler{} systems are not near low-order MMRs (e.g., \citealt{Li2011b, VF2012, F2014, SH2015}), and our model reliably reproduces this observation.  
Nevertheless, the small fraction of systems near MMRs are particularly interesting to dynamicists and for constraining planet formation models.
Therefore, we investigated the ability of our model to reproduce the small spikes in the period ratio distribution slightly wide of first-order MMRs due purely to geometrical effects by assigning planet pairs near MMRs with mutual inclinations also drawn from the population with a narrow distribution of mutual inclinations ($\sigma_{i,\rm low} \sim 1.4^\circ$).
While some but not all of the draws from our clustered models include spikes in the period ratio distribution near MMRs, these spikes are statistically robust (see the quantiles (red curves) in Figures \ref{fig:clustered_P_model} and \ref{fig:clustered_P_R_model}).
Thus, we show that allowing the distribution of mutual inclinations to depend upon the proximity of a planet pair to MMR provides an alternative explanation for the observed spikes in the period ratio distribution. This study did not consider and thus cannot exclude models in which the observed spikes in the period ratio distribution are due to an intrinsically higher rate (e.g., due to migration leading to resonant trapping). Previous studies have noted that most \Kepler{} systems are not in an MMR (e.g., \citealt{VF2012}), and so an additional mechanism would be required to explain why the systems are near, but not in MMR (e.g., \citealt{LW2012, L2013, PMT2013, DL2014, GS2014, X2014, CF2015, I2017}).
Future studies may be able to distinguish between these two models and better constrain the properties of near-MMR systems by making use of additional observational constraints (e.g., transit timing variations and transit duration variations). The results of precision radial velocity surveys may also help constrain the rate of additional non-transiting planet companions.

Our framework of simulating ensembles of planetary systems in a forward model provides a way of directly probing the underlying populations of exoplanets and their properties, including planets that are not observed or do not transit. 
Our clustered models also provide considerable utility for informing follow-up efforts of new exoplanet surveys such as the Transiting Exoplanet Survey Satellite (TESS) mission, which is expected to discover thousands of additional planetary candidates \citep{S2015, B2017, S2017}. 
By matching the observed distributions of period ratios and transit depth ratios for the \Kepler{} exoplanets and assuming that the \Kepler{} population is representative of planetary systems to be observed by TESS, our clustered periods and sizes model can be used to compute conditional probabilities of additional RV detectable planets in systems with already known planets of measured periods and radii.

We have made the core SysSim code (\url{https://github.com/ExoJulia/ExoplanetsSysSim.jl}), inputs collated from numerous datafiles (\url{https://github.com/ExoJulia/SysSimData}), and the code specific to the clustered models (\url{https://github.com/ExoJulia/SysSimExClusters}, Zenodo DOI 10.5281/zenodo.3386372) available to the public via Github. We encourage other researchers to contribute model extensions via pull requests and/or additional public git repositories.  
Additionally, researchers can use our forward modelling pipeline to perform detailed comparisons of the results of planet formation simulations with \Kepler{} observations, so as to improve our understanding of planet formation and the architectures of planetary systems more generally.
For researchers who prefer not to run the SysSim code themselves, we provide hundreds of \textit{physical} and \textit{observed} catalogues in table format, each containing the simulated true and observed properties of $\sim 10^5$ planetary systems, generated from each of our three models using either the best-fitting parameter values or draws from the ABC-posterior distribution for parameter values explained in \S\ref{Model_params}.

\section*{Acknowledgements}
\addcontentsline{toc}{section}{Acknowledgements}
We thank the entire \Kepler{} team for years of work leading to a successful mission and data products critical to this study.  
We acknowledge many valuable contributions with members of the \Kepler{} Science Team's working groups on multiple body systems, transit timing variations, and completeness working groups.  
We thank Keir Ashby, Danley Hsu, Neal Munson, Shane Marcus, and Robert Morehead for contributions to the broader SysSim project.  
We thank Derek Bingham, Earl Lawrence, Ilya Mandell, Oded Aahronson, Ben Bar-Oh, Dan Fabrycky, Jack Lissauer, Gijs Mulders, Aviv Ofir, and Jason Rowe for useful conversations.  
We thank Rebekah Dawson for reading preliminary drafts of this paper and providing detailed suggestions.
MYH acknowledges the support of the Natural Sciences and Engineering Research Council of Canada (NSERC), funding reference number PGSD3-516712-2018.
EBF and DR acknowledge support from NASA Origins of Solar Systems grant \# NNX14AI76G and Exoplanet Research Program grant \# NNX15AE21.
EBF acknowledges support from NASA \Kepler{} Participating Scientist Program Cycle II grant \# NNX14AN76G.
MYH and EBF acknowledge support from the Penn State Eberly College of Science and Department of Astronomy \& Astrophysics, the Center for Exoplanets and Habitable Worlds, and the Center for Astrostatistics.  
EBF acknowledges support and collaborative scholarly discussions during  residency at the Research Group on Big Data and Planets at the Israel Institute for Advanced Studies.  
The citations in this paper have made use of NASA's Astrophysics Data System Bibliographic Services.  
This research has made use of the NASA Exoplanet Archive, which is operated by the California Institute of Technology, under contract with the National Aeronautics and Space Administration under the Exoplanet Exploration Program.
This work made use of the stellar catalogue from \citet{H2019} and thus indirectly the gaia-kepler.fun crossmatch data base created by Megan Bedell.
Several figures in this manuscript were generated using the \texttt{corner.py} package \citep{Fm2016}.
We acknowledge the Institute for CyberScience (\url{http://ics.psu.edu/}) at The Pennsylvania State University, including the CyberLAMP cluster supported by NSF grant MRI-1626251, for providing advanced computing resources and services that have contributed to the research results reported in this paper.
This study benefited from the 2013 SAMSI workshop on Modern Statistical and Computational Methods for Analysis of \Kepler{} Data, the 2016/2017 Program on Statistical, Mathematical and Computational Methods for Astronomy, and their associated working groups.
This material was based upon work partially supported by the National Science Foundation under Grant DMS-1127914 to the Statistical and Applied Mathematical Sciences Institute (SAMSI). Any opinions, findings, and conclusions or recommendations expressed in this material are those of the author(s) and do not necessarily reflect the views of the National Science Foundation.




\bibliographystyle{mnras}





\appendix

\begin{figure*}
\includegraphics[scale=0.38,trim={0 0 0 0},clip]{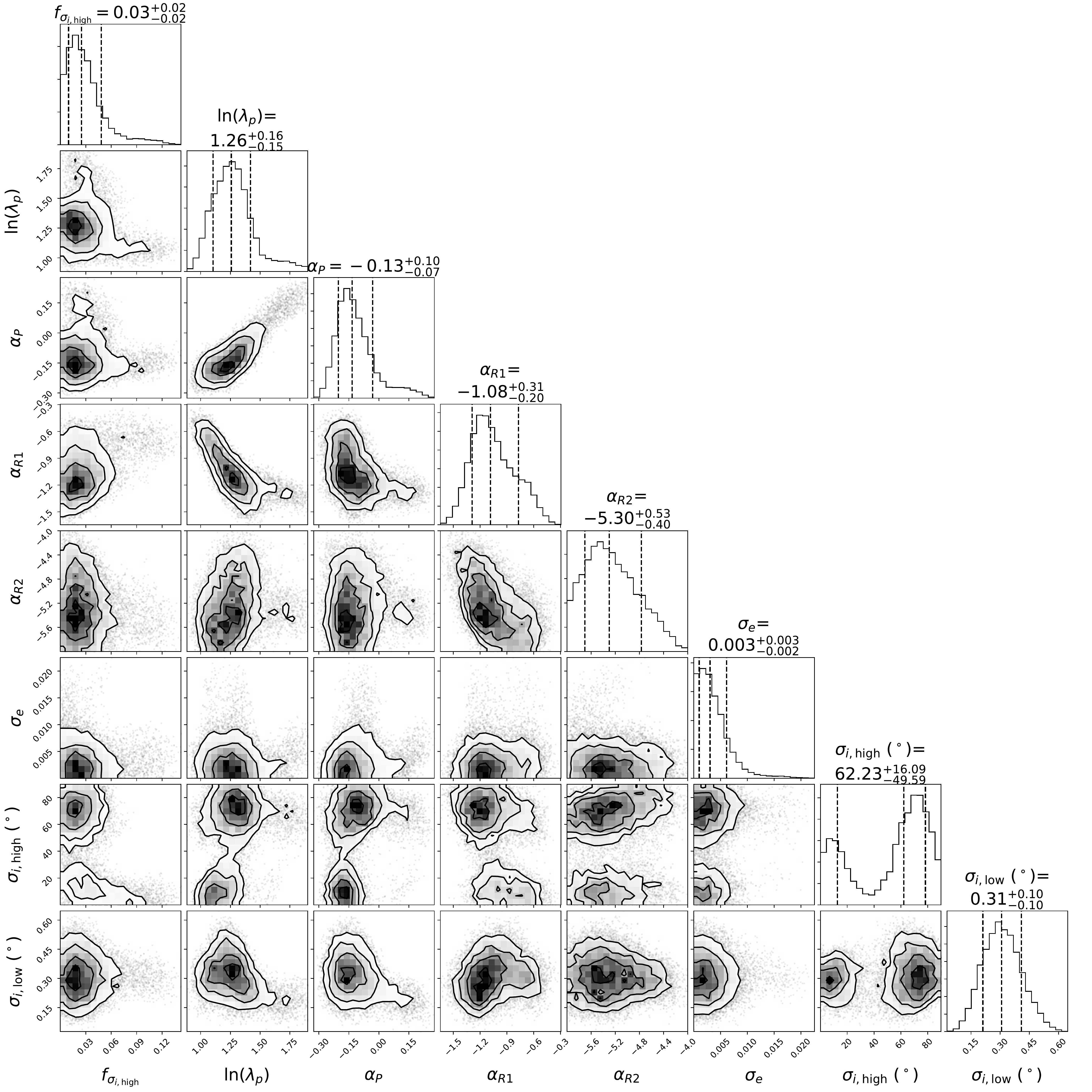}
\caption{Marginal posterior distributions of the model parameters for the non-clustered model, showing the projections of the points that pass a distance threshold of $\mathcal{D}_{W,\rm KS} = 55$ as drawn from the GP emulator. The prior mean function was set to a constant value of 75.}
\label{fig:non_clustered_corner}
\end{figure*}

\begin{figure*}
\includegraphics[scale=0.3,trim={0 0 0 0},clip]{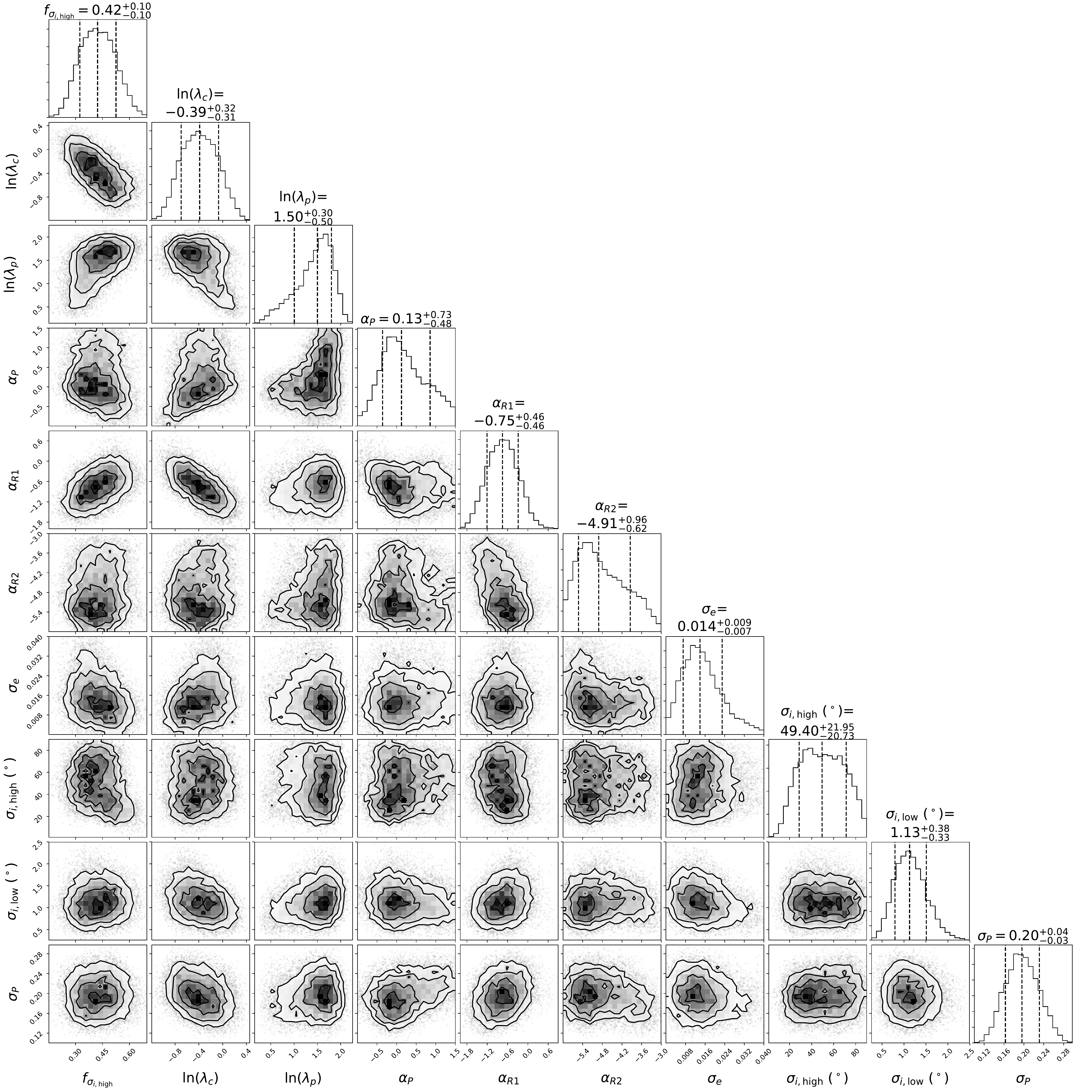}
\caption{Marginal posterior distributions of the model parameters for the clustered periods model, showing the projections of the points that pass a distance threshold of $\mathcal{D}_{W,\rm KS} = 35$ as drawn from the GP emulator. The prior mean function was set to a constant value of 75.}
\label{fig:clustered_P_corner_KS}
\end{figure*}

\begin{figure*}
\includegraphics[scale=0.3,trim={0 0 0 0},clip]{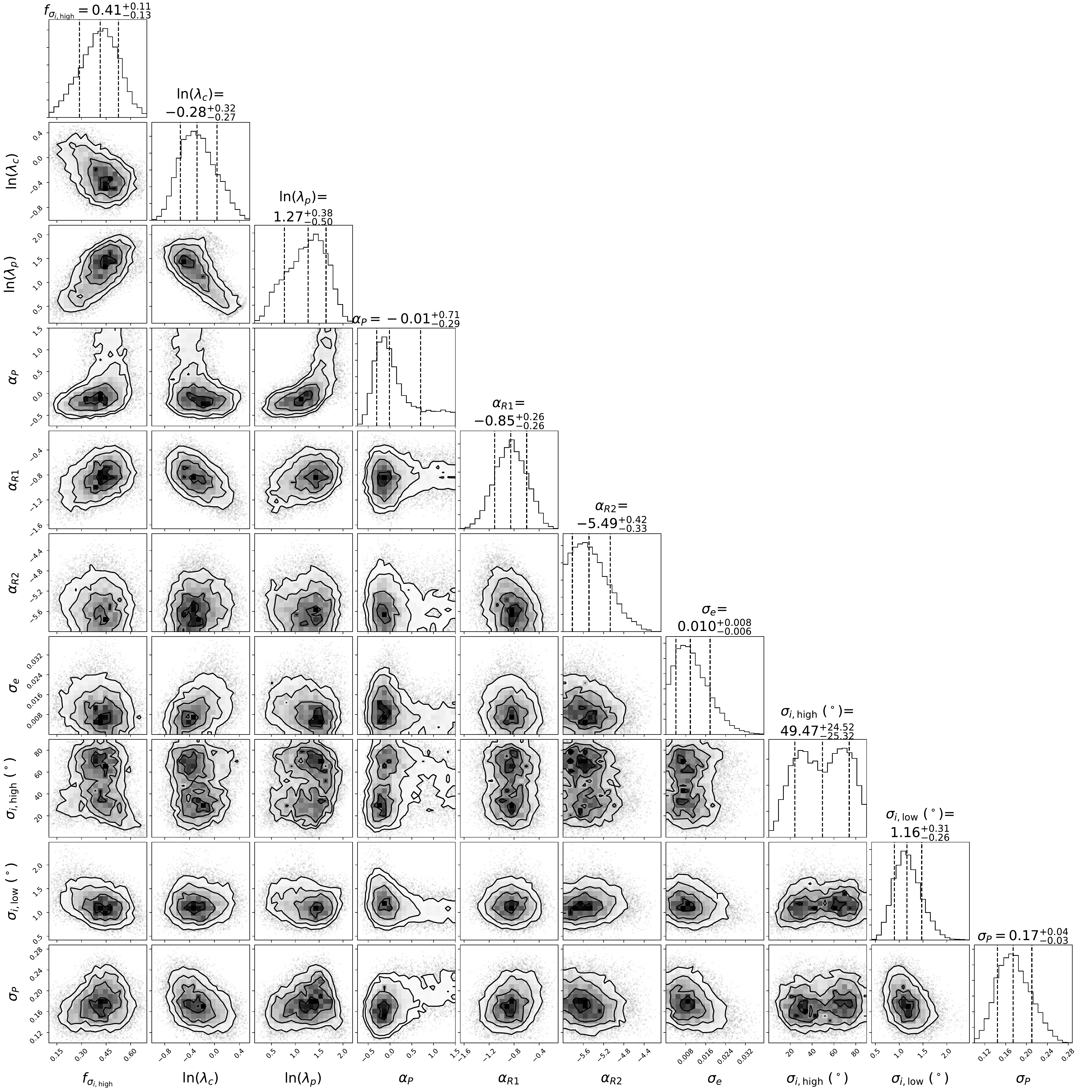}
\caption{Marginal posterior distributions of the model parameters for the clustered periods model, showing the projections of the points that pass a distance threshold of $\mathcal{D}_{W,\rm AD'} = 140$ as drawn from the GP emulator. The prior mean function was set to a constant value of 250.}
\label{fig:clustered_P_corner_AD}
\end{figure*}

\begin{figure*}
\includegraphics[scale=0.28,trim={0 0 0 0},clip]{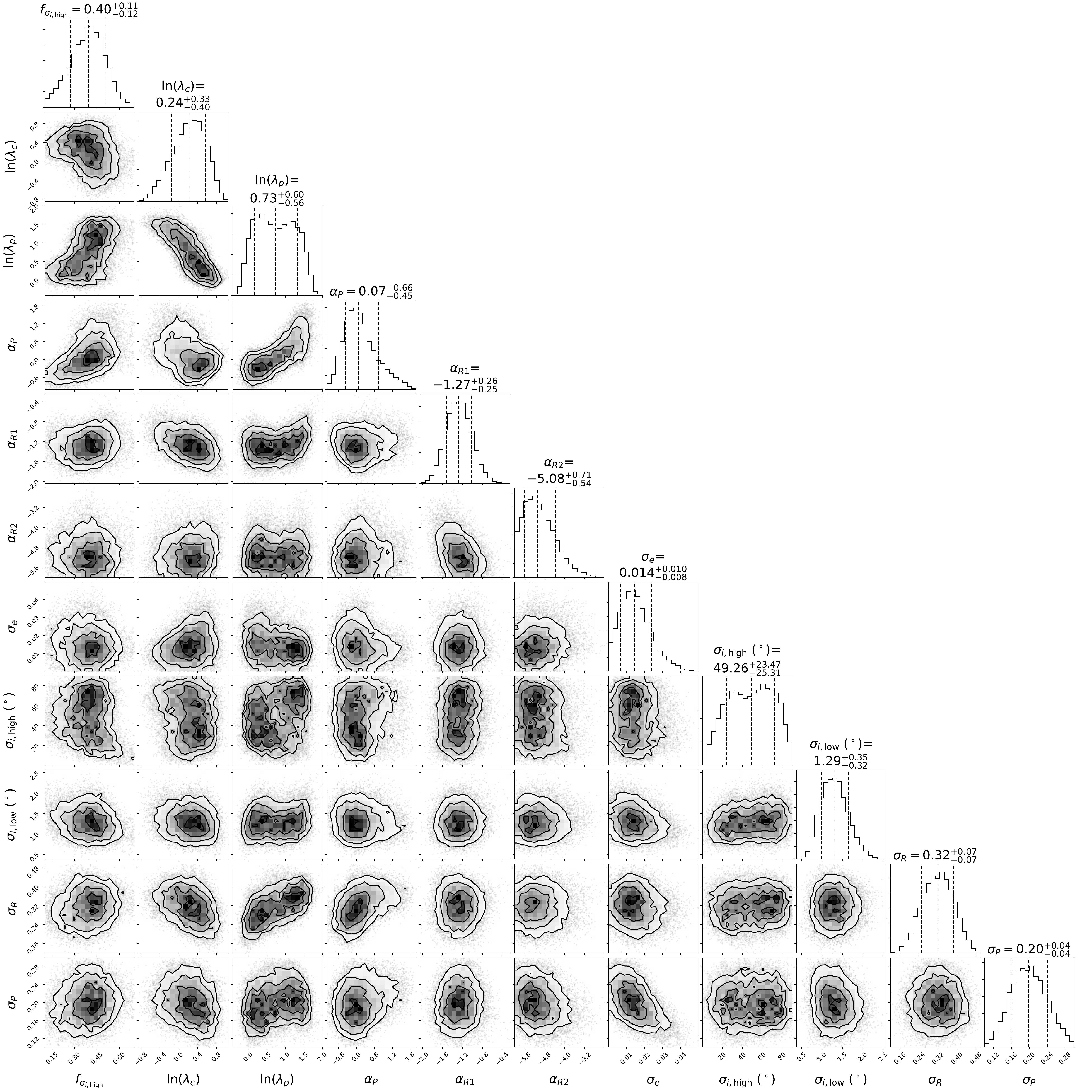}
\caption{Marginal posterior distributions of the model parameters for the clustered periods and sizes model, showing the projections of the points that pass a distance threshold of $\mathcal{D}_{W,\rm AD'} = 140$ as drawn from the GP emulator. The prior mean function was set to a constant value of 250.}
\label{fig:clustered_P_R_corner_AD}
\end{figure*}

\begin{figure*}
\includegraphics[scale=0.28,trim={0 0 0 0},clip]{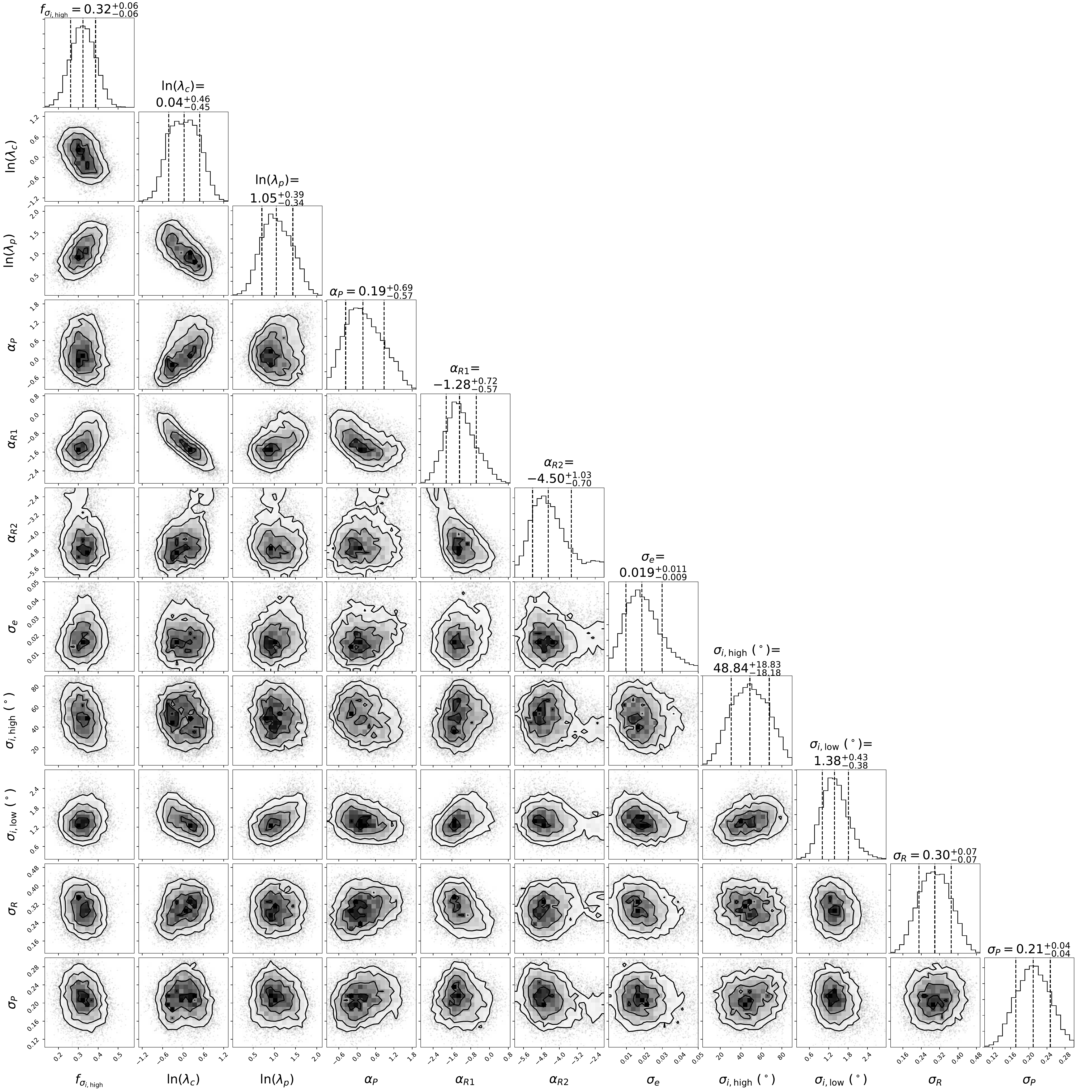}
\caption{Marginal posterior distributions of the model parameters for the alternative MMR inclinations model (clustered periods and sizes model without the lowering of mutual inclinations for planets near an MMR), showing the projections of the points that pass a distance threshold of $\mathcal{D}_{W,\rm KS} = 35$ as drawn from the GP emulator. The prior mean function was set to a constant value of 75.}
\label{fig:clustered_P_R_no_mmr_corner_KS}
\end{figure*}

\begin{figure*}
\includegraphics[scale=0.28,trim={0 0 0 0},clip]{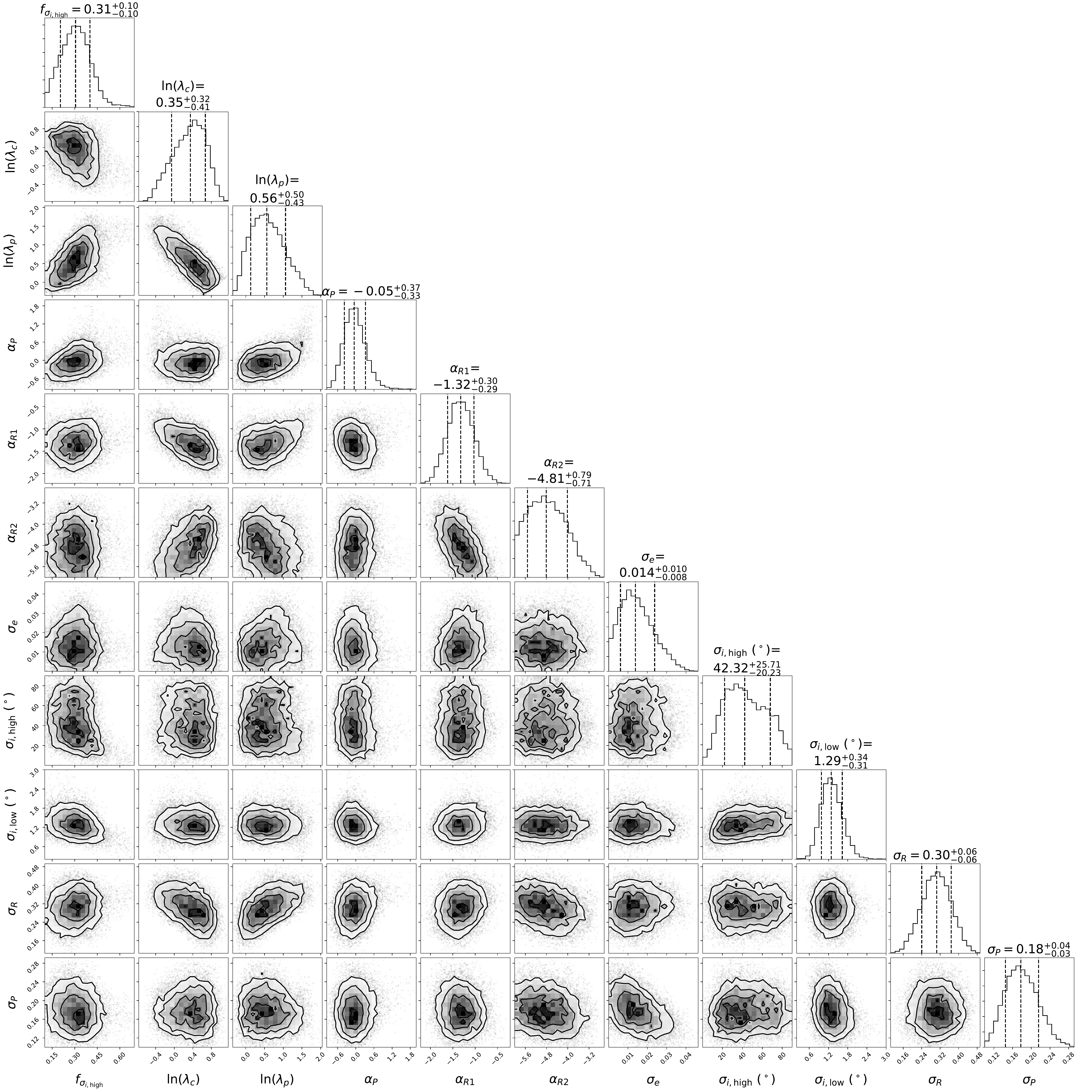}
\caption{Marginal posterior distributions of the model parameters for the alternative MMR inclinations model (clustered periods and sizes model without the lowering of mutual inclinations for planets near an MMR), showing the projections of the points that pass a distance threshold of $\mathcal{D}_{W,\rm AD'} = 140$ as drawn from the GP emulator. The prior mean function was set to a constant value of 250.}
\label{fig:clustered_P_R_no_mmr_corner_AD}
\end{figure*}

\begin{figure*}
\begin{tabular}{cc}
 \includegraphics[scale=0.33,trim={0 0.3cm 0 0.2cm},clip]{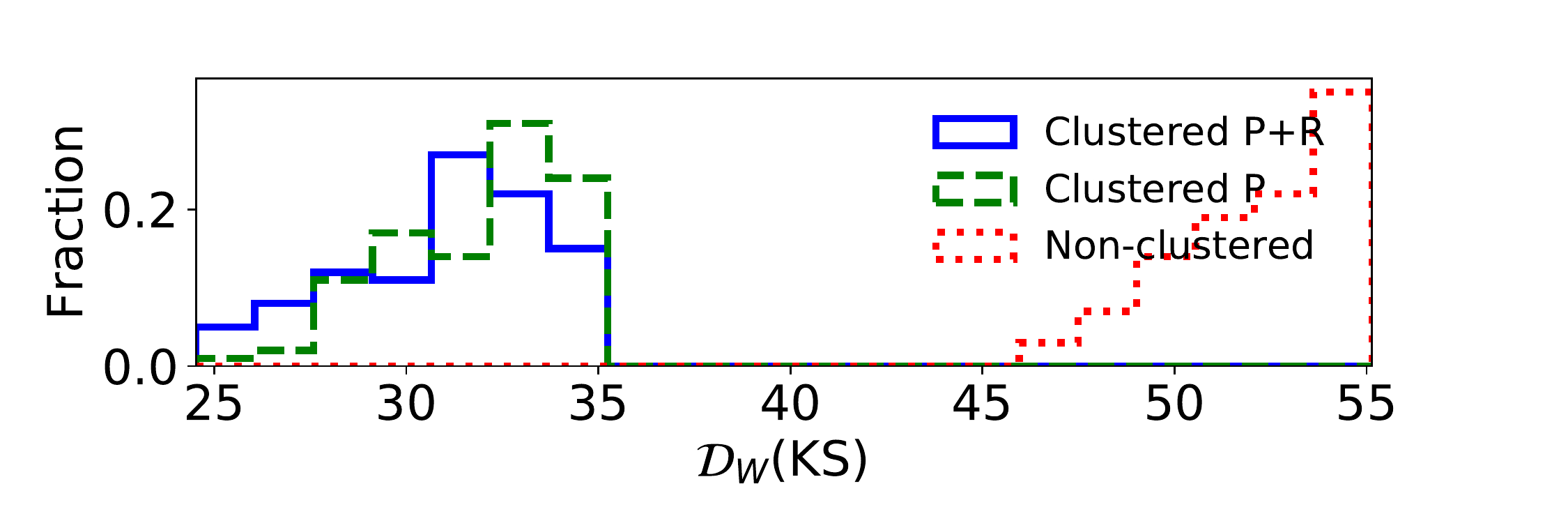} & 
 \includegraphics[scale=0.33,trim={0 0.3cm 0 0.2cm},clip]{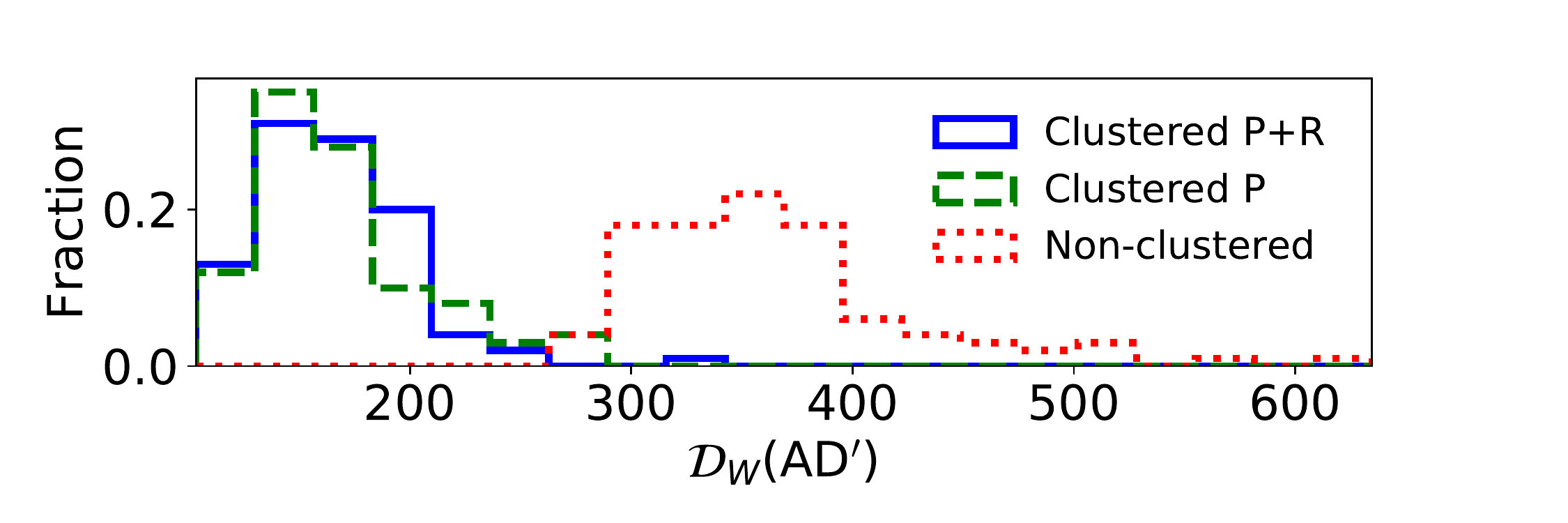} \\
 \includegraphics[scale=0.33,trim={0 0.3cm 0 0.2cm},clip]{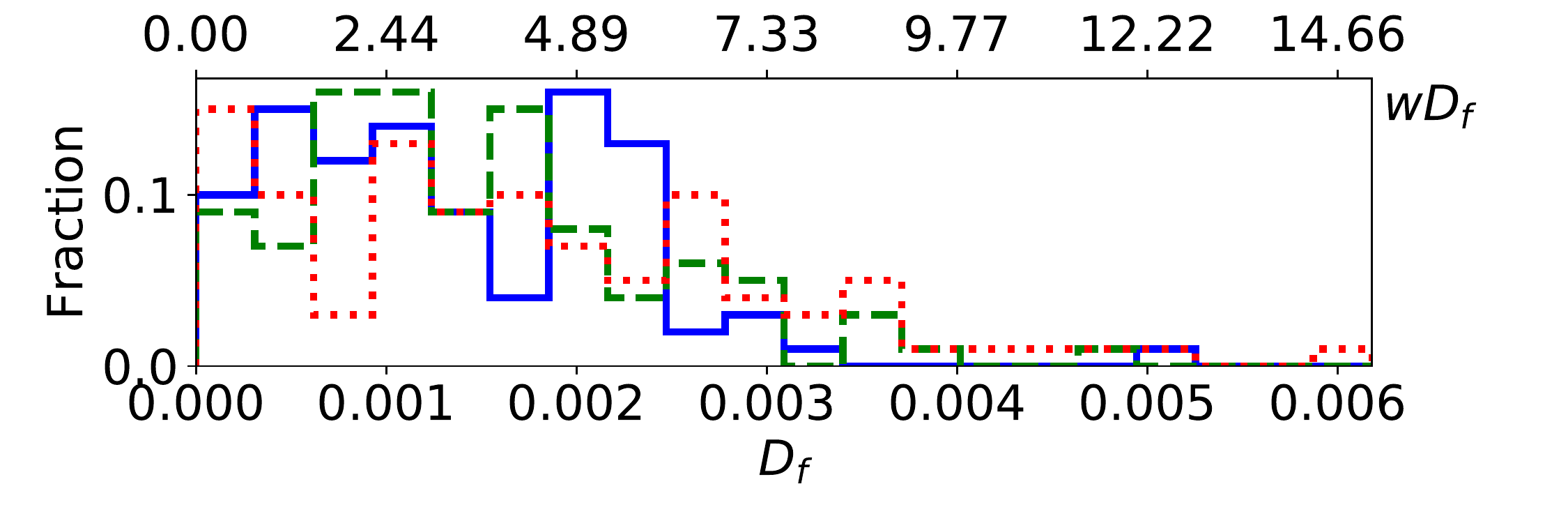} & 
 \includegraphics[scale=0.33,trim={0 0.3cm 0 0.2cm},clip]{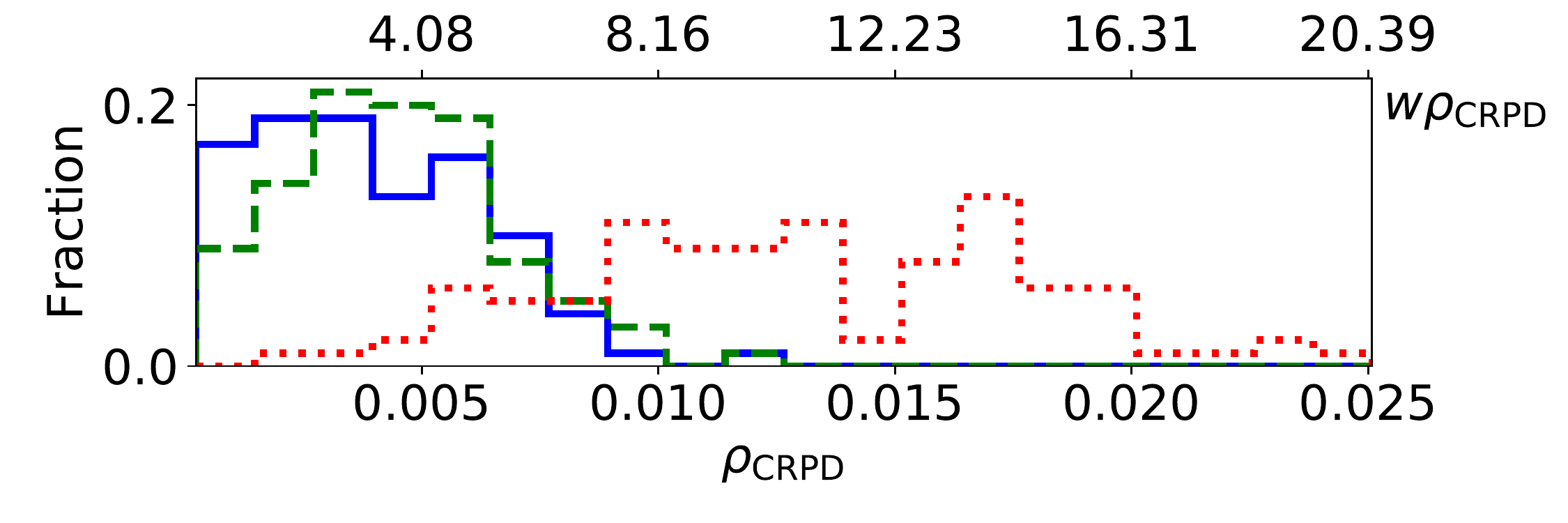} \\
 \includegraphics[scale=0.33,trim={0 0.3cm 0 0.2cm},clip]{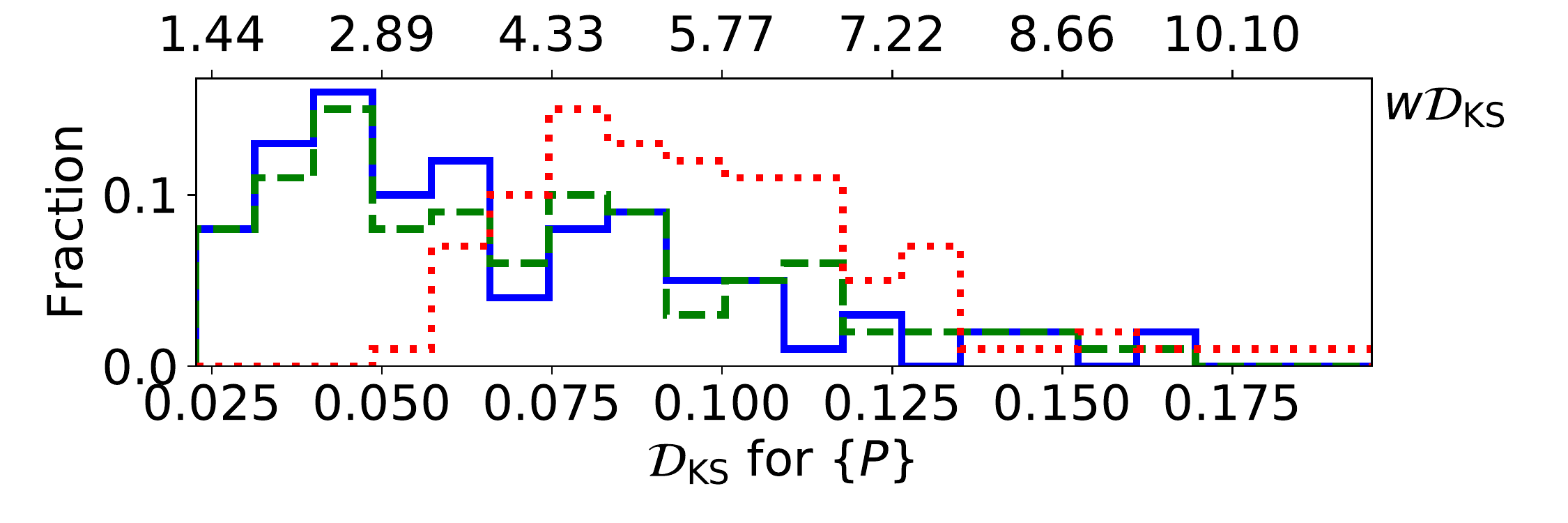} & 
 \includegraphics[scale=0.33,trim={0 0.3cm 0 0.2cm},clip]{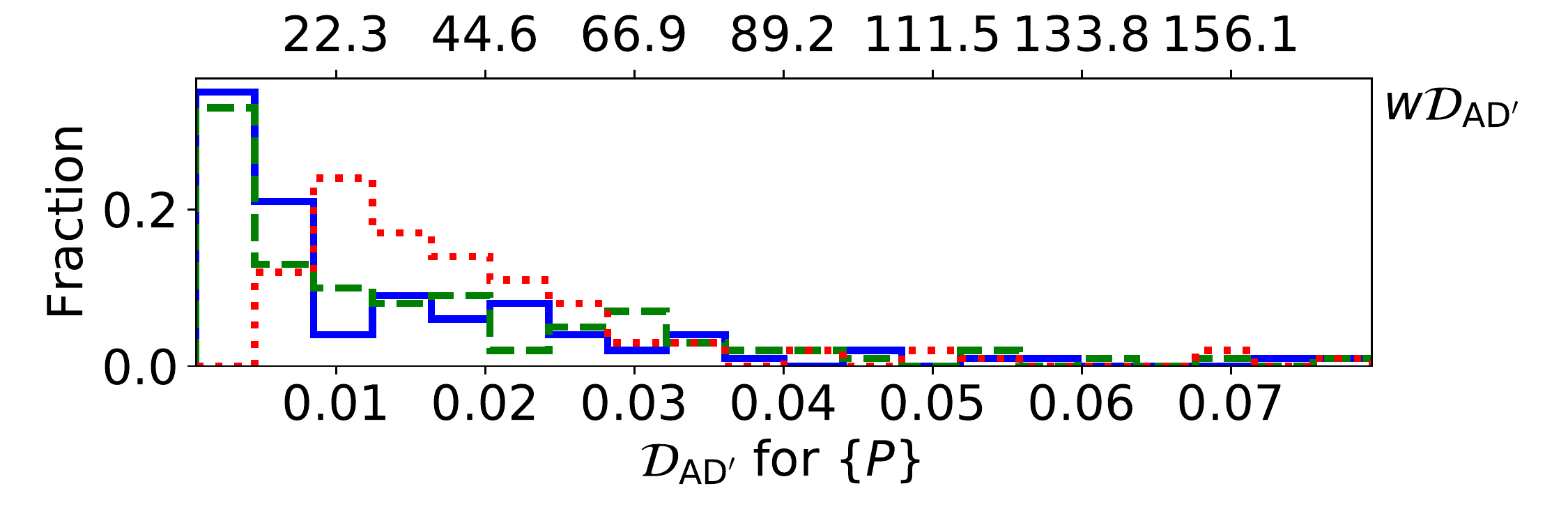} \\
 \includegraphics[scale=0.33,trim={0 0.3cm 0 0.2cm},clip]{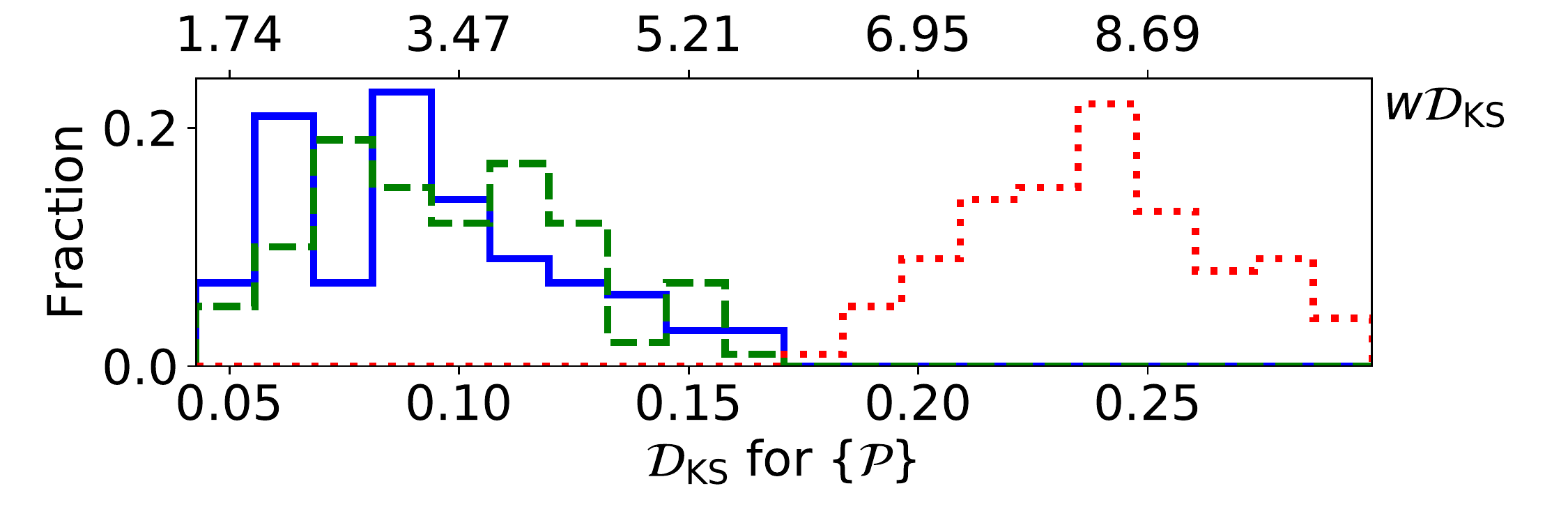} & 
 \includegraphics[scale=0.33,trim={0 0.3cm 0 0.2cm},clip]{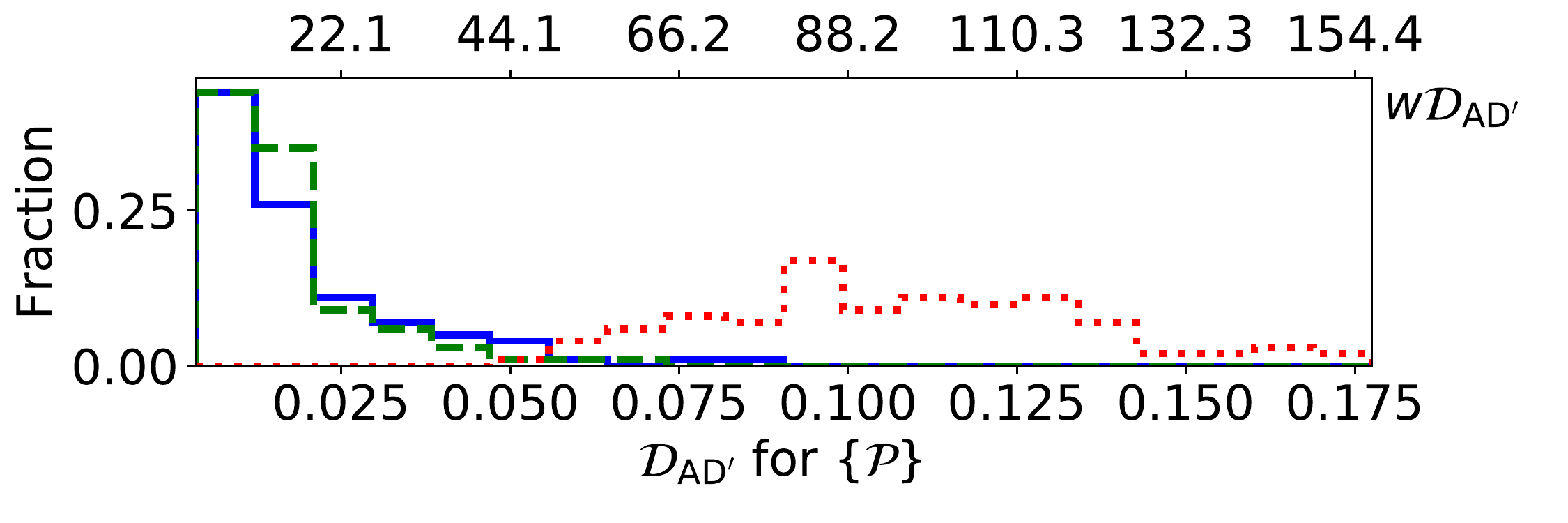} \\
 \includegraphics[scale=0.33,trim={0 0.3cm 0 0.2cm},clip]{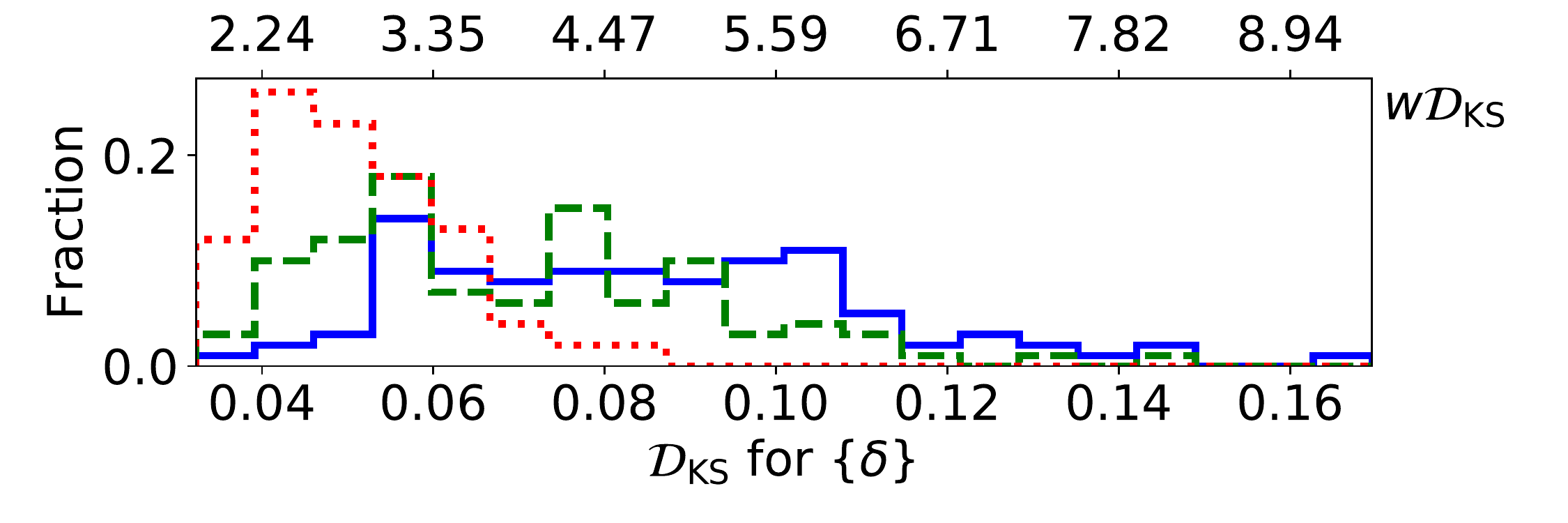} & 
 \includegraphics[scale=0.33,trim={0 0.3cm 0 0.2cm},clip]{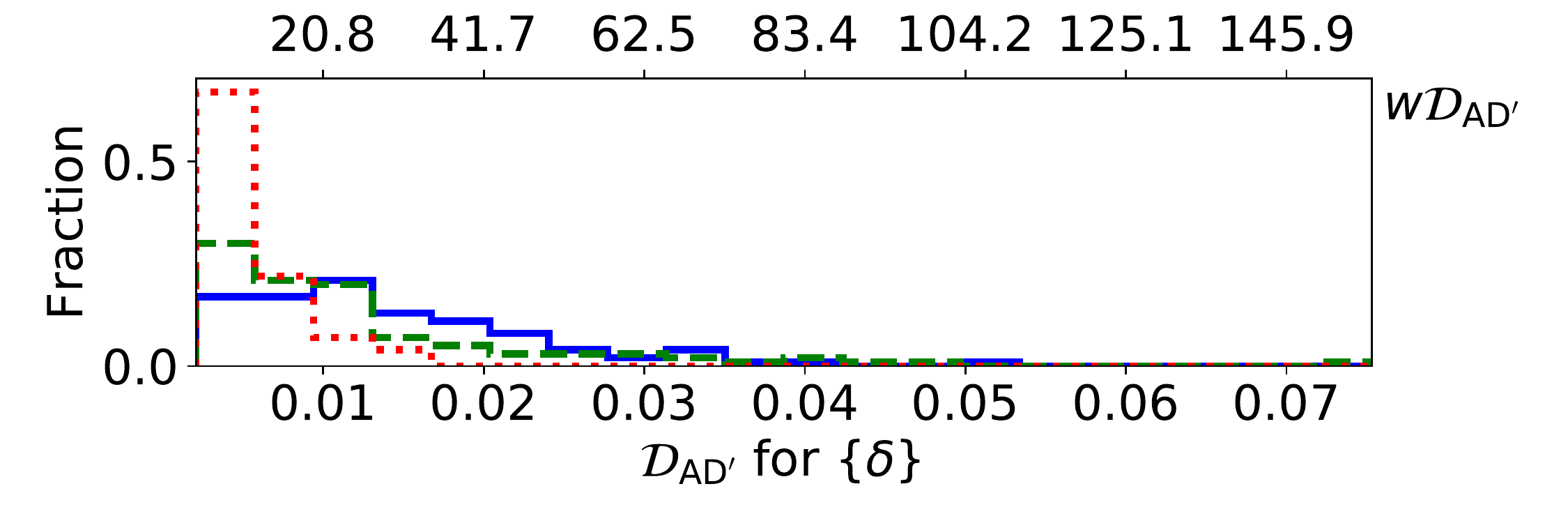} \\
 \includegraphics[scale=0.33,trim={0 0.3cm 0 0.2cm},clip]{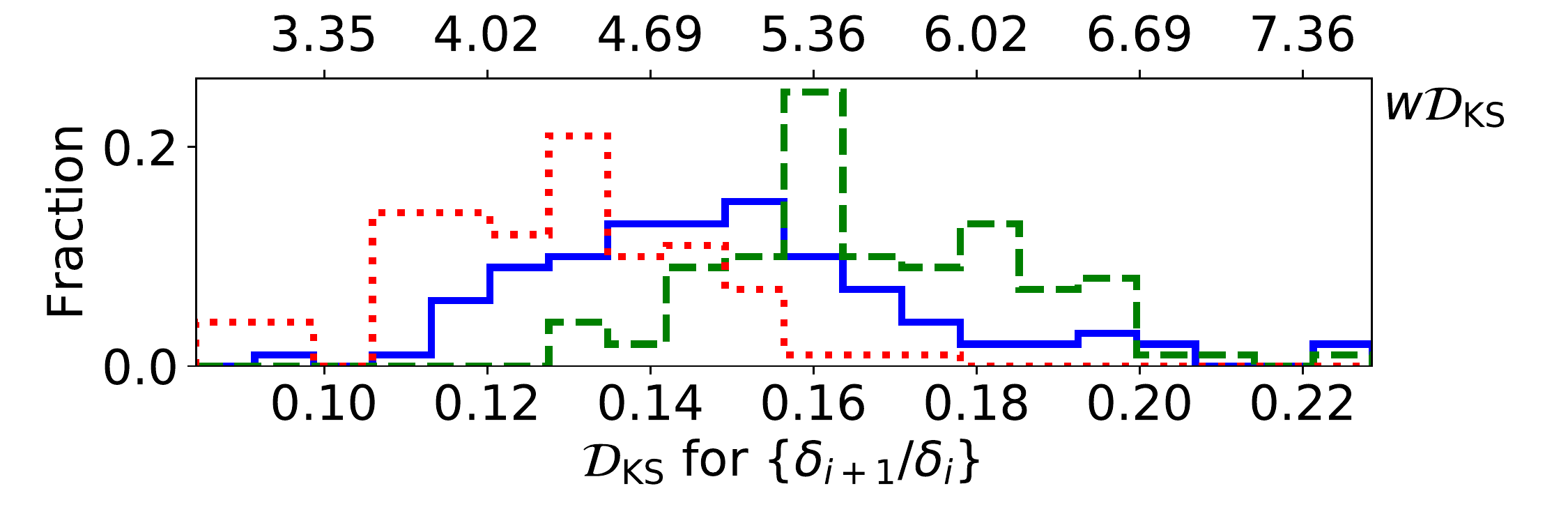} & 
 \includegraphics[scale=0.33,trim={0 0.3cm 0 0.2cm},clip]{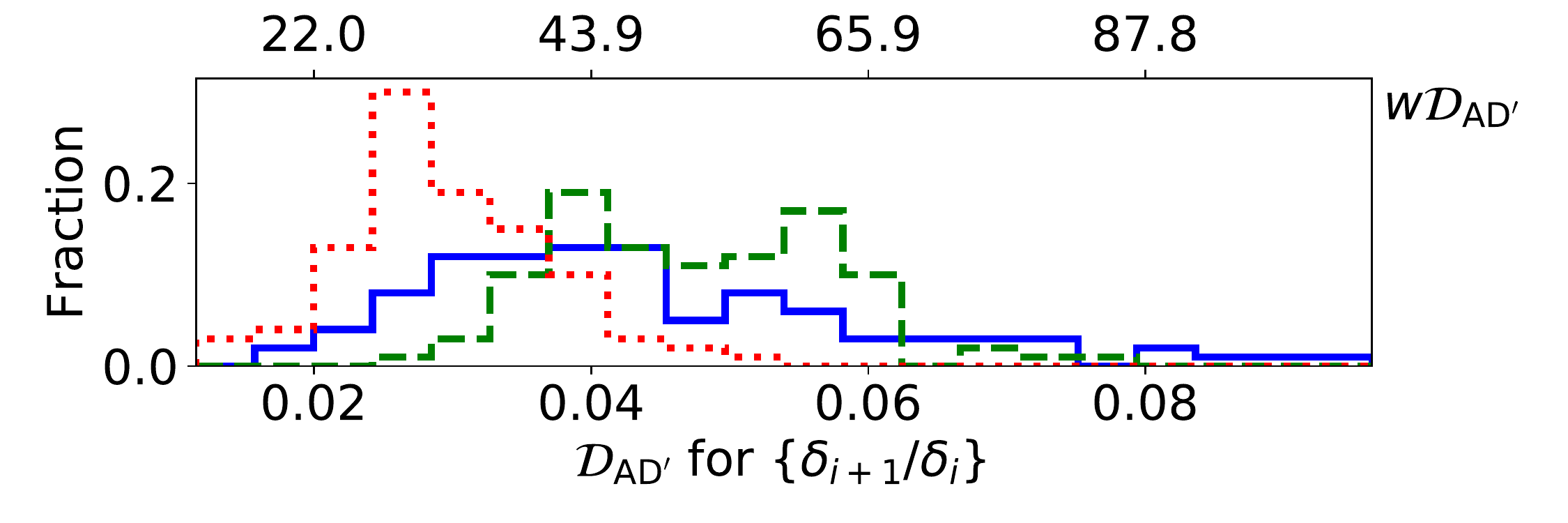} \\
 \includegraphics[scale=0.33,trim={0 0.3cm 0 0.2cm},clip]{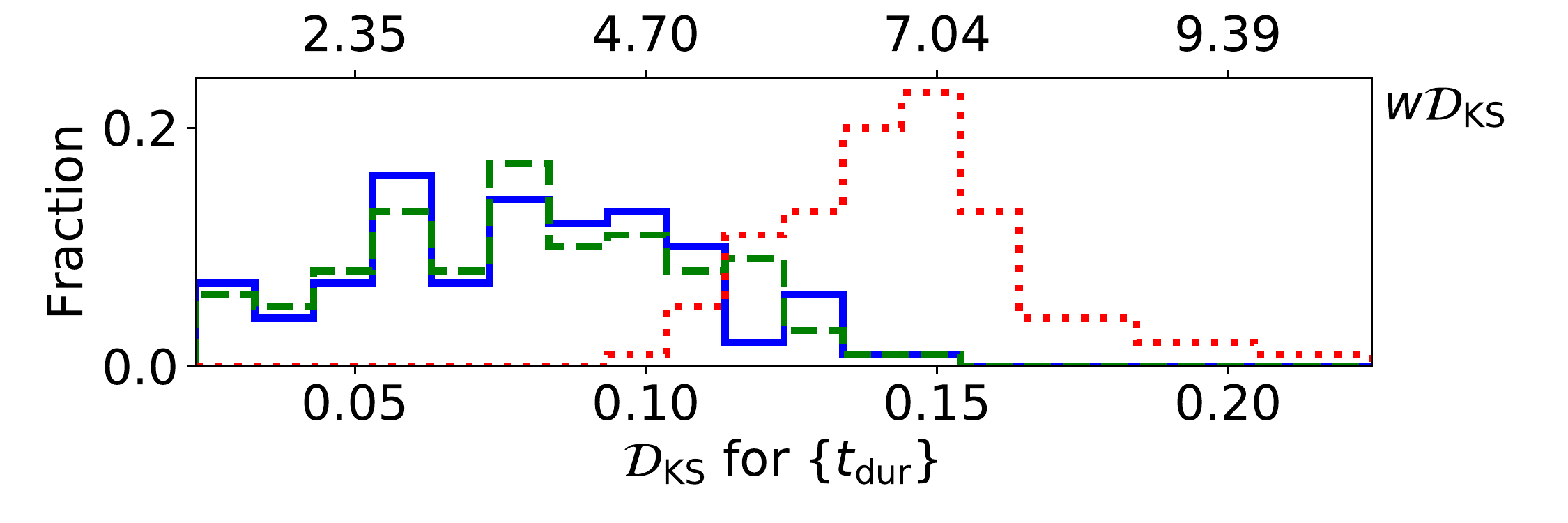} & 
 \includegraphics[scale=0.33,trim={0 0.3cm 0 0.2cm},clip]{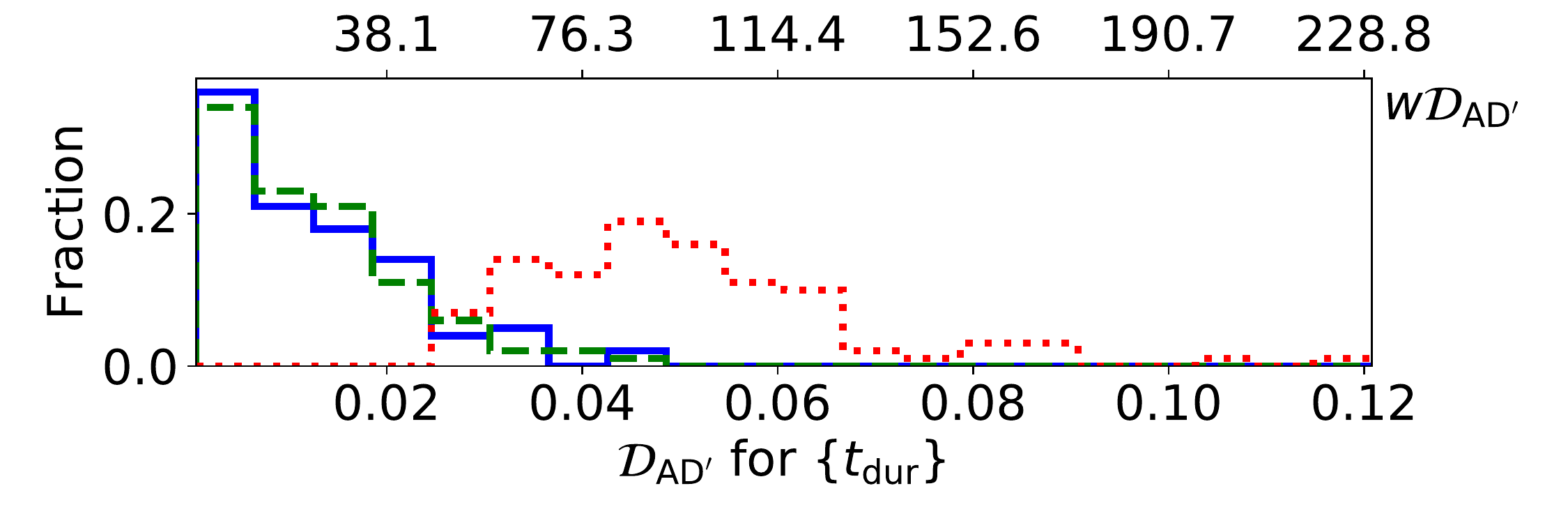} \\
 \includegraphics[scale=0.33,trim={0 0.3cm 0 0.2cm},clip]{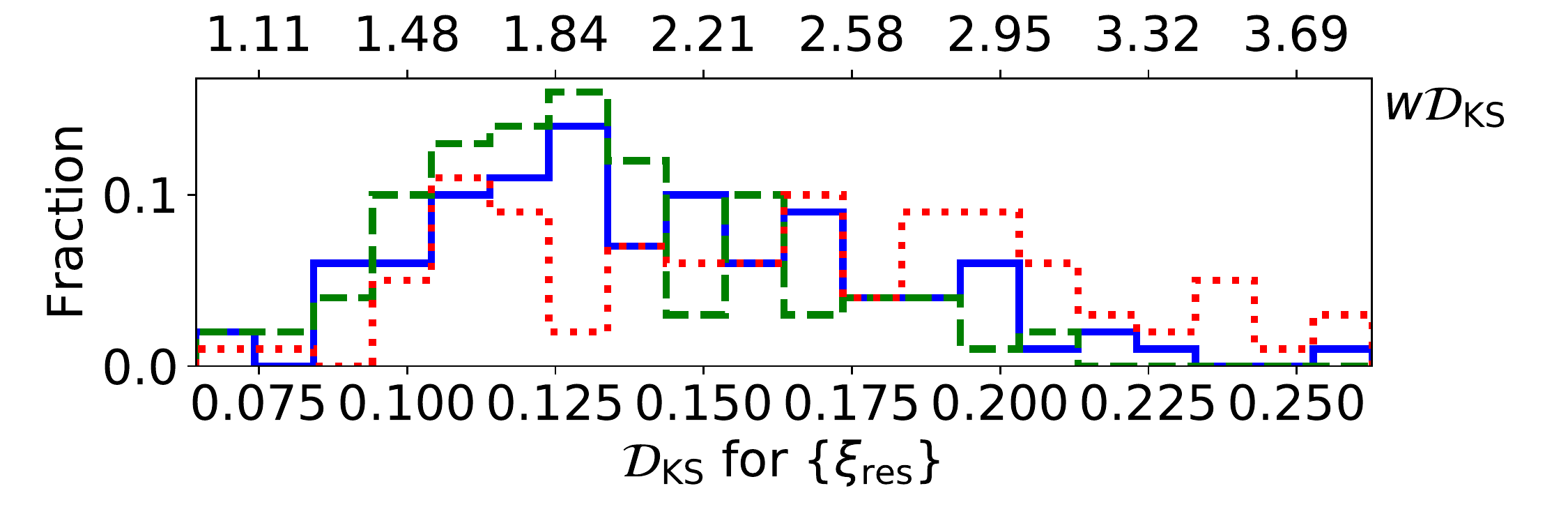} & 
 \includegraphics[scale=0.33,trim={0 0.3cm 0 0.2cm},clip]{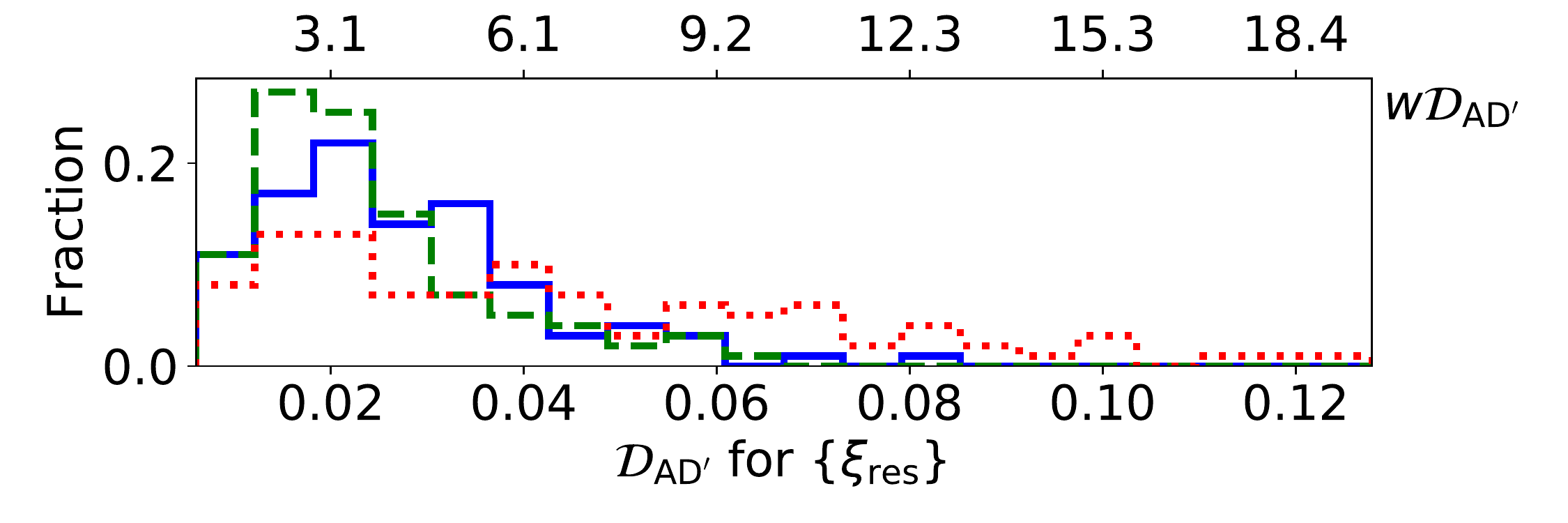} \\
 \includegraphics[scale=0.33,trim={0 0.3cm 0 0.2cm},clip]{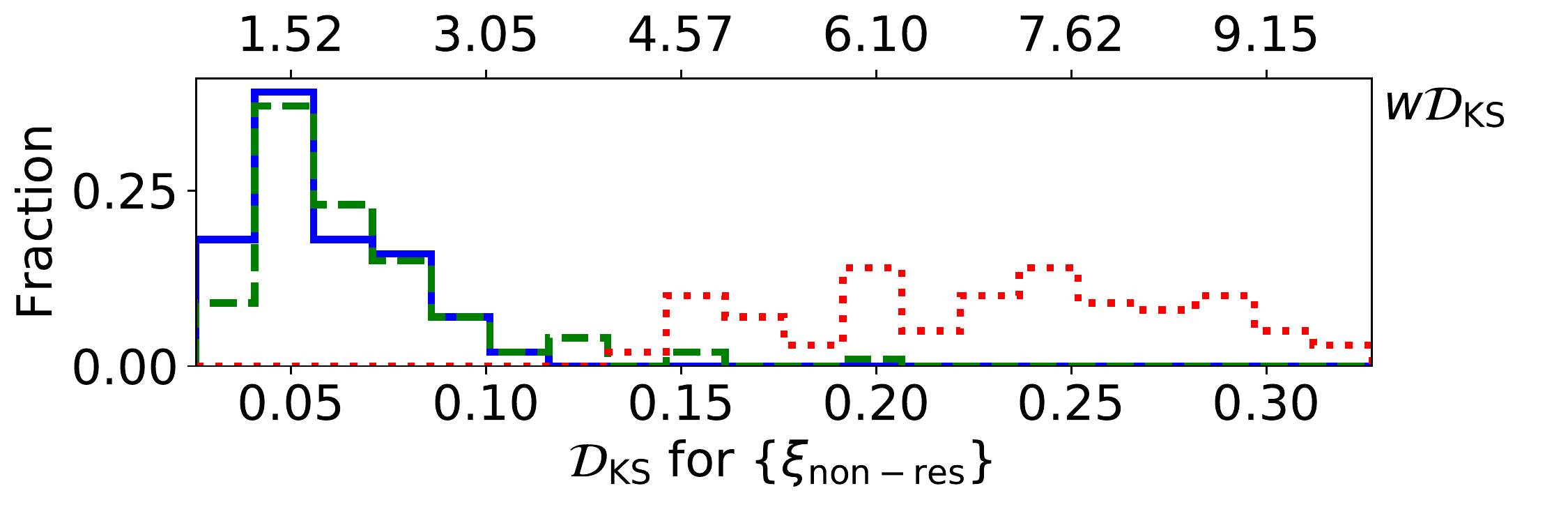} & 
 \includegraphics[scale=0.33,trim={0 0.3cm 0 0.2cm},clip]{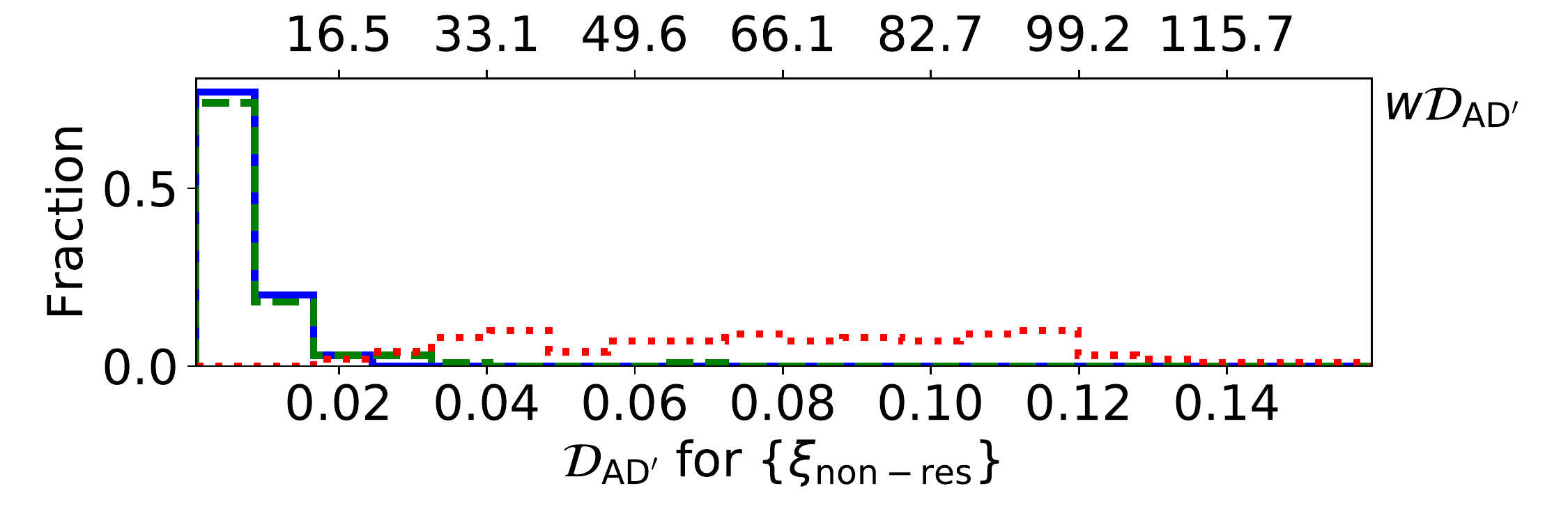} \\
\end{tabular}
\caption{Histograms of the total (weighted; top row) and individual distance terms (second row and below) for each of our models. Each histogram includes 100 simulated catalogues with parameters drawn from the emulator passing our KS distance thresholds for each model. With the exception of the two upper-most rows, the left-hand panels show the individual KS distance terms while the right-hand panels show the individual AD distance terms. For each panel, the bottom x-axis labels the true KS or AD distances ($\mathcal{D}_{\rm KS}$ or $\mathcal{D}_{\rm AD'}$), while the top x-axis labels the weighted distance (i.e. the bottom axis scaled by a weight $w$ as listed in Table \ref{tab:weights}).}
\label{fig:distances}
\end{figure*}


\bsp	
\label{lastpage}
\end{document}